\documentclass[structabstract]{aa} 
\usepackage{graphicx}
\usepackage{txfonts}
\usepackage{natbib}
\usepackage{longtable}
\usepackage{lscape}
\usepackage{multirow}
\usepackage{color}
\usepackage{textcomp}
\usepackage{url}

\bibpunct{(}{)}{;}{a}{}{,}

\newcommand{\kms}{km\,s$^{-1}$}

\newcommand{\degree}{$^{\circ}$}

\begin{document}
\title{A 1.3 cm Line Survey toward Orion KL}
\author{Y. Gong\inst{1,2,3} 
\and
C. Henkel\inst{2,4}
\and
S. Thorwirth\inst{5}
\and\
S. Spezzano\inst{5}
\and
K.~M. Menten\inst{2}
\and
C.~M. Walmsley\inst{6,7}
\and
F. Wyrowski\inst{2}
\and
R.~Q. Mao\inst{1}
\and
B. Klein\inst{2,8}
}
\institute{Purple Mountain Observatory \& Key Laboratory for Radio Astronomy, Chinese Academy of Sciences, 2 West Beijing Road, 210008 Nanjing, PR China
\and Max-Planck Institut f\"ur Radioastronomie, Auf Dem H\"ugel 69, 53121 Bonn, Germany
\and University of Chinese Academy of Sciences, No. 19A Yuquan Road, 100049 Beijing, PR China
\and Astronomy Department, King Abdulaziz University, P.O. Box 80203, Jeddah 21589, Saudi Arabia
\and I. Physikalisches Institut, Universit{\"a}t zu K{\"o}ln, 50937 K{\"o}ln, Germany
\and Osservatorio Astrofisico di Arcetri, Largo E. Fermi, 5, I-50125 Firenze, Italy
\and Dublin Institute of Advanced Studies, Fitzwilliam Place 31, Dublin 2, Ireland
\and University of Applied Sciences Bonn-Rhein-Sieg, Grantham-Allee 20, 53757 Sankt Augustin, Germany
}
    
   \date{}

 \abstract{The nearby Orion Kleinmann-Low nebula is one of the most prolific sources of molecular line emission. It has served as a benchmark for spectral line searches throughout the (sub)millimeter regime.}
{The main goal is to systematically study spectral characteristics of Orion KL in the $\lambda \sim 1.3$~cm band.}
{We carried out a spectral line survey with the Effelsberg-100 m telescope towards Orion KL. It covers the frequency range between 17.9~GHz and 26.2~GHz, i.e., the radio ``K-band''. We also examine ALMA maps to address the spatial origin of molecules detected by our 1.3 cm line survey.}
{In Orion KL, we find 261 spectral lines, yielding an average line density of about 32 spectral features per GHz above 3$\sigma$ (a typical value of 3$\sigma$ is 15 mJy). The identified lines include 164 radio recombination lines (RRLs) and 97 molecular lines. The RRLs, from hydrogen, helium, and carbon, originate from the ionized material of the Orion Nebula, part of which is covered by our beam. The molecular lines are assigned to 13 different molecular species including rare isotopologues. A total of 23 molecular transitions from species known to exist in Orion KL are detected for the first time in the interstellar medium. Non-metastable ($J>K$) $^{15}$NH$_{3}$ transitions are detected in Orion KL for the first time. Based on the velocity information of detected lines and the ALMA images, the spatial origins of molecular emission are constrained and discussed. A narrow feature is found in SO$_{2}$ ($8_{1,7}-7_{2,6}$), but not in other SO$_{2}$ transitions, possibly suggesting the presence of a maser line. Column densities and fractional abundances relative to H$_{2}$ are estimated for 12 molecules with local thermodynamic equilibrium (LTE) methods. Rotational diagrams of non-metastable $^{14}$NH$_{3}$ transitions with $J=K+1$ to $J=K+4$ yield different results; metastable ($J=K$) $^{15}$NH$_{3}$ is found to have a higher excitation temperature than non-metastable $^{15}$NH$_{3}$, also indicating that they may trace different regions. Elemental and isotopic abundance ratios are also estimated: He/H = (8.7$\pm0.7$)\% derived from the ratios between helium RRLs and hydrogen RRLs; $^{12}$C/$^{13}$C =63$\pm$17 from $^{12}$CH$_{3}$OH/$^{13}$CH$_{3}$OH; $^{14}$N/$^{15}$N =100$\pm$51 from $^{14}$NH$_{3}$/$^{15}$NH$_{3}$; D/H = (8.3$\pm$4.5)$\times$10$^{-3}$ from NH$_{2}$D/NH$_{3}$. The dispersion of the He/H ratios derived from H$\alpha$/He$\alpha$ pairs to H$\delta$/He$\delta$ pairs is very small, which is consistent with theoretical predictions that the departure coefficients $b_{\rm n}$ factors for hydrogen and helium are nearly identical. Based on a non-LTE code neglecting excitation by the infrared radiation field and a likelihood analysis, we find that the denser regions have lower kinetic temperature, which favors an external heating of the Hot Core.}
{}
\keywords{astrochemistry -- ISM: individual (Orion KL) -- ISM: molecules -- radio lines: ISM -- surveys}

\maketitle

\section{Introduction}
One of the best ways to study physical and chemical conditions of astronomical objects is via spectral line surveys. In fact, they are the only means to obtain a complete unbiased view of the molecular inventory of a given source. However, despite the long history of powerful large radio telescopes, only very few unbiased frequency surveys have been conducted in the centimeter wave regime. Spectroscopic features which are expected at centimeter wavelengths include radio recombination lines (RRLs) from hydrogen, helium and carbon, and pure rotational transitions of heavy species such as complex organic molecules which can be detected conveniently at centimeter wavelengths due to their small rotational constants. The molecular lines at centimeter wavelengths are likely to be optically thin and thus analysis is easier than at millimeter wavelengths. Furthermore, the ``line density'' is not as high as at millimeter wavelengths, reducing the confusion level. 

To summarize previous surveys at centimeter wavelengths: \citet{1993ApJS...86..211B} surveyed W51 from 17.6 to 22.0 GHz with the NRAO 43m telescope; \citet{2004ApJ...610..329K} investigated the dark cloud TMC-1 at 4--6 and 8--10 GHz with the Arecibo 305m telescope; \citet{2004PASJ...56...69K} observed TMC-1 from 9 to 50 GHz with the Nobeyama 45m telescope; \citet{2015A&A...574A..56G} measured IRC +10216 from 17.8 to 26.3 GHz with the Effelsberg 100m telescope finding 23 new transitions from species detected previously; \citet{2008ApJ...675L..85R} surveyed Sgr B2 near the Galactic center with the Green Bank Telescope (GBT) in the course of their GBT PRIMOS project\footnote[1]{http://www.cv.nrao.edu/{\raise.17ex\hbox{$\scriptstyle\sim$}}aremijan/PRIMOS/index.html} which led to detections of complex organic molecules such as CH$_{2}$CHCHO (propenal), CH$_{3}$CH$_{2}$CHO (propanal) and CNCHO (cyanoformaldehyde)~\citep{2004ApJ...610L..21H,2008ApJ...675L..85R}. They have shown that the centimeter wave range of the electromagnetic spectrum has an enormous potential to detect new molecules and new transitions.

In this study, we present a 1.3 cm line survey toward the Orion Kleinmann-Low nebula (Orion KL). Orion KL has a high luminosity of $\sim 10^{5} L_{\odot}$ \citep{1984ApJ...281..172W}, harboring several luminous embedded IR sources, and is the closest high mass star formation region at a distance of 414$\pm$7 \,pc \citep{2007A&A...474..515M}. The chemistry of this source is particularly rich due to high kinetic temperature of the gas caused by the evaporation of dust grain mantles and the interaction of the newly formed stars with their environment \citep[e.g.,][]{2009ARA&A..47..427H,1997ApJ...481..396C}. It is thus one of the best sources to study the spectral characteristics associated with high-mass star forming regions. So far, many spectral line surveys have been carried out toward Orion KL. The existing surveys, summarized in Table~\ref{Tab:surveys}, cover a wide frequency range from 34.25~GHz to 6.8~THz. These efforts have resulted in the discovery of a large number of molecules, including, recently, the unusual long complex molecule methyl acetate, CH$_{3}$COOCH$_{3}$ \citep{2013ApJ...770L..13T}, as well as a deuterated ammonium cation, NH$_{3}{\rm D}^{+}$ \citep{2013ApJ...771L..10C}. Meanwhile, no comprehensive view exists of our frequency range covered here (17.9--26.2 GHz), although numerous studies have targeted specific lines it contains, e.g., from NH$_{3}$ or CH$_{3}$OH. Here, we therefore systematically study the $\lambda$ $\sim$1.3\,cm (K band) characteristics of Orion KL.

\begin{table*}[!hbt]
\caption{Existing line surveys of Orion KL.}\label{Tab:surveys}             
\small
\centering                                      
\begin{tabular}{ccc}          
\hline\hline                        
Covered Frequencies                 & Telescope       &  reference                   \\
\hline
\multicolumn{3}{l}{Orion KL}\\
\hline
\textbf{17.9--26.2 GHz}     & \textbf{Effelsberg-100m} &  \textbf{This work} \\
34.25--50 GHz \& 83.5--84.5 GHz \& 86--91.5 GHz \tablefootmark{(1)} & Nobeyama-45m & \citet{1986ApSS.118..405O}\\
42.3--43.6 GHz                   & GBT-100m       & \citet{2009ApJ...691.1254G} \\
70--115 GHz                      & NRAO-11m       & \citet{1989ApJS...70..539T,1991ApJS...76..617T} \\
72--91 GHz                       & Onsala-20m     & \citet{1984AA...130..227J}  \\
67--93.6 GHz                     & GBT-100m       & \citet{2015AJ....149..162F} \\
80--280 GHz\tablefootmark{(2)}   & IRAM-30m       & \citet{2010AA...517A..96T,2011AA...528A..26T,2013AA...556A.143E}\\
138.3--150.7 GHz                 & TRAO-14m       & \citet{2001ApJ...551..333L} \\
130--170 GHz                     & NRAO-12m       & \citet{2008arXiv0802.2273R} \\
150--160 GHz                     & FCRAO-14m      & \citet{1993ApJS...89..155Z} \\
159.7--164.7 GHz                 & TRAO-14m       & \citet{2002JKAS...35..187L} \\
190--900 GHz\tablefootmark{(2)}    & CSO-10m        & \citet{1995ApJ...451..238S} \\
215--247 GHz                     & OVRO-10.4m     & \citet{1985ApJS...58..341S} \\
216.2--217.2 GHz \& 219.1--220.1 GHz \& 241--242 GHz \& 244-245 GHz  & OVRO   & \citet{1996ApJ...472L..49B} \\ 
218.9--221.2 GHz \& 228.9--231.2 GHz & IRAM-30m+SMA & \citet{2015arXiv150408012F} \\
247--263 GHz                     & OVRO-10.4m     & \citet{1986ApJS...60..357B} \\
257--273 GHz                     & JCMT-15m       & \citet{1991AAS...91..237G}  \\
325--360 GHz                     & CSO-10m        & \citet{1997ApJS..108..301S} \\
330.5--360.1 GHz \& 200.7--202.3 GHz \& 203.7--205.3 GHz & NRAO-12m & \citet{1989ApJS...70..833J}  \\
334--343 GHz                     & JCMT-15m       & \citet{1995ApJS...97..455S} \\
455--507 GHz                     & JCMT-15m       & \citet{2003AA...407..589W} \\
486--492 GHz \& 541--577 GHz      & Odin satellite &  \citet{2007AA...476..791O,2007AA...476..807P} \\ 
607--725 GHz                     & CSO-10m        & \citet{2001ApJS..132..281S} \\ 
795--903 GHz                     & CSO-10m        & \citet{2005ApJS..156..127C} \\
480--1907 GHz\tablefootmark{(2)} & Herschel/HIFI  & \citet{2014ApJ...787..112C} \\
1595--6816 GHz\tablefootmark{(3)}& ISO-60cm       & \citet{2006MNRAS.370..597L} \\    
\hline
 \end{tabular}
 \tablefoot{\\
 \tablefoottext{1}{The covered frequency range is not entirely displayed in their paper.}\\
 \tablefoottext{2}{Note that the covered frequency range is discontinuous.}\\
\tablefoottext{3}{The wavelength range corresponding to the given frequencies is 44~$\mu$m--188~$\mu$m.}\\
 }
 			        \normalsize
\end{table*}

\section{Observation and data reduction} 
\subsection{Observations with the Effelsberg-100m telescope}
The measurements were carried out in a position switching mode with the primary focus $\lambda$ = 1.3 cm K-band receiver of the 100-m telescope at Effelsberg/Germany\footnote[2]{The 100-m telescope at Effelsberg is operated by the Max-Planck-Institut f{\"u}r Radioastronomie (MPIFR) on behalf of the Max-Planck-Gesellschaft (MPG).}, during 2012 January, April and 2013 January, March, and May. The telescope was pointed at $\alpha_{\rm J2000}=05^{\rm h}35^{\rm m}14.17^{\rm s}$, $\delta_{\rm J2000}=-05^{\circ}22^{\prime}46.5^{\prime\prime}$, about 12\arcsec\,south of the dust peak (see Fig.~\ref{Fig:alma-dust}a). On- and off-source integration times were two minutes per scan. The newly installed Fast Fourier Transform Spectrometer (FFTS) was used as backend. Each of the two orthogonal linear polarizations was covered with a bandwidth of 2 GHz providing 32768 channels and resulting in a channel spacing of 61 kHz, equivalent to 0.7 \kms at 26 GHz. The actual frequency and velocity resolutions are coarser by a factor of 1.16 \citep{2012A&A...542L...3K}. Several frequency setups were tuned to cover the entire frequency range from 17.9 GHz to 26.2 GHz with an overlap of at least 100 MHz between two adjacent setups. Across the whole frequency range, the FWHM beamsize varies from 35\arcsec\,to 50\arcsec ($\sim$40\arcsec at 23~GHz). The survey encompasses a total of $\sim$ 25 observing hours. The focus was checked every few hours, in particular after sunrise and sunset. Pointing was obtained every hour toward the nearby pointing sources PKS~0420$-$01 or 3C161 and was found to be accurate to about 5\arcsec. Strong continuum sources (mostly NGC7027 and 3C286) were used to calibrate the spectral line flux, assuming standard flux densities \citep[5.5 Jy for NGC 7027 and 2.5 Jy for 3C286 at 22 GHz;][]{1994A&A...284..331O}. The typical rms noise is about $3-15$ mJy in 61~kHz wide channels. The conversion factor from Jy on a flux density scale ($S_{\nu}$) to K on a main beam brightness temperature scale ($T_{\rm mb}$) is $T_{\rm mb}/S_{\nu} \sim 1.7~{\rm K/Jy}$ at 18.5~GHz, $1.5~{\rm K/Jy}$ at 22~GHz, and $1.4~{\rm K/Jy}$ at 23.7~GHz. Local standard of rest (LSR) velocities are used throughout the paper.

For the data reduction, the GILDAS\footnote[3]{http://www.iram.fr/IRAMFR/GILDAS} software package including CLASS and GreG was employed. During the data reduction, some inevitable defects were found at the edges of spectra. Therefore 100 channels at each edge of the original spectra were excluded. Fourth to tenth order baselines were subtracted from each spectrum with 2000 to 3000 channels to avoid lines to be truncated at the edge of sub-spectra, and then these sub-spectra were stitched together to resconstruct the complete spectrum. Since some of those sub-spectra also suffer contamination from time variable radio frequency interference (RFI), the channels showing RFI signals have been ``flagged'', i.e. discarded from further analysis, resulting in some limited gaps in the averaged spectra. Furthermore, there were a few individual bad channels, which also have been eliminated. 
\subsection{Archival data}
Here, we also present archival ALMA line survey data covering a frequency range of 214--247 GHz to study the distribution of molecules detected by our 1.3 cm line survey.  The observations were carried out with 16 antennas on 2012 January 20, during ALMA's science verification (SV) phase. Callisto and the quasar J0607-085 were used as the flux and phase calibrator, respectively. The spectral resolution is 0.488 MHz, which corresponds to $\sim$0.6 \kms\,at 230~GHz. Its primary beam size ($\sim$30\arcsec) is slightly smaller than the beam size of the Effelsberg-100m in the $\lambda$ $\sim$1.3\,cm band. The CLEAN algorithm in the Common Astronomy Software Applications package (CASA) was used to deconvolve the images applying natural weighting. The synthesized beam size is 1.9\arcsec$\times$1.4\arcsec. In this work, we chose to use 11 unpublished transitions (see Table~\ref{Tab:alma}) since several other lines and dust continuum emission from the ALMA data have been already studied \citep[e.g.,][]{2013ApJ...770..142N,2013ApJ...777...85N,2014ApJ...787..112C,2014ApJ...791..123W,2015ApJ...801...82H}. The data are available to the public and can be accessed from the ALMA-SV website\footnote[4]{http://almascience.nrao.edu/alma-data/science-verification}.

Since the ALMA images without short spacings are only sensitive to small-scale structure, we also make use of the VLA C band continuum data\footnote[5]{Image credit: NRAO/VLA Archive Survey, (c) 2005-2007 AUI/NRAO} (project AY191A) and the SCUBA--850 $\mu$m dust emission \citep{2008ApJS..175..277D} to trace the large-scale structure of Orion KL. In addition, we also compare the ALMA data with other studies to address the missing flux problem (see Appendix~\ref{app.c}).

\section{Components of Orion KL}\label{sec:comp}
\begin{figure*}[!htbp]
\centering
\includegraphics[width = 0.8 \textwidth]{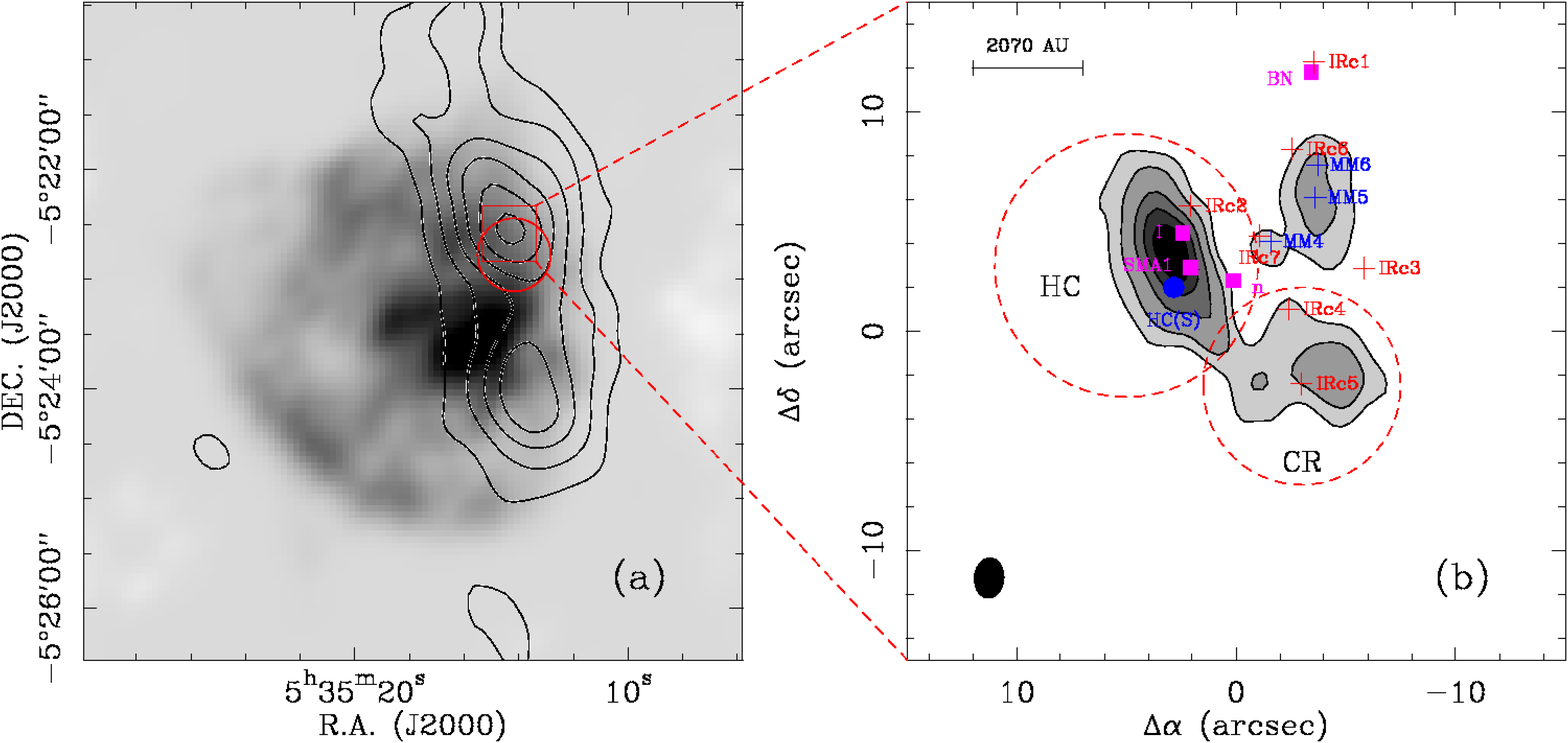}
\caption{{(a) The SCUBA--850~$\mu$m dust emission (contours) overlaid on the 6 cm VLA continuum image. The contours correspond to 5, 10, 20, 40, 80, 160 Jy~beam$^{-1}$. The red circle represents the Effelsberg FWHM beamsize (40\arcsec). The red box represents the mapped region of Fig.~\ref{Fig:alma-dust}b. (b) Continuum map of Orion KL at 230 GHz from the ALMA-SV line survey modified from Figure 8 of \citet{2014ApJ...787..112C}. The contour levels are 10\%, 20\%, 40\%, 60\%, 80\% of the peak intensity of 1.406 Jy~beam$^{-1}$. The positions of source n, source I, SMA1 and the BN object are indicated by the purple squares. The position of the hot core south (HC(S)) is indicated by the blue circle. The positions of Infrared sources IRc1, IRc2, IRc3, IRc4, IRc5, IRc6, and IRc7 are marked by red crosses. Blue crosses indicate millimeter wave continuum sources. The approximate positions of the compact ridge and the hot core are indicated by the red dashed circles and the acronyms CR and HC. (these sources are described in more details in Sect.~\ref{sec:comp}.) The physical scale is given by the horizontal bar. The synthesized beam is shown in the lower left of the panel. The reference point corresponds to ($\alpha_{\rm J2000}$, $\delta_{\rm J2000}$) = (05$^{\rm h}$35$^{\rm m}$14.350$^{\rm s}$, $-$05\degree22\arcmin35.00\arcsec).} \label{Fig:alma-dust}}
\end{figure*}

Fig.~\ref{Fig:alma-dust}a shows the large-scale structure of Orion KL. In Fig.~\ref{Fig:alma-dust}b, we present an ALMA 230~GHz continuum image with prominent sources near or within Orion KL indicated. The positions of a few objects, Source n (a protostar coincident with a peculiar double 8.4 GHz radio continuum source), Source I (an embedded massive protostar without submillimeter and infrared continuum, but only SiO maser emission), SMA1 (a compact submillimeter source not detected in the infrared or cm regime) and the BN object (a massive protostar with a circumstellar disk), are based on previous studies \citep[e.g.,][]{1995ApJ...445L.157M,2005Natur.437..112J,2008ApJ...679L.121B,2011ApJ...727..113B,2011A&A...532A..32F}. The positions of IRc1, IRc2, IRc3, IRc4, IRc5, IRc6, and IRc7 infrared emission peaks are taken from \citet{2004AJ....128..363S}. The millimeter continuum sources MM4, MM5, MM6 are taken from \citet{2014ApJ...791..123W}. In the $\lambda$ $\sim$1.3\,cm band, the spectral features of Orion KL can originate from eight distinct components within our $\sim$40\arcsec\,beam. These components are:
\begin{itemize}
	\item[(1)] ionized gas from the foreground M42 H{\scriptsize II} region (see Fig.~\ref{Fig:alma-dust}a). The contribution can be traced by hydrogen and helium RRLs which have typical velocities $\upsilon_{\rm lsr}\sim$ $-5$~\kms\, and large line widths $\Delta \upsilon>15$~\kms\,\citep[e.g.,][]{1997A&A...327.1177W,1992ApJ...395..484P}. 

\item[(2)] the photon dominated region (PDR). It is the interface region between the molecular cloud and the foreground M42 H{\scriptsize II} region. It can be traced by carbon RRLs  ($\upsilon_{\rm lsr}\sim 9-10$\,\kms\,and $\Delta \upsilon\sim 3-5$~\kms) \citep[e.g.,][]{1994ApJ...428..209N}, which is because carbon ions can exist outside H{\scriptsize II} regions due to their lower ionization energy (11.3 eV) with respect to that of hydrogen (13.6 eV) \citep[e.g.,][]{1997ApJ...487L.171W,2009tra..book.....W}.

\item[(3)] the ``hot core'' (HC) component. The HC is a well-known source, $\sim$ 2\arcsec\,from IRc2 (see Fig.~\ref{Fig:alma-dust}b). It has a high kinetic temperature and a high density \citep[$T_{\rm kin}>150$~K and $n>10^{7}$~cm$^{-3}$, e.g.,][]{1989ARA&A..27...41G,1993ApJS...89..123M,1997ApJS..108..301S,2000ApJ...538..665W,2010ApJ...713.1192W,2011ApJ...739L..13G,2014ApJ...787..112C}. Note that an extremely high kinetic temperature is also indicated by detections of high $J$ metastable NH$_{3}$ transitions with energy levels up to 1900~K and a high rotational temperature of $\sim 400$ K \citep{1993A&A...276L..29W}. The HC is believed to be heated by an explosive event \citep{2011ApJ...727..113B,2011A&A...529A..24Z,2012A&A...540A.119N}. Its line profiles have velocities of $\upsilon_{\rm lsr} \sim 3-6$~\kms\,and widths of $\Delta \upsilon\sim 5-12$~\kms. 

\item[(4)] the ``compact ridge'' (CR) component. It is located $\sim$12\arcsec\,southwest of the well-known HC (see Fig.~\ref{Fig:alma-dust}b). It has densities of $n\sim$10$^{6}$~cm$^{-3}$ and temperatures of $T_{\rm kin}\sim80-280$~K \citep[e.g.,][]{1993ApJS...89..123M,2005ApJ...632..355B,2007AA...476..807P,2010ApJ...713.1192W,2011A&A...532A..32F}. Studies of methanol and methyl formate emission suggest that the CR is externally heated \citep{2011A&A...527A..95W,2011A&A...532A..32F}. Its line profiles have velocities of $\upsilon_{\rm lsr}\sim 7-9$~\kms\, and widths of $\Delta \upsilon\sim 3-6$~\kms. 

\item[(5)] the ``hot core south'' (HC(S)) component. The component was first noticed by studies of HDO \citep{2013ApJ...770..142N,2014ApJ...787..112C} and is likely to originate from 1\arcsec\,south of the well-known HC submillimeter continuum peak (see Fig.~\ref{Fig:alma-dust}b). It shows a spectral feature with $\upsilon_{\rm lsr} \sim 6.5-8$\,\kms\, and $\Delta \upsilon\sim 5-10$~\kms. 


\item[(6)] the ``extended ridge'' (ER) component. It represents the ambient gas of Orion KL. It has densities of $n\sim10^{4}$--10$^{6}$~cm$^{-3}$ and temperatures of $T_{\rm kin}< 60$~K showing spectral features of $\upsilon_{\rm lsr}\sim 8-10$~\kms\,and $\Delta \upsilon \sim 2-5$~\kms \citep[e.g.,][]{1986ApJS...60..357B,1997ApJS..108..301S,2007AA...476..807P}.

\item[(7)] the ``plateau'' component. The component is characterized by velocities $\upsilon_{\rm lsr} \sim 6-12$\,\kms\,and large line widths $\Delta \upsilon > 20$~\kms, which are attributed to outflows. Orion KL is known to contain at least two outflows, historically referred to as the low velocity flow (LVF) and the high velocity flow (HVF) \citep[e.g.,][]{1981ApJ...244..884G}. The LVF is oriented along the SW-NE direction and is thought to be driven by radio source I \citep[e.g.,][]{1989ARA&A..27...41G,1995ApJ...445L.157M,1998Natur.396..650G,2009ApJ...704L..25P}. It has spectral features of $\upsilon_{\rm lsr}\sim 5$~\kms\,and $\Delta \upsilon \sim 18$~\kms. The HVF, more extended than the LVF, is oriented along the SE-NW direction and is thought to be driven by the submillimeter source SMA1 or the dynamical decay of a multi-star system involving radio source I, source n and the BN object that caused the explosive event described in (3) above \citep[e.g.,][]{2008ApJ...679L.121B,2011ApJ...727..113B,2015ApJ...799..102G}. It has spectral features of $\upsilon_{\rm lsr}\sim 10$~\kms\,and the line widths can be very large reaching velocities up to 150~\kms.

  \item[(8)] the millimeter continuum sources MM4, MM5, and MM6 taken from \citet{2014ApJ...791..123W}. Based on their $^{13}$CH$_{3}$CN studies, MM4, MM5, and MM6 are found to have rotational temperatures of 182$^{+81}_{-43}$~K, 157$^{+43}_{-28}$~K, and 119$^{+18}_{-14}$~K, respectively. The $^{13}$CH$_{3}$CN spectral features are: MM4 with velocities of $\upsilon_{\rm lsr}\sim 5.5$~\kms\, and widths of $\Delta \upsilon \sim 5.9$~\kms; MM5 with velocities of $\upsilon_{\rm lsr}\sim 10.1$~\kms\, and widths of $\Delta \upsilon \sim 5.4$~\kms; MM6 with velocities of $\upsilon_{\rm lsr}\sim 7.5$~\kms\, and widths of $\Delta \upsilon \sim 5.7$~\kms. From Fig.~\ref{Fig:alma-dust}b, MM4 is close to IRc7 while MM6 is close to IRc6.
\end{itemize}

\section{Results}
\subsection{Line identifications}\label{sec.id}
The line identification was performed with the help of the JPL\footnote[6]{http://spec.jpl.nasa.gov}, CDMS\footnote[7]{http://www.astro.uni-koeln.de/cdms/catalog}, and splatalogue\footnote[8]{http://www.splatalogue.net} databases as well as the online Lovas line list\footnote[9]{http://www.nist.gov/pml/data/micro/index.cfm} for astronomical spectroscopy \citep{2005JMoSt.742..215M,1998JQSRT..60..883P,2004JPCRD.33..177L}. However, the rest frequencies of RRLs are not contained in these databases, and are thus calculated with the Rydberg formula 
\begin{equation}\label{redberg}
 \nu_{n_{2}\to n_{1}} = R(\frac{1}{n_{1}^{2}}-\frac{1}{n_{2}^{2}}) ,~n_{1} < n_{2} , 
\end{equation}
where $\Delta n = n_{2}-n_{1}$. $n_{2}$ and $n_{1}$ are the principle quantum numbers of the upper and lower state, $R$ is the Rydberg constant which is equal to 3.28805129$\times 10^{15}$ Hz for hydrogen, and 3.28939118$\times 10^{15}$ Hz for helium \citep{2009tra..book.....W}. The nomenclature for RRLs is based on $n_{1}$ and $\Delta n$. For example, H71$\alpha$ is corresponding to a hydrogen RRL with $n_{1}=71$, $n_{2}=72$, and $\Delta n=1$. The Greek alphabet corresponds to $\Delta n$, e.g., $\alpha$ is corresponding to $\Delta n=1$, $\beta$ is corresponding to $\Delta n=2$, etc. 

A line is identified as real if it exhibits a 3$\sigma$ feature in more than 3 adjacent raw channels. On the other hand, there are a few narrow lines which are also identified as real even though their intensity is above $3\sigma$ for fewer than 3 adjacent raw channels. Such lines show up as doublets (e.g., CH$_{3}$OCHO, CH$_{3}$OCH$_{3}$), which confirms our assignment. Identified lines are listed in Table~\ref{Tab:orilines}. Given that line profiles are complex in Orion KL, different methods are employed to fit our observed lines. For RRLs, we simply use one single Gaussian component to derive their observed properties. For molecular lines showing nuclear quadrupole hyperfine structure (hfs, like HC$_{3}$N, CH$_{3}$CN, HNCO, NH$_{3}$, etc.), we used the HFS fitting routine embedded in CLASS to derive the observed properties (including the optical depth) while least-square fits of Gaussians are performed for the other molecular transitions. On the other hand, the observed lines of sulfur-bearing molecules are composites of different distinct components, so here the approach we take is to adopt a range of velocities and line widths of different components to separate their contributions. This is based on general knowledge of the source (see Sect.~\ref{sec:comp}). The NH$_{3}$ emission also originates from several components, but here the situation is even more complex. In this case, multiple component fitting becomes unrealistic. Therefore, we only give the integrated intensities for those NH$_{3}$ transitions which have a heterogeneous origin. Although C$\alpha$ lines are blended with He$\alpha$ lines, we present double Gaussian fits to separate their relative contributions to obtain the observed properties of C$\alpha$ lines since little information about C$\alpha$ lines exists in the literature. For other blended transitions, we do not separate the lines but simply integrate the whole line profiles to obtain upper limits to the integrated intensities of individual spatial subcomponents. 

\begin{figure*}[!htbp]
\centering
\includegraphics[width = 0.8 \textwidth]{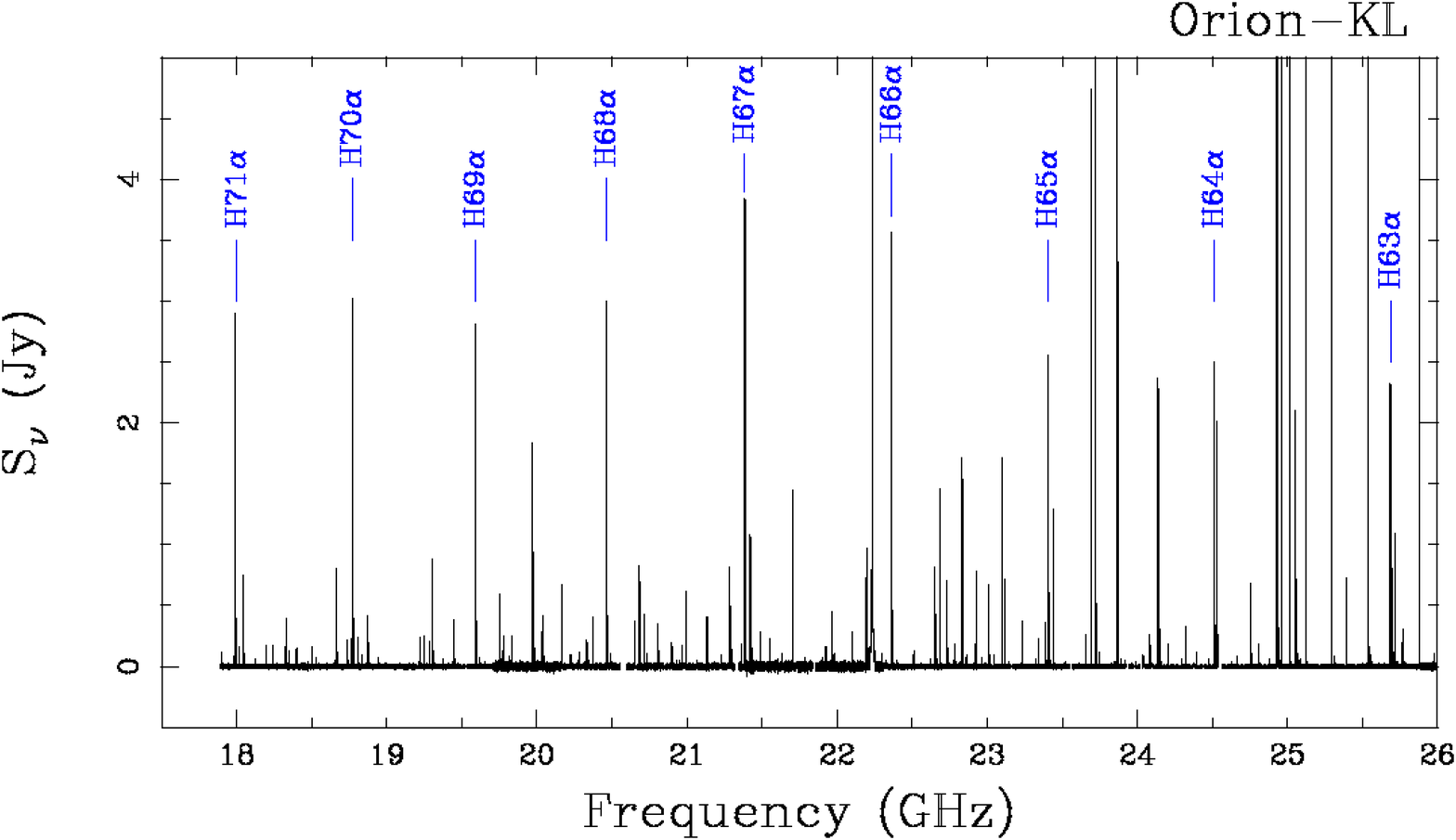}
\includegraphics[width = 0.8 \textwidth]{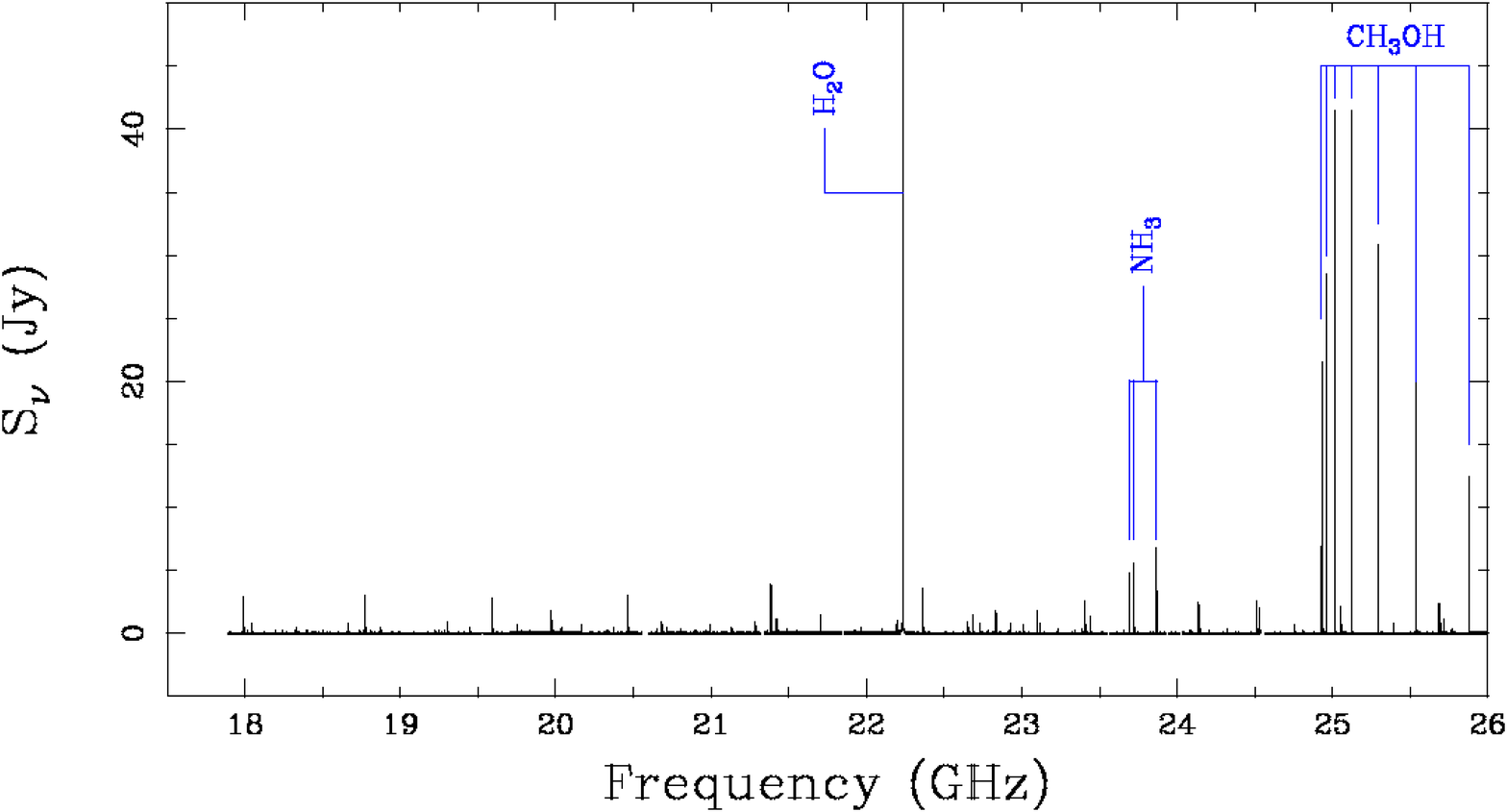}
\caption{{Overview of the 1.3 cm line survey toward Orion KL with strong lines marked. The displayed frequency scale is based on the Local Standard of Rest velocity 0 \kms.} \label{Fig:all}}
\end{figure*}

The observed line parameters for different species are displayed in Tables \ref{Tab:rrl}--\ref{Tab:one}, where the rms noise $\sigma$ given is for 61~kHz wide channels. Fig.~\ref{Fig:all} shows an overview of the 1.3 cm spectral line survey toward Orion KL. In the survey, we find 261 lines, yielding an average line density of about 32 spectral features per GHz above $3\sigma$. The line density is much smaller than that at millimeter wavelengths \citep[e.g.,][]{2010AA...517A..96T}. All lines are identified. The identified lines include 164 RRLs and 97 molecular lines. The RRLs, from hydrogen, helium, and carbon, originate from the ionized material of the Orion Nebula, part of which is covered by our beam. The molecular lines are assigned to 13 different molecules including rare isotopologues. 23 lines from already known species are detected for the first time in the interstellar medium (marked by an ``N'' in the last column of Table~\ref{Tab:orilines}). 19 molecular lines or carbon RRLs are blended with neighbouring hydrogen or helium RRLs. From Fig.~\ref{Fig:all}, we can see that the spectrum is dominated by strong lines arising from four species, the super strong H$_{2}$O (6$_{1,6}$--5$_{2,3}$) maser with $S_{\nu}>30000$~Jy, the series of CH$_{3}$OH maser lines starting at $\approx 25$~GHz with $S_{\nu}>$10~Jy, low $J$ metastable transitions of NH$_{3}$ with $S_{\nu}>5$ Jy and H$\alpha$ lines with $S_{\nu}>$ 2~Jy. Fig.~\ref{Fig:ori} shows the observed spectrum in more detail. Each panel covers $\sim$500~MHz with a 10~MHz overlap between adjacent panels so that lines truncated in one panel will not be truncated in the neighboring panel. In addition, Fig.~\ref{Fig:alpha}--\ref{Fig:one} shows zoom-in plots of all identified lines.

\subsection{Radio recombination lines}\label{sec.rrl}
From Fig.~\ref{Fig:ori} and Table~\ref{Tab:orilines}, we find that the frequency range is mainly occupied by RRLs. Among the 164 RRLs, there are 116 hydrogen RRLs, 39 helium RRLs, and 9 carbon RRLs. Their spectra are shown in Figs.~\ref{Fig:alpha}--\ref{Fig:iota} and their observed line parameters are given in Table \ref{Tab:rrl}. 56 hydrogen RRLs with  $\Delta n \ge 6$ are detected. Furthermore, hydrogen RRLs with $\Delta n$ as high as 11 are detected in this survey while hydrogen RRLs with even higher $\Delta n$ ($\leq 25$) are reported by \citet{2011Ap&SS.333..377B}. Carbon RRLs are all blended with helium RRLs since line widths of RRLs are broad and the separations between the rest frequencies of carbon RRLs and helium RRLs with the same $n_{1}$ and $\Delta n$ are small. Furthermore, the velocity differences shift the carbon RRLs slightly into the He RRLs (See Sect.~\ref{sec:comp} for the differences with respect to radial velocities.).

\begin{figure*}[!htbp]
\centering
\includegraphics[width = 0.4 \textwidth]{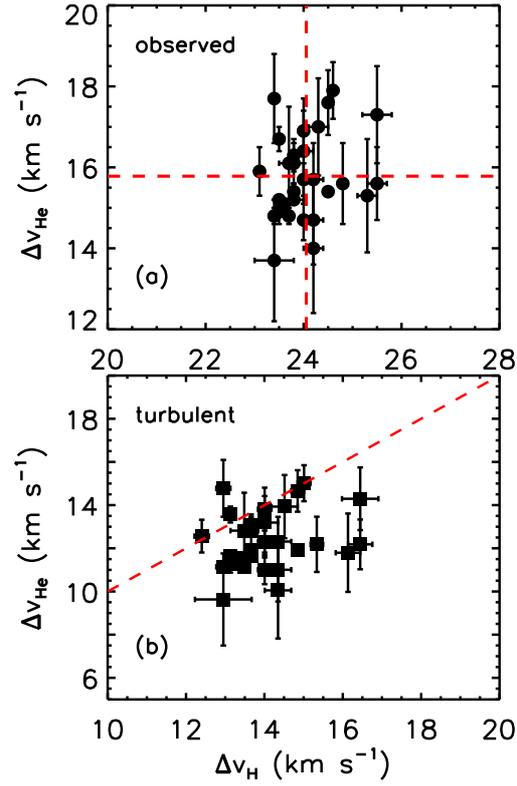}
\caption{{(a) The relationship between the observed line widths of hydrogen (abszissa) and helium (ordinate) RRLs. The red dashed lines represent the unweighted mean values of observed line widths of hydrogen and helium RRLs, respectively. (b) The relationship between the turbulent line widths derived from hydrogen and helium RRLs. The red dashed line connects points where turbulent line widths for hydrogen and helium are equal.} \label{Fig:linwid}}
\end{figure*}

From Fig.~\ref{Fig:linwid}a, we can see that line widths ($\sim 24$~\kms) of hydrogen RRLs are broader than those ($\sim 16$~\kms) of helium RRLs. The line broadening of RRLs is thought to arise from thermal motions and turbulence \citep{2009tra..book.....W}. The observed line width, $\Delta \upsilon$, has a thermal and a turbulent contribution, $\Delta \upsilon = \sqrt{\Delta \upsilon_{\rm th}^{2}+\Delta \upsilon_{\rm tur}^{2}}$, where $\Delta \upsilon_{\rm th}$ is the thermal width, and $\Delta \upsilon_{\rm tur}$ is the line width caused by turbulent flows. Taking the electron temperature of the foreground H{\scriptsize II} region as 8300~K \citep{1997A&A...327.1177W} and following a Maxwell-Boltzmann velocity distribution, the thermal widths of hydrogen and helium are estimated to be 19.5~\kms\, and 9.8~\kms with the formula $\Delta \upsilon_{\rm th} = 2\sqrt{2{\rm ln}2}(\frac{{\rm k}T}{m})^{1/2}$, where k is the Boltzmann constant, $T$ is the electron temperature, and $m$ is the mass of a particle. Consequently, the average hydrogen and helium turbulent line widths are estimated to be $14.1\pm1.0$~\kms\, and $12.4\pm1.4$~\kms, respectively. From Fig.~\ref{Fig:linwid}b, we can see that most points lie below the red dashed line where the turbulent line widths are equal for hydrogen and helium. This suggests that hydrogen turbulent line widths might be slightly larger than those of helium. However, the large errors make it uncertain. If the difference can be confirmed, that may indicate that hydrogen RRLs trace a larger portion of ionized gas than helium RRLs in Orion KL. The exciting stars of the H{\scriptsize II} region have $T_{\rm eff}\sim 4\times 10^{4}$~K \citep{1991ApJ...374..580B}. This is higher than the critical $T_{\rm eff}$ of 37000 K, below which helium is no longer appreciably ionised \citep{2009tra..book.....W}. The innermost parts of their H{\scriptsize II} regions should therefore show ionized He{\scriptsize II}. The outer parts, however, are devoid of He{\scriptsize II}, while still containing ionized hydrogen. The larger volume leads to a larger line width. Nevertheless, the bulk of the discrepancy between observed line widths of hydrogen and helium RRLs originates from their thermal contributions.

\begin{figure*}[!htbp]
\centering
\includegraphics[width = 0.45 \textwidth]{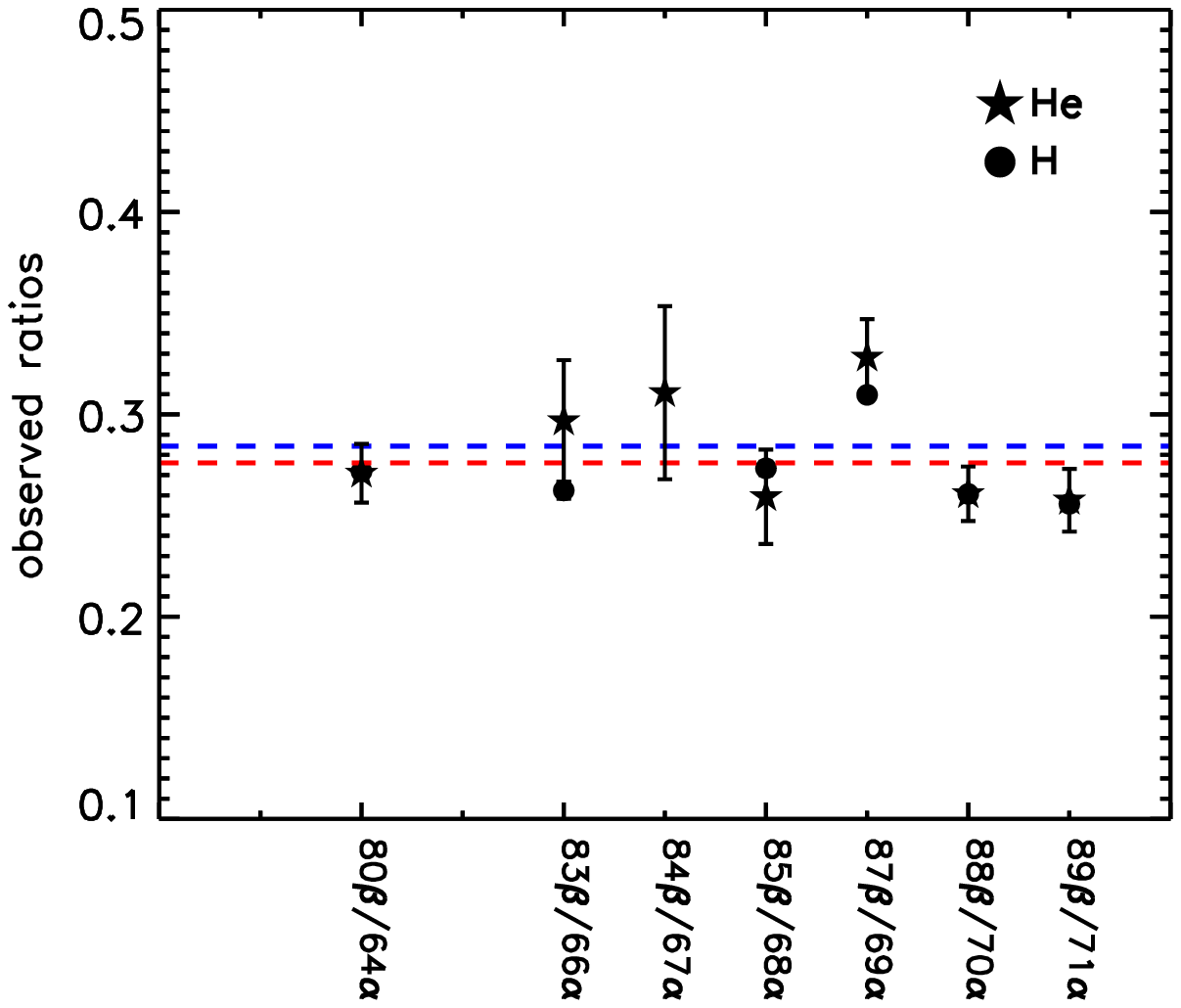}
\includegraphics[width = 0.45 \textwidth]{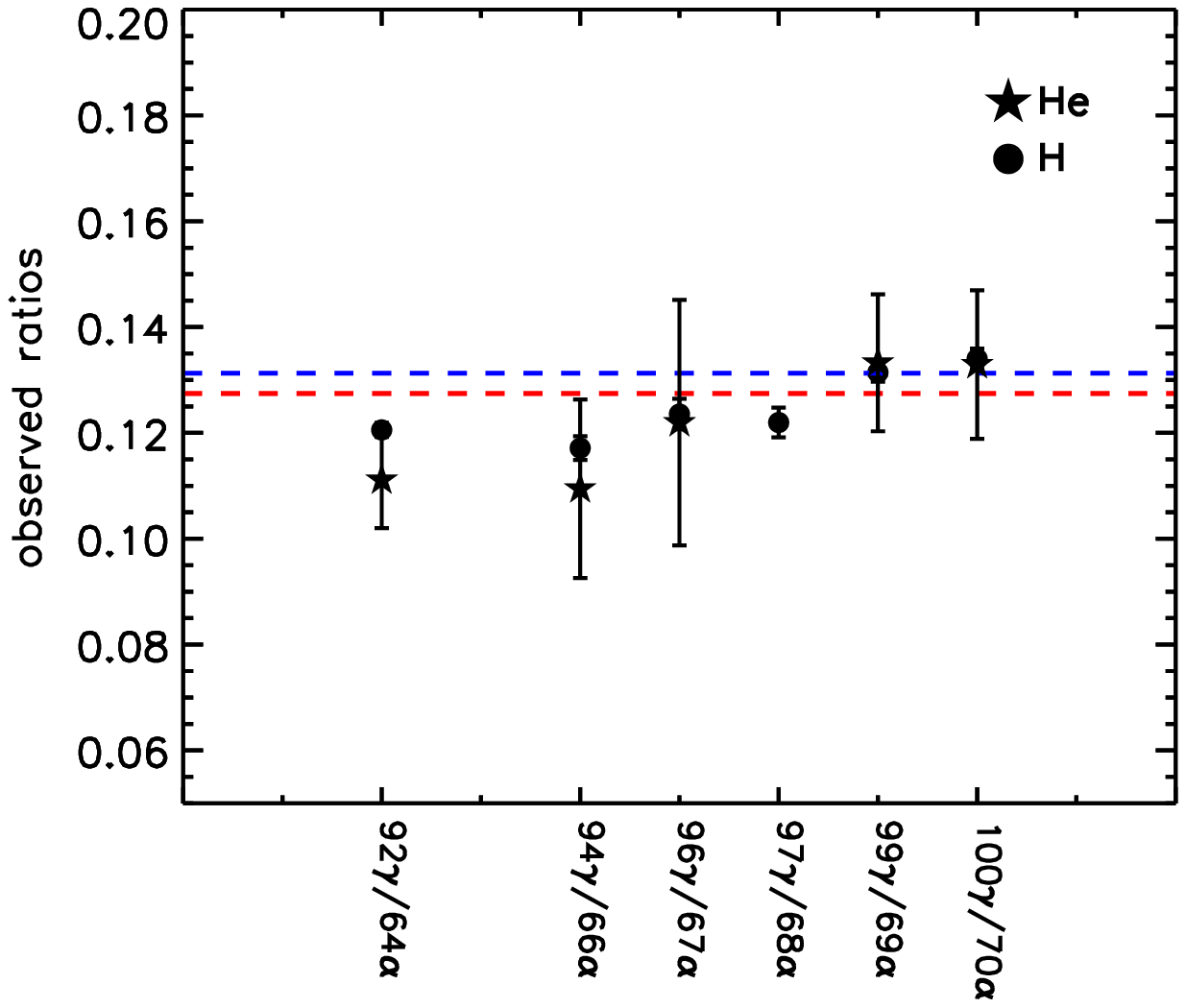}
\includegraphics[width = 0.45 \textwidth]{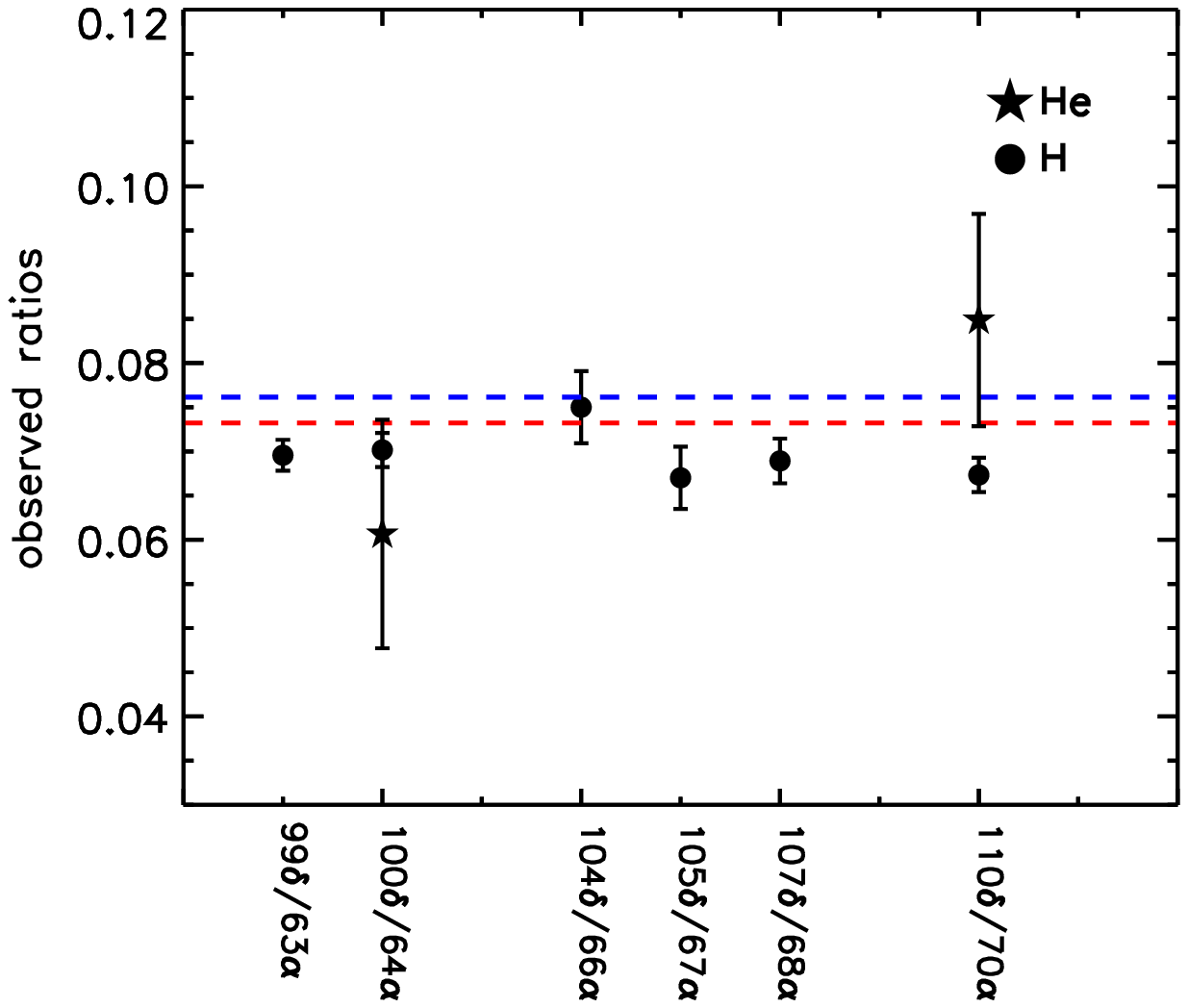}
\caption{{Comparison of observed and LTE ratios of recombination lines. The observed ratios for hydrogen and helium RRLs are marked with circles and pentagrams. The red dashed lines represent the LTE ratios while the blue dashed lines represent the LTE ratios with the departure coefficients $b_{n}$ corrected. Each ratio is given below the respective abszissa.} \label{Fig:ltetest}}
\end{figure*}

Assuming local thermodynamic equilibrium (LTE) and following the formulae (6.24)--(6.27) of \citet{1972MNRAS.157..179B}, the intensity ratios of RRL pairs with different $\Delta n$ from the same atom at neighbouring frequencies (e.g., H88$\beta$/H70$\alpha$) are found to be 
\begin{equation}
\frac{T_{\Delta n_{1}}}{T_{\Delta n_{2}}} = \frac{\Delta n_{1}K(\Delta n_{1})}{\Delta n_{2}K(\Delta n_{2})}, 
\end{equation} 
where $\Delta n_{1}$ and $\Delta n_{2}$ represent different $\Delta n$ defined in formula (\ref{redberg}), and $\Delta nK(\Delta n)$ is given in Table I of  \citet{1972MNRAS.157..179B}. From Fig.~\ref{Fig:ltetest}, we find that observed ratios from H$\beta$/H$\alpha$,H$\gamma$/H$\alpha$,H$\delta$/H$\alpha$, are consistent with LTE ratios. Based on a table of the departure coefficients $b_{n}$ provided by \citet{1979ApJS...39..633S}, we find that all $b_{n}$ factors are larger than 0.94 and smaller than unity, when adopting an electron temperature of 8300 K and an electron density of $1\times10^{4}$ cm$^{-3}$ \citep{1997A&A...327.1177W}. Therefore, we suggest that the LTE deviations are very small for these RRLs in Orion KL.

\subsection{Molecular lines}\label{sec.mol}
In this section, we focus on the origin of detected molecular lines according to the line parameters obtained by Gaussian fits or hfs fits in our survey as well as spatial distributions of their corresponding transitions at higher frequencies obtained from the ALMA-SV data (see Table~\ref{Tab:alma} and Fig.~\ref{Fig:cmap}).  

\subsubsection{NH$_{3}$, $^{15}$NH$_{3}$, and NH$_{2}$D}
In our band, we find 7 metastable NH$_{3}$ transitions, 27 non-metastable NH$_{3}$ transitions, 7 metastable $^{15}$NH$_{3}$ transitions, 3 non-metastable $^{15}$NH$_{3}$ transitions, and 2 NH$_{2}$D transitions (see Table~\ref{Tab:nh3} and Fig.~\ref{Fig:nh3}). It is worth noting that the non-metastable $^{15}$NH$_{3}$ transitions are detected in Orion KL for the first time. The line widths of non-metastable $^{15}$NH$_{3}$ transitions become narrower with increasing $J$, while there is no such trend for the metastable $^{15}$NH$_{3}$ transitions (see Fig.~\ref{Fig:nh32} and Table~\ref{Tab:nh3}). This suggests that higher-$J$ non-metastable transitions may trace denser regions where turbulent motions become less dominant. We note that the errors of the line widths ($\sim$1~\kms) are large compared to the differences between these line widths, so further observations are needed to confirm the trend. Based on previous VLA ammonia observations \citep[e.g.,][]{1982ApJ...259L.103G,2011ApJ...739L..13G}, NH$_{3}$ emission is mainly found in the HC, the Plateau, MM4, MM5, and MM6, but barely in the CR. Velocities around 6\,\kms\,and large line widths suggest that $^{15}$NH$_{3}$ mainly comes from the HC. One of the two NH$_{2}$D transitions belongs to the para, the other to the ortho species \footnote[10]{For NH$_{2}$D, the inversion motion splits each rotational level (denoted by the quantum numbers $J_{k_{\rm a}k_{\rm c}}$) into two inversion states, labeled as ``s'' and ``a'', corresponding to symmetric and antisymmetric states of the two equivalent minimum-energy structures \citep{1983ARA&A..21..239H}. The ortho levels are those in the ``s'' state with odd $k_{\rm a}$ and in the ``a'' state with even $k_{\rm a}$ while other states belong to the para species.}, which have already been reported by \citet{1987A&A...172..311W}. The intensity of NH$_{2}$D ($4_{1,4}{\rm a}-4_{0,4}{\rm s}$) from our observations is smaller than theirs, but is still within their uncertainties. NH$_{2}$D ($3_{1,3}{\rm s}-3_{0,3}{\rm a}$) is blended with NH$_{3}$ (8,5) at 18808.5 MHz and H118$\epsilon$ at 18808.4 MHz (see Fig.~\ref{Fig:nh32}). From Fig.~\ref{Fig:cmap}, NH$_{2}$D emission is found in the HC, the HC(S), and in MM4. NH$_{2}$D ($3_{1,3}s-3_{0,3}a$) and ($4_{1,4}a-4_{0,4}s$) are considered to arise from the HC in the following analysis.

\subsubsection{H$_{2}$O and HDO}
We find one H$_{2}$O line and one HDO line (see Table~\ref{Tab:one}). The H$_{2}$O (6$_{1,6}$-5$_{2,3}$) maser is blended with NH$_{3}$ (3,1), and shows multiple velocity components which have peak intensities of $>$1000 Jy (see Fig.~\ref{Fig:one}). A previous deuterated water study has shown that HDO mainly comes from the HC(S) \citep[see Fig.~6 of][]{2013ApJ...770..142N}. Although HDO ($3_{2,1}-4_{1,4}$) is blended, it clearly peaks at $\sim 7$~\kms\,, which coincides with the velocity of the HC(S). This also points out that the deuterated water arises mainly from the HC(S).

\subsubsection{Sulfur-bearing molecules, SO$_{2}$ and OCS}
We find five transitions of SO$_{2}$ and one transition of OCS in the survey. Most of the unblended lines can be fitted with two velocity components (see Fig.~\ref{Fig:so2} and Table~\ref{Tab:so2}). From Fig.~\ref{Fig:cmap}, we can see that the SO$_{2}$ emission around 6~\kms\, mainly comes from the HC. Thus, the two velocity components are assigned to the HC and the plateau. But for SO$_{2}$ (8$_{1,7}$--7$_{2,6}$), it cannot be fitted with two Gaussian components. There is an additional narrow component which has a velocity of $\sim$8~\kms\,and a line width of $\sim$2~\kms\, when we fit the line with three Gaussian components. It may result from emission from the CR, an outstanding emitter of transitions at low upper level energy, but it is narrower than the typical SO$_{2}$ line width (3--6~\kms) of the CR \citep{2013AA...556A.143E}. Alternatively, it may be affected by population inversion since the narrow component is not detected in other SO$_{2}$ transitions in our band and the 1.3, 2mm, and 3 mm bands \citep{2013AA...556A.143E}. Furthermore, SO$_{2}$ is extremely enhanced by shocks so that it becomes the strongest coolant in the 607--725~GHz survey which outperforms CO by a large factor \citep{2001ApJS..132..281S}.

 OCS (2--1) also shows two components, but the fitted velocity and line width are different from those derived from SO$_{2}$ (see Table~\ref{Tab:so2}). The line width of the plateau component from OCS is much narrower than found for SO$_{2}$. This is probably because OCS (2--1) is only excited in the LVF. Based on the distributions of OCS (see Fig.~\ref{Fig:cmap}), the other component of OCS (2--1) is assigned to the HC(S) according to the fitted velocity and line width.

\subsubsection{Cyanopolyynes, HC$_{3}$N and HC$_{5}$N}
Cyanopolyynes are well studied and abundant in many different astronomical environments such as the late-type carbon star IRC +10216 and the cold starless core TMC-1 \citep[e.g.,][]{2000AAS..142..181C,2004PASJ...56...69K} whereas they are not prominent in Orion KL \citep{2013AA...559A..51E}. BIMA observations show that toward Orion KL, HC$_{3}$N emission originates in the HC \citep{1996ApJ...469..216W} whereas the origin of HC$_{5}$N is still a puzzle.  In our survey, we find one HC$_{3}$N line and two HC$_{5}$N lines (see Fig.~\ref{Fig:one}). Based on the measured velocities and line widths (see Table~\ref{Tab:one}),  HC$_{5}$N seems to originate from the CR or the HC(S) or the ER. From Fig.~\ref{Fig:cmap}, we confirm that HC$_{3}$N is mainly from HC while HC$_{5}$N is not detected in the archival ALMA data. Nevertheless, it would be very surprising if HC$_{3}$N and HC$_{5}$N would not share the same spatial origin in Orion KL. Assuming that HC$_{5}$N originates from the same region as HC$_{3}$N, the HC$_{5}$N/HC$_{3}$N abundance ratio is estimated to be $(2.8\pm0.7)\times10^{-2}$\, according to Table~\ref{Tab:rd}. This is slightly lower than the $(7\pm4)\times 10^{-2}$ of \citet{2013AA...559A..51E}. HC$_{5}$N ($7-6$) is inside the frequency range covered by our survey. With the rotational temperatures and column densities listed in Table~\ref{Tab:rd}, we can estimate the peak intensity of HC$_{5}$N ($7-6$) by assuming a low opacity, a Gaussian profile, a line width of 5~\kms, and a source size of 10\arcsec. The peak intensity is estimated to be about 0.012~Jy, so its non-detection can be expected. On the other hand, we do not detect HC$_{7}$N and HC$_{9}$N transitions in this band which are detected in large numbers toward TMC-1 and IRC +10216 \citep{2004PASJ...56...69K,2015A&A...574A..56G}. HCN $\upsilon_{2}$=1 direct $l$-type transitions detected originally in the protoplanetary nebula CRL 618 \citep{2003ApJ...586..338T} and selected high-mass star-forming regions \citep{2001PhDT........34T} as well as the prototypical starburst galaxy Arp 220 \citep{2008AJ....136..389S} are also not detected in this survey. A tentative detection of the $J=9$ direct $l$-type transition towards Orion KL reported earlier \citep{2001PhDT........34T} could not be confirmed here.

\subsubsection{CH$_{3}$OH and $^{13}$CH$_{3}$OH}
\begin{figure*}[!htbp]
\centering
\includegraphics[width = 0.5 \textwidth]{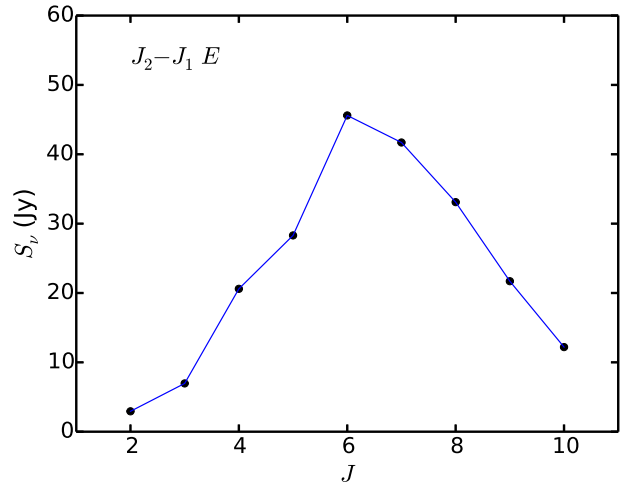}
\caption{{Observed peak intensities of CH$_{3}$OH ($J_{2}-J_{1}$~$E$) masers as a function of rotational quantum number $J$.} \label{Fig:25GHzmaser}}
\end{figure*}

In this survey, we find seventeen CH$_{3}$OH (v$_{\rm t}$=0) transitions, four torsionally excited CH$_{3}$OH (v$_{\rm t}$=1) transitions, and one \hbox{$^{13}{\rm CH}_{3}{\rm OH}$} (v$_{\rm t}$=0) transition. The $J_{2}-J_{1}$~$E$ methanol masers with $J=2...10$, already reported \citep[e.g.,][]{1988A&A...198..267M}, are all detected. From Fig.~\ref{Fig:25GHzmaser}, our measurements show that the peak intensities of these masers increase with principle quantum number $J$ from 2 to 6 and decrease with $J$ from 6 to 10. This is consistent with the result of \citet{1986A&A...157..318M}, although the $J_{2}-J_{1}$~$E$ methanol masers with $J=8,~9$ are not included in their analysis. The newly detected CH$_{3}$OH ($26_{2}-26_{1}$~$E$) line, belonging to the $J_{2}-J_{1}$~$E$ class, might not to be affected by population inversion, since it agrees well with the fit to the rotational diagram to CH$_{3}$OH transitions (see Sect.~\ref{sec.abu}). From Fig.~\ref{Fig:cmap}, we can see that methanol exists in the HC, the HC(S), the CR, MM4, MM5 and MM6. Based on the fitted line widths and velocities (see Table~\ref{Tab:ch3oh}), we suggest that transitions detected by our survey mainly arise from the CR. 

\subsubsection{HNCO}
From Fig.~\ref{Fig:cmap}, HNCO is found to exist mainly in the HC, the HC(S), and MM4 while only weak emission arises from MM5, MM6, and the CR. The HNCO emission in HC(S) stands out around velocities ranging from 6--10~\kms. In combination with the fitted parameters of HNCO (1--0) (see Table~\ref{Tab:one}), we suggest that HNCO (1--0) mainly originates from the HC(S). HNCO is believed to be enhanced in shocked regions \citep{2000A&A...361.1079Z}, which may indicate that the HC(S) component is formed via shocks due to the interaction between the outflowing gas and ambient clouds.
\subsubsection{H$_{2}$CO}
In this survey, we find one transition of H$_{2}$CO, the 9$_{2,7}$--9$_{2,8}$ line, which is from the para state of H$_{2}$CO. From Fig.~\ref{Fig:cmap}, H$_{2}$CO is found to exist in the HC, the HC(S), the CR, the Plateau, MM5, and MM6. Based on the fitted parameters (see Table~\ref{Tab:one}), H$_{2}$CO (9$_{2,7}$--9$_{2,8}$) may arise from the HC(S) or the CR. On the other hand, H$_{2}$CO (9$_{2,7}$--9$_{2,8}$) has an upper energy of 205~K, so we suggest that H$_{2}$CO (9$_{2,7}$--9$_{2,8}$) is mainly from the HC(S).
\subsubsection{CH$_{3}$CN and CH$_{3}$CH$_{2}$CN}
One CH$_{3}$CN transition and one CH$_{3}$CH$_{2}$CN transition are detected in our survey. Based on Fig.~\ref{Fig:cmap}, we find that the CH$_{3}$CN emission arises from many spatial components, among which the HC is the strongest, while CH$_{3}$CH$_{2}$CN emission is mainly from the HC, followed by MM4. Inspecting the fitted parameters for CH$_{3}$CN ($1_{0}-0_{0}$) and CH$_{3}$CH$_{2}$CN ($3_{1,3}-2_{1,2}$) (see Table~\ref{Tab:one}), we suggest that the two transitions are also mainly from the HC. The line profile of CH$_{3}$CH$_{2}$CN ($3_{1,3}-2_{1,2}$) is broad, because its hfs lines are blended. Although many other CH$_{3}$CH$_{2}$CN transitions with similar upper level energies fall in this band, they are expected to be much weaker since their intrinsic strengths are much lower. 
\subsubsection{CH$_{3}$OCHO and CH$_{3}$OCH$_{3}$}
In our survey, we find twelve CH$_{3}$OCHO transitions and two CH$_{3}$OCH$_{3}$ transitions (see Table~\ref{Tab:ch3ocho} and Table~\ref{Tab:one}). Based on the fitted velocities and narrow line widths, we suggest that CH$_{3}$OCHO and CH$_{3}$OCH$_{3}$ originate from the CR. From Fig.~\ref{Fig:cmap}, we find that the emissions of CH$_{3}$OCHO and CH$_{3}$OCH$_{3}$ mainly come from the CR, followed by MM5 and MM6.


\subsubsection{A brief summary}
The origin of molecules detected by our 1.3 cm line survey is briefly summarized in Table~\ref{Tab:origin}, demonstrating that chemical differentiation exists in Orion KL. This differentiation is roughly consistent with the fact that nitrogen-bearing molecules are favored in the HC while oxygen-bearing molecules are more predominant in the CR \citep[e.g.,][]{2005ApJ...632..355B,2008ApJ...672..962F}.

\section{Discussion}
\subsection{Rotational temperatures, column densities and fractional abundances relative to H$_{2}$}\label{sec.abu}
 Given that the density in Orion KL is relatively high (see Sect.~\ref{sec:comp}), LTE should be a good approximation. In the 1.3 cm wavelength range, the continuum emission is dominated by free-free emission. Toward Orion KL, free-free emission originates mainly from its foreground H{\scriptsize II} region M42 and is optically thin in the 18--26 GHz band \citep{1967ApJ...147..471M,1970ApL.....5..261T}. So the free-free continuum emission of Orion KL itself is simply assumed to be zero in the radiative transfer equation to estimate the opacity of molecular transitions at 1.3 cm wavelength. Thus, the opacity at the line center can be estimated via a formula given by \citet{2009tra..book.....W},
\begin{equation}\label{tau}
\tau = - {\rm ln} (1- \frac{T_{\rm mb}}{f(J(T_{\rm rot})-J(T_{\rm bg}))} ) ,
\end{equation}
where $T_{\rm mb}$ is the observed peak intensity, $T_{\rm rot}$ is the rotational temperature, and $T_{\rm bg}$ is the 2.73~K cosmic microwave background brightness temperature. $f$ is the filling factor $f=\theta^{2}_{\rm s}/(\theta^{2}_{\rm s}+\theta^{2}_{\rm beam})$, where we assume a Gaussian source and beam width FWHM of $\theta_{\rm s}$ and $\theta_{\rm beam}$, respectively. $J(T)=\frac{h\nu}{k} \frac{1}{e^{h\nu/kT}-1}$, $h$ is the Planck constant, $\nu$ is the rest frequency of the analyzed transition, and $k$ is the Boltzmann constant. If we take a filling factor of 0.06 (corresponding to an assumed source size of 10\arcsec) and a rotational temperature of 100~K, a peak intensity of 1~K (corresponding to $\sim$~0.7 Jy at 22~GHz) is equivalent to an opacity of 0.18 at the line center. When a higher rotational temperature is taken, the opacity becomes even lower. Thus, most of the detected thermally excited molecular lines appear to be optically thin, with the exception of the low $J$ NH$_{3}$ transitions. On the other hand, lines with hfs allow for determinations of opacity as well. Based on the HFS fitting routine in CLASS, we confirm that HC$_{3}$N (2--1), CH$_{3}$CN (1$_{0}$--0$_{0}$), HNCO (1$_{0,1}$-0$_{0,0}$), and CH$_{3}$CH$_{2}$CN ($3_{1,3}-2_{1,2}$) are optically thin (see Table~\ref{Tab:one}).

 Below, we use rotational diagrams to roughly estimate rotational temperatures and column densities. The standard formula used here is 
\begin{equation}\label{f1}
{\rm ln}(\frac{3kW}{8\pi^{3}\nu\mu^{2}S})~=~{\rm ln}(\frac{N_{\rm tot}}{Q(T_{\rm rot})})-\frac{E_{\rm u}}{kT_{\rm rot}},
\end{equation}
where $k$ is the Boltzmann constant, $W$ is the integrated intensity, $\nu$ is the rest frequency, $\mu$ is the permanent dipole moment, $S$ is the transition's intrinsic strength, $N_{\rm tot}$ is the total column density, $T_{\rm rot}$ is the rotational temperature, $Q$ is the partition function, and $E_{\rm u}$ is the upper level energy of the transition. The values of $Q$ and $\mu$ are taken from the CDMS and JPL catalogs. Note that the partition function for CH$_{3}$OH and $^{13}$CH$_{3}$OH has included the vibrational state $v_{\rm t}$=0 and $v_{\rm t}$=1 (Dr. Christian P. Endres, priv. comm.). For molecules with at least two transitions, we use least-square fits to the rotational diagrams to derive their rotational temperatures and column densities. For molecules with only one transition detected, we fix rotational temperatures, according to the values from literature or chemically related molecules, to derive their column densities (see Table~\ref{Tab:rd}). But for HC$_{5}$N, the fitted rotational temperature is about 1~K, which is not reasonable. This is because we only detect two HC$_{5}$N transitions and their upper level energy difference ($\sim$1~K) is very small. Thus, rotational temperatures of 30~K and 200~K are taken to derive the column density of HC$_{5}$N assuming that HC$_{5}$N originates from the CR and HC, respectively.  

Following previous studies \citep[][]{2013ApJ...777...85N,2014ApJ...787..112C}, we assume that all molecules with the same origin have the same source size to simplify the calculation. Adopted source sizes for each component are also based on Table~2 of \citet{2014ApJ...787..112C} and the parameters of the HC(S) are assumed to be the same as those of the HC. With the adopted source sizes, the observed intensities are corrected for beam dilution by dividing by the filling factor to derive source-averaged column densities. The H$_{2}$ column densities adopted in \citet{2014ApJ...787..112C} are used to calculate fractional abundances of the detected molecules. 

\begin{figure*}[!htbp]
\centering
\includegraphics[width = 0.45 \textwidth]{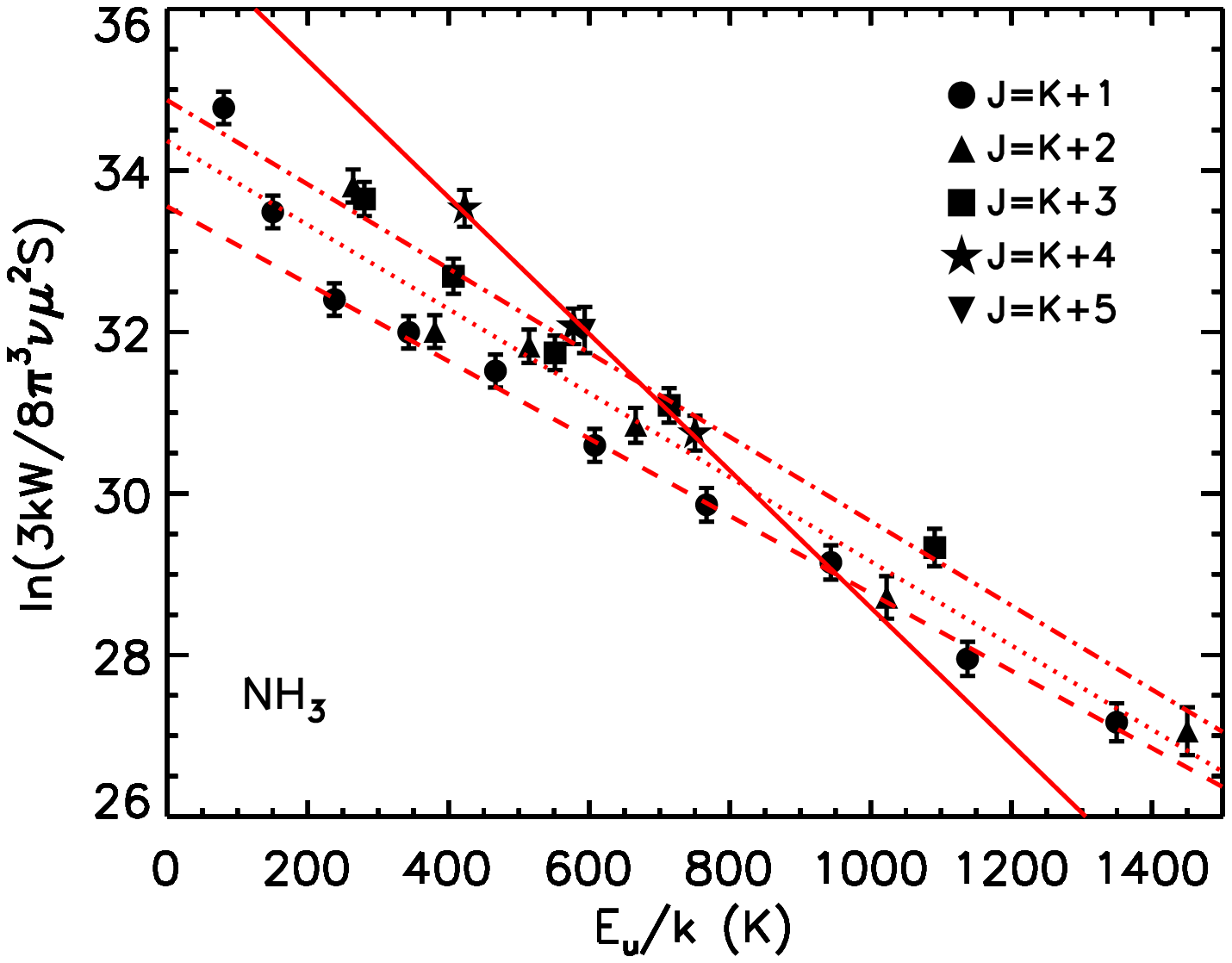}
\includegraphics[width = 0.45 \textwidth]{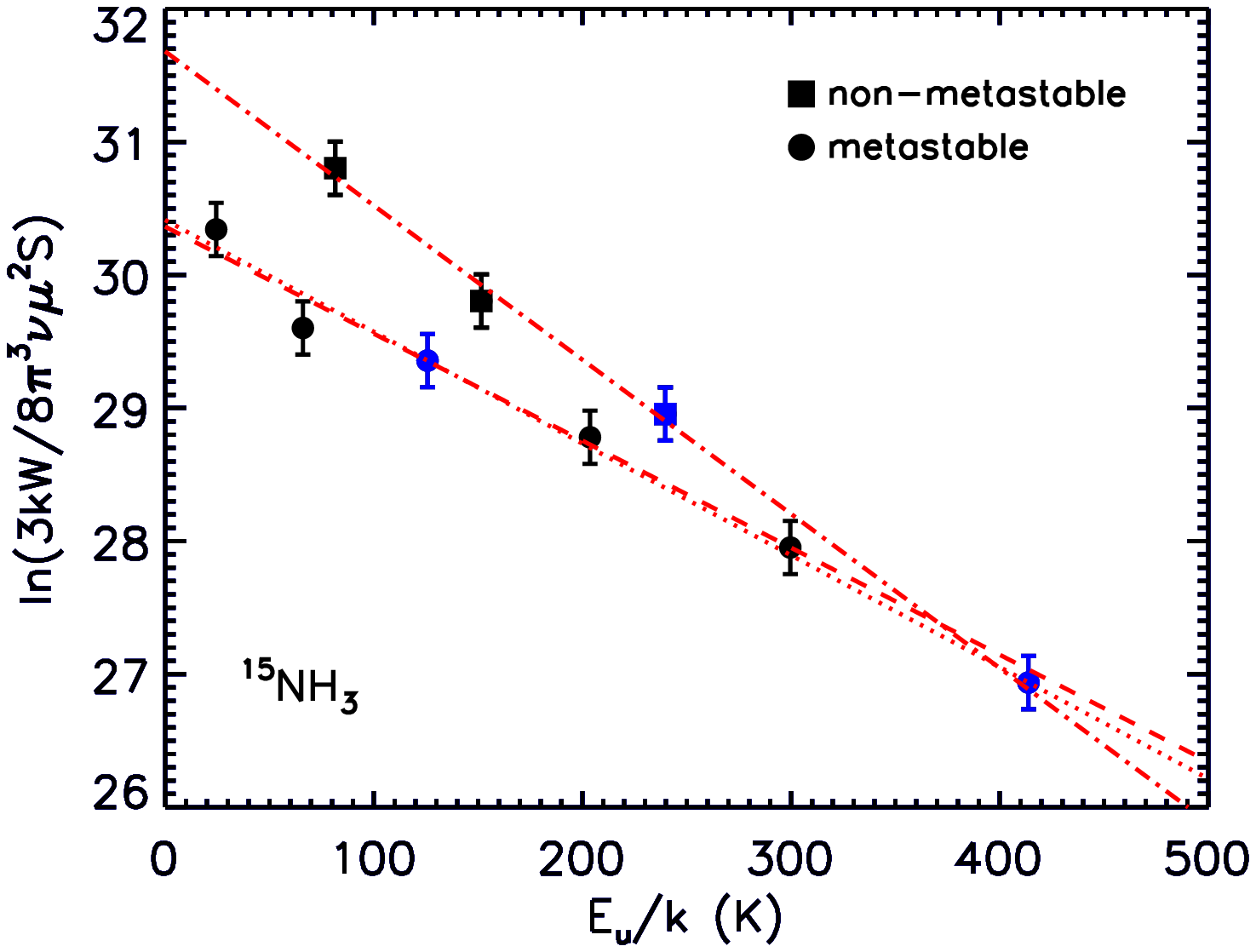}
\includegraphics[width = 0.45 \textwidth]{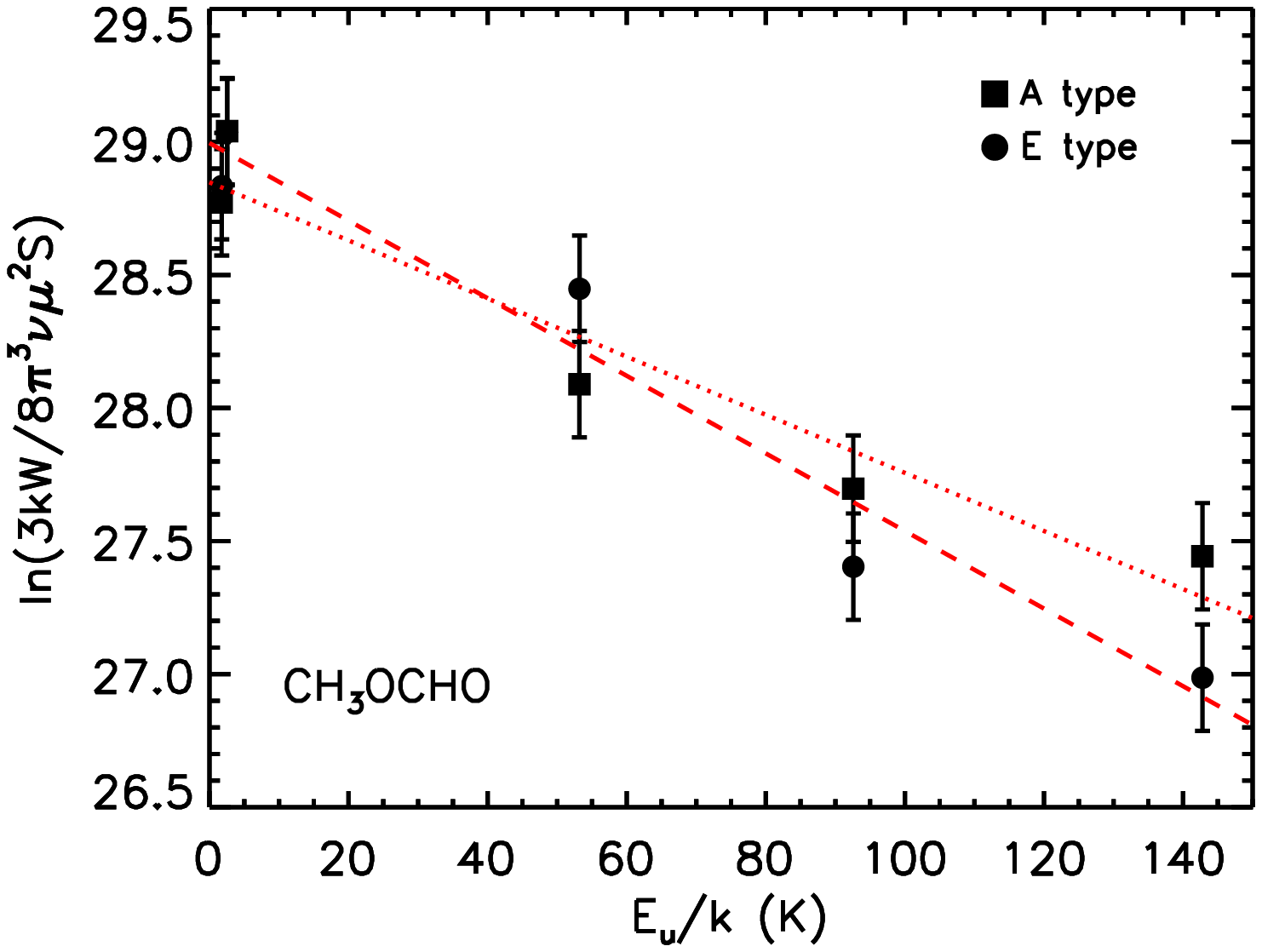}
\includegraphics[width = 0.45 \textwidth]{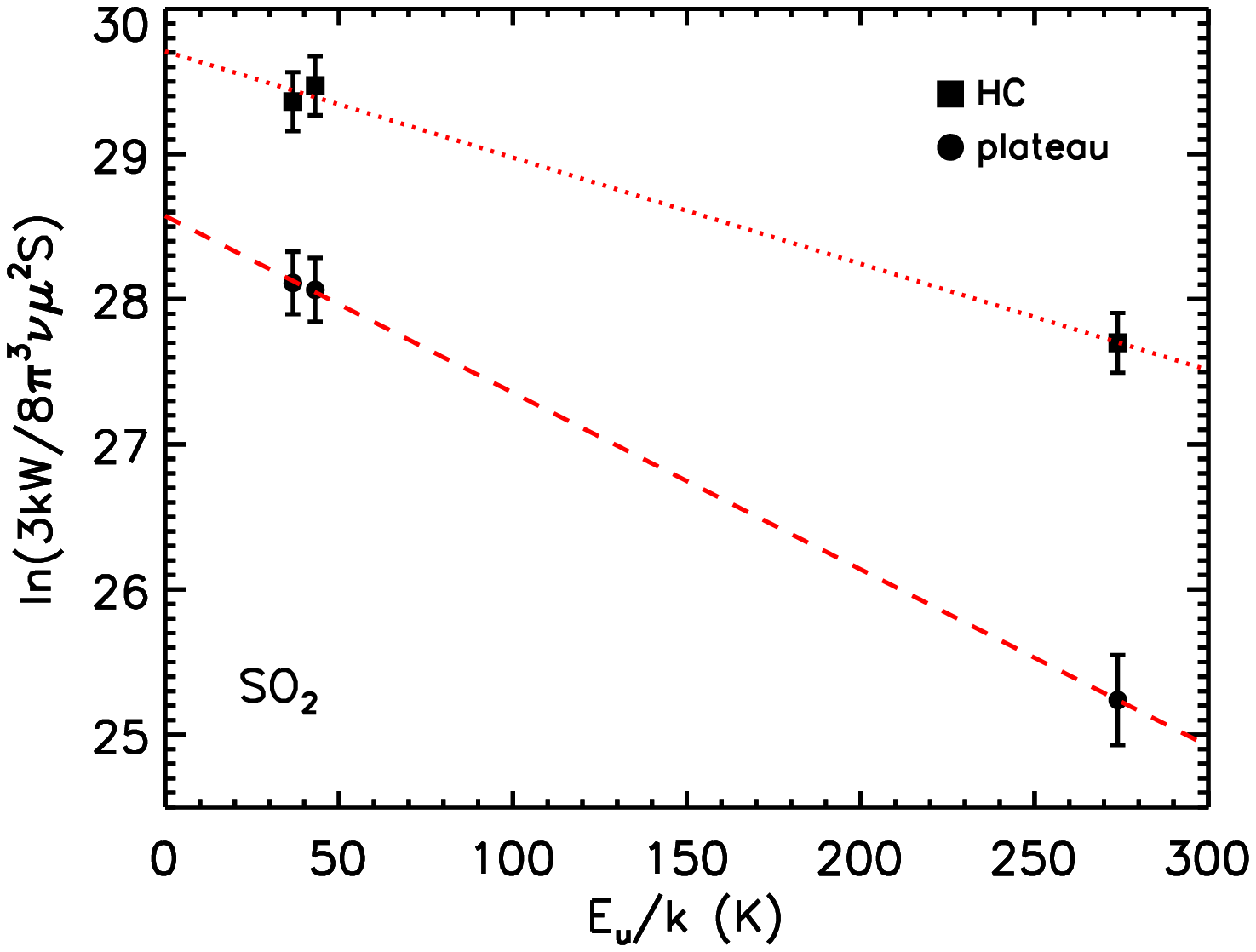}
\includegraphics[width = 0.45 \textwidth]{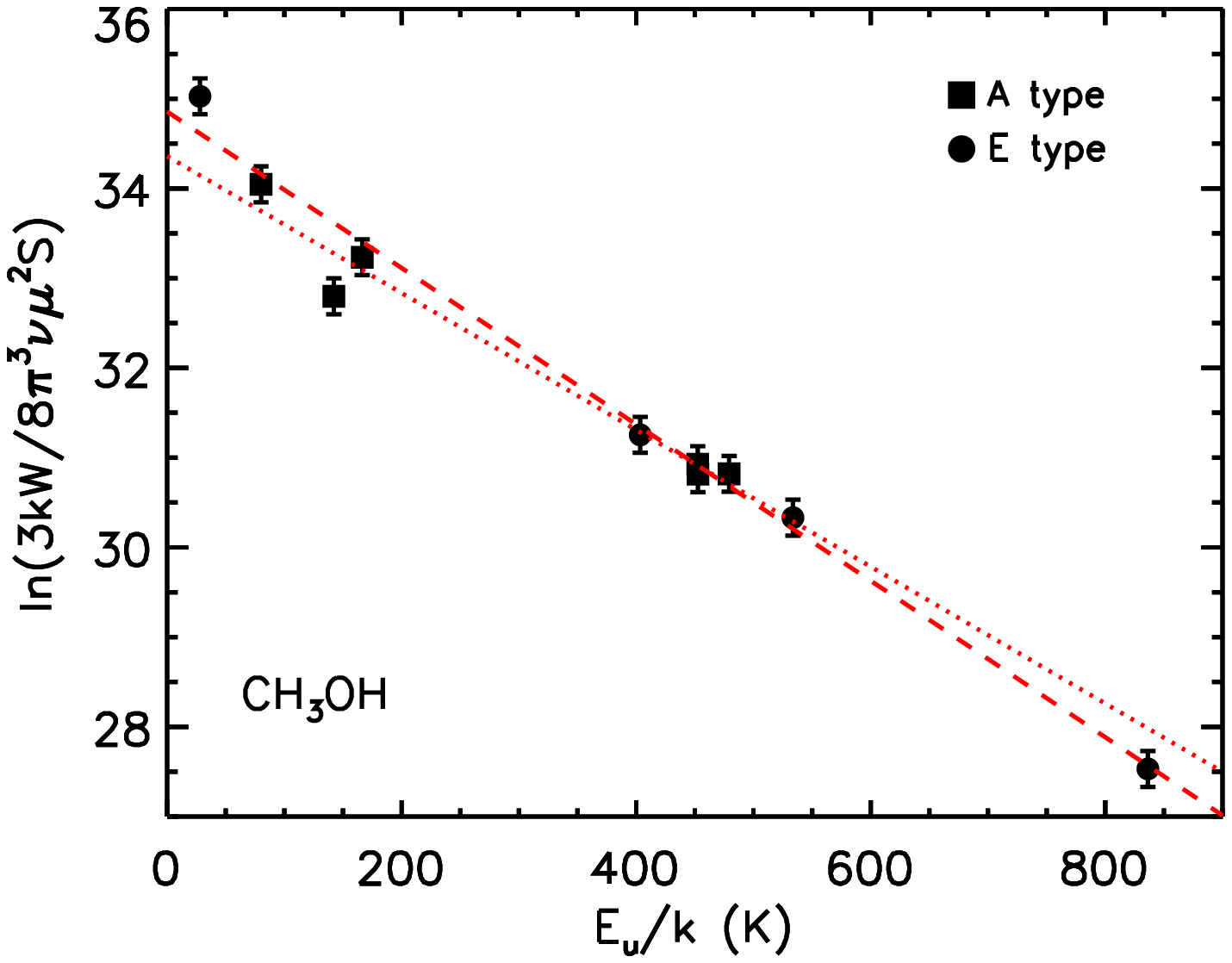}
\caption{{Rotational diagrams for NH$_{3}$, $^{15}$NH$_{3}$, CH$_{3}$OCHO, SO$_{2}$, and CH$_{3}$OH. The circles and squares are explained by the legends in the upper right of each panel. A calibration error of 20\% has been included in the error bars. In the NH$_{3}$ panel, least-square fits to $J+1$, $J+2$, $J+3$, and $J+4$ are indicated by a red dashed line, a red dotted line, a red dot-dashed line, and a red solid line. In the $^{15}$NH$_{3}$ panel, the para and ortho transitions of $^{15}$NH$_{3}$ are black and blue; the least-square fits to para and ortho metastable $^{15}$NH$_{3}$ transitions and non-metastable $^{15}$NH$_{3}$ transitions are indicated by a red dashed line, a red dotted line and a red dot-dashed line, respectively. In the CH$_{3}$OH and CH$_{3}$OCHO panels, the dashed and dotted lines represent fits to $A$ type and $E$ type transitions, respectively. In the SO$_{2}$ panel, dashed and dotted lines represent fits to its plateau and its HC component. } \label{Fig:rd}}
\end{figure*}

Fig.~\ref{Fig:rd} shows rotational diagrams for molecules with more than two transitions detected in our survey. For NH$_{3}$, we only fit the non-metastable transitions since metastable transitions consist of several components and decompositions based on the velocity information may be less accurate. In the fitting process, we separate non-metastable transitions into four classes which are labeled as $J=K+1$, $J=K+2$, $J=K+3$, and $J=K+4$. For $J=K+1$, NH$_{3}$ (2,1), (3,2), (4,3), (5,4), (6,5), and (7,6) are ignored in the fitting process since they are potentially contaminated by emission from other components rather than only from the HC component (see Fig.~\ref{Fig:nh3}). From the fitted result (see Fig.~\ref{Fig:rd}) where the ortho-to-para ratio is set to unity (based on the analysis of $^{15}$NH$_{3}$, see the discussion below), we find that only NH$_{3}$ (2,1) and (3,2) deviate significantly, while all others lie along the fitted line which suggests that other $J=K+1$ transitions are also dominated by the HC component. Nevertheless, if NH$_{3}$ (8,7), (9,8), (10,9), and (11,10) are not optically thin, the derived rotational temperature will be overestimated and the derived column density will be underestimated. We also find that the fits for $J=K+2$ and $J=K+3$ have rotational temperatures similar to $J=K+1$. However, the fit for $J=K+4$ has a much lower rotational temperature. Furthermore, the derived NH$_{3}$ column densities go up from $J=K+1$ to $J=K+4$. This suggests that the four classes may trace different NH$_{3}$ emission regions. However, it is surprising that the $J=K+4$ class has the largest column density among the four since the $J=K+4$ transitions require particularly high excitation. We also note that the fit for the $J=K+4$ class may have large uncertainties since we only have three data points.

To determine the ortho-to-para ratio of ammonia, $^{15}$NH$_{3}$ is more suited since the optical depths are much smaller than those of NH$_{3}$. Assuming that this rare isotopologue was formed in the same region as its main species, we fit its para and ortho states of metastable transitions separately (see Fig.~\ref{Fig:rd}). The ortho-to-para ratio is estimated to be $0.99\pm0.34$, in accord with the result ($0.7^{+0.5}_{-0.3}$) of \citet{1985A&A...146..134H}, agreeing very well with the fact that the ortho-to-para ratio approaches the statistical ratio of unity when ammonia is equilibrated under high kinetic temperature conditions \citep[e.g.,][]{2002PASJ...54..195T}. On the other hand, we find that metastable $^{15}$NH$_{3}$ has a higher excitation temperature than non-metastable $^{15}$NH$_{3}$ although only three points have been fitted for the non-metastable $^{15}$NH$_{3}$ (see Fig.~\ref{Fig:rd} and Table~\ref{Tab:rd}). This is likely due to the fact that the metastable line excitation is dominated by collisions while the excitation of the non-metastable transitions follows the radiation field. In the very dense medium, collisional ($T_{\rm kin}$) and radiative ($T_{\rm rad}$) temperatures could be the same. However, $T_{\rm rad}$ may suffer from beam dilution effects, which results in lower rotational temperatures. Alternatively, non-metastable $^{15}$NH$_{3}$ transitions trace higher densities than its metastable counterparts, and thus represent the temperature of higher excited inner regions than metastable transitions. Meanwhile, the HC is believed to be externally heated by an explosive event \citep{2011ApJ...727..113B,2011A&A...529A..24Z,2012A&A...540A.119N}, so the innermost region could be colder. Therefore, non-metastable transitions of $^{15}$NH$_{3}$ can be expected to have a lower rotational temperature than its metastable transitions. This is also consistent with the fact that the $J=K+4$ transitions of $^{14}$NH$_{3}$, tracing denser regions, have a lower rotational temperature.

No CH$_{3}$OH maser lines were included in the fit due to their deviation from LTE. CH$_{3}$OH ($2_{1}-3_{0} E$) was also not included because this line comprises several velocity components due to its low upper level energy ($E_{\rm u}/k=$28 K). We fit the A type and the E type of CH$_{3}$OH and CH$_{3}$OCHO, separately. Consequently, the CH$_{3}$OCHO A/E abundance ratio is estimated to be $1.38\pm0.26$ slightly higher than the expected value of unity while the CH$_{3}$OH A/E abundance ratio is estimated to be $0.77\pm0.33$, slightly lower than the expected value of unity. Nevertheless, the differences could be caused by uncertainties. For SO$_{2}$, we fit its HC component and its plateau component separately. We find that the HC component has a higher rotational temperature and a higher column density than its plateau component. 

The resulting rotational temperatures and column densities, together with results obtained with single dish telescopes from the literature, are given in Table~\ref{Tab:rd}. We find fitted rotational temperatures ranging from 69 to 209 K and derived molecular column densities ranging from 9.6$\times10^{13}$ to 1.9$\times10^{18}$~cm$^{-2}$. This results in fractional abundances spanning over more than four orders of magnitude. Taking beam dilution into account, we find that most of our results agree well with previous studies based on (sub)millimeter lines except for SO$_{2}$, CH$_{3}$CN, and HNCO. Our SO$_{2}$ column density for the HC is less than that of \citet{2014ApJ...787..112C} by a factor of 6. This is because our fitted  rotational temperature is much lower and the SO$_{2}$ lines of \citet{2014ApJ...787..112C} at higher frequencies, trace higher temperatures. The derived column densities of HC$_{3}$N, HC$_{5}$N, CH$_{3}$CN and HNCO are higher than those from previous studies listed in Table~\ref{Tab:rd}. The upper level energies of their transitions are very low, so they are probably contaminated by other components not being part of the HC. Thus, our assumed source size of 10$\arcsec$ could be an underestimate, which results in an overestimated column density. The derived column densities should then be upper limits for the average column density of those molecules. Alternatively, previously reported column densities were derived from (sub)millimeter lines which may suffer from opacity effects, leading to an underestimate of their column densities. 

\subsection{Elemental and isotopic abundance ratios in Orion KL}\label{sec.ratio}
The He$^{+}$/H$^{+}$, $^{12}$C/$^{13}$C, $^{14}$N/$^{15}$N, and D/H abundance ratios are fundamental parameters for the study of cosmology and the evolution of galaxies \citep[e.g.,][]{1985ARA&A..23..319B,1994ARA&A..32..191W,2007ARNPS..57..463S}. 
\subsubsection{He/H}
 Benefiting from a large number of RRLs, we can make use of hydrogen and helium RRL pairs to estimate the He$^{+}$/H$^{+}$ abundance ratio which can be used to evaluate the He/H abundance ratio. Under LTE conditions, the He/H abundance ratio can be calculated via the integrated intensity of Gaussian fits to the hydrogen and helium RRLs \citep[e.g.,][]{1974A&A....32..283C}:
\begin{equation}\label{f2}
y= \frac{N({\rm He})}{N({\rm H})} \simeq \frac{1}{R} \frac{\int \limits_{\Omega_{\rm s}}{\rm d}\Omega \int S_{\nu}({\rm He^{+}}){\rm d}\upsilon}{\int \limits_{\Omega_{\rm s}}{\rm d}\Omega \int S_{\nu}({\rm H^{+}}){\rm d}\upsilon} \;,
\end{equation}
where $R$ is the ratio of volumes of He$^{+}$ and H$^{+}$ Str\"omgren spheres, weighted by the square of the proton density, $\Omega_{\rm s}$ is the source solid angle, $\int S_{\nu}({\rm He}^{+}){\rm d}\upsilon$ and $\int S_{\nu}({\rm H}^{+}){\rm d}\upsilon$ are the observed integrated intensities of corresponding helium and hydrogen RRLs. Assuming that the H{\scriptsize II} region has the same size as the He{\scriptsize II} region, formula~(\ref{f2}) becomes
\begin{equation}\label{f3}
y= \frac{N({\rm He})}{N({\rm H})}\simeq \frac{\int S_{\nu}({\rm He^{+}}){\rm d}\upsilon}{\int S_{\nu}({\rm H^{+}}){\rm d}\upsilon} \;.
\end{equation}

\begin{figure*}[!htbp]
\centering
\includegraphics[width = 0.8 \textwidth]{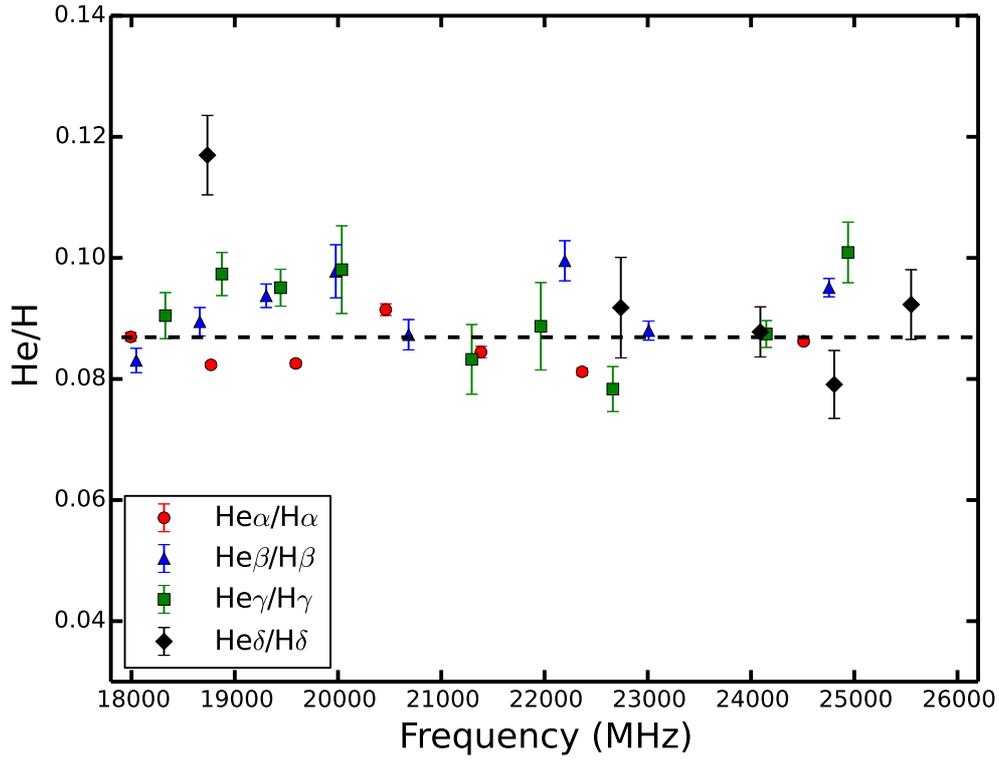}
\caption{{He$^{+}$/H$^{+}$ abundance ratios derived from RRLs as a function of rest frequency. The abundance ratios derived from He$\alpha$/H$\alpha$, He$\beta$/H$\beta$, He$\gamma$/H$\gamma$ and He$\delta$/H$\delta$ are marked with red filled circles, blue filled triangles, green filled squares and black filled diamonds, respectively. The dashed line represents the sigma-weighted mean value of He$^{+}$/H$^{+}$ abundance ratios.} \label{Fig:yr}}
\end{figure*}

In this work, there are 28 unblended pairs which are from H$\alpha$/He$\alpha$, H$\beta$/He$\beta$, H$\gamma$/He$\gamma$, and H$\delta$/He$\delta$. The rest frequencies of these pairs are only about 7 MHz apart, so their ratios should be free of uncertainties related to pointing accuracy, calibration errors, and different beam widths. Fig.~\ref{Fig:yr} shows the derived He/H abundance ratios according to formula~(\ref{f3}). We find that uncertainties increase from H$\alpha$/He$\alpha$ to H$\delta$/He$\delta$. Thus, we take the sigma-weighted mean value as the He/H abundance ratio which is estimated to be (8.7$\pm$0.7)\%. This agrees well with previous studies \citep{1992ApJ...395..484P,1994ARA&A..32..191W,2006ApJ...653.1226Q}, and is slightly lower than the (9.1$\pm$0.5)\% from previous H66$\alpha$ and He66$\alpha$ measurements which were pointed at the H{\scriptsize II} maximum \citep{1980A&A....87..269T}. Therefore, we confirm that the He/H abundance ratio is comparable to the primordial He/H abundance ratio of $\sim$8.3\% due to Big Bang Nucleosynthesis \citep[BBN;][]{2004ApJ...617...29O,2007ARNPS..57..463S} and the solar value \citep[$\sim$9.8\%;][]{1994ARA&A..32..191W}, which indicates that the He/H abundance ratio is mainly determined by BBN but little affected by stellar nucleosynthesis. Furthermore, the dispersion of the ratios derived from H$\alpha$/He$\alpha$ pairs to H$\delta$/He$\delta$ pairs is small (see Fig.~\ref{Fig:yr}), suggesting that the departure coefficients, the $b_{\rm n}$ factors, are nearly identical for hydrogen and helium. This agrees well with theoretical predictions \citep[e.g.,][]{1995MNRAS.272...41S}.

\subsubsection{$^{12}$C/$^{13}$C}
We use the total column density of CH$_{3}$OH and $^{13}$CH$_{3}$OH, including both the A and E type, to estimate the $^{12}$C/$^{13}$C ratio in the CR. For CH$_{3}$OH, we obtain a total column density of $(2.0\pm0.4)\times10^{18}$~cm$^{-2}$. Since the A/E abundance ratio is estimated to be $0.77\pm0.33$ (see Sect.~\ref{sec.abu}), we simply assume the A/E abundance ratio to be unity for $^{13}$CH$_{3}$OH. We arrive at a total column density of $(3.2\pm0.6)\times10^{16}$~cm$^{-2}$ for $^{13}$CH$_{3}$OH. This results in a $^{12}$C/$^{13}$C ratio of $63\pm17$, which agrees very well with the result ($57\pm14$) based on CH$_{3}$OH transitions at higher frequencies \citep{2007AA...476..807P}, and the $50\pm20$ from H$_{2}$CS transitions \citep{2010AA...517A..96T}, while it is slightly higher than the value ($43\pm7$) from CN lines \citep{2002ApJ...578..211S}. This confirms that the $^{12}$C/$^{13}$C ratio in the CR is slightly smaller than the value ($77\pm7$) in the local ISM and the solar value \citep[89,][]{1994ARA&A..32..191W}. Since the solar value represents the local interstellar medium 4.5 Gyrs ago, the difference may arise from the Galactic chemical evolution. Isotope selective photodissociation by UV photons can influence the $^{12}$C/$^{13}$C ratio due to the difference in self-shielding of $^{12}$C and $^{13}$C \citep[e.g.,]{1988ApJ...334..771V,2002ApJ...578..211S,2005ApJ...634.1126M}. $^{13}$C is expected to be more easily photodissociated, which will result in a higher $^{12}$C/$^{13}$C ratio, rather than a lower value that we obtained. This suggests that the effect cannot be significant influencing the $^{12}$C/$^{13}$C ratio in Orion KL. Previous studies have also shown that chemical fractionation does not play a substantial role in influencing such ratios \citep[e.g.,][]{2005ApJ...634.1126M}. 

\subsubsection{$^{14}$N/$^{15}$N}\label{sec.nratio}
Here, we use the average column density of NH$_{3}$ and $^{15}$NH$_{3}$ to estimate the $^{14}$N/$^{15}$N ratio in the HC. For NH$_{3}$, we find that the column density obtained from the NH$_{3}$ ($J=K+4$) transitions is nearly an order of magnitude higher than the values derived from the $J=K+1$, $J=K+2$, and $J=K+3$ transitions, which may be due to different excitation conditions. For NH$_{3}$ ($J=K+1$, $J=K+2$, and $J=K+3$) which share similar excitation conditions, we take a sigma-weighted average value of the three groups as the column density of NH$_{3}$ in the HC, which is estimated to be $(2.4\pm 1.2)\times 10^{17}$~cm$^{-2}$. For $^{15}$NH$_{3}$, we take the average value derived from metastable and non-metastable transitions as the column density of $^{15}$NH$_{3}$, which is estimated to be $(2.4\pm 0.3)\times 10^{15}$~cm$^{-2}$. This results in a $^{14}$N/$^{15}$N ratio of $100\pm51$, which is roughly consistent with the previous value of 170$^{+140}_{-80}$ by \citet{1985A&A...146..134H}, but smaller than the $234\pm47$ derived from CN transitions \citep{2012ApJ...744..194A}. This indicates that the $^{14}$N/$^{15}$N ratio in the HC is smaller that the $450\pm22$ in the local ISM \citep{1994ARA&A..32..191W}, the $1000\pm200$ in the prototypical starless cloud core L1544 \citep{2013A&A...555A.109B}, and the $300\pm50$ in Barnard 1 \citep{2013A&A...560A...3D}. Fractionation in nitrogen is likely to play an important role in cold temperature ($\sim$10 K) regions \citep{2008MNRAS.385L..48R}, but the effect can be neglected in Orion KL due to its high temperature. Meanwhile, the $^{14}$N/$^{15}$N ratio is found to be $361\pm141$ in the prototypical PDR region Orion Bar \citep{2012ApJ...744..194A}. Thus, isotope selective photodissociation by UV photons should not affect the $^{14}$N/$^{15}$N ratio a lot either. Since $^{15}$N may originate from massive stars and is potentially destroyed in lower mass stars \citep[e.g.,][]{1994LNP...439...72H,1994ARA&A..32..191W,1999ApJ...512L.143C,2009ApJ...690..580W}, such an enhanced $^{15}$N in the massive star-forming regions like Orion KL can be expected. Furthermore, our value is similar to the $111\pm17$ in the Large Magellanic Cloud (LMC) massive star-forming region N113, the $\sim100$ in the LMC star-forming region N159HW and the value in the central region of NGC 4945 \citep{1999ApJ...512L.143C}.

\subsubsection{D/H}
The deuterium fraction can be estimated from the ratio between the column densities of para--NH$_{3}$ and para--NH$_{2}$D, which have been discussed above and which are given in Table~\ref{Tab:rd}. We obtain a D/H ratio of $(8.3\pm4.5)\times 10^{-3}$. This is slightly higher than the value $3\times10^{-3}$ from a previous deuterated ammonia \citep{1987A&A...172..311W} and a deuterated water study \citep{2013ApJ...770..142N}, and is consistent with $(2-8)\times10^{-3}$ from a study of other deuterated molecules \citep{2013ApJ...777...85N}. This confirms that the D/H ratio for NH$_{3}$ is strongly affected by fractionation and is nearly two orders of magnitude higher than the abundance ratio in the interstellar medium and the primordial D/H ratio \citep[$\sim 1.5\times 10^{-5}$; e.g.][]{1994ARA&A..32..191W,2003ApJ...587..235O,2007ARNPS..57..463S}. Our value is also consistent with the ratio derived from HCN and DCN in the Orion Bar \citep{2006A&A...454L..47L,2009A&A...508..737P}. The enhancement could be attributed to a fossil record of the deuteration of icy dust grain mantles \citep[e.g.,][]{2000A&A...361..388R,2005A&G....46b..29M} or warm deuterium chemistry driven by CH$_{2}$D$^{+}$ \citep{2009A&A...508..737P}.

\subsection{A RADEX non-LTE model for $^{15}$NH$_{3}$}\label{sec.radex}
The inversion lines of ammonia are widely used as a Galactic and extragalactic molecular temperature tracer \citep[e.g.,][]{1983A&A...122..164W,1983ARA&A..21..239H,2003A&A...403..561M}. However, the $^{14}$NH$_{3}$ transitions tend to become optically thick in high density regions containing hot cores. Furthermore, Orion KL contains several spatial components which make contributions to the $^{14}$NH$_{3}$ emission, resulting in difficulties of disentangling the different velocity components. On the contrary, the $^{15}$NH$_{3}$ transitions can be expected to be optically thin since the $^{14}$N/$^{15}$N abundance ratio is of order 100 in Orion KL (see Sect.~\ref{sec.nratio}); the transitions of $^{15}$NH$_{3}$ only show one velocity component from the HC, thus they can be used to constrain the physical properties of the HC. Furthermore, $^{15}$NH$_{3}$ is proved to be a good tracer for high density environments such as ultra-compact H{\scriptsize II} regions \citep[e.g.,][]{1996A&A...314..265W}. Here, we present a non-LTE analysis of $^{15}$NH$_{3}$ using the RADEX code \citep{2007A&A...468..627V}, where the Einstein A coefficients of $^{15}$NH$_{3}$ are from the CDMS database while its collision rates are taken to be the same as those of $^{14}$NH$_{3}$ \citep{1988MNRAS.235..229D}. A spherical cloud geometry is assumed. In the models, we do not account for any external radiation fields with the notable exception of the cosmic microwave background. However, infrared pumping is likely to affect the populations of $^{15}$NH$_{3}$ in the innermost regions of Orion KL. Therefore, spatial densities derived from purely collisional excitation can be considered as upper limits. To evaluate the models, we compare the observed line ratios with the modeled line ratios to minimize $\chi^{2}$ which is defined as
\begin{equation}\label{f:chi}
\chi^{2} = \Sigma_{i} \frac{(R_{{\rm obs}(i)}-R_{{\rm model}(i)})^{2}}{\sigma_{{\rm obs}(i)}^{2}},
\end{equation}
where $R_{{\rm obs}(i)}$, and $R_{{\rm model}(i)}$ represent the observed and modeled line ratios, and $\sigma_{{\rm obs}(i)}$ represents the uncertainty in $R_{{\rm obs}(i)}$. We note that the uncertainties of observed line ratios, ($\frac{^{15}{\rm NH}_{3} (2,2)}{^{15}{\rm NH}_{3} (1,1)}$, $\frac{^{15}{\rm NH}_{3} (4,4)}{^{15}{\rm NH}_{3} (2,2)}$, $\frac{^{15}{\rm NH}_{3} (5,5)}{^{15}{\rm NH}_{3} (4,4)}$, $\frac{^{15}{\rm NH}_{3} (3,2)}{^{15}{\rm NH}_{3} (2,1)}$, etc.) should be free of calibration errors and beam dilution effects due to being close in frequency. Following previous $\chi^{2}$ studies \citep[e.g.,][]{2014A&A...568A.122Z}, we only take solutions with a likelihood $L>0.6$ into account, where $L$ is defined as 
\begin{equation}\label{f:l}
L={\rm exp}(-\chi^{2}/2)/L_{\rm max} .
\end{equation}
$L_{\rm max}$ is the maximum likelihood, corresponding to the minimum value of $\chi^{2}$.

\begin{figure*}[!htbp]
\centering
\includegraphics[height = 0.4 \textwidth]{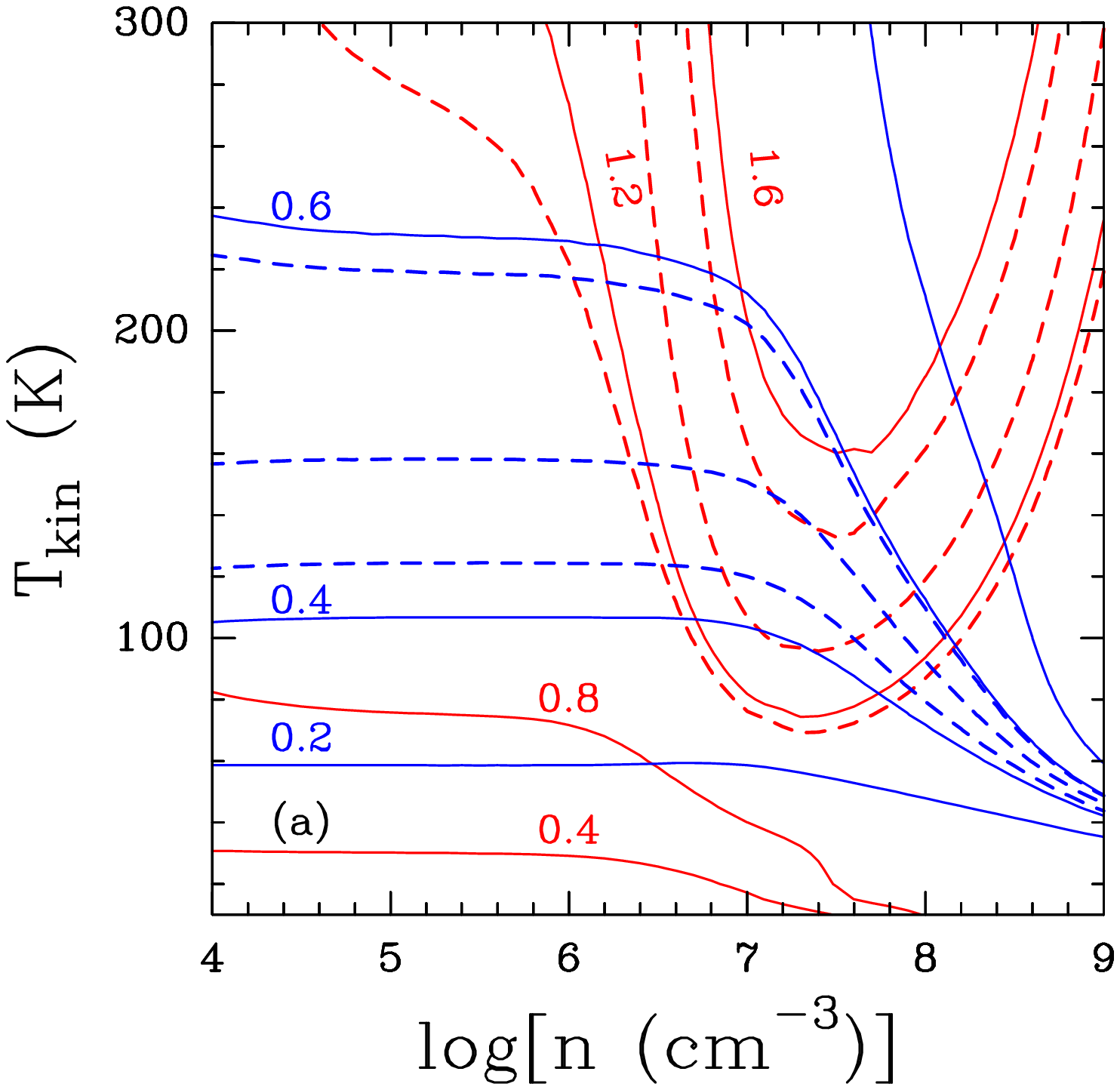}
\includegraphics[height = 0.4 \textwidth]{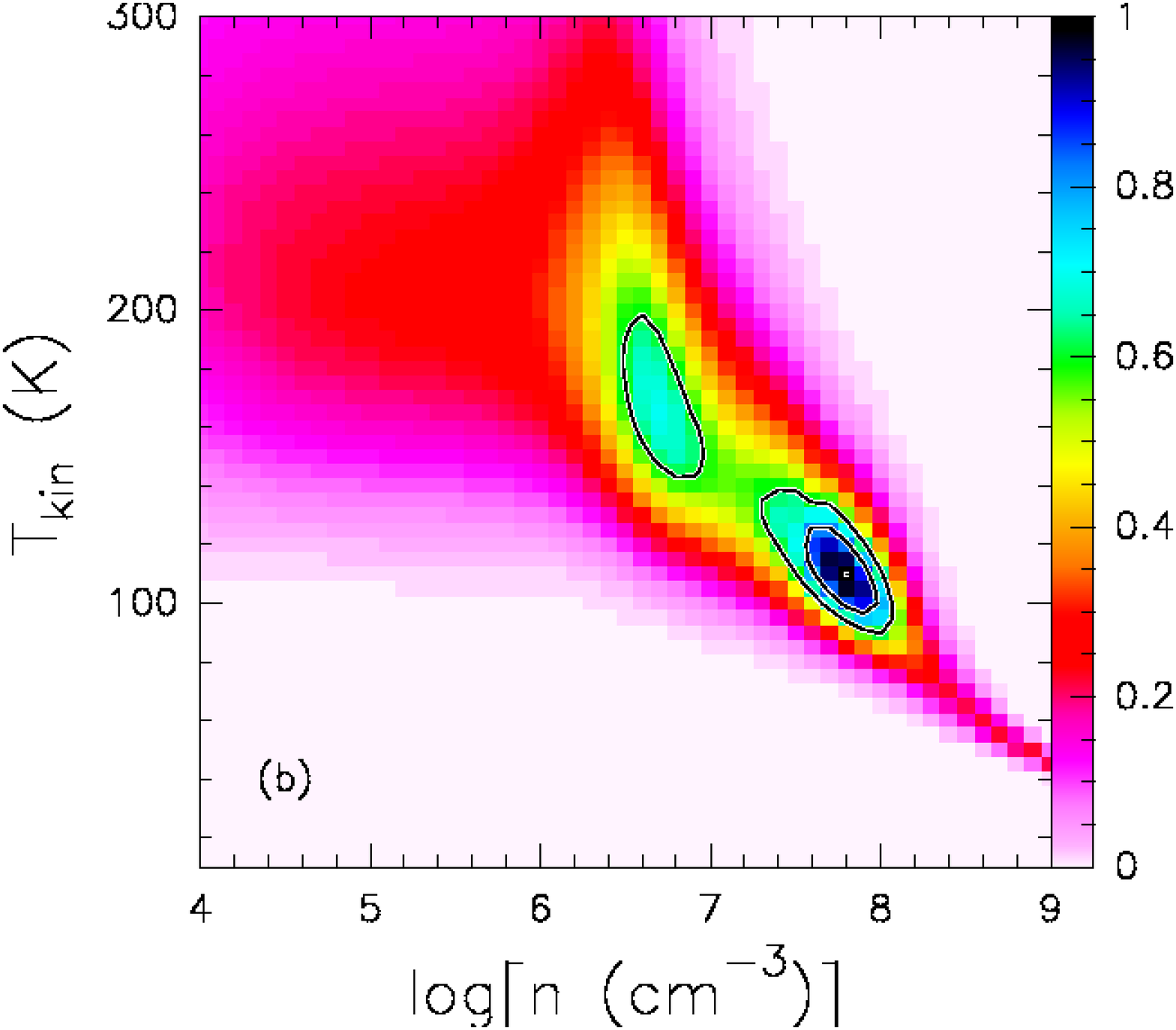}
\includegraphics[height = 0.4 \textwidth]{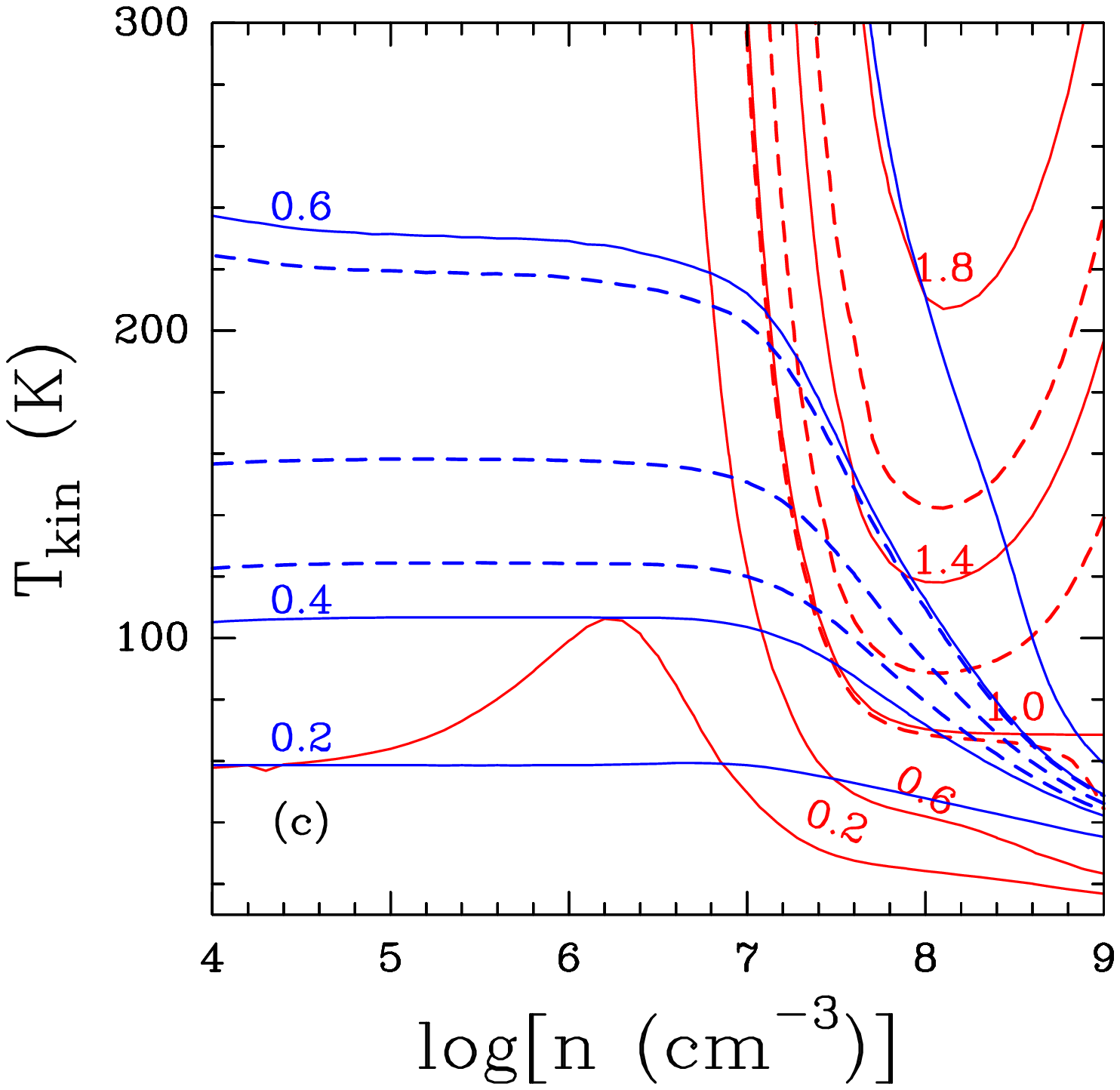}
\includegraphics[height = 0.4 \textwidth]{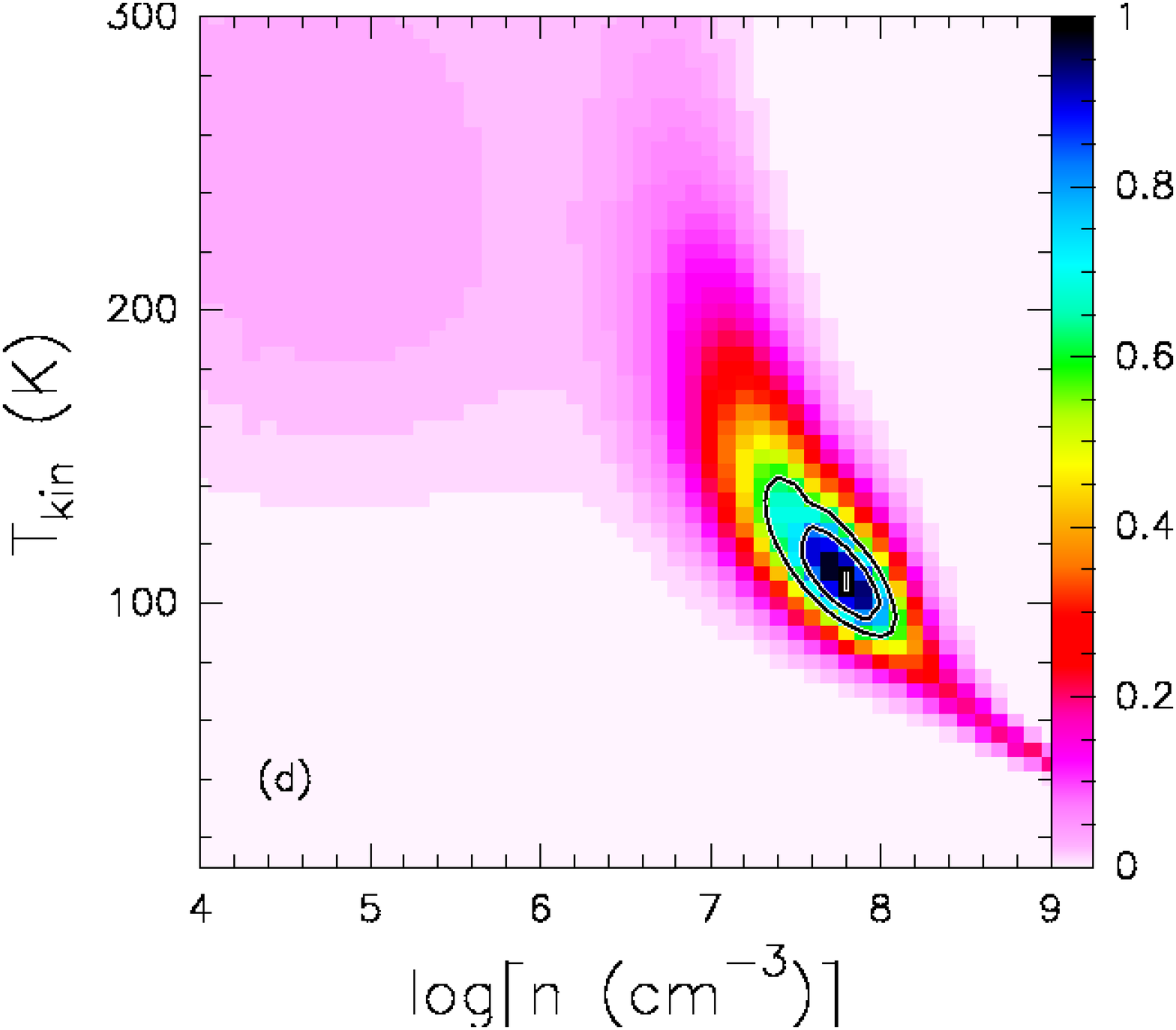}
\includegraphics[height = 0.4 \textwidth]{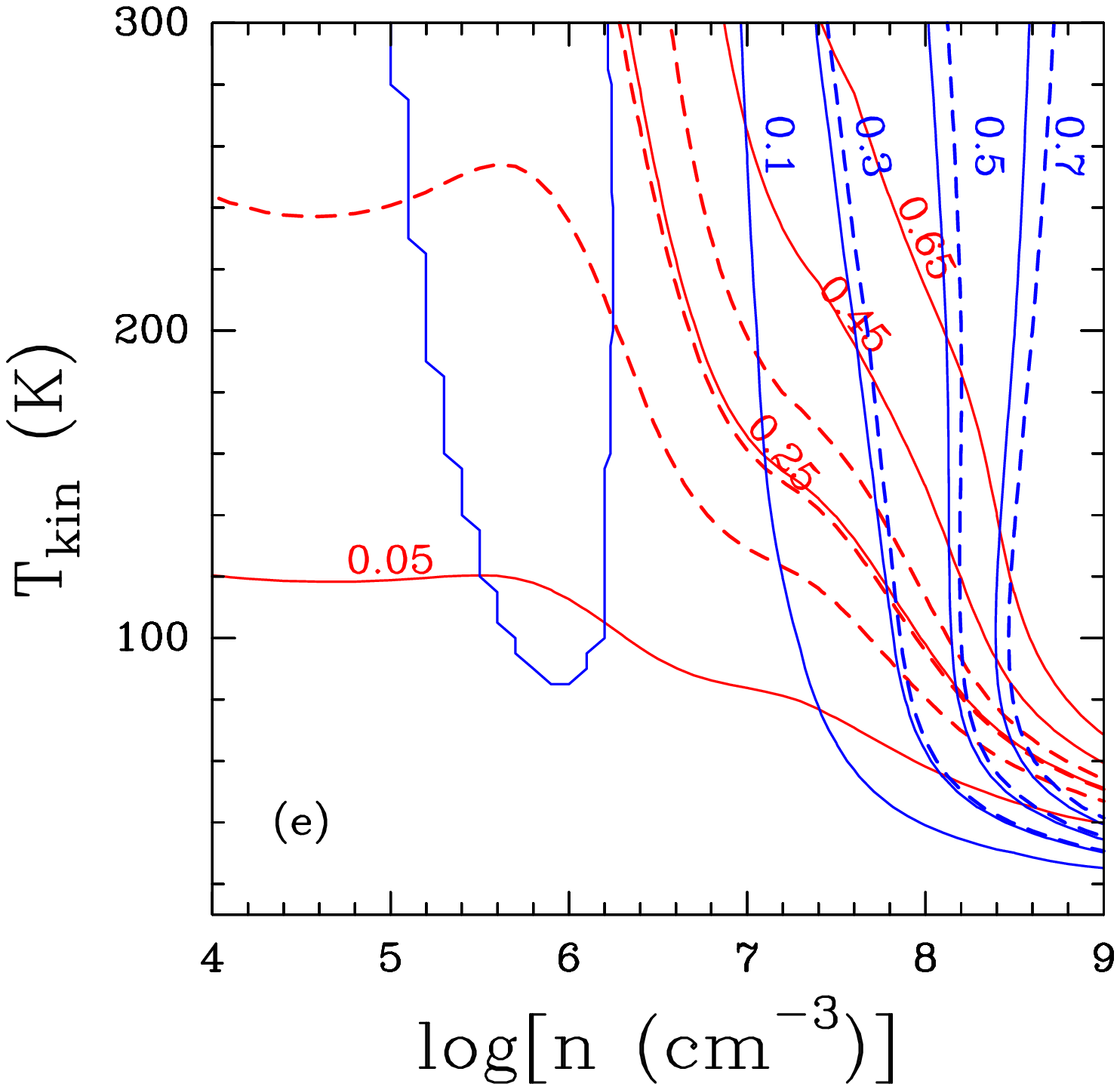}
\includegraphics[height = 0.4 \textwidth]{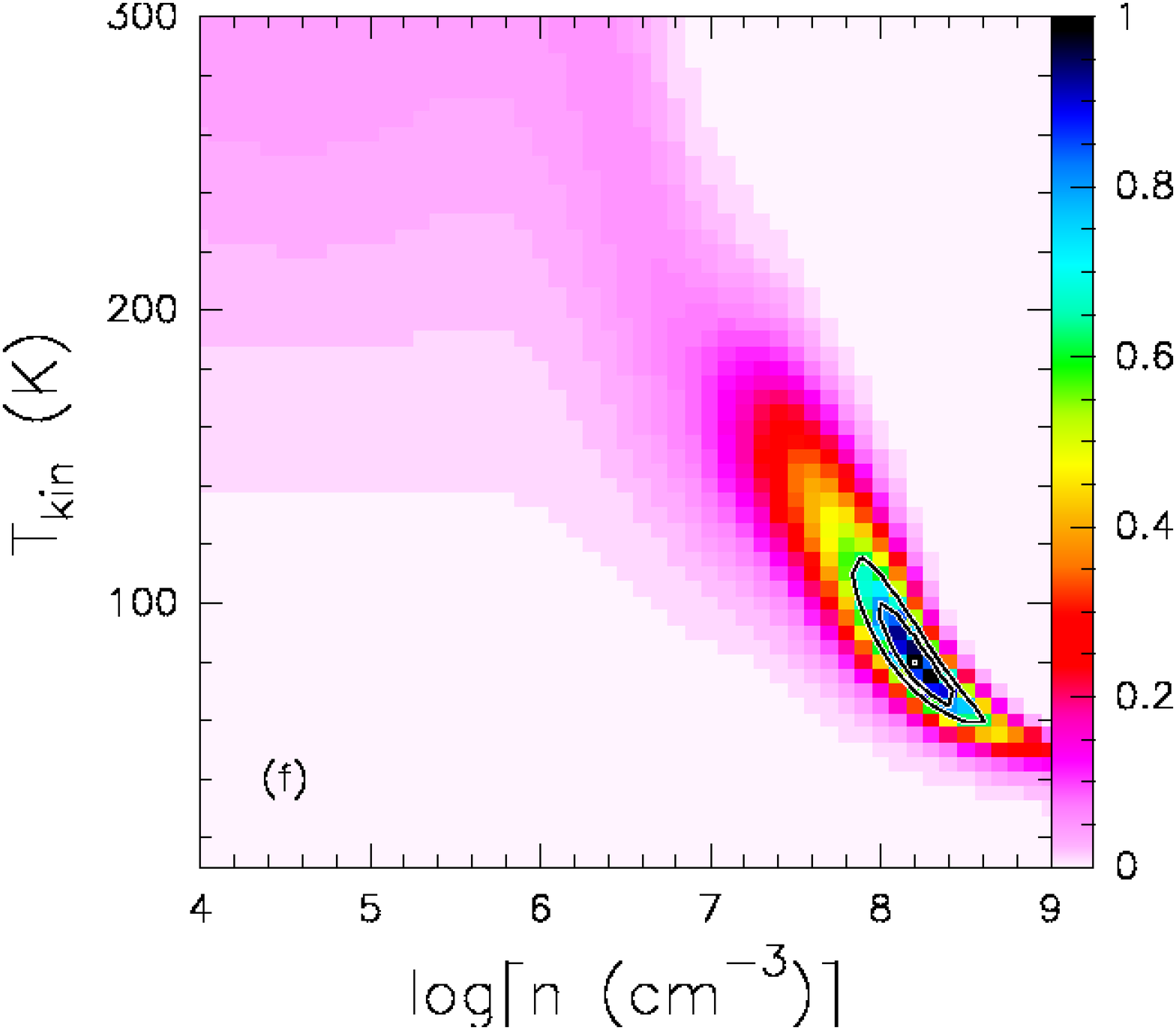}
\caption{{Radex radiative transfer calculations on the excitation of $^{15}$NH$_{3}$. (a) The modeled line ratios $\frac{^{15}{\rm NH}_{3} (2,2)}{^{15}{\rm NH}_{3} (1,1)}$ (red lines) and $\frac{^{15}{\rm NH}_{3} (5,5)}{^{15}{\rm NH}_{3} (4,4)}$ (blue lines) as a function of $n_{\rm H_{2}}$ and $T_{\rm kin}$. The solid red and blue lines represent the calculated $\frac{^{15}{\rm NH}_{3} (2,2)}{^{15}{\rm NH}_{3} (1,1)}$ and $\frac{^{15}{\rm NH}_{3} (5,5)}{^{15}{\rm NH}_{3} (4,4)}$ line ratios which are labeled. The dashed lines represent observed line ratios and their upper and lower limits. (b) The derived likelihood as a function of $n_{\rm H_{2}}$ and $T_{\rm kin}$. Only metastable para--$^{15}$NH$_{3}$ transitions are included in the modeling. The likelihood scale is indicated by the color bar. The contours represent the likelihoods 0.6, and 0.8. (c) Same as Fig.~\ref{Fig:radex1}a, but the solid red lines represent the line ratios of $\frac{^{15}{\rm NH}_{3} (3,2)}{^{15}{\rm NH}_{3} (2,1)}$. (d) Same as Fig.~\ref{Fig:radex1}b, but the non-metastable transitions are also included in the modeling. (e) Same as Fig.~\ref{Fig:radex1}a, but the solid red and blue lines represent the line ratios of $\frac{^{15}{\rm NH}_{3} (6,6)}{^{15}{\rm NH}_{3} (3,3)}$ and $\frac{^{15}{\rm NH}_{3} (4,3)}{^{15}{\rm NH}_{3} (3,3)}$. (f) Same as Fig.~\ref{Fig:radex1}d, but for ortho--$^{15}$NH$_{3}$ transitions. } \label{Fig:radex1}}
\end{figure*}

Assuming that the HC has a typical size of 10\arcsec\,and a line width of 7~\kms, we obtain a velocity gradient of 350~\kms\,pc$^{-1}$. Adopting a fixed para-$^{15}$NH$_{3}$ abundance of 8$\times 10^{-9}$ (see Table~\ref{Tab:rd}), we arrive at a para-$^{15}$NH$_{3}$ abundance per velocity gradient [$X$]/($dv/dr$) of 2.3$\times 10^{-11}$ pc~(\kms)$^{-1}$. The modeled kinetic temperatures ($T_{\rm kin}$) range from 10 to 500~K with a step size of 5~K. The H$_{2}$ number density log($\frac{n({\rm H_{2}})}{{\rm cm}^{-3}}$) varies from 3.0 to 9.0 with a step size of 0.1. We modeled the metastable para-$^{15}$NH$_{3}$ line ratios, and the results are shown in Fig.~\ref{Fig:radex1}a and \ref{Fig:radex1}b. We can see that there are two groups of solutions: (1) a warmer group with $T_{\rm kin}=145-190$~K and $n_{\rm H_{2}}= 10^{6.6}-10^{7.1}$~cm$^{-3}$ and (2) a colder group with $T_{\rm kin}=95-150$~K and $n_{\rm H_{2}}= 10^{7.2}-10^{8.0}$~cm$^{-3}$. The first group of solutions is roughly consistent with previous studies \citep{1989ARA&A..27...41G,2010ApJ...713.1192W,2011ApJ...739L..13G}. The second group of solutions indicates that the $^{15}$NH$_{3}$ emission region may have a colder kinetic temperature in an even denser region. This can be expected since the HC in Orion KL is thought to be externally heated \citep{2011ApJ...727..113B,2011A&A...529A..24Z,2012A&A...540A.119N}, which may result in an inner denser region with a lower kinetic temperature. On the other hand, we find that the line ratio $\frac{^{15}{\rm NH}_{3} (5,5)}{^{15}{\rm NH}_{3} (4,4)}$ is only sensitive to kinetic temperature in a region with a spatial density less than 10$^{7}$ cm$^{-3}$. When densities are higher than 10$^{7}$ cm$^{-3}$ at the abundance given above, these lines become optically thicker, and thus the ratios are no longer a good kinetic temperature tracer. Including the non-metastable para-$^{15}$NH$_{3}$ in the modeling (see Figs.~\ref{Fig:radex1}c and \ref{Fig:radex1}d), we obtain a solution with $T_{\rm kin}=95-145$~K and $n_{\rm H_{2}}= 10^{7.4}-10^{8.0}$~cm$^{-3}$, similar to the second group of solutions based on metastable transitions, which is supportive of a colder and denser region existing in the HC. 

We also modeled the line ratios from ortho--$^{15}$NH$_{3}$. Since the ortho--to--para ratio is found to be around unity (see Sect.~\ref{sec.abu}), the ortho--$^{15}$NH$_{3}$ abundance per velocity gradient is assumed to be the same as that for para--$^{15}$NH$_{3}$. With the line ratios $\frac{^{15}{\rm NH}_{3} (6,6)}{^{15}{\rm NH}_{3} (3,3)}$ and $\frac{^{15}{\rm NH}_{3} (4,3)}{^{15}{\rm NH}_{3} (3,3)}$, we use the same method to model these ratios from ortho--$^{15}$NH$_{3}$, and the results are shown in Figs.~\ref{Fig:radex1}e and \ref{Fig:radex1}f. We find that ortho--$^{15}$NH$_{3}$ emission is likely to come from a region with $T_{\rm kin}=60-105$~K and $n_{\rm H_{2}}= 10^{8.0}-10^{8.6}$~cm$^{-3}$, which is indicative of an even colder and denser region. This further suggests that the HC is externally heated. 

\begin{figure*}[!htbp]
\centering
\includegraphics[height = 0.4 \textwidth]{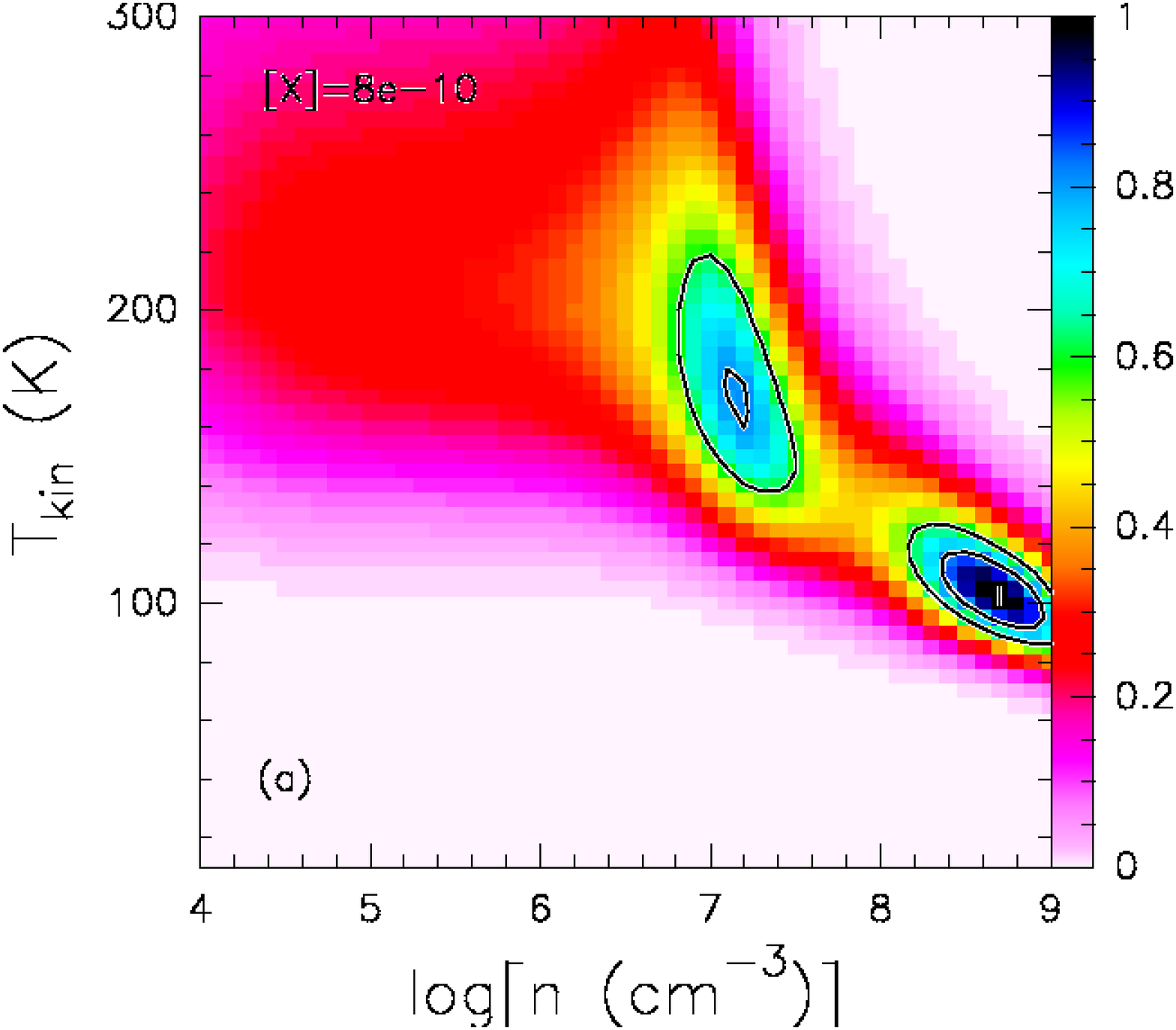}
\includegraphics[height = 0.4 \textwidth]{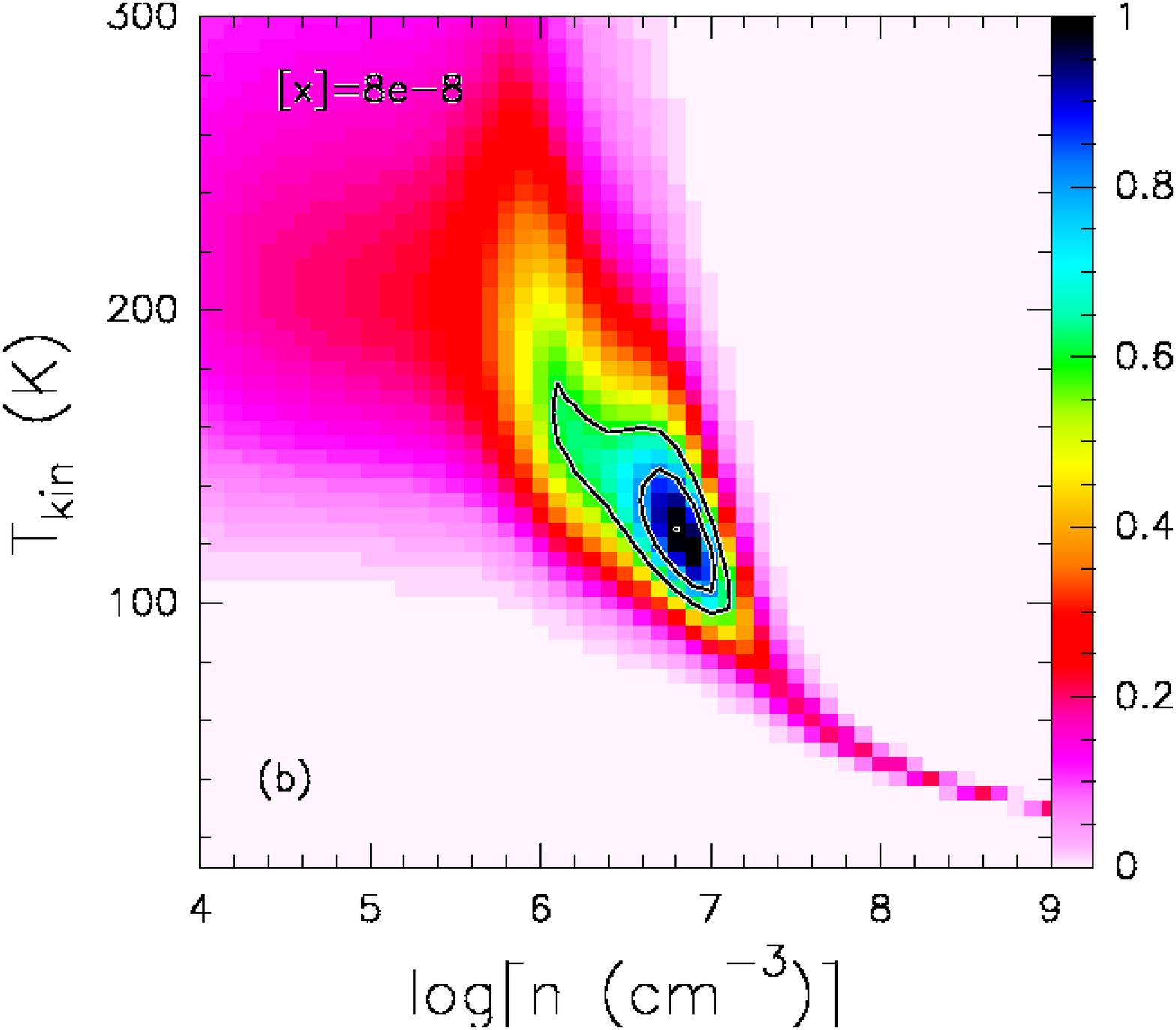}
\includegraphics[height = 0.4 \textwidth]{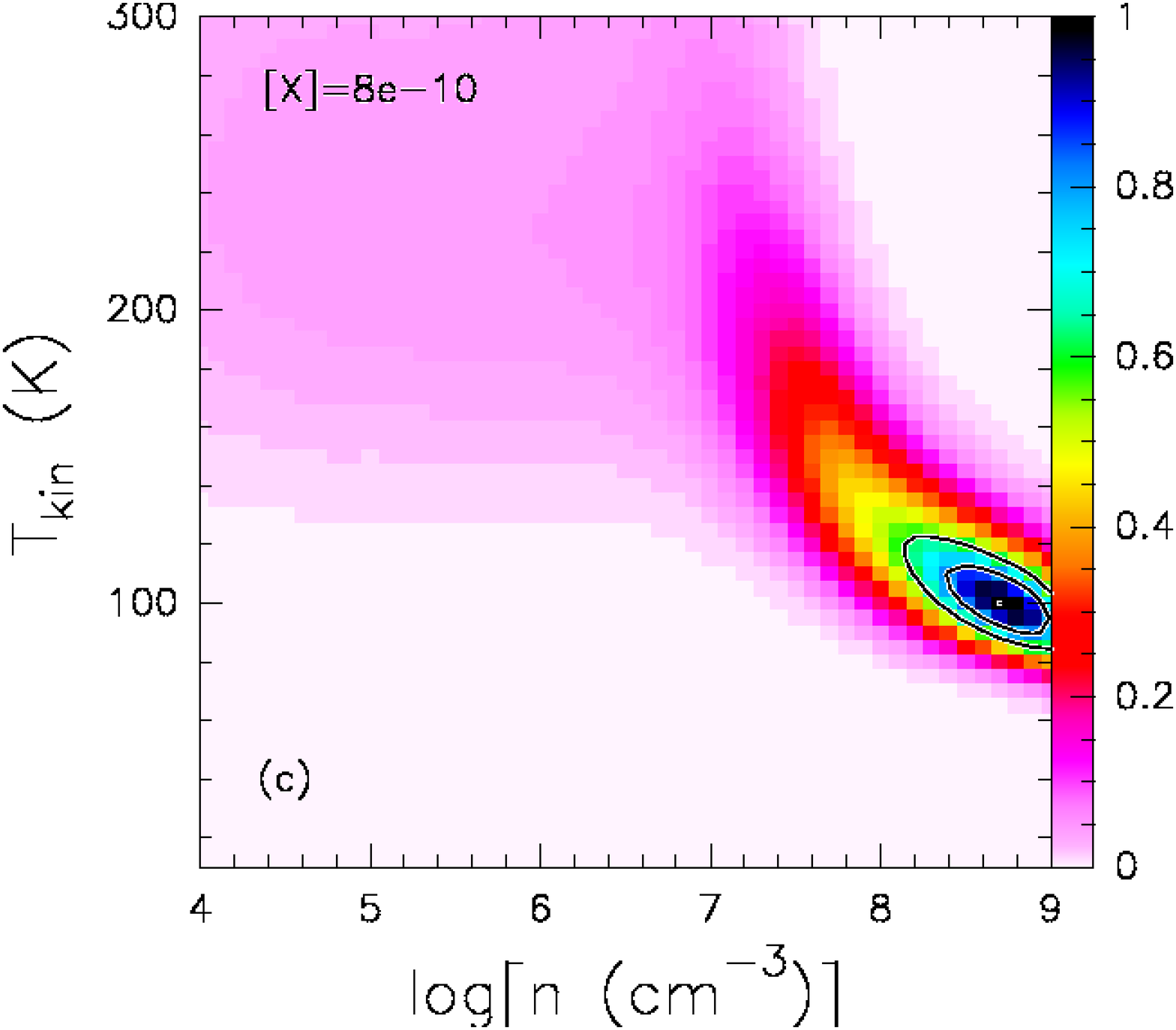}
\includegraphics[height = 0.4 \textwidth]{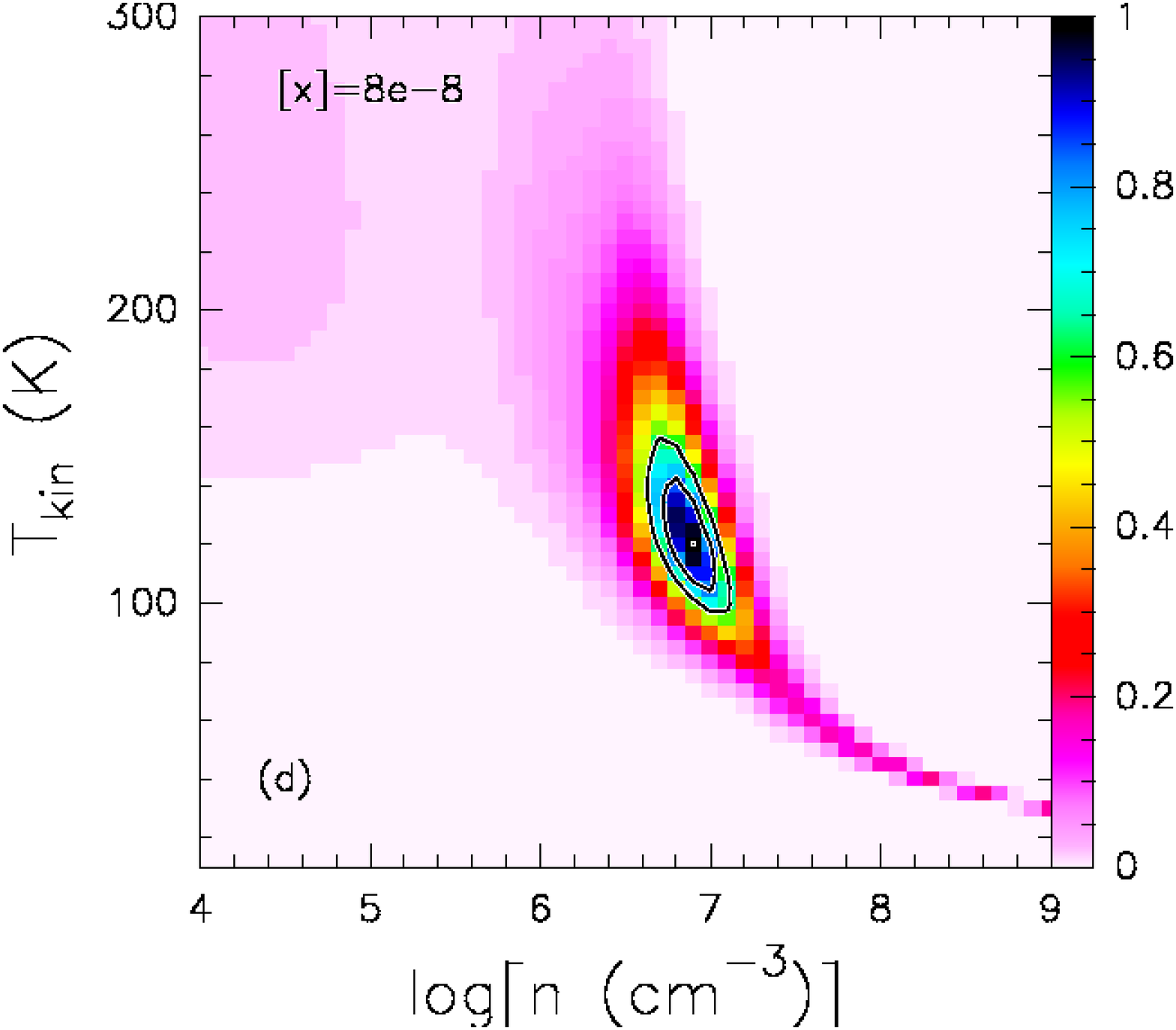}
\includegraphics[height = 0.4 \textwidth]{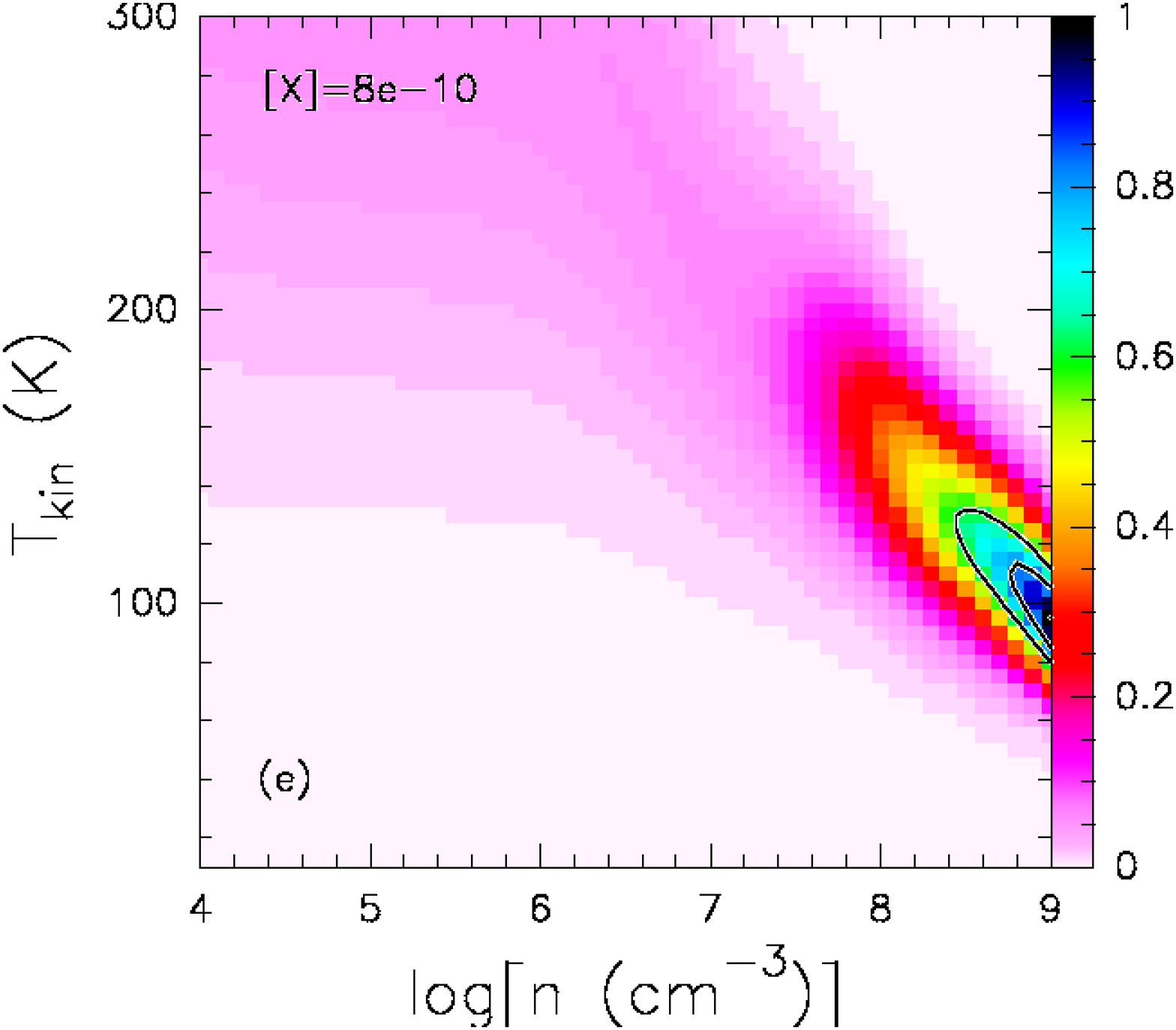}
\includegraphics[height = 0.4 \textwidth]{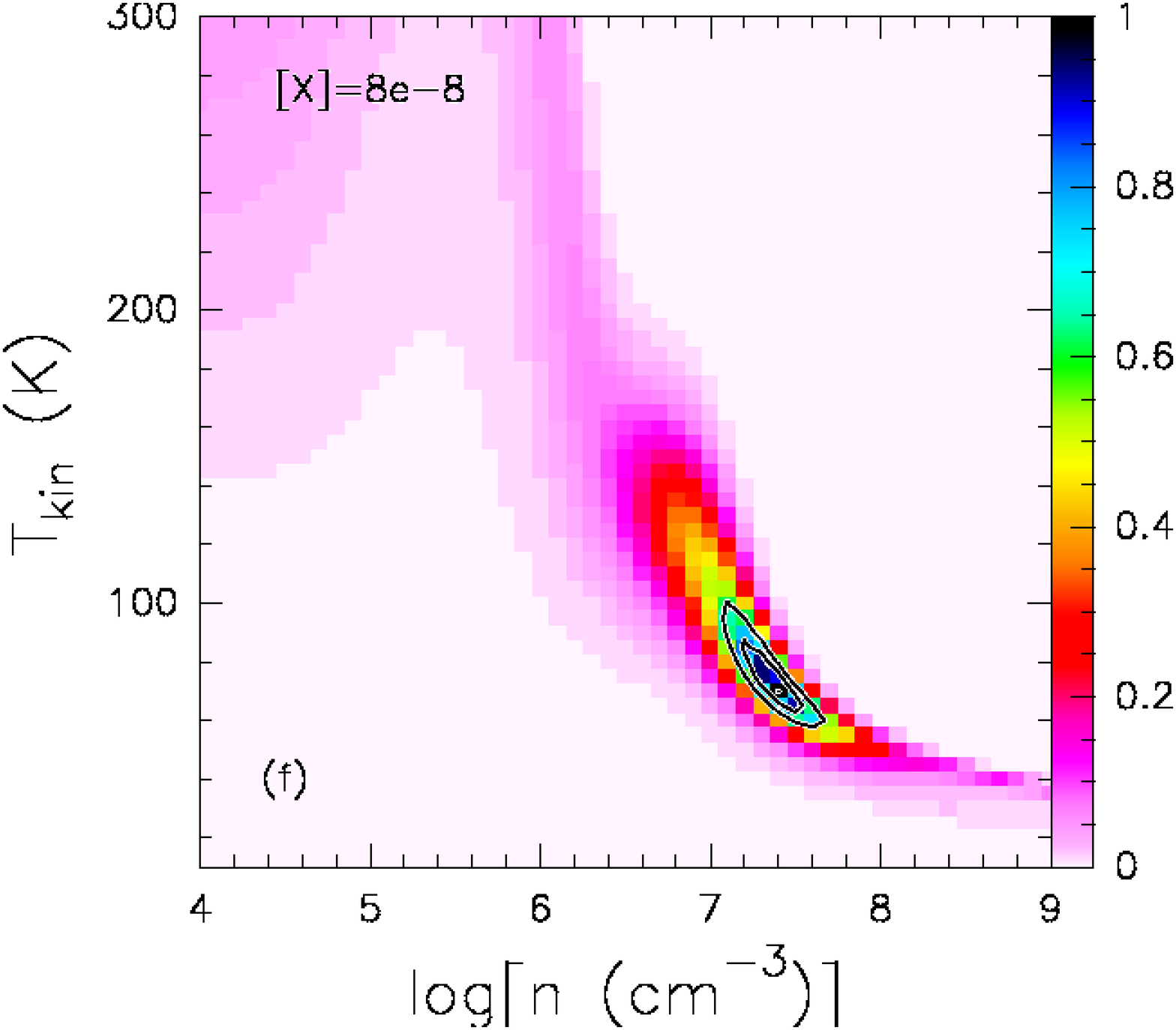}
\caption{{Same as Fig.~\ref{Fig:radex1}b, Fig.~\ref{Fig:radex1}d, and Fig.~\ref{Fig:radex1}f, but with different $^{15}$NH$_{3}$ abundances of $8\times10^{-10}$ and $8\times10^{-8}$ indicated in the upper left of each panel.} \label{Fig:radex2}}
\end{figure*}

The $^{15}$NH$_{3}$ abundance calculated with the LTE column density and an assumed H$_{2}$ column density may have a large uncertainty, which probably results in a large error in $T_{\rm kin}$ and $n({\rm H_{2}})$. Thus, we carried out another RADEX modelling with two different $^{15}$NH$_{3}$ abundances of 8$\times 10^{-10}$ and 8$\times 10^{-8}$. The results are shown in Fig.~\ref{Fig:radex2}. By comparison with Fig.~\ref{Fig:radex1}, we find that the derived kinetic temperature depends little on the assumed $^{15}$NH$_{3}$ abundances while the modeled spatial density depends strongly on the the assumed $^{15}$NH$_{3}$ abundances. Fig.~\ref{Fig:radex2} also shows that denser region has a lower kinetic temperature even when changing abundances. Therefore, external heating of the HC still holds for the $^{15}$NH$_{3}$ abundances ranging from 8$\times 10^{-10}$ to 8$\times 10^{-8}$. 

As mentioned above, an external radiation field is not taken into account in this modeling. In principle, infrared pumping can affect the population of ammonia via vibrationally excited lines \citep[e.g., see Fig.~1 of][]{1988A&A...205..235M}. When the infrared radiation field becomes very intense, the population can be significantly affected, and such effect may have led to a large number ($\sim$20) of NH$_{3}$ masers in W51-IRS2 \citep{2013A&A...549A..90H}. The lack of NH$_{3}$ masers and $T_{\rm kin}$ values below 200 K indicate that vibrational excitation of ammonia does not play a major role for the HC in Orion KL. However, our modeling without excitation by the infrared radiation field only gives upper limits for the spatial density.

\section{Summary and Conclusions}
We have carried out a 1.3 cm spectral line survey toward Orion KL with the Effelsberg-100 m telescope. We detect a total of 261 lines, yielding an average line density of about 23 spectral features per GHz above 3$\sigma$ (a typical value of 3$\sigma$ is 15 mJy). Among them, 164 lines are radio recombination lines (RRLs) from hydrogen, helium and carbon, 97 lines can be assigned to 13 different molecules including rare isotopologues. A total of 23 molecular transitions from species known to exist in Orion KL are detected for the first time in the interstellar medium. Non-metastable $^{15}$NH$_{3}$ transitions are detected in Orion KL for the first time. An analysis of line widths suggests that the bulk of the discrepancy between observed line widths of hydrogen and helium RRLs originates from their thermal contributions. The intensity ratios of RRL pairs with different $\Delta n$ from the same atom at neighbouring frequencies are consistent with the local thermodynamic equilibrium (LTE) ratios, suggesting that the LTE deviations are very small in this band. The origin of molecules detected by our survey is discussed according to observed lines and ALMA images. A narrow feature is found in SO$_{2}$ ($8_{1,7}-7_{2,6}$), but not in other SO$_{2}$ transitions, possibly suggesting the presence of a maser line. Column densities and fractional abundances relative to H$_{2}$ are estimated for 12 molecules with rotational diagrams. We find that the four classes of non-metastable NH$_{3}$ from $J=K+1$ to $J=K+4$ have different excitation conditions and metastable $^{15}$NH$_{3}$ has a higher excitation temperature than non-metastable $^{15}$NH$_{3}$. The elemental and isotopic abundance ratios are calculated: He$^{+}$/H$^{+}$ = (8.7$\pm0.7$)\% derived from the ratios between helium RRLs and hydrogen RRLs; $^{12}$C/$^{13}$C =63$\pm$17 from $^{12}$CH$_{3}$OH/$^{13}$CH$_{3}$OH; $^{14}$N/$^{15}$N =100$\pm$51 from $^{14}$NH$_{3}$/$^{15}$NH$_{3}$; D/H = (8.3$\pm$4.5)$\times$10$^{-3}$ from NH$_{2}$D/NH$_{3}$. The dispersion of the He/H ratios derived from H$\alpha$/He$\alpha$ pairs to H$\delta$/He$\delta$ pairs is very small, which is consistent with theoretical predictions that the departure coefficients $b_{\rm n}$ factors for hydrogen and helium are nearly identical. A non-LTE model of $^{15}$NH$_{3}$ neglecting external radiation fields and a likelihood analysis supports the view that the hot core is externally heated.

\section*{ACKNOWLEDGMENTS}\label{sec.ack}
We would like to thank an anonymous referee for a helpful report that led to improvements in the paper. We greatly thank Christian P. Endres for providing $^{15}$NH$_{3}$ spectroscopic information. We wish to thank Zhiyu Zhang for discussions of the $\chi^{2}$ analysis and appreciate the assistance of the Effelsberg-100m operators during the observations. Y. Gong acknowledges support by the MPG-CAS Joint Doctoral Promotion Programme (DPP), and NSFC Grants 11127903, 11233007 and 10973040. S. Thorwirth gratefully acknowledges funding by the Deutsche Forschungsgemeinschaft (DFG) through grant TH 1301/3-2. S. Spezzano wishes to thank the DFG SFB956 and the ``Fondazione Angelo della Riccia'' for funding. This paper makes use of the following ALMA data: ADS/JAO.ALMA\#2011.0.00009.SV. ALMA is a partnership of ESO (representing its member states), NSF (USA) and NINS (Japan), together with NRC (Canada) and NSC and ASIAA (Taiwan), in cooperation with the Republic of Chile. The Joint ALMA Observatory is operated by ESO, AUI/NRAO and NAOJ. This research has made use of NASA's Astrophysics Data System.

\bibliographystyle{aa}
\bibliography{references}

\clearpage

\renewcommand{\tabcolsep}{0.15 cm}
\normalsize \longtab{2}{
\begin{longtable}{clll}
	\caption{Lines detected in the survey of Orion-KL.}\label{Tab:orilines}\\
\hline
Rest frequency      & Transition  &   $E_{\rm u}$/k         &       \\
(MHz)	       &	     &   (K)                 & notes \\
\hline
\endfirsthead
\caption{continued.} \\
\hline
Rest frequency  & Transition   &  $E_{\rm u}$/k            &              \\
(MHz)	   &		  &  (K)	            &  notes        \\ 
\hline
\endhead
\hline  \\
\endfoot
\hline
\endlastfoot

17901.6     &    H120$\varepsilon$                                   &      &             \\ 
17978.5     &	 H127$\zeta$			                     &      &             \\ 
17992.6     &	 H71$\alpha$			                     &      &             \\ 
17999.9     &	 He71$\alpha$			                     &      &   1         \\ 
18001.5     &	 C71$\alpha$			                     &      &   1         \\ 
18017.3     &	 NH$_{3}$ (7,3)			                     & 750  &             \\ 
18019.0     &    H139$\theta$                                        &      &             \\
18045.9     &	 H89$\beta$			                     &      &             \\ 
18053.2     &	 He89$\beta$			                     &      &             \\ 
18123.4     &	 H133$\eta$			                     &      &             \\ 
18196.2     &	 HC$_{3}$N (2--1)		                     & 1    &             \\ 
18241.7     &	 H111$\delta$			                     &      &             \\ 
18327.5     &	 H101$\gamma$			                     &      &             \\ 
18335.0     &	 He101$\gamma$			                     &      &             \\ 
18347.5     &	 H119$\varepsilon$			             &      &             \\ 
18391.5     &	 NH$_{3}$ (6,1)			                     & 593  &             \\ 
18398.0     &	 CH$_{3}$CN (1$_{0}$--0$_{0}$)		             & 1    &  2          \\ 
18400.1     &	 H126$\zeta$			                     &      &  2, 3       \\
18402.7     &    H138$\theta$                                        &      &  3          \\
18499.3     &	 NH$_{3}$ (9,6)			                     & 1091 &             \\ 
18528.0     &	 H132$\eta$				             &      &             \\ 
18661.1     &	 H88$\beta$			                     &      &             \\ 
18668.7     &	 He88$\beta$			                     &      &             \\ 
18734.9     &	 H110$\delta$			                     &      &             \\ 
18742.5     &	 He110$\delta$			                     &      &             \\ 
18769.2     &	 H70$\alpha$			                     &      &             \\ 
18776.8     &	 He70$\alpha$			                     &      &  1          \\ 
18778.5     &	 C70$\alpha$			                     &      &  1          \\ 
18807.9     &	 NH$_{2}$D (3$_{1,3}$s--3$_{0,3}$a)                     & 94   &  4          \\ 
18808.4     &    H118$\varepsilon$                                   &      &  4          \\
18808.5     &	 NH$_{3}$ (8,5)			                     & 893  &  4          \\ 
18835.1     &	 H125$\zeta$			                     &      &             \\ 
18859.0     &    H142$\iota$                                         &      &             \\
18874.8     &	 H100$\gamma$			                     &      &             \\ 
18882.5     &	 He100$\gamma$			                     &      &             \\ 
18884.6     &	 NH$_{3}$ (6,2)			                     & 578  &             \\ 
18944.7     &	 H131$\eta$			                     &      &             \\ 
19218.4     &	 NH$_{3}$ (7,4)			                     & 713  &             \\ 
19246.1     &	 H109$\delta$			                     &      &             \\ 
19283.8     &    H124$\zeta$                                         &      &  5          \\
19284.8     &	 H117$\varepsilon$			             &      &  5          \\ 
19304.7     &	 H87$\beta$			                     &      &             \\ 
19312.6     &	 He87$\beta$			                     &      &             \\ 
19374.0     &	 H130$\eta$				             &      &             \\ 
19444.0     &	 H99$\gamma$			                     &      &             \\ 
19451.9     &	 He99$\gamma$			                     &      &             \\ 
19591.1     &	 H69$\alpha$			                     &      &             \\ 
19599.1     &	 He69$\alpha$			                     &      &  1          \\ 
19600.9     &	 C69$\alpha$			                     &      &  1          \\ 
19621.6     &    H135$\theta$                                        &      &             \\
19654.0     &    H140$\iota$                                         &      &             \\
19757.5     &	 NH$_{3}$ (6,3)			                     & 551  &             \\ 
19776.0     &	 H108$\delta$			                     &      &  6          \\ 
19777.4     &	 H116$\varepsilon$		                     &      &  6          \\ 
19816.4     &	 H129$\eta$			                     &      &             \\ 
19838.3     &	 NH$_{3}$ (5,1)			                     & 423  &             \\ 
19967.4     &	 CH$_{3}$OH ($2_{1}$--$3_{0}$ E)                       & 28   &             \\ 
19978.2     &	 H86$\beta$			                     &      &             \\ 
19986.3     &	 He86$\beta$			                     &      &             \\ 
20036.3     &	 H98$\gamma$			                     &      &             \\ 
20044.5     &	 He98$\gamma$			                     &      &             \\ 
20051.7     &	 H134$\theta$				             &      &             \\ 
20171.1     &	 CH$_{3}$OH (11$_{1}$ --10$_{2}$ A$^{+}$)               & 166  &             \\ 
20225.1     &	 H122$\zeta$			                     &      &             \\ 
20272.4     &	 H128$\eta$				             &      &             \\ 
20287.0     &	 H115$\varepsilon$			             &      &             \\ 
20325.5     &	 H107$\delta$			                     &      &             \\ 
20333.8     &    He107$\delta$                                       &      &   7         \\
20335.4     &	 SO$_{2}$ (12$_{3,9}$--13$_{2,12}$)	             & 94   &   7, N      \\ 
20346.8     &	 CH$_{3}$OH (17$_{6}$--18$_{5}$ E)	             & 534  &   N         \\ 
20371.5     &	 NH$_{3}$ (5,2)			                     & 407  &             \\ 
20460.0     &    HDO ($3_{2,1}-4_{1,4}$)                                & 226  &   8        \\ 
20461.8     &	 H68$\alpha$			                     &      &  8          \\ 
20470.1     &	 He68$\alpha$			                     &      &  1          \\ 
20472.0     &	 C68$\alpha$			                     &      &  1          \\ 
20494.5     &	 H133$\theta$			                     &      &  9          \\ 
20494.3     &	 H138$\iota$			                     &      &  9          \\ 
20653.0     &	 H97$\gamma$			                     &      &             \\ 
20683.3     &	 H85$\beta$			                     &      &             \\ 
20691.8     &	 He85$\beta$			                     &      &             \\ 
20718.7     &    H121$\zeta$                                         &      & 10          \\
20719.2     &	 NH$_{3}$ (8,6)			                     & 835  & 10          \\ 
20735.4     &	 NH$_{3}$ (9,7)			                     & 1023 &             \\ 
20804.9     &	 NH$_{3}$ (7,5)			                     & 666  &             \\ 
20814.3     &	 H114$\varepsilon$			             &      &             \\ 
20895.6     &	 H106$\delta$			                     &      &             \\ 
20908.9     &	 CH$_{3}$OH (16$_{-4}$--15$_{-5}$ E)                    & 403  &             \\ 
20932.5     &	 H137$\iota$			   	             &      &             \\ 
20950.5     &	 H132$\theta$				             &      &             \\ 
20970.7     &	 CH$_{3}$OH v$_{t}$=1 (10$_{1}$--11$_{2}$ A$^{+}$)       & 452  &             \\ 
20994.7     &	 NH$_{3}$ (6,4)			                     & 514  &             \\ 
21070.7     &	 NH$_{3}$ (11,9)			             & 1450 &             \\ 
21134.3     &	 NH$_{3}$ (4,1)			                     & 280  &             \\ 
21227.2     &    H126$\eta$                                          &      &  11         \\
21228.6     &	 H120$\zeta$			                     &      &  11         \\ 
21285.3     &	 NH$_{3}$ (5,3)			                     & 380  &             \\ 
21295.2     &	 H96$\gamma$			                     &      &             \\ 
21301.3     &    HC$_{5}$N ($8-7$)                                    & 2    &             \\
21303.9     &	 He96$\gamma$			                     &      &             \\ 
21359.9     &	 H113$\varepsilon$			             &      &             \\ 
21383.2     &    H136$\iota$                                         &      &  12         \\
21384.8     &	 H67$\alpha$			                     &      &  12         \\ 
21393.5     &	 He67$\alpha$			                     &      &  1          \\ 
21395.5     &	 C67$\alpha$			                     &      &  1          \\ 
21420.0     &    H131$\theta$                                        &      &  13         \\
21422.1     &	 H84$\beta$			                     &      &  13         \\ 
21430.8     &	 He84$\beta$			                     &      &             \\ 
21487.2     &	 H105$\delta$			                     &      &             \\ 
21550.3     &	 CH$_{3}$OH v$_{t}$=1 (12$_{2}$--11$_{1}$ A$^{-}$)       & 479  &             \\ 
21637.9     &	 $^{15}$NH$_{3}$ (4,3)		                     & 240  &             \\ 
21703.4     &	 NH$_{3}$ (4,2)			                     & 264  &             \\ 
21784.0     &	 $^{15}$NH$_{3}$ (3,2)			             & 152  &  N          \\ 
21903.8     &	 H130$\theta$				             &      &             \\ 
21924.8     &	 H112$\varepsilon$			             &      &             \\ 
21964.3     &	 H95$\gamma$			                     &      &             \\ 
21973.3     &	 He95$\gamma$			                     &      &             \\ 
21981.6     &	 HNCO (1$_{0,1}$--0$_{0,0}$)		             & 1    &             \\ 
22044.2     &	 $^{15}$NH$_{3}$ (2,1)			             & 82   &  N          \\ 
22101.4     &	 H104$\delta$			                     &      &             \\ 
22196.5     &	 H83$\beta$			                     &      &             \\ 
22205.5     &	 He83$\beta$			                     &      &             \\ 
22235.1     &	 H$_{2}$O (6$_{1,6}$--5$_{2,3}$)		             & 643  &  14         \\ 
22299.6     &	 H118$\zeta$			                     &      &             \\ 
22324.4     &	 H134$\iota$			                     &      &             \\ 
22364.2     &	 H66$\alpha$			                     &      &             \\ 
22373.3     &	 He66$\alpha$			                     &      &  1          \\ 
22375.3     &	 C66$\alpha$			                     &      &  1          \\ 
22402.1     &	 H129$\theta$				             &      &             \\ 
22509.8     &	 H111$\varepsilon$			             &      &             \\ 
22624.9     &	 $^{15}$NH$_{3}$ (1,1)			             & 24   &             \\ 
22649.8     &	 $^{15}$NH$_{3}$ (2,2)			             & 66   &             \\ 
22653.0     &	 NH$_{3}$ (5,4)			                     & 343  &             \\ 
22661.8     &	 H94$\gamma$			                     &      &             \\ 
22671.0     &	 He94$\gamma$			                     &      &             \\ 
22688.3     &	 NH$_{3}$ (4,3)			                     & 238  &             \\ 
22732.5     &	 NH$_{3}$ (6,5)			                     & 467  &             \\ 
22739.2     &	 H103$\delta$			                     &      &             \\ 
22748.5     &	 He103$\delta$			                     &      &             \\ 
22775.1     &	 H123$\eta$				             &      &             \\ 
22789.4     &	 $^{15}$NH$_{3}$ (3,3)			             & 126  &             \\ 
22815.7     &	 H133$\iota$			                     &      &             \\ 
22827.7     &	 CH$_{3}$OCHO (2$_{1,2}$--1$_{1,1}$ E)                  & 2    & 15          \\ 
22828.1     &	 CH$_{3}$OCHO (2$_{1,2}$--1$_{1,1}$ A)                  & 2    & 15          \\ 
22834.2     &	 NH$_{3}$ (3,2)			                     & 150  &             \\ 
22862.3     &	 H117$\zeta$			                     &      &             \\ 
22915.8     &	 H128$\theta$				             &      &             \\ 
22925.0     &	 NH$_{3}$ (7,6)			                     & 608  &             \\ 
22965.6     &	 H$_{2}$CO (9$_{2,7}$--9$_{2,8}$)                       & 205  &  N          \\ 
23008.6     &	 H82$\beta$			                     &      &             \\ 
23018.0     &	 He82$\beta$			                     &      &             \\ 
23046.0     &	 $^{15}$NH$_{3}$ (4,4)			             & 204  &             \\ 
23098.8     &	 NH$_{3}$ (2,1)			                     & 80   &             \\ 
23115.9     &	 H110$\varepsilon$			             &      &             \\ 
23121.0     &	 CH$_{3}$OH (9$_{2}$--10$_{1}$ A$^{+}$)	             & 142  &             \\ 
23145.4     &	 $^{13}$CH$_{3}$OH (4$_{0}$--3$_{1}$ E)              & 36   &  N          \\ 
23232.3     &	 NH$_{3}$ (8,7)			                     & 767  &             \\ 
23321.6     &	 H132$\iota$			                     &      & 16          \\ 
23324.4     &	 H122$\eta$			                     &      & 16          \\ 
23347.0     &	 CH$_{3}$OH (7$_{1}$--7$_{1}$ A$^{+}$)                  & 80   & N           \\ 
23389.1     &	 H93$\gamma$			                     &      & 17          \\ 
23390.0     &    CH$_{3}$OCH$_{3}$ (12$_{3,10}$--11$_{4,7}$ AA)          & 83.7 &  17, N      \\
23393.1     &	 CH$_{3}$OCH$_{3}$ (12$_{3,10}$--11$_{4,7}$ EE)          & 83.7 &  N          \\ 
23398.6     &	 He93$\gamma$			                     &      &  18         \\
23401.8     &    H102$\delta$                                        &      &  18, 19     \\ 
23404.3     &	 H65$\alpha$			                     &      &  19         \\ 
23411.4     &    He102$\delta$                                       &      &  20         \\
23413.8     &	 He65$\alpha$			                     &      &  1, 20, 21  \\ 
23414.2     &	 SO$_{2}$ ($5_{2,4}$--$6_{1,5}$)                        & 24   & N, 21       \\ 
23416.0     &    C65$\alpha$                                         &      &  1          \\
23422.0     &	 $^{15}$NH$_{3}$ (5,5)			             & 300  &             \\ 
23444.0     &    H116$\zeta$                                         &      &  22         \\
23444.8     &	 CH$_{3}$OH (10$_{1}$--9$_{2}$ A$^{-}$)                 & 143  &  22         \\ 
23445.3     &    H127$\theta$                                        &      &  22         \\
23518.2     &    H136$\kappa$                                        &      &             \\
23657.6     &	 NH$_{3}$ (9,8)			                     & 943  &             \\ 
23694.5     &	 NH$_{3}$ (1,1)			                     & 23   &             \\ 
23722.6     &	 NH$_{3}$ (2,2)			                     & 64   &             \\ 
23743.8     &	 H109$\varepsilon$			             &      &             \\ 
23842.5     &	 H131$\iota$			                     &      &             \\ 
23860.9     &	 H81$\beta$			                     &      &             \\ 
23870.1     &	 NH$_{3}$ (3,3)			                     & 124  & 23          \\ 
23870.6     &	 He81$\beta$			                     &      & 23          \\ 
23891.6     &	 H121$\eta$				             &      &             \\ 
23922.3     &	 $^{15}$NH$_{3}$ (6,6)			             & 414  &             \\ 
23963.9     &	 HC$_{5}$N (9--8)			             & 6    &             \\ 
23991.2     &	 H126$\theta$				             &      &             \\ 
24039.6     &	 SO$_{2}$ (21$_{5,17}$--22$_{4,18}$)                     & 274  &             \\ 
24045.6     &	 H115$\zeta$			                     &      &             \\ 
24083.5     &	 SO$_{2}$ (8$_{2,6}$--9$_{1,9}$)	                     & 43   &             \\ 
24090.4     &	 H101$\delta$			                     &      &             \\ 
24100.2     &	 He101$\delta$			                     &      &             \\ 
24139.4     &	 NH$_{3}$ (4,4)			                     & 201  &             \\ 
24147.9     &	 H92$\gamma$			                     &      &             \\ 
24157.7     &	 He92$\gamma$			                     &      &             \\ 
24205.4     &	 NH$_{3}$ (10,9)			             & 1138 &             \\ 
24296.5     &	 CH$_{3}$OCHO (2$_{0,2}$--1$_{0,1}$ E)                  & 2    &             \\ 
24298.5     &	 CH$_{3}$OCHO (2$_{0,2}$--1$_{0,1}$ A)                  & 2    &             \\  
24325.9     &	 OCS (2--1)				             & 2    &             \\
24379.1     &	 H130$\iota$			                     &      &             \\ 
24394.8     &	 H108$\varepsilon$			             &      &             \\ 
24477.3     &	 H120$\eta$				             &      &             \\ 
24509.9     &	 H64$\alpha$			                     &      &             \\ 
24519.9     &	 He64$\alpha$			                     &      &  1          \\ 
24522.1     &	 C64$\alpha$			                     &      &  1          \\ 
24532.9     &	 NH$_{3}$ (5,5)			                     & 295  &             \\ 
24625.2     &	 CH$_{3}$OCHO (12$_{3,9}$--12$_{3,10}$ E)	             & 53   &  N          \\ 
24649.4     &    CH$_{3}$OCHO (12$_{3,9}$--12$_{3,10}$ A)	             & 53   &  N          \\ 
24668.0     &	 H114$\zeta$			                     &      &             \\ 
24755.7     &	 H80$\beta$			                     &      &             \\ 
24765.8     &	 He80$\beta$			                     &      &             \\ 
24806.3     &	 H100$\delta$			                     &      &             \\ 
24816.4     &	 He100$\delta$			                     &      &             \\ 
24882.0     &	 NH$_{3}$ (11,10)			             & 1350 &  N          \\ 
24928.7     &	 CH$_{3}$OH (3$_{2}$--3$_{1}$ E)                       & 36   & 24          \\ 
24931.9     &    H129$\iota$                                        &       & 24          \\
24933.5     &	 CH$_{3}$OH (4$_{2}$--4$_{1}$ E)                       & 45   &             \\ 
24934.4     &	 CH$_{3}$OH (2$_{2}$--2$_{1}$ E)                       & 29   &             \\ 
24939.8     &	 H91$\gamma$			                     &      &             \\ 
24950.0     &	 He91$\gamma$			                     &      &             \\ 
24959.1     &	 CH$_{3}$OH (5$_{2}$--5$_{1}$ E)                       & 57   &             \\ 
25018.1     &	 CH$_{3}$OH (6$_{2}$--6$_{1}$ E)                       & 71   &             \\ 
25023.8     &	 NH$_{2}$D (4$_{1,4}$a--4$_{0,4}$s)                     & 152  &             \\ 
25056.0     &	 NH$_{3}$ (6,6)			                     & 408  &             \\ 
25069.7     &	 H107$\varepsilon$		                     &      &             \\ 
25079.9     &    He107$\varepsilon$		                     &      &  25         \\ 
25082.3     &	 H119$\eta$			                     &      &  25         \\ 
25124.9     &	 CH$_{3}$OH (7$_{2}$--7$_{1}$ E)                       & 87   &             \\ 
25134.9     &	 H124$\theta$				             &      &             \\ 
25294.4     &	 CH$_{3}$OH (8$_{2}$--8$_{1}$ E)                       & 106  &             \\ 
25312.0     &	 H113$\zeta$			                     &      &             \\ 
25322.8     &	 CH$_{3}$OH v$_{t}$=1 (9$_{9,1}$--9$_{8,2}$ A)           & 788  &   26, 27, N \\ 
            &	 CH$_{3}$OH v$_{t}$=1 (9$_{9,0}$--9$_{8,1}$ A)           & 788  &   26, 27, N \\ 
25323.5     &	 $^{15}$NH$_{3}$ (8,8)			             & 688  &   27, N     \\ 
25392.8     &	 SO$_{2}$ (8$_{1,7}$--7$_{2,6}$)			     & 37   &   N         \\ 
25497.5     &    CH$_{3}$OCHO (20$_{5,15}$--20$_{5,16}$ E)	             & 143  &   N         \\ 
25501.5     &	 H128$\iota$			                     &      &	          \\ 
25530.7     &	 CH$_{3}$OCHO (20$_{5,15}$--20$_{5,16}$ A)	             & 143  &   N         \\ 
25541.4     &	 CH$_{3}$OH (9$_{2,7}$--9$_{1,8}$ E)	             & 127  &             \\ 
25550.8     &	 H99$\delta$			                     &      &             \\ 
25561.2     &	 He99$\delta$			                     &      &             \\ 
25609.7     &	 H136$\lambda$			                     &      &             \\ 
25642.8     &	 H132$\kappa$			                     &      &             \\ 
25686.3     &	 H63$\alpha$			                     &      &             \\ 
25695.9     &	 H79$\beta$			                     &      &  28         \\ 
25696.7     &	 He63$\alpha$			                     &      &  1, 28      \\ 
25699.1     &    C63$\alpha$                                         &      &  1          \\
25706.4     &	 He79$\beta$			                     &      &  29         \\ 
25707.4     &    H118$\eta$                                          &      &  29         \\
25715.1     &	 NH$_{3}$ (7,7)			                     & 539  &             \\ 
25730.2     &	 CH$_{3}$OCHO (16$_{4,12}$--16$_{4,13}$ E)	             & 93   & N           \\ 
25734.1     &	 H123$\theta$				             &      &             \\ 
25759.8     &	 CH$_{3}$OCHO (16$_{4,12}$--16$_{4,13}$ A)	             & 93   & N           \\ 
25766.8     &	 H90$\gamma$			                     &      & 30          \\ 
25769.8     &	 H106$\varepsilon$			             &      & 30          \\ 
25777.3     &	 He90$\gamma$			                     &      &             \\ 
25787.1     &	 CH$_{3}$OH (26$_{2}$--26$_{1}$ E)	             & 836  & N           \\ 
25878.3     &	 CH$_{3}$OH (10$_{2}$--10$_{1}$ E)	             & 150  &             \\ 
25978.7     &	 H112$\zeta$			                     &      &             \\ 
26044.8     &	 CH$_{3}$OCHO (2$_{1,1}$--1$_{1,0}$ E)	             & 3    &             \\ 
26048.5     &	 CH$_{3}$OCHO (2$_{1,1}$--1$_{1,0}$ A)	             & 3    &             \\ 
26120.6     &	 CH$_{3}$OH v$_{t}$=1 (10$_{1}$-11$_{2}$ A$^{+}$)         & 453  & N           \\ 
26124.6     &	 CH$_{3}$CH$_{2}$CN ($3_{1,3}-2_{1,2}$)                  &  4    & N            \\ 
\end{longtable}   
\tablefoot{Notes-- (1) C$\alpha$ lines all blended with nearby He$\alpha$ lines in our band. (2) CH$_{3}$CN ($1_{0}-0_{0}$) at 18398.0 MHz is blended with H126$\zeta$ at 18400.1 MHz. (3) H126$\zeta$ at 18400.1 MHz is blended with H138$\theta$ at 18402.7 MHz. (4) NH$_{2}$D (3$_{1,3}$--3$_{0,3}$) at 18807.9 MHz is blended with NH$_{3}$ (8,5) at 18808.5 MHz and H118$\varepsilon$ at 18808.4 MHz. (5) H124$\zeta$ at 19283.8 MHz is blended with H117$\varepsilon$ at 19284.8 MHz. (6) H108$\delta$ at 19776.0 MHz is blended with H116$\varepsilon$ at 19777.4 MHz. (7) He107$\delta$ at 20333.8 MHz is blended with SO$_{2}$ ($12_{3,9}-13_{2,12}$) at 20335.4 MHz. (8) HDO ($3_{2,1}-4_{1,4}$) at 20460.0 MHz is blended with H68$\alpha$ at 20461.8 MHz. (9) H133$\theta$ at 20494.5 MHz is blended with H138$\iota$ at 20494.3 MHz. (10) H121$\zeta$ at 20718.7 MHz is blended with NH$_{3}$ (8,6) at 20719.2 MHz. (11) H126$\eta$ at 21227.2 MHz is blended with H120$\zeta$ at 21228.6 MHz. (12) H136$\iota$ at 21383.2 MHz is blended with H67$\alpha$ at 21384.8 MHz. (13) H131$\theta$ at 21420.0 MHz is blended with H84$\beta$ at 21422.1 MHz. (14) H$_{2}$O ($6_{1,6}-5_{2,3}$) at 22235.1 MHz is blended with NH$_{3}$ (3,1) at 22234.6 MHz. (15) CH$_{3}$OCHO ($2_{1,2}-1_{1,1}$ E) at 22827.7 MHz is blended with CH$_{3}$OCHO ($2_{1,2}-1_{1,1}$ A) at 22828.1 MHz. (16) H132$\iota$ at 23321.6 MHz is blended with H122$\eta$ at 23324.4 MHz. (17) H93$\gamma$ at 23389.1 MHz is blended with CH$_{3}$OCH$_{3}$ (12$_{3,10}$--11$_{4,7}$ AA) at 23390.0 MHz. (18) He93$\gamma$ at 23398.6 MHz is blended with H102$\delta$ at 23401.8 MHz. (19) H102$\delta$ at 23401.8 MHz is blended with H65$\alpha$ at 23404.3 MHz. (20) He102$\delta$ at 23411.4 MHz is blended with He65$\delta$ at 23413.8 MHz. (21) He65$\alpha$ at 23413.8 MHz is blended with SO$_{2}$ ($5_{2,4}$--$6_{1,5}$) at 23414.2 MHz. (22) H116$\zeta$ at 23444.0 MHz is blended with CH$_{3}$OH (10$_{1,9}$--9$_{2,8}$ A) at 23444.8 MHz and H127$\theta$ at 23445.3 MHz. (23) NH$_{3}$ (3,3) at 23870.1 MHz is blended with He81$\beta$ at 23870.6 MHz. (24) CH$_{3}$OH ($3_{2,1}-3_{1,2}$ E) at 24928.7 MHz is blended with H129$\iota$ at 24931.9 MHz. (25) He107$\varepsilon$ at 25079.9 MHz is blended with H119$\eta$ at 25082.3 MHz. (26) CH$_{3}$OH v$_{t}$=1 (9$_{9,1}$--9$_{8,2}$ A) and CH$_{3}$OH v$_{t}$=1 (9$_{9,0}$--9$_{8,1}$ A) at 25322.8 MHz are nearly degeneracy. (27) CH$_{3}$OH v$_{t}$=1 (9$_{9,1}$--9$_{8,2}$ A) at 25322.8 MHz is blended with $^{15}$NH$_{3}$ (8,8) at 25323.5 MHz. (28) H79$\beta$ at 25695.9 MHz is blended with He63$\alpha$ at 25696.7 MHz. (29) He79$\beta$ at 25706.4 MHz is blended with H118$\eta$ at 25707.4 MHz. (20) H90$\gamma$ at 25766.8 MHz is blended with H106$\varepsilon$ at 25769.8 MHz. (N) Newly detected transitons in the interstellar medium from known astronomical molecules.}
                                                  
} 

\renewcommand{\tabcolsep}{0.1 cm}
\scriptsize \longtab{3}{
\footnotesize
\begin{longtable}{cccccccccccc}
	\caption{Column densities and rotational temperatures of the detected molecules compared with previous studies.}\label{Tab:rd}\\
\hline
          & \multicolumn{5}{c}{This work}                                     & & \multicolumn{4}{c}{Other studies} &              \\
\cline{2-6} \cline{8-12}
          & $T_{\rm rot}$        &   $N$                & $\chi$($N/N_{\rm H_{2}}$) & $\theta_{\rm s}$  &       & &$T_{\rm rot}$ &   $N$     & $\theta_{\rm s}$  &   &         \\
Species   & (K)                & (cm$^{-2}$ )          &                 & (\arcsec)       & Note  & &(K)         & (cm$^{-2}$ ) & (\arcsec) & Note\tablefootmark{(b)}  & Ref.      \\
\hline
\endfirsthead
\caption{continued.} \\
\hline
          & \multicolumn{5}{c}{This work}                                     &  & \multicolumn{4}{c}{Other study} &              \\
\cline{2-6} \cline{8-12}
          & $T_{\rm rot}$        &   $N$                & $\chi$($N/N_{\rm H_{2}}$) & $\theta_{\rm s}$  &       & &$T_{\rm rot}$ &   $N$   &  $\theta_{\rm s}$    &   &         \\
Species   & (K)                & (cm$^{-2}$ )          &                 & (\arcsec)       & Note  & &(K)         & (cm$^{-2}$ ) & (\arcsec)& Note\tablefootmark{(b)} & Ref.      \\
\hline
\endhead
\hline  \\
\endfoot
\hline
\endlastfoot
NH$_{3}$ ($J=K+1$)                & $209\pm22$ & $(1.2\pm0.7)\times10^{17}$ & $(4.0\pm2.1)\times 10^{-7}$ &10 & HC       & & $160\pm25$ & $2.9^{+0.9}_{-0.7}\times 10^{16}$&  40 & beam & Her88        \\   
NH$_{3}$ ($J=K+2$)                & $192\pm13$ & $(2.4\pm0.8)\times10^{17}$ & $(7.9\pm2.5)\times 10^{-7}$ &10 & HC       & & 250        & $1.3\times10^{17}$            &  10 & HC   & Cro14        \\
NH$_{3}$ ($J=K+3$)                & $191\pm13$ & $(4.1\pm0.9)\times10^{17}$ & $(1.3\pm0.3)\times 10^{-6}$ &10 & HC       & &            &                              &     &      &              \\
NH$_{3}$ ($J=K+4$)                & $118\pm13$ & $(1.9\pm1.1)\times10^{18}$ & $(6.0\pm3.5)\times 10^{-6}$ &10 & HC       & &            &                              &     &      &              \\
$^{15}$NH$_{3}$ (metastable, para) & $124\pm14$ & $(2.6\pm0.4)\times10^{15}$ & $(8.4\pm0.1)\times 10^{-9}$ &10 & HC       & & $110\pm10$ & $6.6\times 10^{13}$           &  40 & beam & Her85        \\
$^{15}$NH$_{3}$ (metastable, ortho)& $119\pm14$ & $(2.5\pm0.8)\times10^{15}$ & $(8.1\pm0.3)\times 10^{-9}$ &10 & HC       & & $115\pm15$ & $9.4\times 10^{13}$           &  40 & beam & Her85        \\
$^{15}$NH$_{3}$ (non-metastable)   & $86\pm13$  & $(1.6\pm1.8)\times10^{15}$ & $(5.2\pm5.8)\times 10^{-9}$ &10 & HC       & & 200        & $1.0\times10^{15}$            &  10 & HC   & Cro14        \\
NH$_{2}$D (para)     & 120\tablefootmark{(a)}  & $(2.0\pm0.4)\times10^{15}$ & $(6.4\pm1.3)\times 10^{-9}$ &10 & HC       & &   115      &  $1.7\times 10^{14}$          &  40 & beam & Wal87        \\
                    &                         &                           &                            &   &          & &            &  $(2.8\pm0.7)\times 10^{15}$   &  10 & HC   & Nei13        \\
                    &                         &                           &                            &   &          & &  $75\pm11$ &  $(9.7\pm3.8)\times 10^{15}$   &  11 & beam & Sch01        \\
                    &                         &                           &                            &   &          & &  250       &  $3.9\times 10^{15}$           &  10 & HC   & Cro14        \\
                    &                         &                           &                            &   &          & &  160       &  $8.7\times 10^{15}$           &  11 & beam & Whi03        \\
CH$_{3}$OH (E)       & $115\pm 8$              & $(1.1\pm0.4)\times10^{18}$ & $(2.8\pm1.1)\times 10^{-6}$ &10 & CR       & &$599\pm295$ & $(9.3\pm4.8)\times10^{16}$    &  11 & beam &  Whi03       \\
CH$_{3}$OH (A)       & $131\pm8$               & $(8.6\pm1.4)\times10^{17}$ & $(2.2\pm0.4)\times 10^{-6}$ &10 & CR       & &   $188\pm3$ & $(7.0\pm0.2)\times10^{16}$    & 20 & beam &  Sch97       \\
                    &                         &                           &                            &   &          & &  $303\pm6$  & $(5.0\pm0.2)\times 10^{16}$   &  11 & beam &  Sch01       \\
                    &                         &                           &                            &   &          & &  $140$      & $4.5\times10^{16}$            & $65-107$& beam &  Tur91       \\
                    &                         &                           &                            &   &          & &   $114$     & $7.7\times10^{15}$            & 40  & beam &  Ziu93       \\
                    &                         &                           &                            &   &          & &   $140$     & $4.5\times10^{16}$            & 46  & beam &  Jon84       \\
$^{13}$CH$_{3}$OH (E)  &  120\tablefootmark{(a)} & $(1.6\pm0.3)\times10^{16}$ & $(4.0\pm0.8)\times 10^{-8}$ & 10& CR       & & $33\pm5$    & $(2.0\pm0.6)\times 10^{15}$  & 25 & CR & Men88        \\
                    &                         &                           &                            &   &          & & $190\pm150$ & $(5.0\pm4.0)\times 10^{15}$   & 25 & CR & Men88        \\
                    &                         &                           &                            &   &          & & $140$       & $1.0\times 10^{16}$           & 10 & CR & Cro14        \\
                    &                         &                           &                            &   &          & & $128$       & $1.5\times 10^{16}$           & 10 & HC(S) & Cro14        \\
                    &                         &                           &                            &   &          & &  $229\pm14$ &  $(9.6\pm1.1)\times 10^{15}$  & 11 & beam & Sch01        \\
                    &                         &                           &                            &   &          & &  115        & $5.9\times 10^{16}$           & 6  & CR &  Per07       \\
SO$_{2}$             &   $137\pm20$            & $(1.5\pm0.2)\times10^{16}$ & $(4.7\pm0.8)\times 10^{-8}$ & 10& HC       & &   $136\pm9$ & $(1.2\pm0.1)\times10^{17}$   & 11 & beam & Whi03        \\
                    &   $82\pm10$             & $(2.2\pm0.4)\times10^{15}$ & $(7.8\pm1.5)\times 10^{-9}$ & 30& Plateau  & &  $187\pm4$  & $(6.0\pm0.3)\times 10^{16}$   & 11 & beam &  Sch01       \\
                    &                         &                           &                            &   &          & &  240        & $9.0\times10^{16}$            & 10 & HC & Cro14        \\
                    &                         &                           &                            &   &          & & $100$       & $1.8\times 10^{15}$           & 10 & CR & Cro14        \\
                    &                         &                           &                            &   &          & &  140        & $2\times10^{16}$              & 40 & beam & Ziu93        \\
                    &                         &                           &                            &   &          & &$138^{+14.7}_{-12.0}$&$1.5^{0.35}_{-0.2}\times10^{16}$& $65-107$& beam &  Tur91   \\
                    &                         &                           &                            &   &          & &   $124\pm3$ & $(7.7\pm0.5)\times10^{16}$    & 20 & beam  &  Sch97      \\
                    &                         &                           &                            &   &          & &   93        & $8.0\times10^{16}$            & 30 & beam  &  Ser95       \\
                    &                         &                           &                            &   &          & &  136        & $1.8\times10^{16}$            & 46 & beam  & Lee02        \\
                    &                         &                           &                            &   &          & &             & $(1.0\pm0.4)\times10^{17}$    & 10 & HC  & Esp13b        \\
                    &                         &                           &                            &   &          & &             & $(1.0\pm0.3)\times10^{16}$    & 30 & Plateau  & Esp13b        \\
OCS                 &  100\tablefootmark{(a)} &  $(5.8\pm1.2)\times10^{16}$& $(1.9\pm0.4)\times 10^{-7}$ & 10& HC(S)    & &   106       & $9.0\times10^{16}$            & 11 & beam  & Whi03        \\
                    &  100\tablefootmark{(a)} & $(9.0\pm1.8)\times10^{15}$ & $(5.0\pm1.0)\times 10^{-8}$ & 30& Plateau  & &  $118$      & $8.6\times10^{14}$            & $65-107$& beam  &  Tur91       \\
                    &                         &                           &                            &   &          & &   $83\pm30$ & $(1.8\pm1.8)\times10^{16}$    & 20  & beam &  Sch97       \\
                    &                         &                           &                            &   &          & &   $190$     & $3.2\times10^{16}$              & 10 & HC  &  Cro14     \\
                    &                         &                           &                            &   &          & &   $165$     & $4.3\times10^{15}$              & 10 & CR  &  Cro14     \\
                    &                         &                           &                            &   &          & &   $110$     & $1.2\times10^{16}$              & 30 & Plateau  &  Cro14     \\
                    &                         &                           &                            &   &          & &             & $(1.6\pm0.4)\times10^{16}$      & 15 & CR  &  Ter10     \\
                    &                         &                           &                            &   &          & &             & $(1.2\pm0.3)\times10^{16}$      & 30 & Plateau  &  Ter10     \\
                    &                         &                           &                            &   &          & &             & $(5.3\pm1.0)\times10^{16}$      & 10 & HC  &  Ter10     \\
HNCO                &   150\tablefootmark{(a)}& $(2.3\pm0.5)\times10^{17}$ & $(7.5\pm1.5)\times 10^{-7}$ &10 & HC(S)    & &  $150\pm14$ & $(4.9\pm0.4)\times10^{15}$    & 11  & beam & Whi03        \\
                    &                         &                           &                            &   &          & &   $160\pm14$ & $(1.3\pm2.3)\times10^{15}$   & 20  & beam &  Sch97       \\
                    &                         &                           &                            &   &          & &  $240\pm13$  & $(3.1\pm0.5)\times 10^{15}$  & 11  & beam &  Sch01       \\
H$_{2}$CO            &   150\tablefootmark{(a)}& $(6.5\pm1.3)\times10^{16}$ & $(2.1\pm0.4)\times 10^{-7}$ &10 & HC(S)    & &  166         & $1.6\times10^{16}$          & 11  & beam & Whi03        \\
                    &                         &                           &                            &   &          & &  $190\pm9$   & $(3.2\pm0.5)\times 10^{15}$  & 11  & beam &  Sch01       \\
                    &                         &                           &                            &   &          & &  $39.9$      & $2.4\times10^{14}$           & $65-107$& beam &  Tur91       \\
HC$_{3}$N           &   200\tablefootmark{(a)} & $(2.5\pm0.5)\times10^{16}$ & $(8.0\pm1.6)\times 10^{-8}$ & 10& HC       & &  $225\pm200$ & $(5.2\pm7.8)\times10^{15}$  & 20  & beam &  Sch97       \\
                   &                          &                           &                            &   &          & &$32.4^{+60}_{-12}$&$4.6^{+3.0}_{-1.8}\times10^{13}$& $65-107$& beam &  Tur91       \\
                   &                          &                           &                            &   &          & &  200          & $1.8\times 10^{15}$         & 10  & HC &  Per07       \\
                   &                          &                           &                            &   &          & &  164          & $1.5\times 10^{15}$         & 11  & beam &  Whi03       \\
                    &                         &                           &                            &   &          & &   $210$       & $1.5\times 10^{15}$         & 10 & HC    &  Cro14     \\
                    &                         &                           &                            &   &          & &   $115$       & $1.3\times 10^{15}$         & 30 & Plateau &  Cro14     \\
                    &                         &                           &                            &   &          & &   $110$       & $(2.0\pm0.5)\times 10^{14}$         & 15 & CR &  Esp13a     \\
                    &                         &                           &                            &   &          & &   $220$       & $(1.0\pm0.3)\times 10^{15}$         & 10 & HC &  Esp13a     \\
                    &                         &                           &                            &   &          & &   $150$       & $(7\pm2)\times 10^{14}$         & 20 & Plateau &  Esp13a     \\
HC$_{5}$N           &   30\tablefootmark{(a)} & $(1.2\pm0.2)\times10^{14}$ & $(3.1\pm0.5)\times 10^{-10}$& 10& CR       & &$31.2^{+3.5}_{-3.4}$&$1.0^{+0.46}_{-0.27}\times10^{13}$& $65-107$& beam &  Tur91      \\
                   &   200\tablefootmark{(a)} & $(6.9\pm1.1)\times10^{14}$ & $(2.2\pm0.4)\times 10^{-9}$& 10& HC       & &   220          & $(7\pm2)\times10^{13}$         & 10     &  HC  &  Esp13a      \\
                   &                          &                           &                             &   &          & & 110          & $(3.0\pm0.9)\times 10^{12}$    & 15      & CR   &  Esp13a      \\
                   &                          &                           &                             &   &          & & 60          & $(5\pm2)\times 10^{12}$         & 120     & ER   &  Esp13a      \\
CH$_{3}$OCH$_{3}$    &  100\tablefootmark{(a)}  & $(4.4\pm0.9)\times10^{16}$ &  $(1.1\pm0.2)\times 10^{-8}$&10 & CR       & &  160          & $2\times 10^{16}$          & 10  &  &  Com05       \\
                   &                          &                           &                             &   &          & &  $89\pm5$     & $(1.8\pm0.2)\times 10^{16}$& 20  & beam &  Sch97       \\
                   &                          &                           &                             &   &          & &  $360\pm26$   & $(2.8\pm0.4)\times 10^{15}$& 11  & beam &  Sch01       \\
                   &                          &                           &                             &   &          & &  112          & $1.3\times 10^{17}$        & 6   & CR   &  Per07       \\
                   &                          &                           &                             &   &          & &  110          & $6.5\times 10^{16}$        & 10  & CR   &  Cro14       \\
                   &                          &                           &                             &   &          & &  100          & $2.1\times 10^{16}$        & 10  & HC(S)&  Cro14       \\
                   &                          &                           &                             &   &          & &  $157\pm30$   & $(1.4\pm0.2)\times 10^{16}$& 11  & beam &  Whi03       \\
                   &                          &                           &                             &   &          & &  197.4        & $1.1\times 10^{16}$        & 46  & beam &  Lee02       \\
                   &                          &                           &                             &   &          & &  75           & $2.5\times 10^{15}$        & 46  & beam &  Joh84       \\
                   &                          &                           &                             &   &          & &  109          & $3.0\times 10^{15}$        & 40  & beam &  Ziu93       \\
                   &                          &                           &                             &   &          & &$91.0^{+33.7}_{-19.0}$&$1.3^{+0.3}_{-0.1}\times10^{15}$& $65-107$ & beam &Tur91    \\
CH$_{3}$OCHO (E)    &   $69\pm8$               & $(1.3\pm0.2)\times10^{17}$ & $(3.4\pm0.4)\times 10^{-7}$  &10 & CR       & & $301\pm95$    & $(5.1\pm1.0)\times 10^{16}$& 11 & beam &  Whi03      \\
CH$_{3}$OCHO (A)    &   $92\pm14$              & $(1.8\pm0.2)\times10^{17}$ & $(4.7\pm0.6)\times 10^{-7}$  &10 & CR       & &   $98\pm3$    & $(1.5\pm0.1)\times10^{16}$& 20  & beam &  Sch97       \\
                   &                          &                           &                             &   &          & &  $110$        & $1.3\times 10^{17}$        & 10  & CR   &  Cro14       \\
                   &                          &                           &                             &   &          & &  $316\pm9$    & $(1.3\pm0.1)\times 10^{17}$& 11  & beam &  Sch01       \\
                   &                          &                           &                             &   &          & & $52$          & $7.7\times 10^{14}$        & 40  & beam &  Ziu93      \\
                   &                          &                           &                             &   &          & &               & $2.5\times 10^{16}$        & 30  & beam &  Ser95      \\
                   &                          &                           &                             &   &          & & $75.2$        & $1.5\times 10^{15}$        & 46  & beam &  Lee02      \\
CH$_{3}$CN          &   200\tablefootmark{(a)} & $(1.9\pm0.4)\times10^{17}$ & $(6.1\pm1.2)\times 10^{-7}$  & 10& HC       & &  $227\pm21$   & $(3.6\pm0.4)\times10^{15}$& 11  & beam & Whi03        \\
                   &                          &                           &                             &   &          & &   $445\pm36$  & $(1.7\pm0.7)\times10^{15}$ & 46  & beam &  Sch97       \\
                   &                          &                           &                             &   &          & &   $274$       & $6.5\times10^{14}$         & $65-107$& beam &  Tur91       \\
                   &                          &                           &                             &   &          & &  $260$        & $3.7\times 10^{15}$        & 10  & HC   &  Cro14       \\
                   &                          &                           &                             &   &          & &  $230$        & $3.5\times 10^{15}$        & 10  & CR   &  Cro14       \\
                   &                          &                           &                             &   &          & &  $120$        & $1.0\times 10^{16}$        & 30  & Plateau   &  Cro14       \\
CH$_{3}$CH$_{2}$CN   &   150\tablefootmark{(a)} & $(4.1\pm0.8)\times10^{16}$ & $(1.3\pm0.3)\times 10^{-7}$  & 10& HC       & &  $136$        & $2.1\times10^{16}$        & 10  & HC   & Cro14        \\
                   &                          &                           &                              &  &          & &  $99\pm3$     & $(1.3\pm0.2)\times10^{16}$  & 20 & beam  &  Sch97       \\
                   &                          &                           &                              &  &          & &  $239\pm4$    & $(3.1\pm0.2)\times10^{16}$  & 11 & beam  &  Sch01       \\
                   &                          &                           &                              &  &          & &               & $(7\pm2)\times10^{16}$\tablefootmark{(c)}      &    & HC    &  Lop14       \\
\end{longtable}   
\normalsize
\tablefoot{\\(a) the rotational temperature is fixed. (b) ``beam'' in the note represents that the column density is beam-averaged. (c) The value is the sum of the three hot core components in \citet{2014AA...572A..44L}.\\
References for rotational temperatures and column densities from the literature. --Her85: \citet{1985A&A...146..134H}; Her88: \citet{1988A&A...201..285H}; Wal87: \citet{1987A&A...172..311W}; Nei13: \citet{2013ApJ...777...85N}; Sch97: \citet{1997ApJS..108..301S}; Sch01: \citet{2001ApJS..132..281S}; Men88: \citet{1988A&A...198..253M}; Com05: \citet{2005ApJS..156..127C}; Per07: \citet{2007AA...476..807P}; Cro14: \citet{2014ApJ...787..112C}; Lee02: \citet{2002JKAS...35..187L}; Joh84: \citet{1984AA...130..227J}; Whi03: \citet{2003AA...407..589W}; Ter10: \citet{2010AA...517A..96T}; Ser95: \citet{1995ApJ...451..238S}; Tur91: \citet{1991ApJS...76..617T}; Ziu93: \citet{1993ApJS...89..155Z}; Lop14: \citet{2014AA...572A..44L}; Esp13a: \citet{2013AA...559A..51E}; Esp13b: \citet{2013AA...556A.143E}.\\}
\normalsize                                                  
} 
\begin{table*}[!hbt]
\caption{The spatial origin of molecules detected by our 1.3 cm line survey.}\label{Tab:origin}             
\normalsize
\centering                                      
\begin{tabular}{ccccccccc}          
\hline\hline                        
                          &  & \multicolumn{7}{c}{origins\tablefootmark{(1)}}                   \\
                            \cline{3-9} 
                          &  & HC       &  HC(S)  & Plateau & CR       & MM4     & MM5    & MM6     \\
\hline
\multicolumn{9}{l}{molecules}\\
\hline
NH$_{3}$                 &   &$\surd$ & $\surd$ & $\surd$ &         & $\surd$ & $\surd$ & $\surd$ \\
$^{15}$NH$_{3}$          &    &$\surd$ &         &         &         &         &         &         \\
NH$_{2}$D                &   &$\surd$ & $\surd$ &         &         & $\surd$ & $\surd$ &         \\
HDO                      &  &        &  $\surd$&         &         &         &         &         \\
SO$_{2}$                 &   &$\surd$ & $\surd$ & $\surd$ & $\surd$&         &         &         \\
OCS                      &  &$\surd$ & $\surd$ & $\surd$ & $\surd$ & $\surd$ & $\surd$ &         \\
HC$_{3}$N                &   &$\surd$ & $\surd$ &         &         & $\surd$ & $\surd$ & $\surd$ \\
HC$_{5}$N                &   &        &   ?     &         &  ?      &         &         &         \\
CH$_{3}$OH               &   & $\surd$  & $\surd$ &         &$\surd$& $\surd$ & $\surd$ & $\surd$ \\
$^{13}$CH$_{3}$OH        &    &           &         &         &$\surd$&         &         &         \\
HNCO                     &  &  $\surd$  & $\surd$ &         & $\surd$ & $\surd$ & $\surd$ & $\surd$ \\
H$_{2}$CO                &   &  $\surd$  & $\surd$ & $\surd$ & $\surd$ &         & $\surd$ & $\surd$ \\
CH$_{3}$CN               &   &$\surd$ & $\surd$   &         & $\surd$ & $\surd$ & $\surd$ & $\surd$ \\
CH$_{3}$CH$_{2}$CN       &    &$\surd$ & $\surd$ &         &         & $\surd$ &         &         \\
CH$_{3}$OCHO             &   &           &         &         &$\surd$&         &         & $\surd$ \\
CH$_{3}$OCH$_{3}$        &    &           &         &         &$\surd$&         & $\surd$ & $\surd$ \\
\hline
 \end{tabular}
 \tablefoot{\\
 \tablefoottext{1}{A ``?'' mark indicates that the origin is still questionable. $\surd$ denotes that the molecule has emission originating from the component.}\\
 }
 			        \normalsize
\end{table*}

\clearpage

\begin{appendix}
\section{The observed properties of detected lines in the survey}\label{app.a}
\renewcommand{\tabcolsep}{0.15 cm}
\normalsize \longtab{1}{
\begin{longtable}{cccccccc}
\caption{Transitions of recombination lines.}\label{Tab:rrl}\\
\hline
Transition   & Rest frequency      & $\int S_{\nu} {\rm d}\upsilon$  & $\upsilon_{\rm lsr}$    & $\Delta \upsilon$ & $S_{\nu}$   &  $\sigma$   &  \\
             & (MHz)      & (Jy~\kms)        & (\kms)      & (\kms)     & (Jy)         &  (Jy)       & notes \\
\hline
\endfirsthead
\caption{continued.} \\
\hline
Transition   & Rest frequency      & $\int S_{\nu} {\rm d}\upsilon$  & $\upsilon_{\rm lsr}$    & $\Delta \upsilon$ & $S_{\nu}$   &  $\sigma$   &  \\
             & (MHz)      & (Jy~\kms)        & (\kms)      & (\kms)     & (Jy)         &  (Jy)       & notes \\
\hline
\endhead
\hline  \\
\endfoot
\hline
\endlastfoot

H$\alpha$      &\multicolumn{7}{c}{}\\
\hline
H71$\alpha$    & 17992.6 & $72.67\pm0.05$ & $-5.4\pm0.1$ & $23.8\pm0.1$ & $2.870\pm0.007$ & $0.006$ &     \\
H70$\alpha$    & 18769.2 & $76.19\pm0.07$ & $-4.4\pm0.1$ & $23.8\pm0.1$ & $3.000\pm0.009$ & $0.006$ &     \\
H69$\alpha$    & 19591.1 & $70.67\pm0.04$ & $-3.4\pm0.1$ & $23.8\pm0.1$ & $2.800\pm0.005$ & $0.005$ &     \\
H68$\alpha$    & 20461.8 & $73.24\pm0.13$ & $-3.6\pm0.1$ & $23.3\pm0.1$ & $2.960\pm0.019$ & $0.008$ &  1   \\
H67$\alpha$    & 21384.8 & $96.27\pm0.11$ & $-5.1\pm0.1$ & $23.7\pm0.1$ & $3.820\pm0.017$ & $0.014$ & 2   \\
H66$\alpha$    & 22364.2 & $92.61\pm0.07$ & $-4.9\pm0.1$ & $24.4\pm0.1$ & $3.560\pm0.010$ & $0.011$ &     \\
H65$\alpha$    & 23404.3 & $<64.39$       &              &              &                 & $0.009$ & b   \\
H64$\alpha$    & 24509.9 & $62.25\pm0.02$ & $-3.6\pm0.1$ & $23.5\pm0.1$ & $2.480\pm0.004$ & $0.004$ &     \\
H63$\alpha$    & 25686.3 & $57.00\pm0.03$ & $-2.6\pm0.1$ & $23.3\pm0.1$ & $2.300\pm0.005$ & $0.005$ &     \\
\hline
He$\alpha$     &\multicolumn{7}{c}{}\\
\hline
He71$\alpha$   & 17999.9 & $6.32\pm0.05$ & $-6.0\pm0.1$ & $15.4\pm0.2$ & $0.386\pm0.007$ & $0.005$ &     \\
He70$\alpha$   & 18776.8 & $6.27\pm0.05$ & $-4.9\pm0.1$ & $15.2\pm0.1$ & $0.388\pm0.008$ & $0.007$ &     \\
He69$\alpha$   & 19599.1 & $5.84\pm0.04$ & $-3.9\pm0.1$ & $15.2\pm0.1$ & $0.361\pm0.006$ & $0.005$ &     \\
He68$\alpha$   & 20470.1 & $6.70\pm0.07$ & $-4.0\pm0.1$ & $15.6\pm0.2$ & $0.405\pm0.011$ & $0.008$ &      \\
He67$\alpha$   & 21393.5 & $8.13\pm0.09$ & $-5.3\pm0.1$ & $14.8\pm0.2$ & $0.515\pm0.015$ & $0.016$ &     \\
He66$\alpha$   & 22373.3 & $7.52\pm0.06$ & $-4.7\pm0.1$ & $15.5\pm0.1$ & $0.455\pm0.010$ & $0.010$ &     \\
He65$\alpha$   & 23413.8 & $<12.96$      &              &              &                 & $0.005$ &  b  \\
He64$\alpha$   & 24519.9 & $5.37\pm0.02$ & $-4.1\pm0.1$ & $15.2\pm0.1$ & $0.333\pm0.004$ & $0.003$ &     \\
He63$\alpha$   & 25696.7 & $<21.85$      &              &              &                 & $0.005$ &  b  \\
\hline
C$\alpha$     &\multicolumn{7}{c}{}\\
\hline
C71$\alpha$   & 18001.5  & $0.40\pm0.03$  & $6.7\pm0.1$ & $4.9\pm0.4$ & $0.077\pm0.007$ & $0.006$ &  t   \\
C70$\alpha$   & 18778.5  & $0.30\pm0.02$  & $8.1\pm0.1$ & $4.0\pm0.3$ & $0.070\pm0.006$ & $0.006$ &  t   \\
C69$\alpha$   & 19600.9  & $0.30\pm0.02$  & $8.8\pm0.2$ & $5.8\pm0.5$ & $0.048\pm0.005$ & $0.005$ &  t   \\
C68$\alpha$   & 20472.0  & $0.51\pm0.12$  & $7.6\pm0.8$ & $11.0\pm3.6$& $0.044\pm0.010$ & $0.008$ &  t   \\
C67$\alpha$   & 21395.5  & $0.54\pm0.07$  & $7.1\pm0.3$ & $5.9\pm0.7$ & $0.085\pm0.015$ & $0.016$ &  t   \\
C66$\alpha$   & 22375.3  & $0.39\pm0.04$  & $7.6\pm0.2$ & $5.8\pm0.6$ & $0.063\pm0.009$ & $0.007$ &  t   \\
C65$\alpha$   & 23416.0  & $0.17\pm0.04$  & $8.0\pm0.3$ & $3.5\pm0.8$ & $0.047\pm0.013$ & $0.005$ &  t  \\
C64$\alpha$   & 24522.1  & $0.27\pm0.02$  & $8.1\pm0.1$ & $4.7\pm0.3$ & $0.054\pm0.005$ & $0.004$ &  t   \\
C63$\alpha$   & 25699.1  & $0.20\pm0.02$  & $9.2\pm0.2$ & $4.3\pm0.4$ & $0.045\pm0.006$ & $0.004$ &  t  \\
\hline
H$\beta$       &\multicolumn{7}{c}{}\\
\hline
H89$\beta$     & 18045.9 & $18.73\pm0.07$ & $-5.2\pm0.1$ & $24.0\pm0.1$ & $0.734\pm0.008$ & $0.007$ &     \\
H88$\beta$     & 18661.1 & $19.98\pm0.04$ & $-4.5\pm0.1$ & $24.0\pm0.1$ & $0.782\pm0.006$ & $0.006$ &     \\
H87$\beta$     & 19304.7 & $21.95\pm0.05$ & $-3.1\pm0.1$ & $23.8\pm0.1$ & $0.867\pm0.007$ & $0.007$ &     \\
H86$\beta$     & 19978.2 & $24.25\pm0.10$ & $-2.6\pm0.1$ & $24.8\pm0.1$ & $0.920\pm0.013$ & $0.013$ &     \\
H85$\beta$     & 20683.3 & $20.49\pm0.06$ & $-3.1\pm0.1$ & $23.8\pm0.1$ & $0.809\pm0.008$ & $0.009$ &     \\
H84$\beta$     & 21422.1 &       $<28.64$ &              &              &                 & $0.026$ & b   \\
H83$\beta$     & 22196.5 & $23.00\pm0.10$ & $-3.5\pm0.1$ & $23.1\pm0.1$ & $0.934\pm0.015$ & $0.014$ &     \\
H82$\beta$     & 23008.6 & $16.32\pm0.03$ & $-4.2\pm0.1$ & $23.5\pm0.1$ & $0.653\pm0.005$ & $0.005$ &     \\
H81$\beta$     & 23860.9 & $17.94\pm0.03$ & $-3.1\pm0.1$ & $23.5\pm0.1$ & $0.716\pm0.005$ & $0.005$ &     \\
H80$\beta$     & 24755.7 & $16.88\pm0.03$ & $-3.9\pm0.1$ & $23.5\pm0.1$ & $0.674\pm0.005$ & $0.005$ &     \\
H79$\beta$     & 25695.9 &       $<22.09$ &              &              &                 & $0.007$ & b   \\
\hline
He$\beta$      &\multicolumn{7}{c}{}\\
\hline
He89$\beta$    & 18053.2 & $1.56\pm0.04$ & $-5.8\pm0.2$ & $14.7\pm0.5$ & $0.099\pm0.006$ & $0.005$ &     \\ 
He88$\beta$    & 18668.7 & $1.79\pm0.05$ & $-4.8\pm0.2$ & $15.7\pm0.4$ & $0.107\pm0.005$ & $0.005$ &     \\ 
He87$\beta$    & 19312.6 & $2.06\pm0.04$ & $-4.1\pm0.2$ & $16.3\pm0.4$ & $0.119\pm0.007$ & $0.006$ &     \\ 
He86$\beta$    & 19986.3 & $2.37\pm0.11$ & $-2.7\pm0.3$ & $15.6\pm1.0$ & $0.142\pm0.015$ & $0.013$ &     \\ 
He85$\beta$    & 20691.8 & $1.79\pm0.05$ & $-3.4\pm0.2$ & $16.1\pm0.5$ & $0.105\pm0.009$ & $0.008$ &     \\ 
He84$\beta$    & 21430.8 & $1.87\pm0.09$ & $-5.7\pm0.3$ & $11.0\pm0.6$ & $0.160\pm0.022$ & $0.020$ & 3   \\ 
He83$\beta$    & 22205.5 & $2.29\pm0.08$ & $-4.3\pm0.2$ & $15.9\pm0.6$ & $0.135\pm0.013$ & $0.013$ &     \\ 
He82$\beta$    & 23018.0 & $1.44\pm0.03$ & $-4.3\pm0.1$ & $14.9\pm0.3$ & $0.090\pm0.005$ & $0.005$ &     \\ 
He81$\beta$    & 23870.6 & $<2.77      $ &              &              &                 & $0.006$ & b   \\
He80$\beta$    & 24765.8 & $1.61\pm0.03$ & $-4.6\pm0.1$ & $16.7\pm0.3$ & $0.090\pm0.005$ & $0.004$ &     \\ 
He79$\beta$    & 25706.4 & $<105.14    $ &              &              &                 & $0.005$ & b   \\ 
\hline
H$\gamma$      &\multicolumn{7}{c}{}\\
\hline
H101$\gamma$   & 18327.5 & $10.12\pm0.05$& $-4.6\pm0.1$ & $25.5\pm0.2$ & $0.373\pm0.007$ & $0.005$ &     \\ 
H100$\gamma$   & 18874.8 & $10.49\pm0.04$& $-3.9\pm0.1$ & $24.5\pm0.1$ & $0.402\pm0.006$ & $0.006$ &     \\ 
H99$\gamma$    & 19444.0 & $9.64\pm0.04$ & $-3.7\pm0.1$ & $24.6\pm0.1$ & $0.368\pm0.005$ & $0.005$ &     \\  
H98$\gamma$    & 20036.3 & $10.82\pm0.09$& $-1.5\pm0.1$ & $25.3\pm0.2$ & $0.402\pm0.012$ & $0.009$ &     \\ 
H97$\gamma$    & 20653.0 & $9.16\pm0.06$ & $-3.2\pm0.1$ & $23.9\pm0.2$ & $0.361\pm0.008$ & $0.007$ &     \\  
H96$\gamma$    & 21295.2 & $12.35\pm0.07$& $-4.4\pm0.1$ & $24.6\pm0.2$ & $0.472\pm0.011$ & $0.011$ &     \\  
H95$\gamma$    & 21964.3 & $11.18\pm0.09$& $-4.0\pm0.1$ & $24.2\pm0.2$ & $0.434\pm0.014$ & $0.014$ &     \\ 
H94$\gamma$    & 22661.8 & $11.17\pm0.05$& $-3.9\pm0.1$ & $25.1\pm0.1$ & $0.417\pm0.008$ & $0.008$ &     \\  
H93$\gamma$    & 23389.1 & $8.50\pm0.05$ & $-3.9\pm0.1$ & $24.2\pm0.2$ & $0.330\pm0.007$ & $0.004$ & 4   \\  
H92$\gamma$    & 24147.9 & $7.63\pm0.02$ & $-3.5\pm0.1$ & $24.0\pm0.1$ & $0.299\pm0.003$ & $0.003$ &     \\  
H91$\gamma$    & 24939.8 & $7.62\pm0.02$ & $-3.7\pm0.1$ & $23.4\pm0.1$ & $0.305\pm0.004$ & $0.004$ &     \\  
H90$\gamma$    & 25766.8 & $<10.64     $ &              &              &                 & $0.004$ & b    \\  
\hline
He$\gamma$     &\multicolumn{7}{c}{}\\
\hline
He101$\gamma$  & 18335.0 & $0.92\pm0.04$ & $-5.2\pm0.3$ & $15.6\pm0.9$ & $0.055\pm0.006$ & $0.005$ &     \\
He100$\gamma$  & 18882.4 & $1.02\pm0.04$ & $-5.4\pm0.3$ & $17.6\pm0.8$ & $0.055\pm0.006$ & $0.006$ &     \\
He99$\gamma$   & 19451.9 & $0.92\pm0.03$ & $-4.7\pm0.3$ & $17.9\pm0.7$ & $0.048\pm0.005$ & $0.005$ &     \\
He98$\gamma$   & 20044.5 & $1.06\pm0.08$ & $-3.4\pm0.5$ & $15.3\pm1.4$ & $0.065\pm0.012$ & $0.011$ &     \\
He96$\gamma$   & 21303.9 & $1.03\pm0.07$ & $-4.7\pm0.5$ & $15.4\pm1.3$ & $0.063\pm0.012$ & $0.010$ &     \\
He95$\gamma$   & 21973.3 & $0.99\pm0.08$ & $-4.0\pm0.5$ & $14.0\pm1.6$ & $0.066\pm0.014$ & $0.014$ &     \\
He94$\gamma$   & 22671.0 & $0.88\pm0.04$ & $-5.3\pm0.4$ & $16.5\pm0.9$ & $0.050\pm0.008$ & $0.008$ &     \\
He93$\gamma$   & 23398.6 & $<0.47      $ &              &              &                 & $0.005$ & b   \\
He92$\gamma$   & 24157.7 & $0.67\pm0.02$ & $-4.2\pm0.2$ & $16.9\pm0.5$ & $0.037\pm0.003$ & $0.003$ &     \\
He91$\gamma$   & 24950.0 & $0.77\pm0.04$ & $-4.0\pm0.3$ & $17.7\pm1.1$ & $0.041\pm0.004$ & $0.004$ &     \\
He90$\gamma$   & 25777.3 & $0.53\pm0.03$ & $-2.8\pm0.3$ & $13.0\pm0.8$ & $0.038\pm0.006$ & $0.005$ &     \\
\hline
H$\delta$      &\multicolumn{7}{c}{}\\
\hline
H111$\delta$   & 18241.7 & $4.58\pm0.06$ & $-4.5\pm0.2$ & $26.8\pm0.4$ & $0.160\pm0.006$ & $0.006$ &     \\     
H110$\delta$   & 18734.9 & $5.48\pm0.05$ & $-4.3\pm0.1$ & $25.5\pm0.3$ & $0.202\pm0.006$ & $0.005$ &     \\     
H109$\delta$   & 19246.1 & $6.80\pm0.05$ & $-2.8\pm0.1$ & $26.7\pm0.2$ & $0.239\pm0.005$ & $0.006$ &     \\     
H108$\delta$   & 19776.0 & $<  9.51    $ &              &              &                 & $0.012$ & b   \\     
H107$\delta$   & 20325.5 & $5.52\pm0.06$ & $-3.7\pm0.1$ & $25.5\pm0.4$ & $0.204\pm0.007$ & $0.007$ &     \\
H106$\delta$   & 20895.6 & $4.99\pm0.07$ & $-3.1\pm0.2$ & $24.8\pm0.4$ & $0.189\pm0.009$ & $0.009$ &     \\     
H105$\delta$   & 21487.2 & $7.34\pm0.13$ & $-5.1\pm0.2$ & $26.9\pm0.6$ & $0.256\pm0.013$ & $0.014$ &     \\     
H104$\delta$   & 22101.4 & $7.00\pm0.12$ & $-3.7\pm0.2$ & $24.6\pm0.5$ & $0.267\pm0.015$ & $0.014$ &     \\     
H103$\delta$   & 22739.2 & $3.63\pm0.05$ & $-4.7\pm0.2$ & $23.4\pm0.4$ & $0.145\pm0.007$ & $0.006$ &     \\
H102$\delta$   & 23401.8 & $<69.10     $ &              &              &                 & $0.005$ & b   \\  
H101$\delta$   & 24090.4 & $4.19\pm0.03$ & $-3.7\pm0.1$ & $24.2\pm0.2$ & $0.163\pm0.003$ & $0.003$ &     \\
H100$\delta$   & 24806.3 & $4.38\pm0.03$ & $-3.7\pm0.1$ & $23.7\pm0.2$ & $0.174\pm0.005$ & $0.005$ &     \\     
 H99$\delta$   & 25550.8 & $4.08\pm0.03$ & $-3.3\pm0.1$ & $24.0\pm0.2$ & $0.160\pm0.004$ & $0.004$ &     \\     
\hline              
He$\delta$     &\multicolumn{7}{c}{}\\
\hline
He110$\delta$  & 18742.5 & $0.64\pm0.04$ & $-4.0\pm0.4$ & $17.3\pm1.2$ & $0.035\pm0.005$ & $0.005$ &     \\    
He107$\delta$  & 20333.8 & $<3.60      $ &              &              &                 & $0.009$ & b   \\   
He103$\delta$  & 22748.5 & $0.33\pm0.03$ & $-5.0\pm0.6$ & $13.7\pm1.5$ & $0.023\pm0.006$ & $0.006$ &     \\    
He102$\delta$  & 23411.4 & $< 12.96$     &              &              &                 & $0.005$ & b   \\  
He101$\delta$  & 24100.2 & $0.37\pm0.02$ & $-5.2\pm0.3$ & $15.7\pm0.9$ & $0.022\pm0.003$ & $0.003$ &     \\    
He100$\delta$  & 24816.4 & $0.35\pm0.02$ & $-5.0\pm0.5$ & $16.1\pm1.4$ & $0.020\pm0.004$ & $0.004$ &     \\    
 He99$\delta$  & 25561.2 & $0.38\pm0.02$ & $-5.0\pm0.5$ & $16.4\pm1.3$ & $0.022\pm0.004$ & $0.004$ &     \\    
\hline
H$\varepsilon$     &\multicolumn{7}{c}{}\\
\hline
H120$\varepsilon$   & 17901.6 & $3.38\pm0.08$ & $-6.5\pm0.3$ & $29.0\pm0.8$ & $0.110\pm0.006$ & $0.006$ &     \\
H119$\varepsilon$   & 18347.5 & $3.64\pm0.07$ & $-5.1\pm0.2$ & $27.3\pm0.6$ & $0.125\pm0.006$ & $0.006$ &     \\
H118$\varepsilon$   & 18808.4 & $<   5.84   $ &              &              &                 & $0.005$ &  b  \\
H117$\varepsilon$   & 19284.8 & $<   7.25   $ &              &              &                 & $0.005$ &  b  \\
H116$\varepsilon$   & 19777.4 & $<   9.29   $ &              &              &                 & $0.012$ &  b  \\
H115$\varepsilon$   & 20287.0 & $2.83\pm0.06$ & $-3.6\pm0.2$ & $23.9\pm0.6$ & $0.111\pm0.007$ & $0.007$ &     \\
H114$\varepsilon$   & 20814.3 & $3.20\pm0.08$ & $-3.8\pm0.3$ & $26.3\pm0.8$ & $0.114\pm0.009$ & $0.009$ &     \\
H113$\varepsilon$   & 21359.9 & $4.25\pm0.14$ & $-6.0\pm0.4$ & $23.9\pm1.0$ & $0.167\pm0.015$ & $0.016$ &     \\
H112$\varepsilon$   & 21924.8 & $4.31\pm0.11$ & $-3.4\pm0.3$ & $25.5\pm0.8$ & $0.159\pm0.013$ & $0.013$ &     \\
H111$\varepsilon$   & 22509.8 & $2.91\pm0.06$ & $-4.1\pm0.2$ & $25.6\pm0.6$ & $0.107\pm0.007$ & $0.007$ &     \\
H110$\varepsilon$   & 23115.9 & $2.71\pm0.03$ & $-4.4\pm0.2$ & $25.1\pm0.4$ & $0.101\pm0.005$ & $0.005$ &     \\
H109$\varepsilon$   & 23743.8 & $2.92\pm0.04$ & $-3.1\pm0.2$ & $24.4\pm0.4$ & $0.112\pm0.006$ & $0.006$ &     \\
H108$\varepsilon$   & 24394.8 & $2.78\pm0.03$ & $-3.4\pm0.1$ & $24.3\pm0.3$ & $0.107\pm0.004$ & $0.004$ &     \\
H107$\varepsilon$   & 25069.7 & $2.45\pm0.04$ & $-4.1\pm0.2$ & $23.5\pm0.4$ & $0.098\pm0.005$ & $0.005$ &     \\
H106$\varepsilon$   & 25769.8 & $<  10.82   $ &              &              &                 & $0.004$ & b   \\
\hline
He$\varepsilon$     &\multicolumn{7}{c}{}\\
\hline
He107$\varepsilon$  & 25079.9 & $<   1.43   $ &              &              &                 & $0.005$ & b    \\
\hline
H$\zeta$     &\multicolumn{7}{c}{}\\
\hline
H127$\zeta$         & 17978.5 & $2.01\pm0.05$ & $-6.5\pm0.3$ & $24.2\pm0.7$ & $0.078\pm0.005$ & $0.004$ &     \\
H126$\zeta$         & 18400.1 & $<   8.62   $ &              &              &                 & $0.006$ & b   \\
H125$\zeta$         & 18835.1 & $2.54\pm0.07$ & $-3.4\pm0.3$ & $27.8\pm0.9$ & $0.086\pm0.005$ & $0.006$ &     \\
H124$\zeta$         & 19283.8 & $<   7.25   $ &              &              &                 & $0.008$ & b   \\
H122$\zeta$         & 20225.1 & $2.12\pm0.06$ & $-3.2\pm0.3$ & $23.6\pm0.9$ & $0.084\pm0.007$ & $0.006$ &     \\
H121$\zeta$         & 20718.7 & $<   5.26   $ &              &              &                 & $0.009$ & b   \\
H120$\zeta$         & 21228.6 & $<   3.75   $ &              &              &                 & $0.006$ & b   \\
H118$\zeta$         & 22299.6 & $2.67\pm0.11$ & $-4.8\pm0.4$ & $25.3\pm1.3$ & $0.099\pm0.011$ & $0.012$ &     \\
H117$\zeta$         & 22862.3 & $2.16\pm0.07$ & $-4.2\pm0.3$ & $27.0\pm1.1$ & $0.075\pm0.006$ & $0.006$ &     \\
H116$\zeta$         & 23444.0 & $<   8.09   $ &              &              &                 & $0.005$ & b   \\
H115$\zeta$         & 24045.6 & $2.21\pm0.04$ & $-2.3\pm0.2$ & $26.0\pm0.6$ & $0.080\pm0.004$ & $0.004$ &     \\
H114$\zeta$         & 24668.0 & $1.79\pm0.03$ & $-4.0\pm0.2$ & $24.9\pm0.6$ & $0.068\pm0.004$ & $0.004$ &     \\
H113$\zeta$         & 25312.0 & $2.05\pm0.04$ & $-3.5\pm0.2$ & $24.8\pm0.7$ & $0.078\pm0.005$ & $0.005$ &     \\
H112$\zeta$         & 25978.7 & $2.24\pm0.06$ & $-3.0\pm0.3$ & $24.0\pm0.7$ & $0.088\pm0.009$ & $0.009$ &     \\
\hline
H$\eta$     &\multicolumn{7}{c}{}\\
\hline
H133$\eta$          & 18123.4 & $1.38\pm0.09$ & $-5.3\pm0.7$ & $27.9\pm2.3$ & $0.047\pm0.007$ & $0.007$ &     \\
H132$\eta$          & 18528.0 & $1.81\pm0.08$ & $-5.3\pm0.5$ & $30.0\pm1.7$ & $0.057\pm0.006$ & $0.006$ &     \\
H131$\eta$          & 18944.7 & $1.71\pm0.08$ & $-4.9\pm0.5$ & $28.9\pm1.6$ & $0.056\pm0.006$ & $0.006$ &     \\
H130$\eta$          & 19374.0 & $1.60\pm0.06$ & $-3.5\pm0.4$ & $26.5\pm1.2$ & $0.057\pm0.005$ & $0.005$ &     \\
H129$\eta$          & 19816.4 & $1.83\pm0.21$ & $-1.6\pm1.2$ & $30.3\pm4.5$ & $0.057\pm0.014$ & $0.014$ &     \\
H128$\eta$          & 20272.4 & $1.16\pm0.09$ & $-6.1\pm0.9$ & $28.9\pm3.0$ & $0.038\pm0.007$ & $0.007$ &     \\
H126$\eta$          & 21227.2 & $<   3.74   $ &              &              &                 & $0.007$ & b    \\
H123$\eta$          & 22775.1 & $1.02\pm0.03$ & $-4.7\pm0.3$ & $23.0\pm0.9$ & $0.042\pm0.004$ & $0.004$ &     \\
H122$\eta$          & 23324.4 & $<   2.12   $ &              &              &                 & $0.006$ & b    \\
H121$\eta$          & 23891.6 & $1.40\pm0.04$ & $-3.3\pm0.3$ & $24.1\pm0.9$ & $0.055\pm0.005$ & $0.005$ &     \\
H120$\eta$          & 24477.3 & $1.25\pm0.03$ & $-3.4\pm0.3$ & $23.7\pm0.8$ & $0.050\pm0.004$ & $0.004$ &     \\
H119$\eta$          & 25082.3 & $<   1.65   $ &              &              &                 & $0.005$ & b    \\
H118$\eta$          & 25707.4 & $<   2.75   $ &              &              &                 & $0.005$ & b   \\                       
\hline
H$\theta$      &\multicolumn{7}{c}{}\\
\hline
H139$\theta$   & 18019.0 & $1.46\pm0.12$ & $-8.4\pm1.0$ & $31.5\pm3.1$ & $0.044\pm0.006$ & $0.006$ &     \\
H138$\theta$   & 18402.8 & $< 6.80     $ &              &              &                 & $0.008$ & b   \\
H135$\theta$   & 19621.6 & $2.02\pm0.06$ & $-3.8\pm0.3$ & $31.0\pm1.2$ & $0.061\pm0.004$ & $0.004$ &     \\
H134$\theta$   & 20051.7 & $2.77\pm0.36$ & $-6.7\pm1.7$ & $41.6\pm6.4$ & $0.063\pm0.013$ & $0.013$ & 5   \\
H133$\theta$   & 20494.5 & $<  3.82    $ &              &              &                 & $0.006$ & b   \\
H132$\theta$   & 20950.5 & $1.27\pm0.15$ & $-0.7\pm1.4$ & $35.3\pm5.3$ & $0.034\pm0.008$ & $0.008$ & 5   \\
H131$\theta$   & 21420.0 & $<  29.45   $ &              &              &                 & $0.051$ & b   \\
H130$\theta$   & 21903.8 & $0.97\pm0.10$ & $-3.4\pm0.9$ & $18.5\pm2.2$ & $0.049\pm0.015$ & $0.015$ & 5   \\
H129$\theta$   & 22402.1 & $1.27\pm0.06$ & $-3.4\pm0.5$ & $26.7\pm1.6$ & $0.045\pm0.006$ & $0.006$ &     \\
H128$\theta$   & 22915.8 & $0.97\pm0.05$ & $-5.1\pm0.5$ & $24.7\pm1.6$ & $0.037\pm0.006$ & $0.006$ &     \\
H127$\theta$   & 23445.2 & $< 8.11     $ &              &              &                 & $0.005$ & b   \\
H126$\theta$   & 23991.2 & $1.06\pm0.03$ & $-3.6\pm0.2$ & $28.2\pm1.0$ & $0.035\pm0.003$ & $0.003$ &     \\
H124$\theta$   & 25134.9 & $1.10\pm0.04$ & $-4.0\pm0.4$ & $28.1\pm1.3$ & $0.037\pm0.004$ & $0.004$ &     \\
H123$\theta$   & 25734.1 & $0.85\pm0.04$ & $-1.9\pm0.6$ & $23.4\pm1.6$ & $0.034\pm0.006$ & $0.006$ &     \\
\hline
H$\iota$       &\multicolumn{7}{c}{}\\
\hline
H142$\iota$    & 18859.0 & $0.62\pm0.04$ & $-7.2\pm0.7$ & $21.9\pm1.9$ & $0.026\pm0.005$ & $0.005$ &     \\
H140$\iota$    & 19654.0 & $0.96\pm0.09$ & $-6.6\pm1.2$ & $35.1\pm4.2$ & $0.026\pm0.005$ & $0.005$ &     \\
H138$\iota$    & 20494.3 & $ <   3.37  $ &              &              &                 & $0.008$ & b   \\
H137$\iota$    & 20932.5 & $0.83\pm0.09$ & $0.3\pm1.2$  & $27.0\pm4.0$ & $0.029\pm0.009$ & $0.009$ & 5   \\
H136$\iota$    & 21383.2 & $ < 97.36   $ &              &              &                 & $0.014$ & b   \\
H134$\iota$    & 22324.4 & $1.02\pm0.07$ & $-5.4\pm0.7$ & $27.8\pm2.4$ & $0.034\pm0.006$ & $0.006$ &     \\
H133$\iota$    & 22815.7 & $0.98\pm0.10$ & $-8.4\pm1.2$ & $32.0\pm3.8$ & $0.029\pm0.006$ & $0.006$ &     \\
H132$\iota$    & 23321.6 & $1.00\pm0.08$ &              &              &                 & $0.006$ & b   \\
H131$\iota$    & 23842.5 & $0.97\pm0.06$ & $-3.6\pm0.6$ & $28.2\pm2.2$ & $0.033\pm0.005$ & $0.005$ &     \\
H130$\iota$    & 24379.1 & $0.67\pm0.03$ & $-2.5\pm0.4$ & $24.0\pm1.2$ & $0.026\pm0.004$ & $0.004$ &     \\
H129$\iota$    & 24931.9 & $ < 63.73   $ &               &             &                 & $0.050$ & b   \\
H128$\iota$    & 25501.5 & $0.68\pm0.03$ & $-4.4\pm0.5$ & $23.6\pm1.3$ & $0.027\pm0.004$ & $0.004$ &     \\
\hline
H$\kappa$      &\multicolumn{7}{c}{}\\
\hline
H136$\kappa$   & 23518.2 & $1.14\pm0.05$ & $-5.1\pm0.8$ & $35.0\pm2.0$ & $0.031\pm0.004$ & $0.004$ & 5   \\
H132$\kappa$   & 25642.8 & $0.50\pm0.04$ & $-3.0\pm0.6$ & $28.4\pm2.5$ & $0.017\pm0.006$ & $0.004$ &     \\
\hline
H$\lambda$     &\multicolumn{7}{c}{}\\
\hline
H136$\lambda$  & 25609.7 & $0.72\pm0.03$ & $-2.8\pm0.6$ & $31.2\pm1.3$ & $0.022\pm0.006$ & $0.004$ & 5   \\
\hline
\end{longtable} 
\tablefoot{-- (1) Although H68$\alpha$ is blended with HDO ($3_{2,1}-4_{1,4}$) (see Fig.~\ref{Fig:one}), the integrated intensity of HDO ($3_{2,1}-4_{1,4}$) can be ignored compared with H68$\alpha$. (2) Although H67$\alpha$ is blended with H136$\iota$ (see Fig.~\ref{Fig:iota}), the integrated intensity of H136$\iota$ can be neglected. (3) Note that the observed line width is narrower than those of other He RRL transitions. (4) Although H93$\gamma$ is blended with CH$_{3}$OCH$_{3}$ ($12_{3,10}-11_{4,7}$) (see Fig.~\ref{Fig:one}), the integrated intensity of CH$_{3}$OCH$_{3}$ ($12_{3,10}-11_{4,7}$) can be ignored compared with H93$\gamma$. (5) Note that the fitted parameters are possibly wrong due to low signal-to-noise ratios. (b) it indicates that this line is blended (see Table~\ref{Tab:orilines}). (t) The fit for the C$\alpha$ line is tentative because it is blended.}              
} 

\renewcommand{\tabcolsep}{0.15 cm}
\normalsize \longtab{2}{
\begin{longtable}{cccccccc}
\caption{The observed properties of NH$_{3}$, $^{15}$NH$_{3}$ and NH$_{2}$D transitions. }\label{Tab:nh3}\\
\hline
Rest frequency    &               & $\int S_{\nu} {\rm d}\upsilon$  & $\upsilon_{\rm lsr}$     & $\Delta \upsilon$ & $S_{\nu}$   &  $\sigma$   &  \\
 (MHz)    & $(J,K)$ & (Jy~\kms)        & (\kms)      & (\kms)     & (Jy)         &  (Jy)       & notes \\
\hline
\endfirsthead
\caption{continued.} \\
\hline
Rest frequency    &               & $\int S_{\nu} {\rm d}\upsilon$  & $\upsilon_{\rm lsr}$     & $\Delta \upsilon$ & $S_{\nu}$   &  $\sigma$   &  \\
 (MHz)    & $(J,K)$ & (Jy~\kms)        & (\kms)      & (\kms)     & (Jy)         &  (Jy)       & notes \\
\hline
\endhead
\hline  \\
\endfoot
\hline
\endlastfoot
\multicolumn{8}{l}{NH$_{3}$ non-metastable}\\
\hline
 18017.3 &  (7,3)           & $1.12\pm0.02$ & $3.4\pm0.1$ & $6.4\pm0.1$ & $0.165\pm0.005$ & $0.005$ & HC    \\
 18391.5 &  (6,1)           & $0.28\pm0.02$ & $3.8\pm0.3$ & $6.7\pm0.6$ & $0.039\pm0.006$ & $0.006$ & HC    \\
 18499.3 &  (9,6)           & $0.94\pm0.03$ & $3.8\pm0.1$ & $6.1\pm0.2$ & $0.145\pm0.008$ & $0.008$ & HC    \\
 18808.5 &  (8,5)           & $<5.89$       &             &             &                 & $0.006$ & b     \\
 18884.6 &  (6,2)           & $1.27\pm0.03$ & $3.9\pm0.1$ & $6.5\pm0.2$ & $0.185\pm0.007$ & $0.006$ & HC    \\
 19218.4 &  (7,4)           & $1.76\pm0.02$ & $4.9\pm0.1$ & $7.3\pm0.1$ & $0.227\pm0.006$ & $0.006$ & HC    \\
 19757.5 &  (6,3)           & $4.82\pm0.07$ & $4.9\pm0.1$ & $8.3\pm0.1$ & $0.543\pm0.014$ & $0.014$ & HC    \\
 19838.3 &  (5,1)           & $1.93\pm0.05$ & $5.8\pm0.1$ & $8.2\pm0.3$ & $0.220\pm0.011$ & $0.011$ & HC    \\
 20371.5 &  (5,2)           & $3.65\pm0.07$ & $7.2\pm0.1$ & $9.4\pm0.2$ & $0.365\pm0.009$ & $0.006$ & HC    \\
 20719.2 &  (8,6)           & $<5.1410$     &             &             &                 & $0.008$ & HC, b \\
 20735.4 &  (9,7)           & $0.51\pm0.03$ & $5.9\pm0.2$ & $4.7\pm0.3$ & $0.102\pm0.011$ & $0.010$ & HC    \\
 20804.9 &  (7,5)           & $2.83\pm0.05$ & $7.5\pm0.1$ & $7.9\pm0.2$ & $0.334\pm0.008$ & $0.008$ & HC    \\
 20994.7 &  (6,4)           & $5.76\pm0.04$ & $7.4\pm0.1$ & $9.1\pm0.1$ & $0.595\pm0.010$ & $0.010$ & HC    \\
 21070.7 &  (11,9)          & $0.28\pm0.03$ & $6.9\pm0.3$ & $6.2\pm0.7$ & $0.043\pm0.007$ & $0.008$ & HC    \\
 21134.3 &  (4,1)           & $3.32\pm0.04$ & $7.2\pm0.1$ & $8.2\pm0.1$ & $0.381\pm0.009$ & $0.008$ & HC    \\
 21285.3 &  (5,3)           & $9.67\pm0.04$ & $5.9\pm0.1$ & $11.8\pm0.1$ &$0.773\pm0.007$ & $0.008$ & a     \\
 21703.4 &  (4,2)           & $17.16\pm0.09$&             &              &                & $0.016$ & a     \\
 22653.0 &  (5,4)           & $10.65\pm0.02$&             &              &                & $0.003$ & a     \\
 22688.3 &  (4,3)           & $22.20\pm0.02$&             &              &                & $0.004$ & a     \\
 22732.5 &  (6,5)           & $8.80\pm0.03$ &             &              &                & $0.005$ & a     \\
 22834.2 &  (3,2)           & $19.39\pm0.04$&             &              &                & $0.006$ & a     \\
 22925.0 &  (7,6)           & $8.99\pm0.04$ &             &              &                & $0.007$ & a     \\
 23098.8 &  (2,1)           & $26.22\pm0.04$&             &              &                & $0.006$ & a     \\
 23232.3 &  (8,7)           & $2.71\pm0.02$ & $5.8\pm0.1$ & $7.1\pm0.1$ & $0.358\pm0.006$ & $0.006$ & HC    \\
 23657.6 &  (9,8)           & $1.65\pm0.02$ & $7.5\pm0.1$ & $6.4\pm0.1$ & $0.244\pm0.005$ & $0.005$ & HC    \\
 24205.4 &  (10,9)          & $1.25\pm0.01$ & $7.1\pm0.1$ & $6.9\pm0.1$ & $0.171\pm0.004$ & $0.004$ & HC    \\
 24882.0 &  (11,10)         & $0.35\pm0.01$ & $7.3\pm0.1$ & $5.7\pm0.2$ & $0.058\pm0.004$ & $0.004$ & HC    \\
\hline
\multicolumn{8}{l}{NH$_{3}$ metastable}\\
\hline
 23694.5 & (1,1)            & $85.91\pm0.04$&             &             &                   & 0.004 & a   \\
 23722.6 & (2,2)            & $82.79\pm0.06$&             &             &                   & 0.006 & a   \\
 23870.1 & (3,3)            & $<104.32$     &             &             &                   & 0.008 & a, b\\
 24139.4 & (4,4)            & $41.12\pm0.03$&             &             &                   & 0.003 & a   \\
 24532.9 & (5,5)            & $31.84\pm0.04$&             &             &                   & 0.005 & a   \\
 25056.0 & (6,6)            & $32.80\pm0.04$&             &             &                   & 0.004 & a   \\
 25715.1 & (7,7)            & $13.29\pm0.03$&             &             &                   & 0.003 & a   \\
\hline
\multicolumn{8}{l}{$^{15}$NH$_{3}$ non-metastable}\\
\hline                                      
 21637.9 & (4,3)            & 0.59$\pm$0.04 & 5.3$\pm$0.2 & 5.8$\pm$0.5 & 0.095$\pm$0.013 & 0.013 & HC  \\
 21784.0 & (3,2)            & 0.41$\pm$0.06 & 6.5$\pm$0.4 & 6.3$\pm$1.3 & 0.062$\pm$0.013 & 0.013 & HC  \\
 22044.2 & (2,1)            & 0.40$\pm$0.05 & 5.9$\pm$0.4 & 7.3$\pm$1.2 & 0.052$\pm$0.011 & 0.011 & HC  \\
 \hline
 \multicolumn{8}{l}{$^{15}$NH$_{3}$ metastable}\\
 \hline
 22624.9 & (1,1)	    & 0.52$\pm$0.04 & 6.1$\pm$0.2 & 8.7$\pm$0.9 & 0.056$\pm$0.006 & 0.006 & HC	 \\
 22649.8 & (2,2)	    & 0.55$\pm$0.02 & 5.8$\pm$0.1 & 6.8$\pm$0.4 & 0.076$\pm$0.007 & 0.006 & HC	 \\
 22789.4 & (3,3)	    & 1.38$\pm$0.02 & 5.7$\pm$0.1 & 7.4$\pm$0.1 & 0.175$\pm$0.005 & 0.005 & HC	 \\
 23046.0 & (4,4)	    & 0.55$\pm$0.02 & 5.5$\pm$0.1 & 6.3$\pm$0.3 & 0.082$\pm$0.006 & 0.006 & HC	 \\
 23422.0 & (5,5)	    & 0.33$\pm$0.02 & 6.1$\pm$0.2 & 7.2$\pm$0.4 & 0.043$\pm$0.005 & 0.005 & HC	 \\
 23922.3 & (6,6)	    & 0.31$\pm$0.01 & 5.7$\pm$0.1 & 7.1$\pm$0.3 & 0.041$\pm$0.003 & 0.003 & HC	 \\
 25323.5 & (8,8)	    & $<0.19$	    &		  &		&		  & 0.004 & HC, b  \\
\hline
NH$_{2}$D &  $J_{k_{\rm a},k_{\rm c}}$                            &\multicolumn{6}{c}{}\\
\hline
 18807.9 &  $3_{1,3}{\rm s}-3_{0,3}{\rm a}$ &    $<5.84$    &             &             &                 & 0.006 & ortho, HC, b\\
 25023.8 &  $4_{1,4}{\rm a}-4_{0,4}{\rm s}$ & 0.19$\pm$0.01 & 6.9$\pm$0.2 & 5.7$\pm$0.5 & 0.031$\pm$0.004 & 0.005 & para, HC  \\
\end{longtable} 
\tablefoot{-- (a) the line originates not only from the HC but also contains emission from other componenets such as the CR, plateau, ER and HC(S). (b) this line is blended (see Table.~\ref{Tab:orilines}).}              
} 

\renewcommand{\tabcolsep}{0.15 cm}
\normalsize \longtab{3}{
\begin{longtable}{cccccccc}
\caption{The observed properties of CH$_{3}$OH transitions.}\label{Tab:ch3oh}                                                     \\
\hline
Rest frequency    &               & $\int S_{\nu} {\rm d}\upsilon$  & $\upsilon_{\rm lsr}$    & $\Delta \upsilon$          & $S_{\nu}$   &  $\sigma$   &                       \\
 (MHz)    & $J_{k_{\rm a}}$  & (Jy~\kms)        & (\kms)      & (\kms)     & (Jy)         &  (Jy)       & notes                \\
\hline
\endfirsthead
\caption{continued.}                                                                                                             \\
\hline
Rest frequency    &               & $\int S_{\nu} {\rm d}\upsilon$  & $\upsilon_{\rm lsr}$     & $\Delta \upsilon$         & $S_{\nu}$   &  $\sigma$   &                       \\
 (MHz)    & $J_{k_{\rm a}}$  & (Jy~\kms)        & (\kms)      & (\kms)     & (Jy)         &  (Jy)       & notes                \\
\hline
\endhead
\hline                                                                                                                           \\
\endfoot
\hline
\endlastfoot
CH$_{3}$OH     &\multicolumn{7}{c}{}                                                                                              \\
\hline                                      
 19967.4 & $2_{1} -3_{0} E$                  & 5.09$\pm$0.03  & 8.8$\pm$0.1 & 2.7$\pm$0.1 & 1.780$\pm$0.013  & 0.011 &  	 \\
 20171.1 & $11_{1} -10_{2} A^{+}$             & 3.36$\pm$0.03  & 8.1$\pm$0.1 & 4.6$\pm$0.1 & 0.688$\pm$0.008  & 0.007 &CR	 \\
 20346.8 & $17_{6} -18_{5} E$	            &0.23$\pm$0.03   & 8.1$\pm$0.2 & 4.9$\pm$0.7 & 0.045$\pm$0.007  & 0.006 &CR   \\
 20908.9 & $16_{-4} -15_{-5}E$		    &0.60$\pm$0.03   & 8.9$\pm$0.1 & 3.8$\pm$0.2 & 0.147$\pm$0.009  & 0.008 &CR   \\
 23121.0 & $9_{2}-10_{1} A^{+}$               & 3.17$\pm$0.02  & 7.0$\pm$0.1 & 4.6$\pm$0.1 & 0.651$\pm$0.008  & 0.006 &CR   \\
 23347.0 & $7_{1}-7_{1}A^{+}$ 	            &0.77$\pm$0.02   & 7.4$\pm$0.1 & 3.4$\pm$0.1 & 0.214$\pm$0.006  & 0.005 &CR	 \\
 23444.8 & $10_{1}-9_{2}A^{-}$	            & $<4.95$        &             &             &                  & 0.004 &CR, b\\
 24928.7 & $3_{2}-3_{1}E$	            & $<20.42$       &             &             &                  & 0.008 & M, b\\
 24933.5 & $4_{2}-4_{1}E$	            & 48.84$\pm$0.11 & 7.8$\pm$0.1 & 2.2$\pm$0.1 & 20.600$\pm$0.0558 & 0.064 & M   \\
 24934.4 & $2_{2}-2_{1}E$		    & 10.05$\pm$0.17 & 8.0$\pm$0.1 & 3.2$\pm$0.1 & 2.940$\pm$0.0691  & 0.054 &  M	 \\
 24959.1 & $5_{2}-5_{1}E$		    & 71.01$\pm$0.12 & 7.9$\pm$0.1 & 2.4$\pm$0.1 & 28.300$\pm$0.061 & 0.009 & M	 \\
 25018.1 & $6_{2}-6_{1}E$		    & 89.20$\pm$0.18 & 7.7$\pm$0.1 & 1.8$\pm$0.1 & 45.600$\pm$0.107 & 0.011 & M	 \\
 25124.9 & $7_{2}-7_{1}E$		    & 82.39$\pm$0.16 & 7.9$\pm$0.1 & 1.9$\pm$0.1 & 41.700$\pm$0.092 & 0.011 & M	 \\
 25294.4 & $8_{2}-8_{1}E$		    & 63.70$\pm$0.19 & 8.1$\pm$0.1 & 1.8$\pm$0.1 & 33.100$\pm$0.110 & 0.010 & M	 \\
 25541.4 & $9_{2}-9_{1}E$                    & 43.21$\pm$0.19  & 8.5$\pm$0.1 & 1.9$\pm$0.1 & 21.700$\pm$0.107 & 0.008 & M	 \\
 25787.1 & $26_{2}-26_{1}E$	            & 0.06$\pm$0.02  & 8.4$\pm$0.2 & 2.6$\pm$0.8 & 0.022$\pm$0.004  & 0.004 & CR  \\
 25878.3 & $10_{2}-10_{1}E$	            & 26.64$\pm$0.16 & 9.0$\pm$0.1 & 2.0$\pm$0.1 & 12.200$\pm$0.087 & 0.005 & M	 \\
\hline
CH$_{3}$OH (v$_{t}$=1)     &\multicolumn{7}{c}{}                                                                                   \\
\hline
 20970.7 & $10_{1}-11_{2}A^{+}$            & 0.72$\pm$0.02 & 8.8$\pm$0.1 & 4.2$\pm$0.2 & 0.160$\pm$0.008 & 0.007 & CR      \\
 21550.3 & $12_{2}-11_{1}A^{-}$            & 0.76$\pm$0.05 & 6.2$\pm$0.1 & 3.4$\pm$0.2 & 0.208$\pm$0.012 & 0.012 & CR      \\
 25322.8 & $9_{9}-9_{8}A$                 & $<0.18$       &             &             &                 & 0.004 & CR, b   \\
 26120.6 & $10_{1}-11_{2}A^{-}$            & 1.44$\pm$0.04 & 8.3$\pm$0.1 & 3.7$\pm$0.1 & 0.364$\pm$0.016 & 0.016 & CR      \\
\hline
$^{13}$CH$_{3}$OH     &\multicolumn{7}{c}{}                                                                                        \\
\hline
 23145.4 & $4_{0}-3_{1}E$                 & 0.11$\pm$0.01 & 6.8$\pm$0.2 & 3.5$\pm$0.4 & 0.030$\pm$0.005 & 0.005  & CR      \\
\end{longtable} 
\tablefoot{-- (M) these transitions are masers \citep[e.g.][]{1988A&A...198..267M,1980ApJ...236..481M}. (b) this line is blended (see Table.~\ref{Tab:orilines}).}              
} 

\renewcommand{\tabcolsep}{0.15 cm}
\normalsize \longtab{4}{
\begin{longtable}{cccccccc}
\caption{Observed properties of SO$_{2}$ and OCS transitions.}\label{Tab:so2}\\
\hline
Rest frequency    &               & $\int S_{\nu} {\rm d}\upsilon$  & $\upsilon_{\rm lsr}$   & $\Delta \upsilon$    & $S_{\nu}$   &  $\sigma$   &  \\
 (MHz)    & $J_{k_{a},k_{c}}$ & (Jy~\kms)        & (\kms)      & (\kms)     & (Jy)         &  (Jy)       & notes \\
\hline
\endfirsthead
\caption{continued.} \\
\hline
Rest frequency    &               & $\int S_{\nu} {\rm d}\upsilon$  & $\upsilon_{\rm lsr}$   & $\Delta \upsilon$    & $S_{\nu}$   &  $\sigma$   &  \\
 (MHz)    & $J_{k_{a},k_{c}}$ & (Jy~\kms)        & (\kms)      & (\kms)     & (Jy)         &  (Jy)       & notes \\
\hline
\endhead
\hline  \\
\endfoot
\hline
\endlastfoot
SO$_{2}$     &\multicolumn{7}{c}{}\\
\hline    
 20335.404 & $12_{3,9}-13_{2,12}$ & $<3.72$         &              &               &                   & 0.007 &  b            \\
 23414.248 & $5_{2,4}-6_{1,5}$   & $<12.96$        &               &              &                    & 0.005 &  b            \\
 24039.629 & $21_{5,17}-22_{4,18}$& $0.90\pm0.05$ & $10.6\pm0.7$ & $26.3\pm1.1$  & $0.032\pm0.004$ & 0.003 & plateau       \\
 24039.629 &                   & $0.60\pm0.05$ & $5.3\pm0.1$  & $8.9\pm0.4$   & $0.063\pm0.004$ & 0.003 &   HC          \\
 24083.479 & $8_{2,6}-9_{1,9}$   & $4.38\pm0.04$ & $9.2\pm0.1$  & $26.9\pm0.2$  & $0.153\pm0.003$ & 0.003 & plateau       \\
 24083.479 &                  & $0.91\pm0.03$ & $6.2\pm0.1$   & $8.2\pm0.2$   & $0.105\pm0.003$ & 0.003 & HC            \\
 25392.818 & $8_{1,7}-7_{2,6}$   & $9.62\pm0.01$ & $10.0\pm0.1$ & $27.7\pm0.1$  & $0.326\pm0.005$ & 0.005 & plateau        \\
 25392.818 &                  & $2.03\pm0.03$ & $6.1\pm0.1$   & $9.7\pm0.1$   & $0.198\pm0.005$ & 0.005 &  HC            \\
 25392.818 &                  & $0.58\pm0.01$ & $8.4\pm0.1$   & $2.3\pm0.1$   & $0.241\pm0.005$ & 0.005 & maser?         \\   
\hline
OCS        &               \multicolumn{7}{c}{}\\
\hline
 24325.9   & 2-1              & $1.08\pm0.04$ & $5.9\pm0.2$ & $14.8\pm0.6$  & $0.069\pm0.004$ & 0.004 &  plateau           \\
           &                  & $1.18\pm0.03$ & $7.7\pm0.1$ & $4.2\pm0.1$   & $0.263\pm0.004$ & 0.004 &  HC(S)             \\
\hline
\end{longtable} 
\tablefoot{-- (b) the line is blended (see Table.~\ref{Tab:orilines}).}              
} 

\renewcommand{\tabcolsep}{0.15 cm}
\normalsize \longtab{5}{
\begin{longtable}{cccccccc}
\caption{The observed properties of CH$_{3}$OCHO transitions.}\label{Tab:ch3ocho}\\
\hline
Rest frequency    &               & $\int S_{\nu} {\rm d}\upsilon$  & $\upsilon_{\rm lsr}$ & $\Delta \upsilon$ & $S_{\nu}$   &  $\sigma$   &  \\
 (MHz)    & $J_{k_{\rm a},k_{\rm c}}$ & (Jy~\kms)        & (\kms)      & (\kms)     & (Jy)         &  (Jy)       & notes \\
\hline
\endfirsthead
\caption{continued.} \\
\hline
Rest frequency    &               & $\int S_{\nu} {\rm d}\upsilon$  & $\upsilon_{\rm lsr}$ & $\Delta \upsilon$ & $S_{\nu}$   &  $\sigma$   &  \\
 (MHz)    & $J_{k_{\rm a},k_{\rm c}}$ & (Jy~\kms)        & (\kms)      & (\kms)     & (Jy)         &  (Jy)       & notes \\
\hline
\endhead
\hline  \\
\endfoot
\hline
\endlastfoot
CH$_{3}$OCHO     &\multicolumn{7}{c}{}\\
\hline                                      
22827.7 & $2_{1,2}-1_{1,1}$ E    & $<0.26      $ &             &             &                   & $0.006$ & CR, b\\
22828.1 & $2_{1,2}-1_{1,1}$ A    & $<0.26      $ &             &             &                   & $0.006$ & CR, b\\
24296.5 & $2_{0,2}-1_{0,1}$ E    & $0.24\pm0.01$ & $7.6\pm0.1$ & $3.8\pm0.2$ & $0.060\pm0.004$ & $0.004$ & CR  \\
24298.5 & $2_{0,2}-1_{0,1}$ A    & $0.23\pm0.01$ & $8.4\pm0.1$ & $3.5\pm0.2$ & $0.061\pm0.004$ & $0.004$ & CR  \\
24625.2 & $12_{3,9}-12_{3,10}$ E & $0.09\pm0.01$ & $7.8\pm0.1$ & $2.8\pm0.3$ & $0.032\pm0.004$ & $0.004$ & CR  \\
24649.4 & $12_{3,9}-12_{3,10}$ A & $0.07\pm0.01$ & $7.4\pm0.3$ & $3.3\pm0.6$ & $0.019\pm0.004$ & $0.004$ & CR  \\
25497.5 & $20_{5,15}-20_{5,16}$ E& $0.04\pm0.01$ & $8.6\pm0.3$ & $2.8\pm0.5$ & $0.013\pm0.004$ & $0.004$ & CR  \\
25530.7 & $20_{5,15}-20_{5,16}$ A& $0.06\pm0.01$ & $7.8\pm0.2$ & $3.0\pm0.8$ & $0.020\pm0.005$ & $0.005$ & CR  \\
25730.2 & $16_{4,12}-16_{4,13}$ E& $0.05\pm0.01$ & $6.7\pm0.1$ & $1.4\pm0.2$ & $0.033\pm0.004$ & $0.005$ & CR  \\
25759.8 & $16_{4,12}-16_{4,13}$ A& $0.07\pm0.01$ & $8.5\pm0.2$ & $3.0\pm0.5$ & $0.021\pm0.004$ & $0.004$ & CR  \\
26044.8 & $2_{1,1}-1_{1,0}$ E    & $0.29\pm0.02$ & $8.7\pm0.1$ & $2.5\pm0.2$ & $0.110\pm0.009$ & $0.009$ & CR  \\
26048.5 & $2_{1,1}-1_{1,0}$ A    & $0.29\pm0.02$ & $9.3\pm0.1$ & $2.6\pm0.2$ & $0.105\pm0.010$ & $0.010$ & CR  \\

\end{longtable} 
\tablefoot{--(b) this line is blended (see Table~\ref{Tab:orilines}).}              
} 

\renewcommand{\tabcolsep}{0.15 cm}
\normalsize \longtab{6}{
\begin{longtable}{cccccccc}
\caption{Observed properties of H$_{2}$O, HDO, CH$_{3}$CN, HC$_{3}$N, HC$_{5}$N, CH$_{3}$OCH$_{3}$, H$_{2}$CO and HNCO transitions.}\label{Tab:one}\\
\hline
Rest frequency    &               & $\int S_{\nu} {\rm d}\upsilon$  & $\upsilon_{\rm lsr}$ & $\Delta \upsilon$ & $S_{\nu}$   &  $\sigma$   &  \\
 (MHz)    & transitions   & (Jy~\kms)               & (\kms)      & (\kms)     & (Jy)       &  (Jy)       & notes \\
\hline
\endfirsthead
\caption{continued.} \\
\hline
Rest frequency    &               & $\int S_{\nu} {\rm d}\upsilon$  & $\upsilon_{\rm lsr}$ & $\Delta \upsilon$ & $S_{\nu}$   &  $\sigma$   &  \\
 (MHz)    & transitions   & (Jy~\kms)               & (\kms)      & (\kms)     & (Jy)      &  (Jy)       & notes \\
\hline
\endhead
\hline  \\
\endfoot
\hline
\endlastfoot
\hline
  \multicolumn{8}{c}{transitions without hfs}\\
\hline    
20460.0  & HDO ($3_{2,1}-4_{1,4}$)       & $<0.54$         &             &            &                    & 0.007 &  HC(S), b\\
21301.3  & HC$_{5}$N (8--7)            & $0.20\pm0.03$   & $8.6\pm0.4$ & $5.3\pm0.9$& $0.034\pm0.008$  & 0.009 &  CR      \\
22235.1  & H$_{2}$O ($6_{1,6}-5_{2,3}$)   & $>75000.00$    &             &            &                    & 0.014 & maser   \\
22965.6  & H$_{2}$CO ($9_{2,7}-9_{2,8}$)  & $0.41\pm0.02$  & $7.1\pm0.1$ & $6.0\pm0.4$& $0.064\pm0.005$  & 0.005 & HC(S)   \\
23390.0  & CH$_{3}$OCH$_{3}$ ($12_{3,10}-11_{4,7}$ AA)  & $<8.77$    &                &                    & 0.007 & CR, b   \\
23393.1  & CH$_{3}$OCH$_{3}$ ($12_{3,10}-11_{4,7}$ EE)  & $0.10\pm0.01$ & $7.3\pm0.2$ & $2.8\pm0.4$& $0.035\pm0.038$  & 0.007 & CR      \\
23963.9  & HC$_{5}$N (9--8)            & $0.14\pm0.01$   & $8.1\pm0.2$ & $4.8\pm0.5$ & $0.028\pm0.004$ & 0.004 & CR      \\
\hline
     \multicolumn{8}{c}{transitions with hfs}\\
\hline
 $\nu$    &               & $\int S_{\nu} {\rm d}\upsilon$\tablefootmark{(1)}  & $\upsilon_{\rm lsr}$ & $\Delta \upsilon$ & $\tau$   &  $\sigma$   &  \\
 (MHz)    & transitions   & (Jy~\kms)               & (\kms)      & (\kms)     &          &  (Jy)       & notes \\ 
\hline  
18196.2   &  HC$_{3}$N (2--1)          & $1.81\pm0.04$ & $6.6\pm0.1$   & $4.3\pm0.1$   & $< 0.1$ & 0.005 & HC         \\ 
18398.0   &  CH$_{3}$CN ($1_{0}-0_{0}$) & $1.57\pm0.05$ & $6.1\pm0.1$   & $6.4\pm0.4$   & $< 0.1$ & 0.006 & HC, b         \\
21981.6   &  HNCO ($1_{0,1}-0_{0,0}$)   & $0.99\pm0.09$ & $8.4\pm0.3$   & $6.4\pm0.7$   & $<0.1$  & 0.014 & HC(S)         \\
26124.6   &  CH$_{3}$CH$_{2}$CN ($3_{1,3}-2_{1,2}$) & $1.09\pm0.09$ & $5.9\pm0.5$   & $7.4\pm0.9$   & $0.2\pm0.1$  & 0.014 & HC \\
\hline
\end{longtable} 
\tablefoot{-- (1) Integrated intensities are calculated from integrating the spectral range with errors derived from $\sigma$. Note that integrated intensities for CH$_{3}$CN ($1_{0}-0_{0}$) are calculated by integrating the spectral range of the fitted hfs model because its observed spectrum is blended. (b) the line is blended (see Table~\ref{Tab:orilines})}              
} 

\clearpage
\section{Zoom-in plots of observed spectra}\label{app.b}
\begin{figure*}[!htbp]
\centering
\includegraphics[width = 0.8 \textwidth]{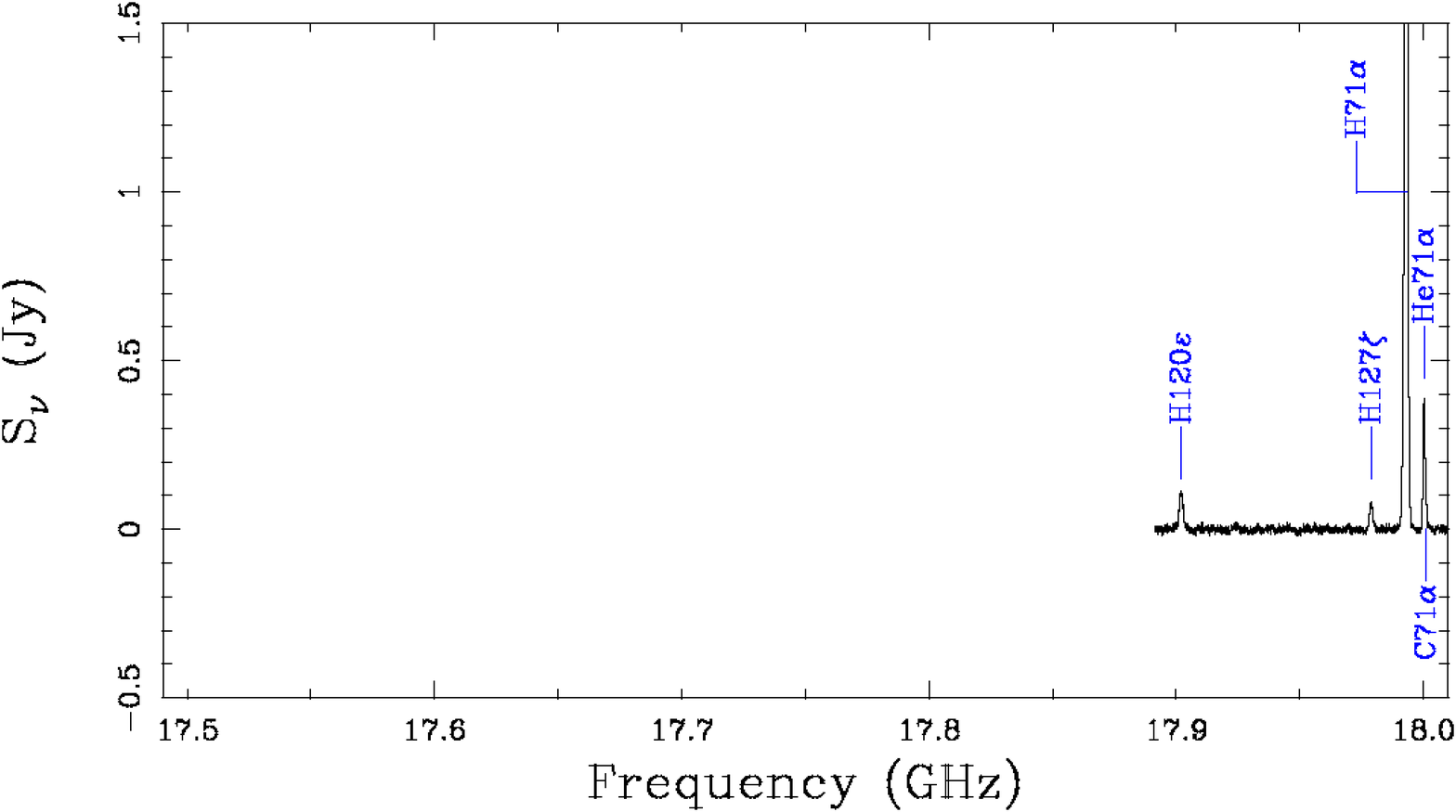}
\includegraphics[width = 0.8 \textwidth]{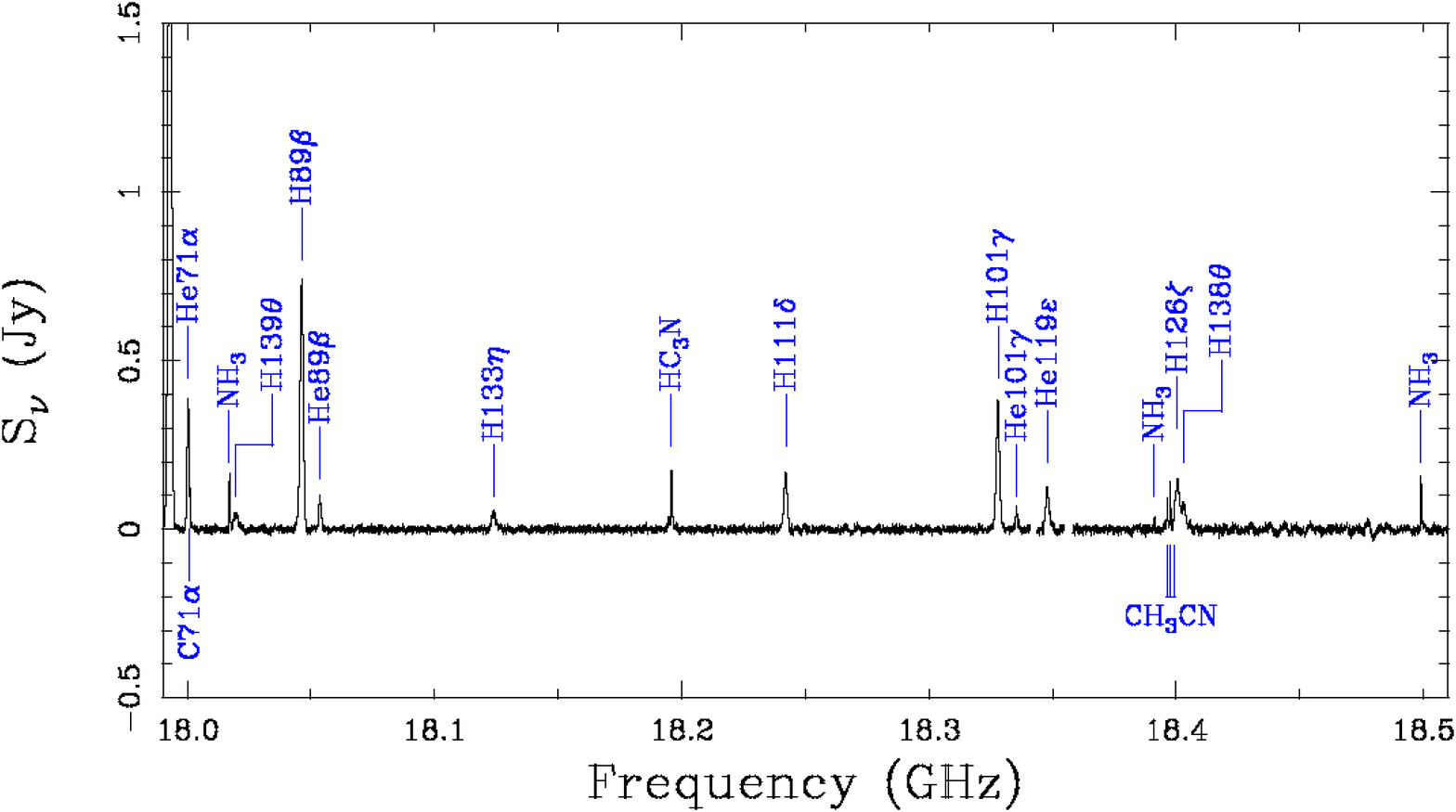}
\includegraphics[width = 0.8 \textwidth]{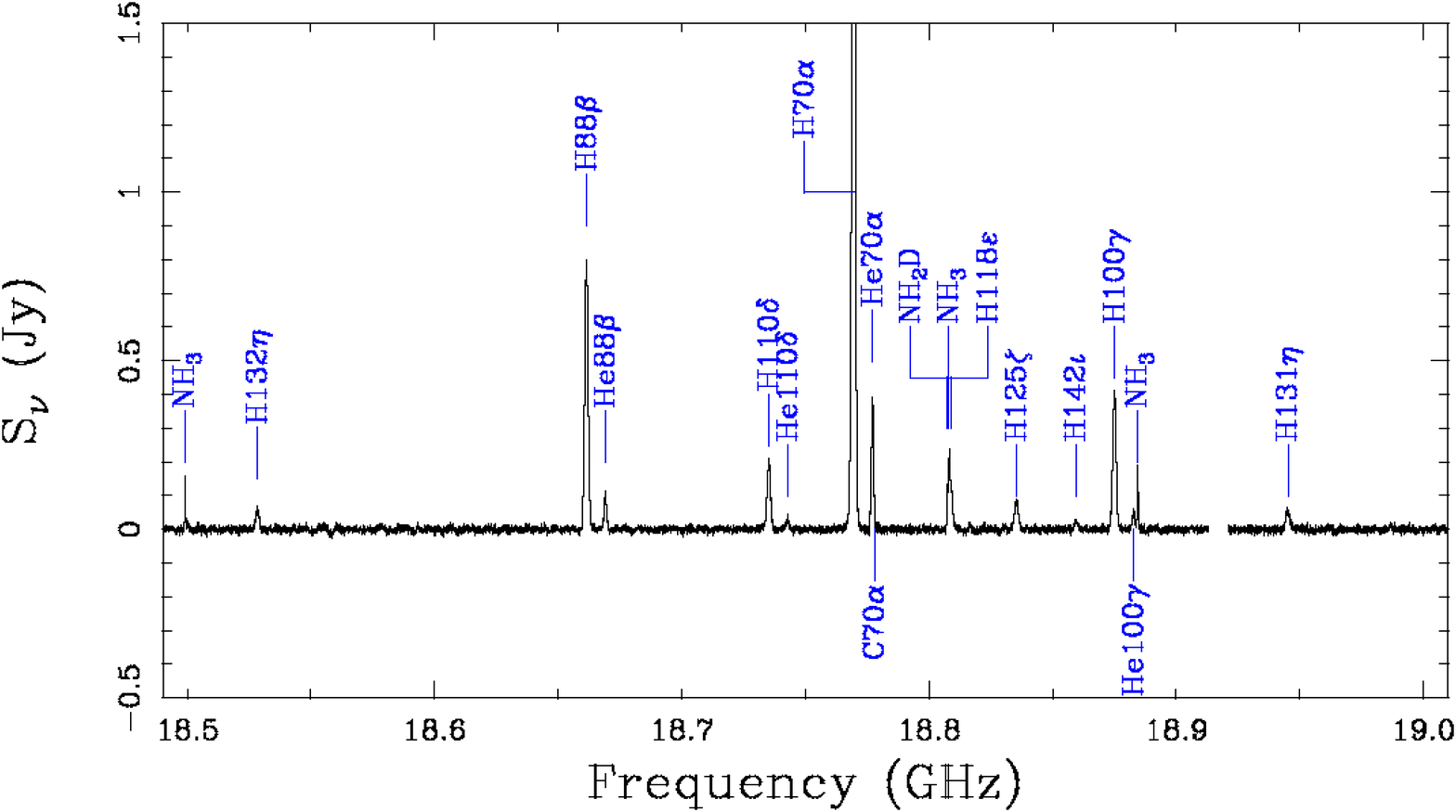}
\caption{{Observed spectrum of Orion KL from 17.9 to 26.2 GHz. The displayed frequency scale is based on the Local Standard of Rest velocity 0 \kms.} \label{Fig:ori}}
\end{figure*}

\begin{figure*}[!htbp]
\centering
\includegraphics[width = 0.8 \textwidth]{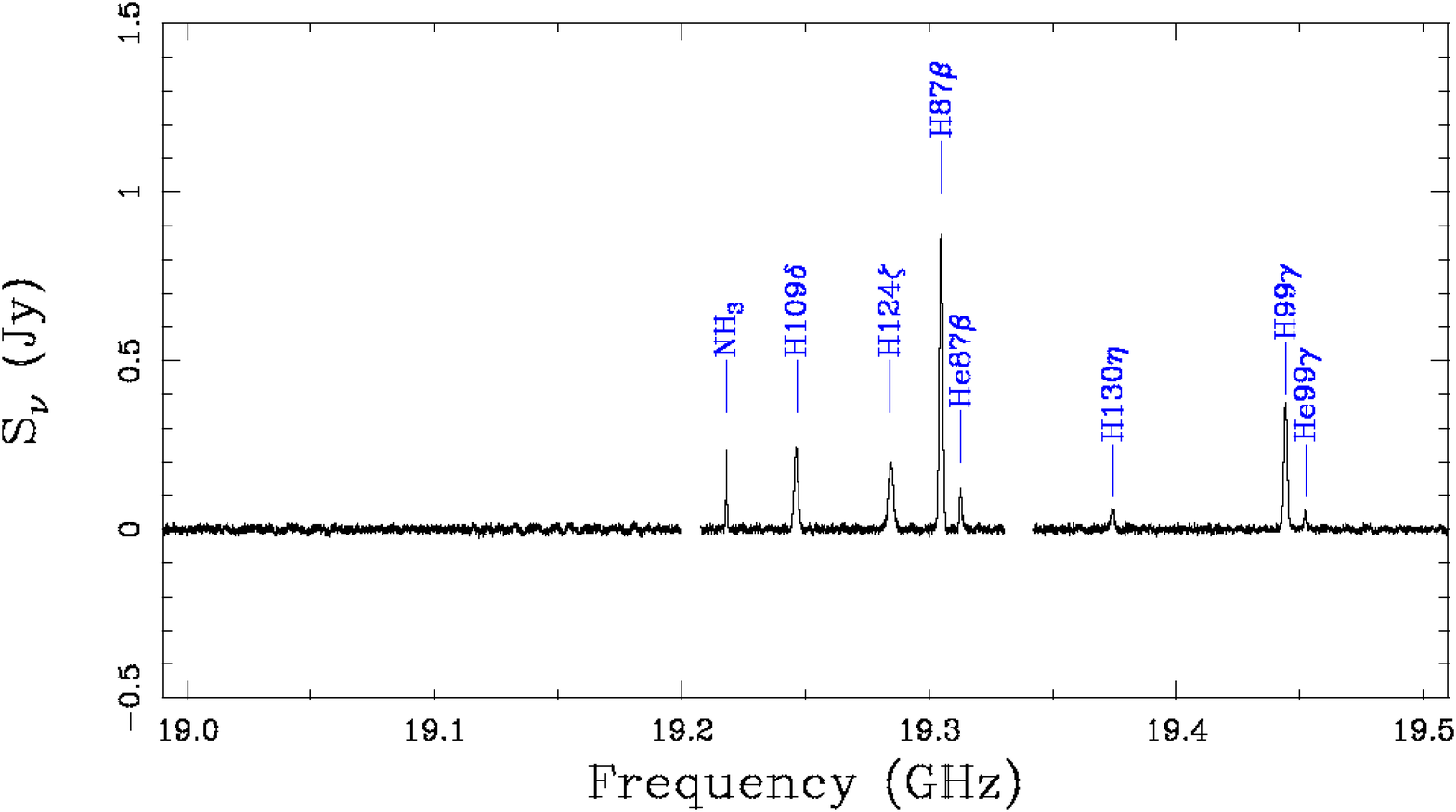}
\includegraphics[width = 0.8 \textwidth]{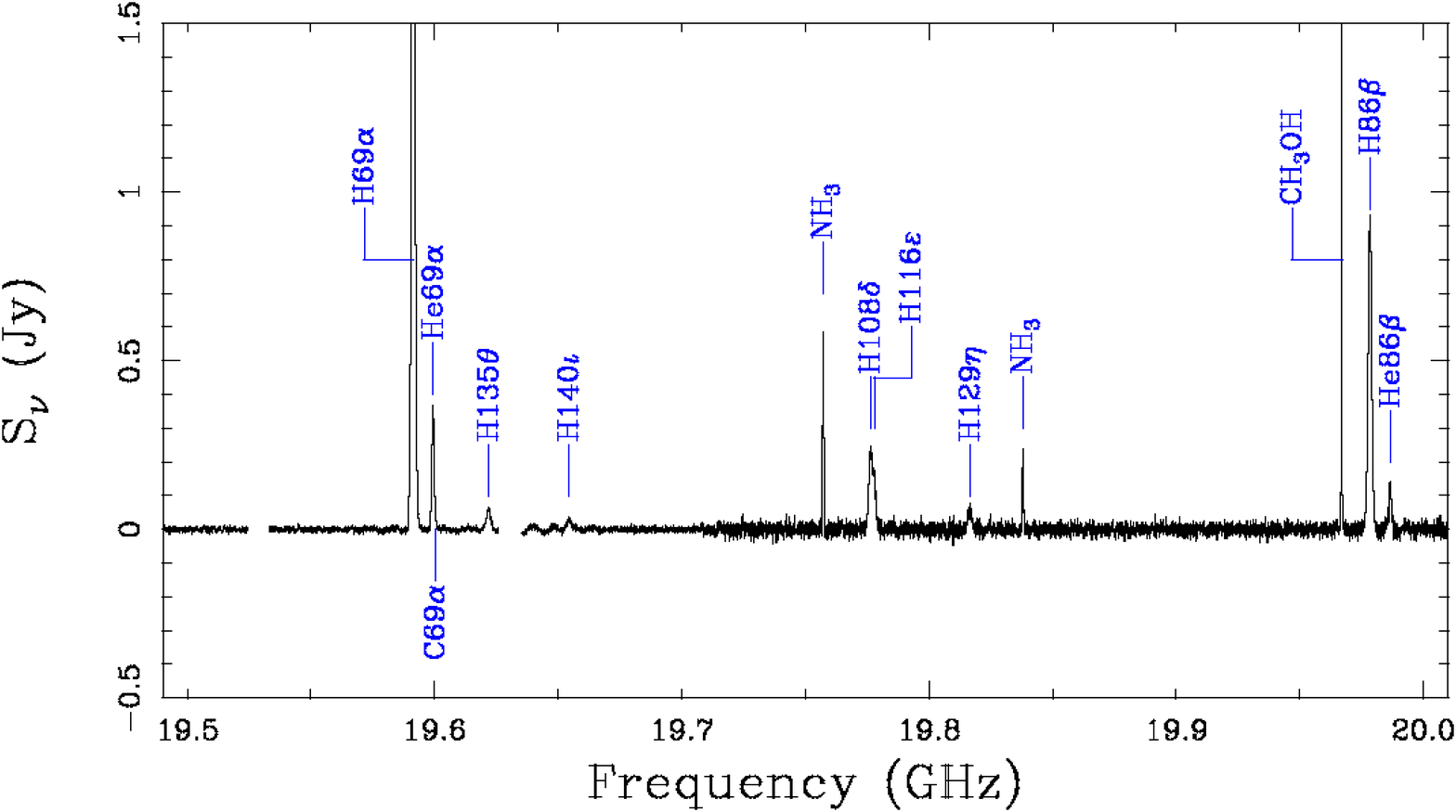}
\includegraphics[width = 0.8 \textwidth]{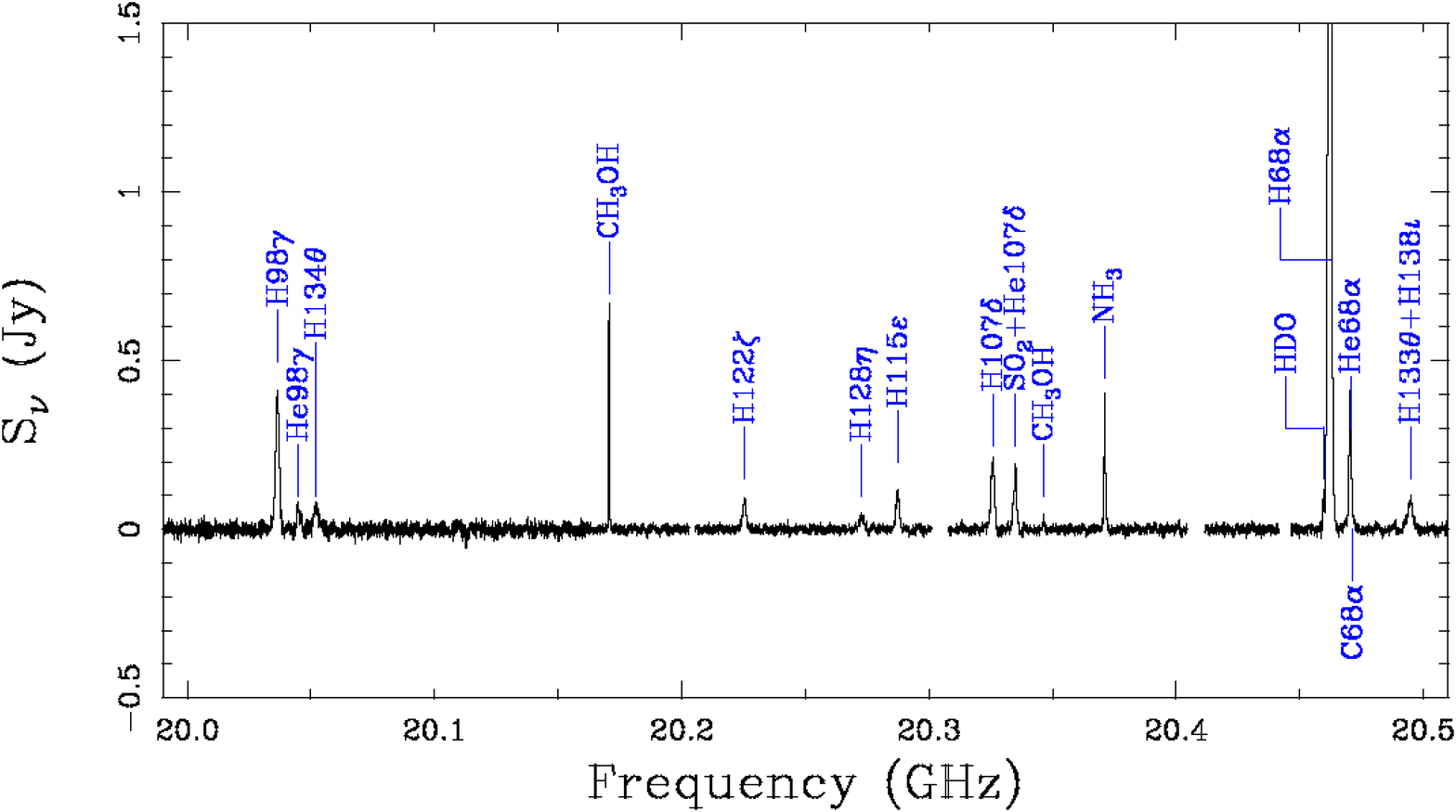}
\centerline{Fig. \ref{Fig:ori}. --- Continued.}
\end{figure*}

\begin{figure*}[!htbp]
\centering
\includegraphics[width = 0.8 \textwidth]{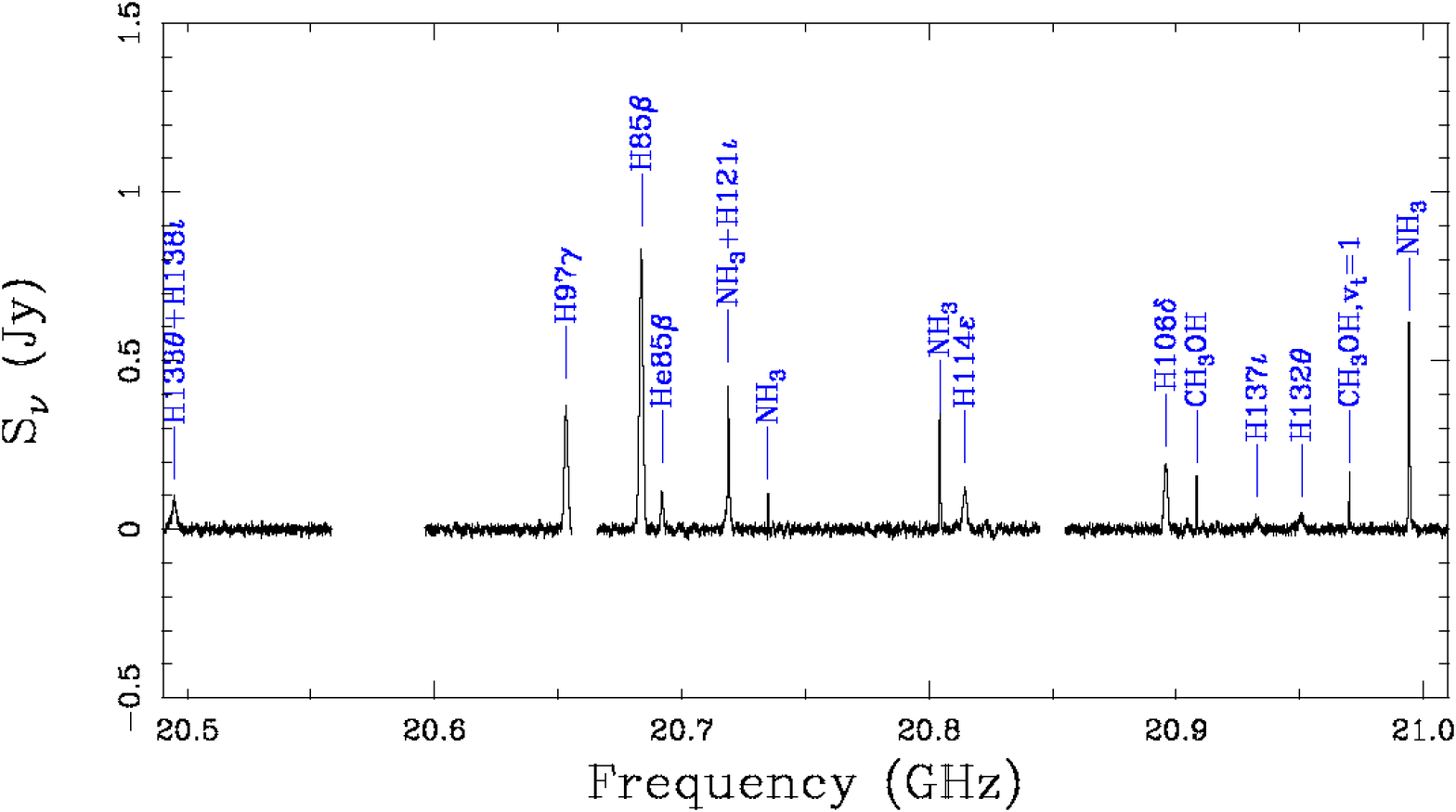}
\includegraphics[width = 0.8 \textwidth]{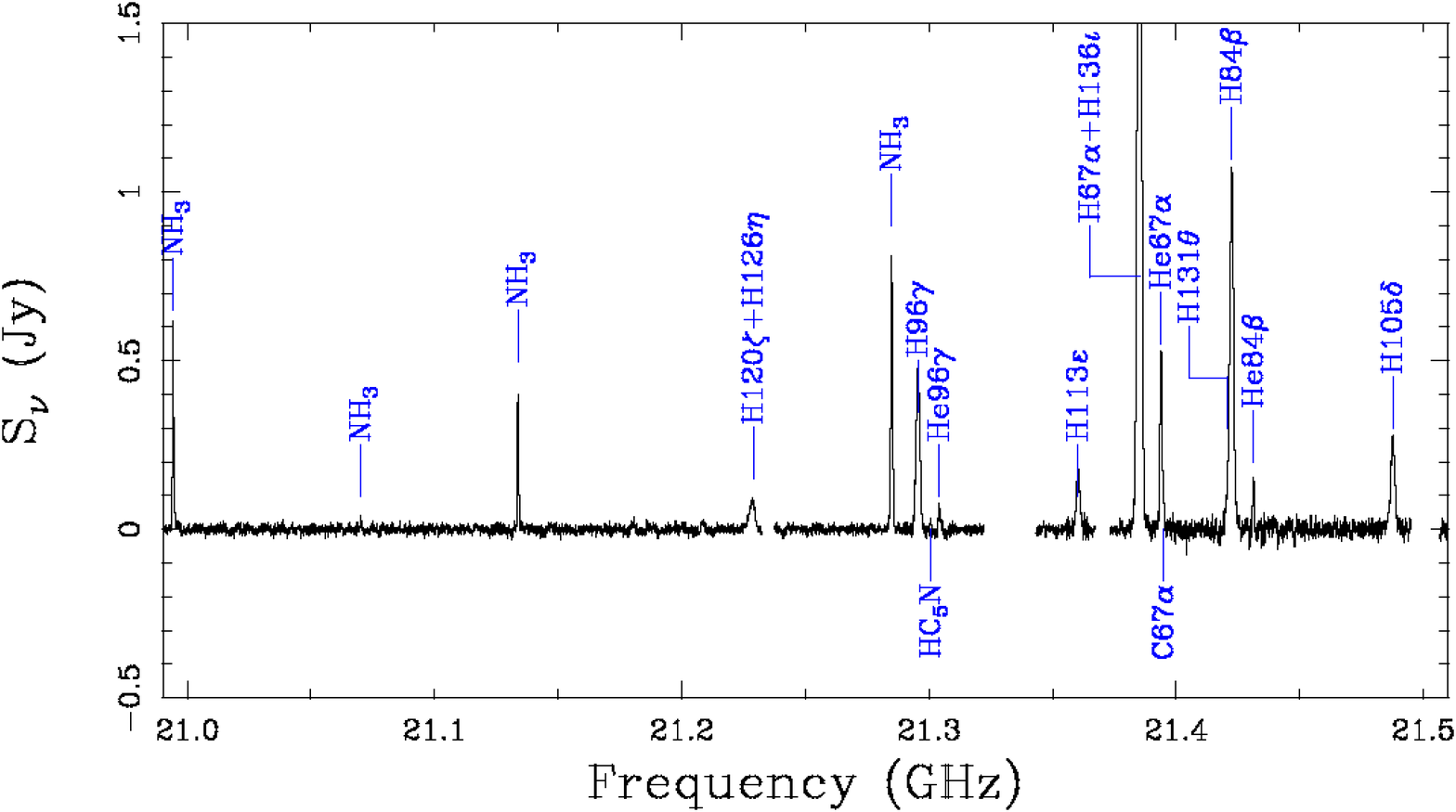}
\includegraphics[width = 0.8 \textwidth]{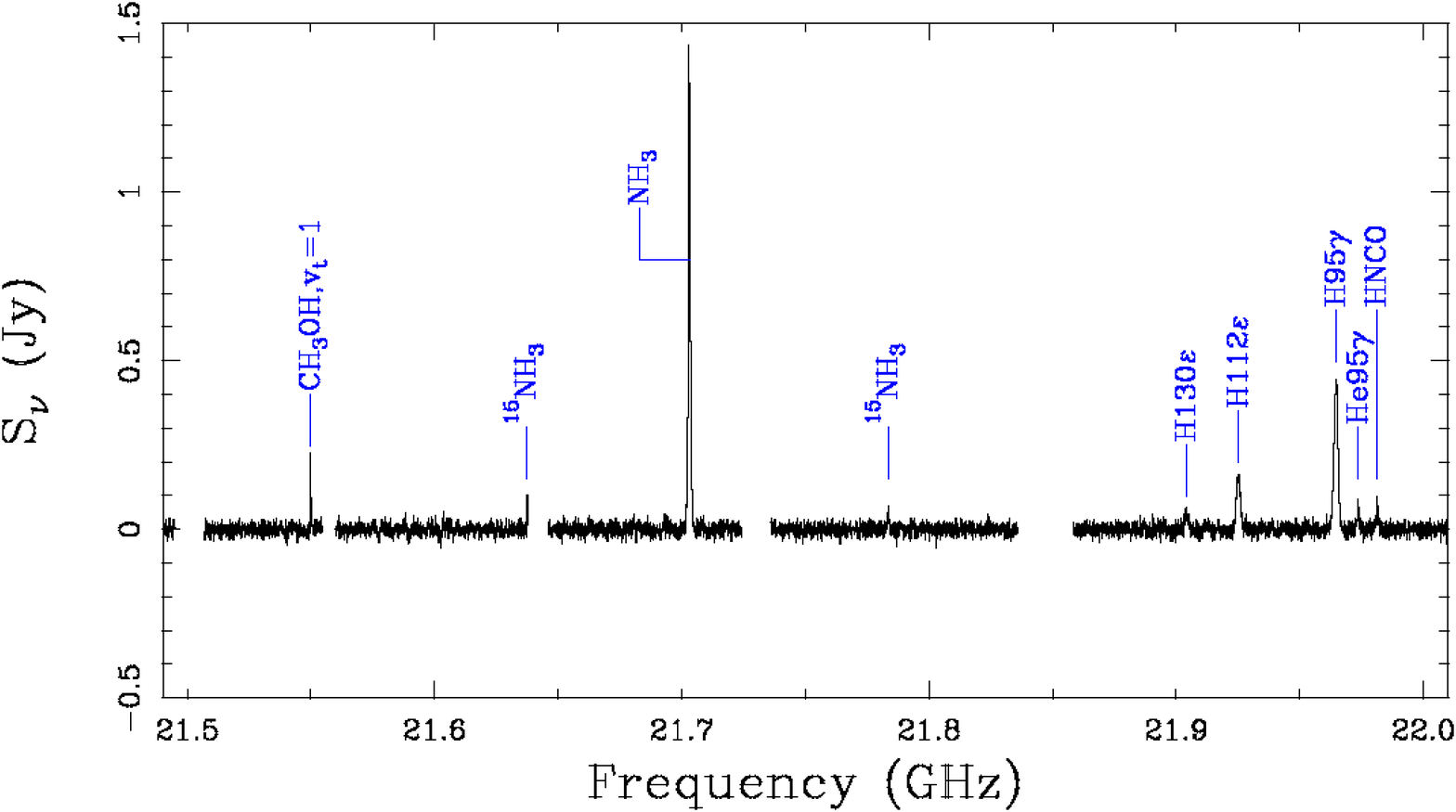}
\centerline{Fig. \ref{Fig:ori}. --- Continued.}
\end{figure*}

\begin{figure*}[!htbp]
\centering
\includegraphics[width = 0.8 \textwidth]{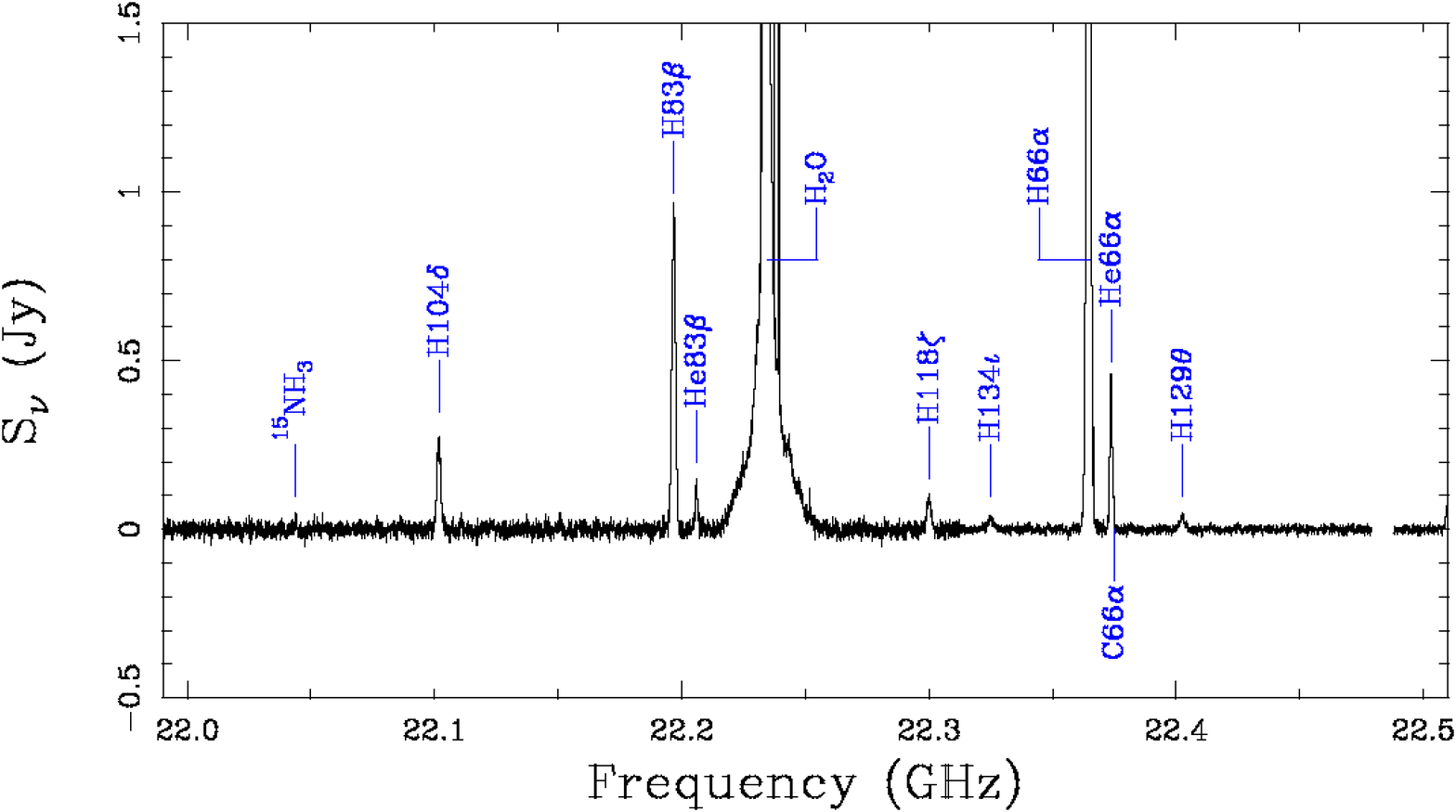}
\includegraphics[width = 0.8 \textwidth]{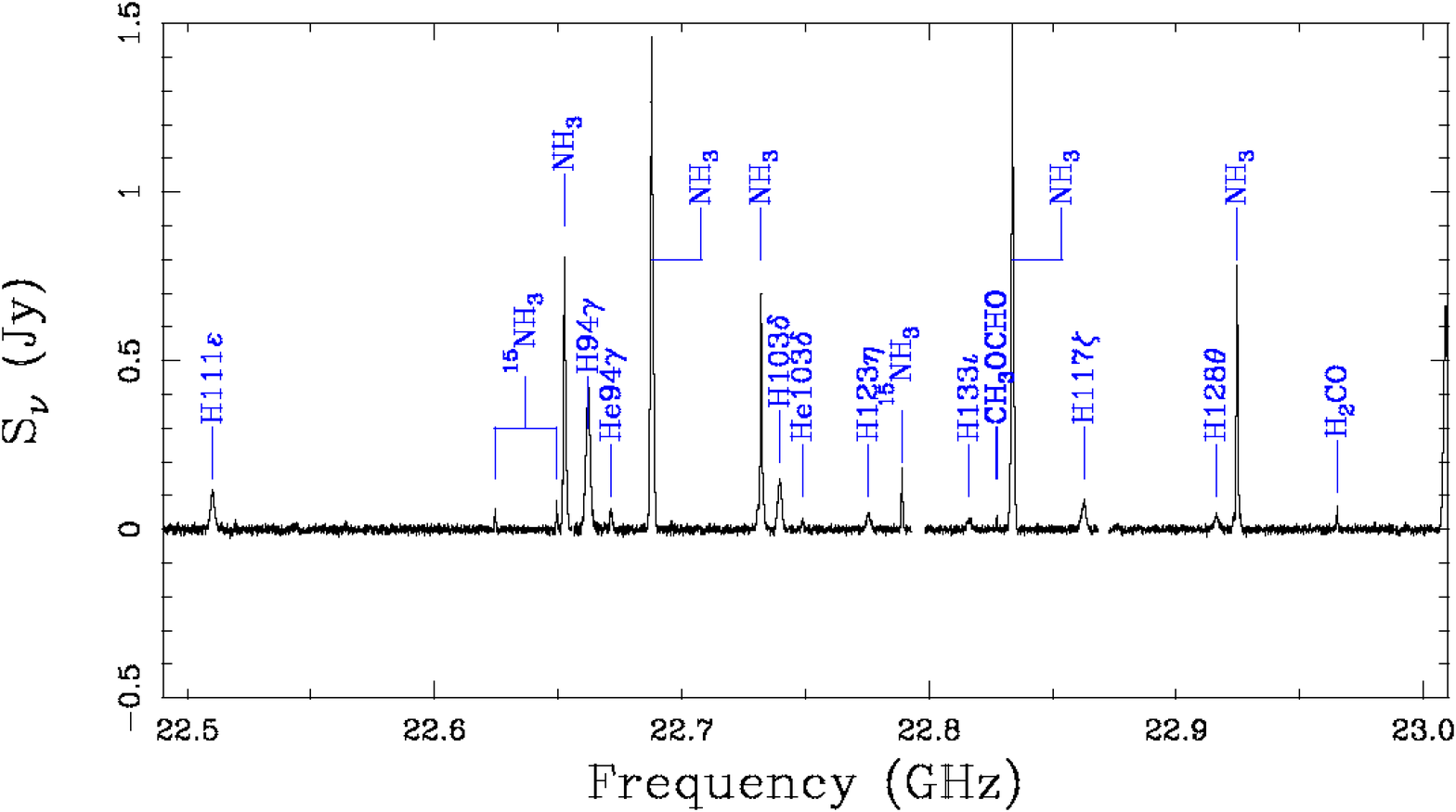}
\includegraphics[width = 0.8 \textwidth]{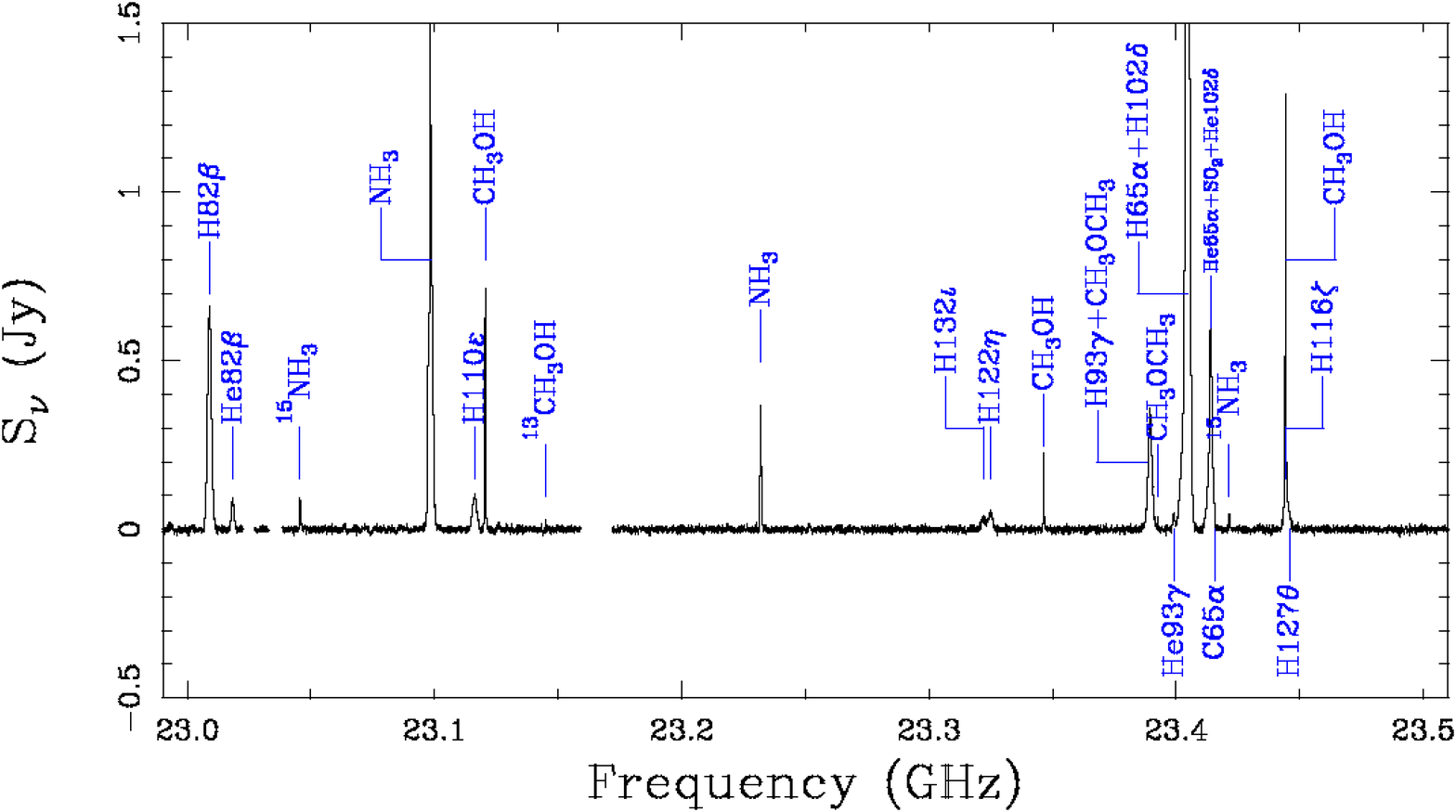}
\centerline{Fig. \ref{Fig:ori}. --- Continued.}
\end{figure*}

\begin{figure*}[!htbp]
\centering
\includegraphics[width = 0.8 \textwidth]{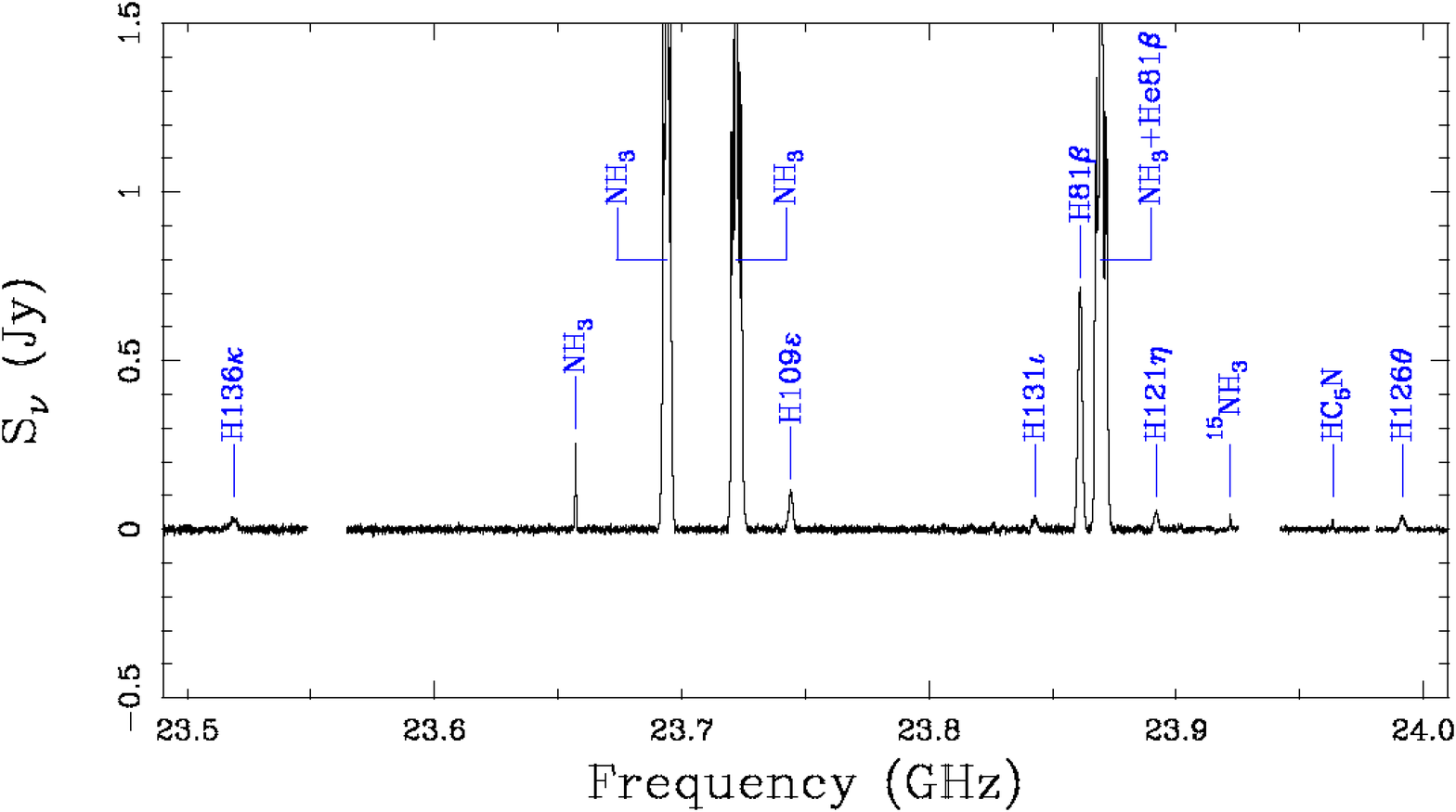}
\includegraphics[width = 0.8 \textwidth]{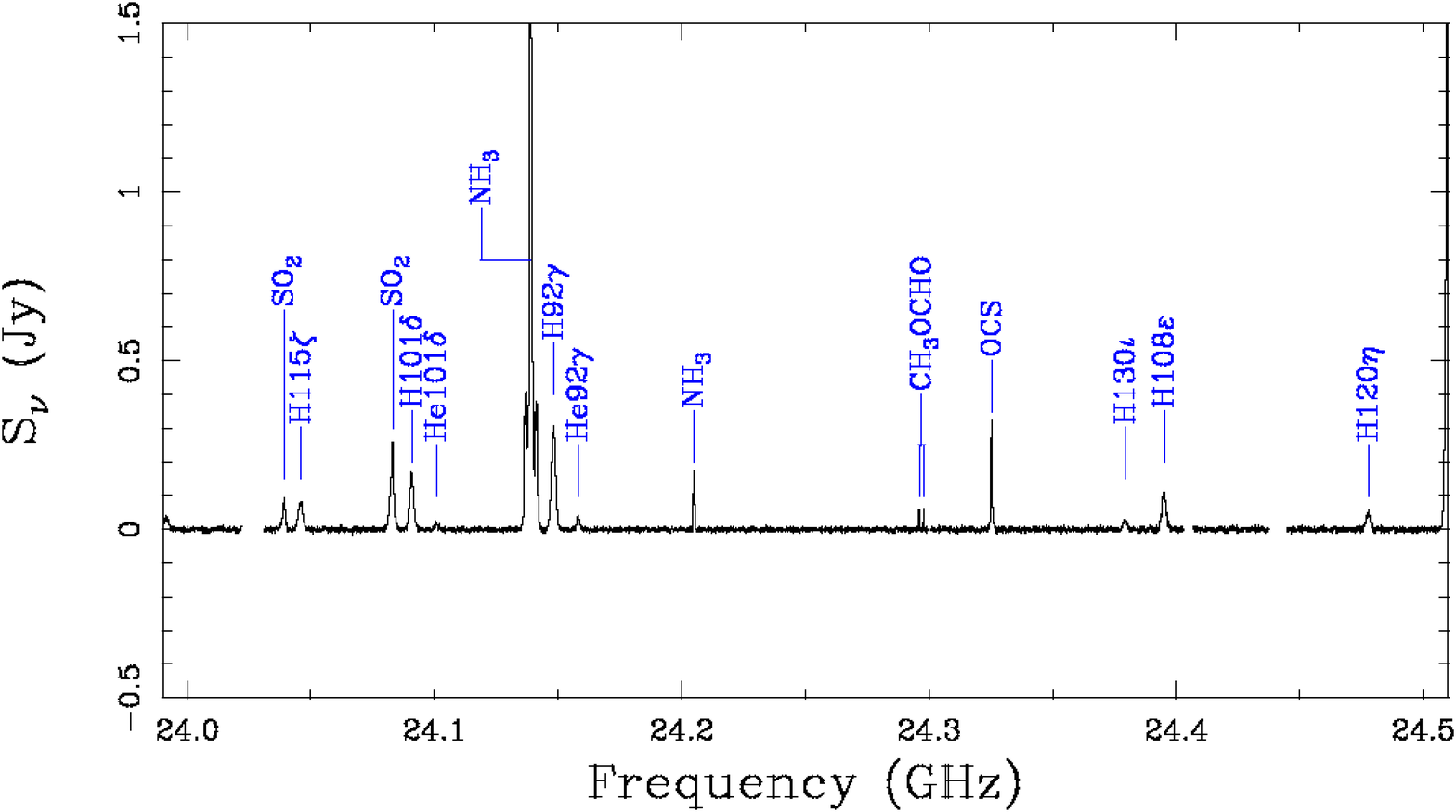}
\includegraphics[width = 0.8 \textwidth]{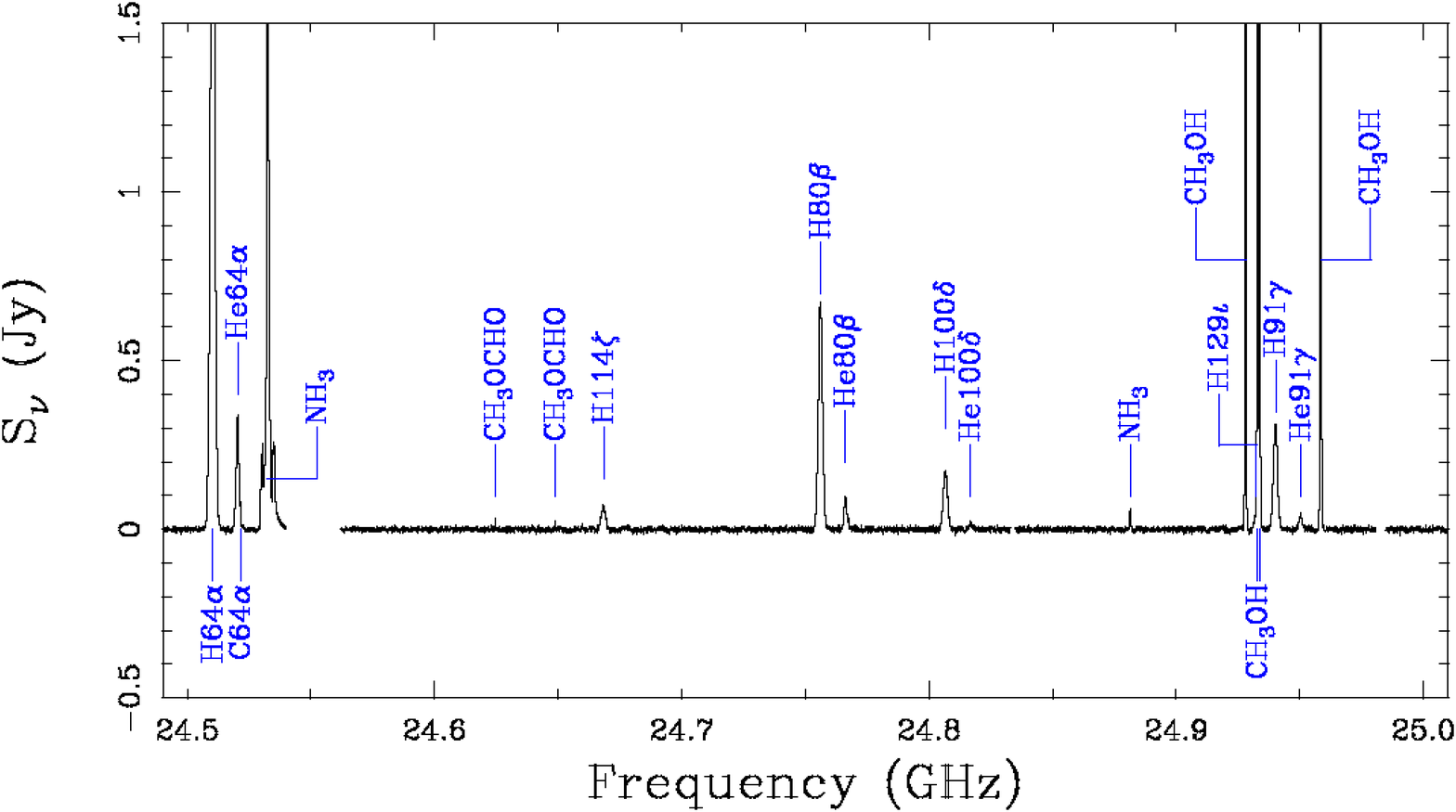}
\centerline{Fig. \ref{Fig:ori}. --- Continued.}
\end{figure*}

\begin{figure*}[!htbp]
\centering
\includegraphics[width = 0.8 \textwidth]{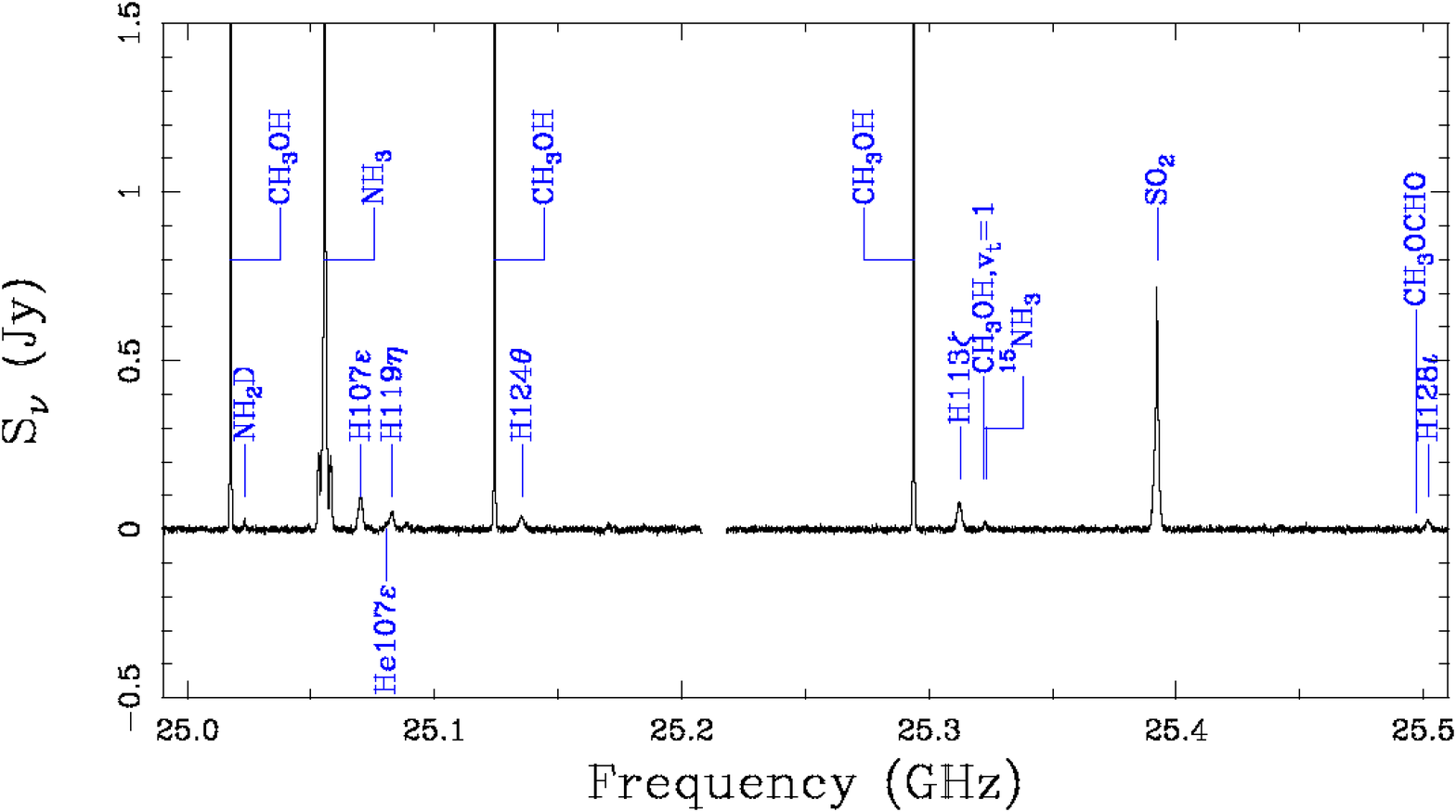}
\includegraphics[width = 0.8 \textwidth]{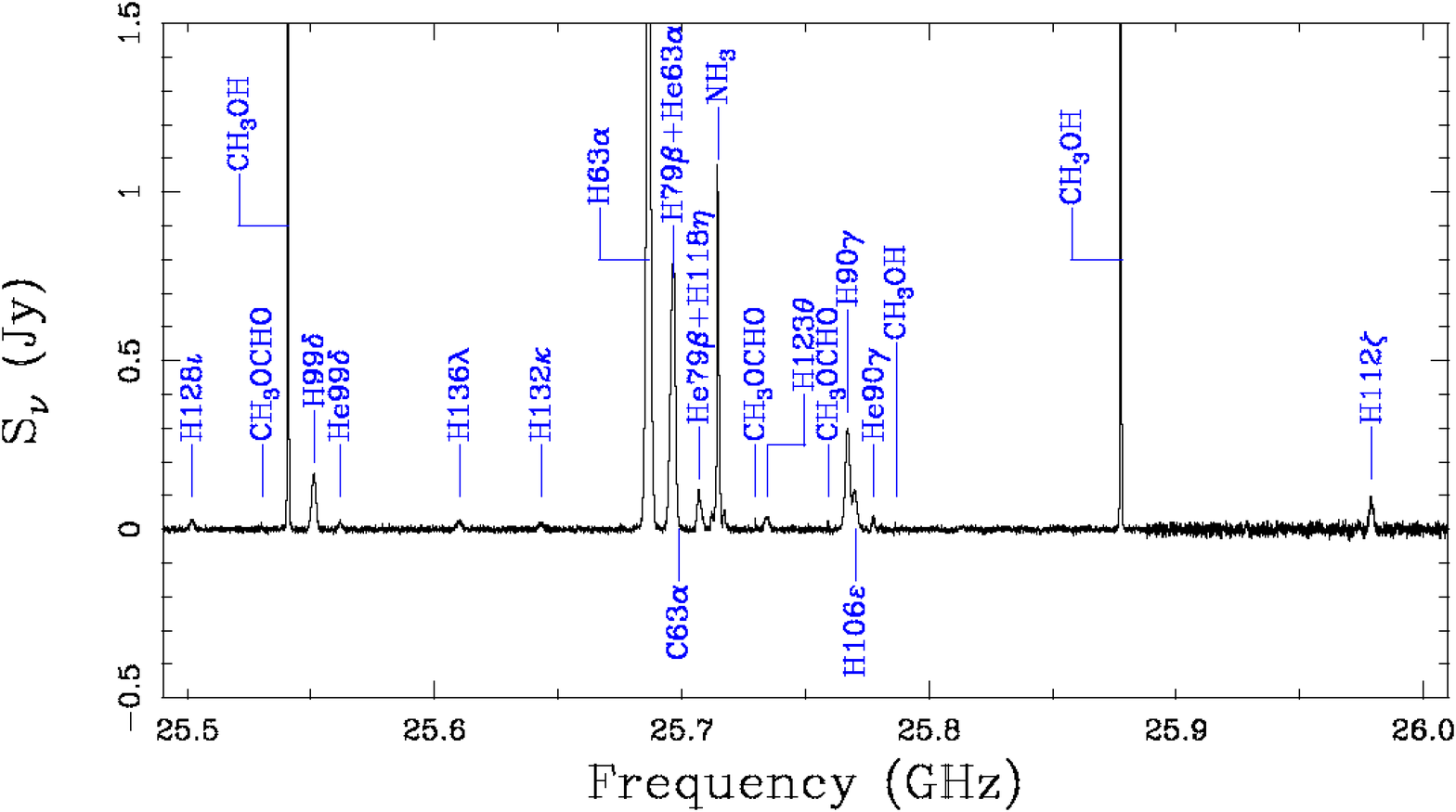}
\includegraphics[width = 0.8 \textwidth]{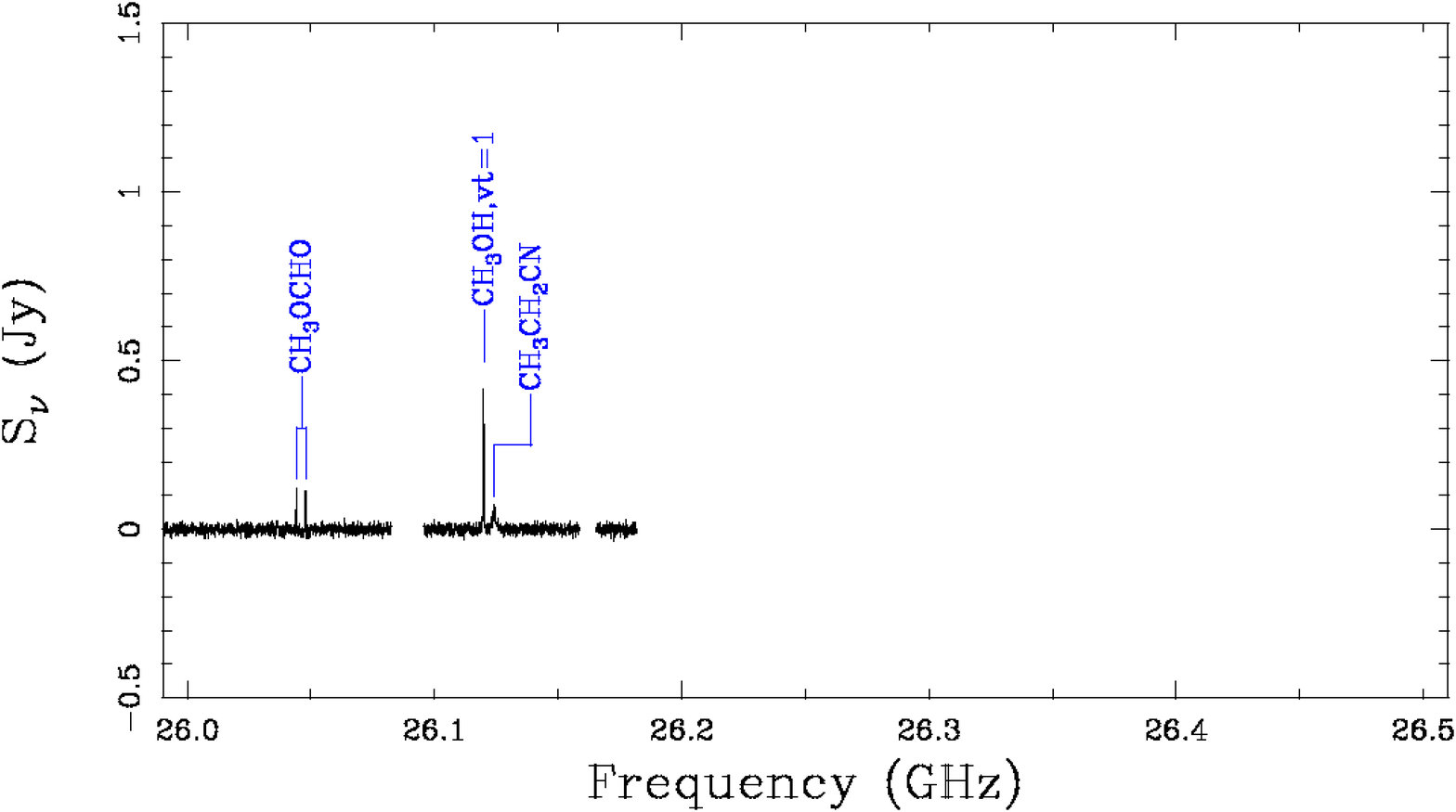}
\centerline{Fig. \ref{Fig:ori}. --- Continued.}
\end{figure*}

\clearpage
\begin{figure*}[!htbp]
\centering
\includegraphics[width = 0.35\textwidth]{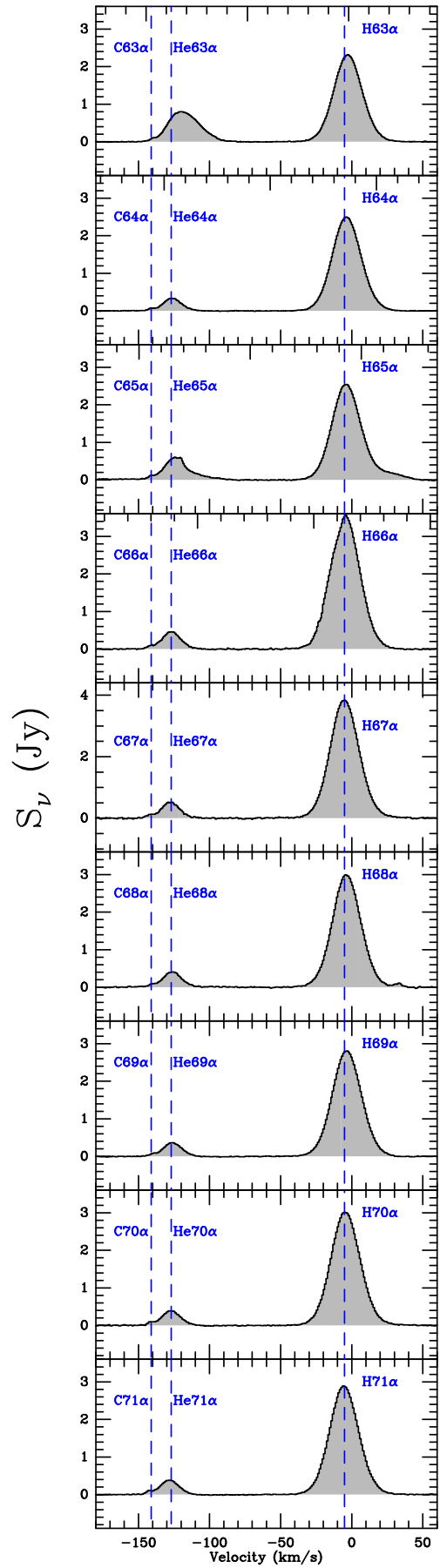}
\caption{{The observed H$\alpha$, He$\alpha$ and C$\alpha$ transitions indicated by dashed lines with $n$ and $\Delta n$ (see Sect.~\ref{sec.id}) also given. He63$\alpha$ is blended with H79$\beta$ while He65$\alpha$ is blended with SO$_{2}$ ($5_{2,4}-6_{1,5}$). The velocity scale refers to the respective H$\alpha$ line in each panel.} \label{Fig:alpha}}
\end{figure*}

\begin{figure*}[!htbp]
\centering
\includegraphics[width = 0.8\textwidth]{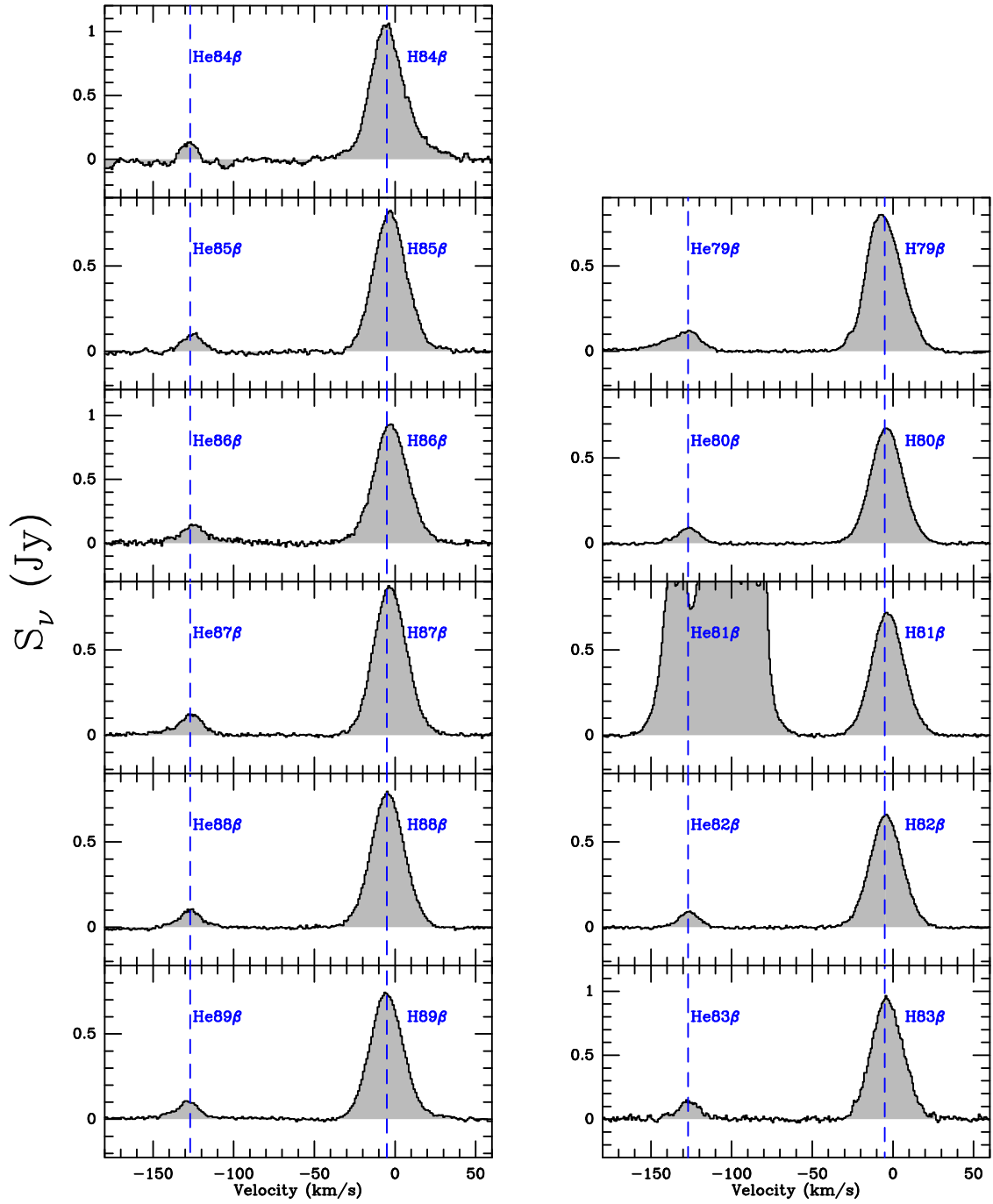}
\caption{{The observed H$\beta$ and He$\beta$ transitions indicated by dashed lines with $n$ and $\Delta n$ also given. H79$\beta$ is blended with He63$\alpha$ while H81$\beta$ is blended with NH$_{3}$ (3,3). The velocity scale refers to the respective H$\beta$ line in each panel.} \label{Fig:beta}}
\end{figure*}

\begin{figure*}[!htbp]
\centering
\includegraphics[width = 0.8\textwidth]{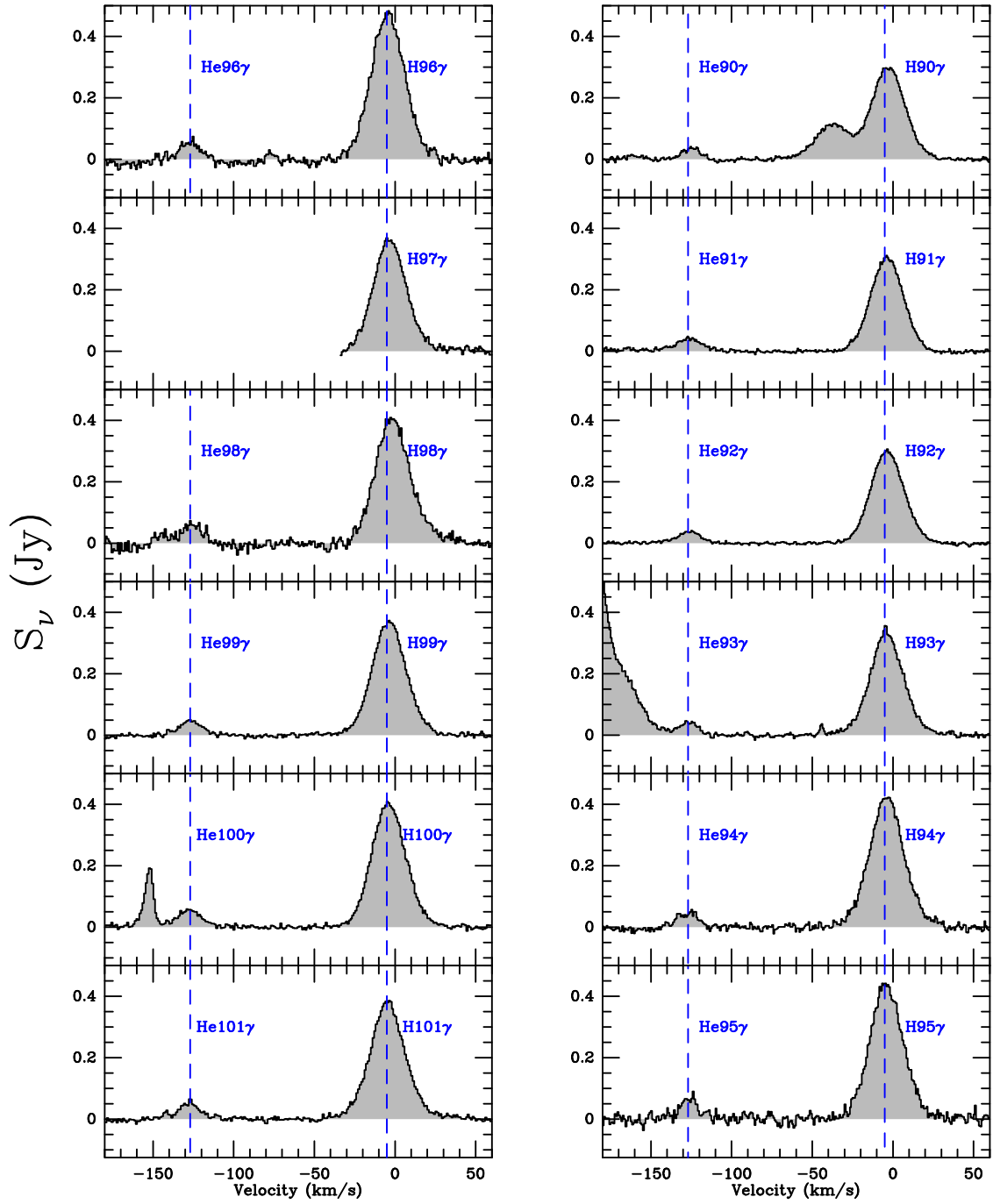}
\caption{{The observed H$\gamma$ and He$\gamma$ transitions indicated by dashed lines with $n$ and $\Delta n$ also given. The velocity scale refers to the respective H$\gamma$ line in each panel. The spectrum near He100$\gamma$ is NH$_{3}$ (6,2).} \label{Fig:gamma}}
\end{figure*}

\begin{figure*}[!htbp]
\centering
\includegraphics[width = 0.8\textwidth]{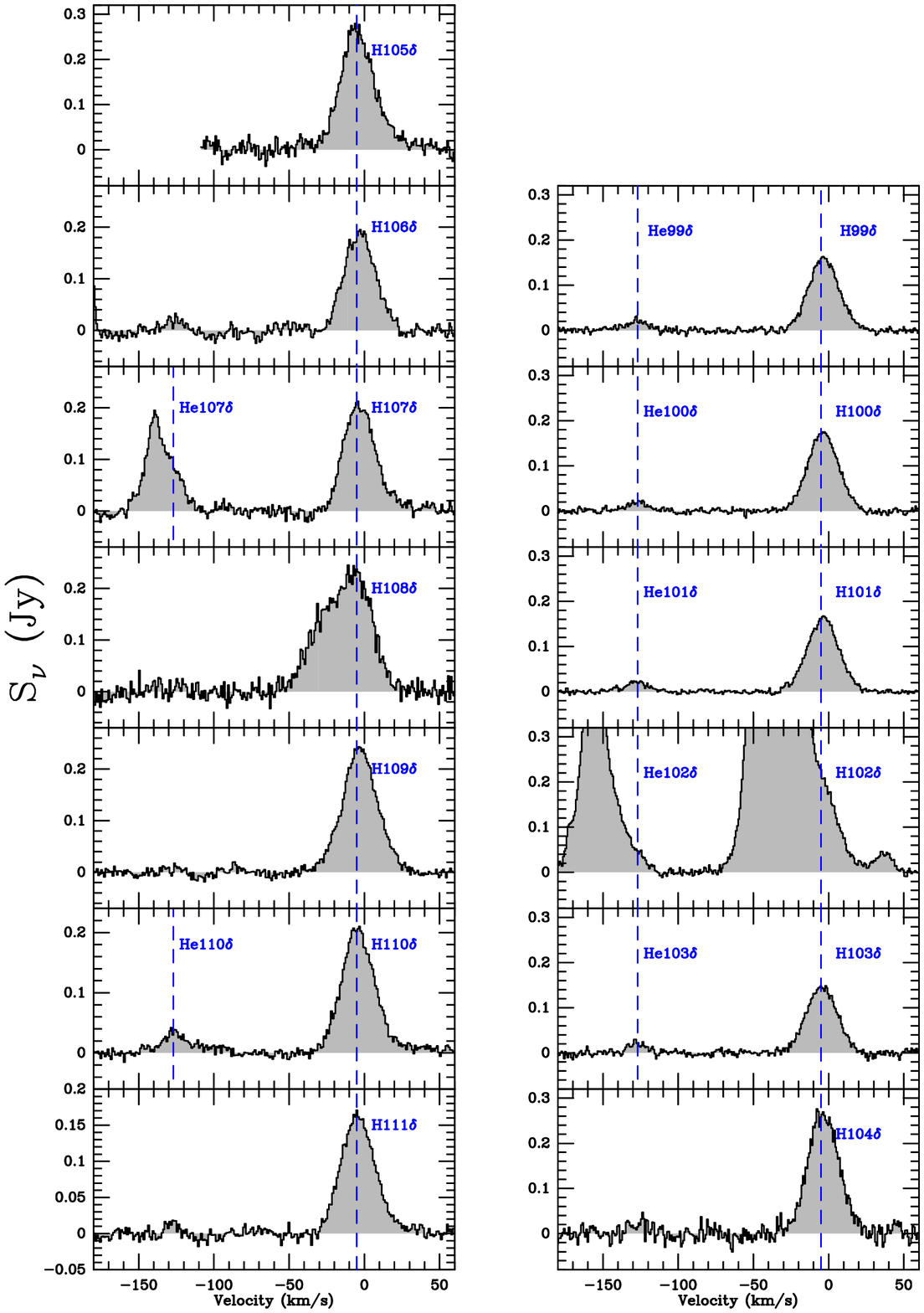}
\caption{{The observed H$\delta$ and He$\delta$ transitions indicated by dashed lines with $n$ and $\Delta n$ also given. The velocity scale refers to the respective H$\delta$ line in each panel.} \label{Fig:delta}}
\end{figure*}

\begin{figure*}[!htbp]
\centering
\includegraphics[width = 0.8\textwidth]{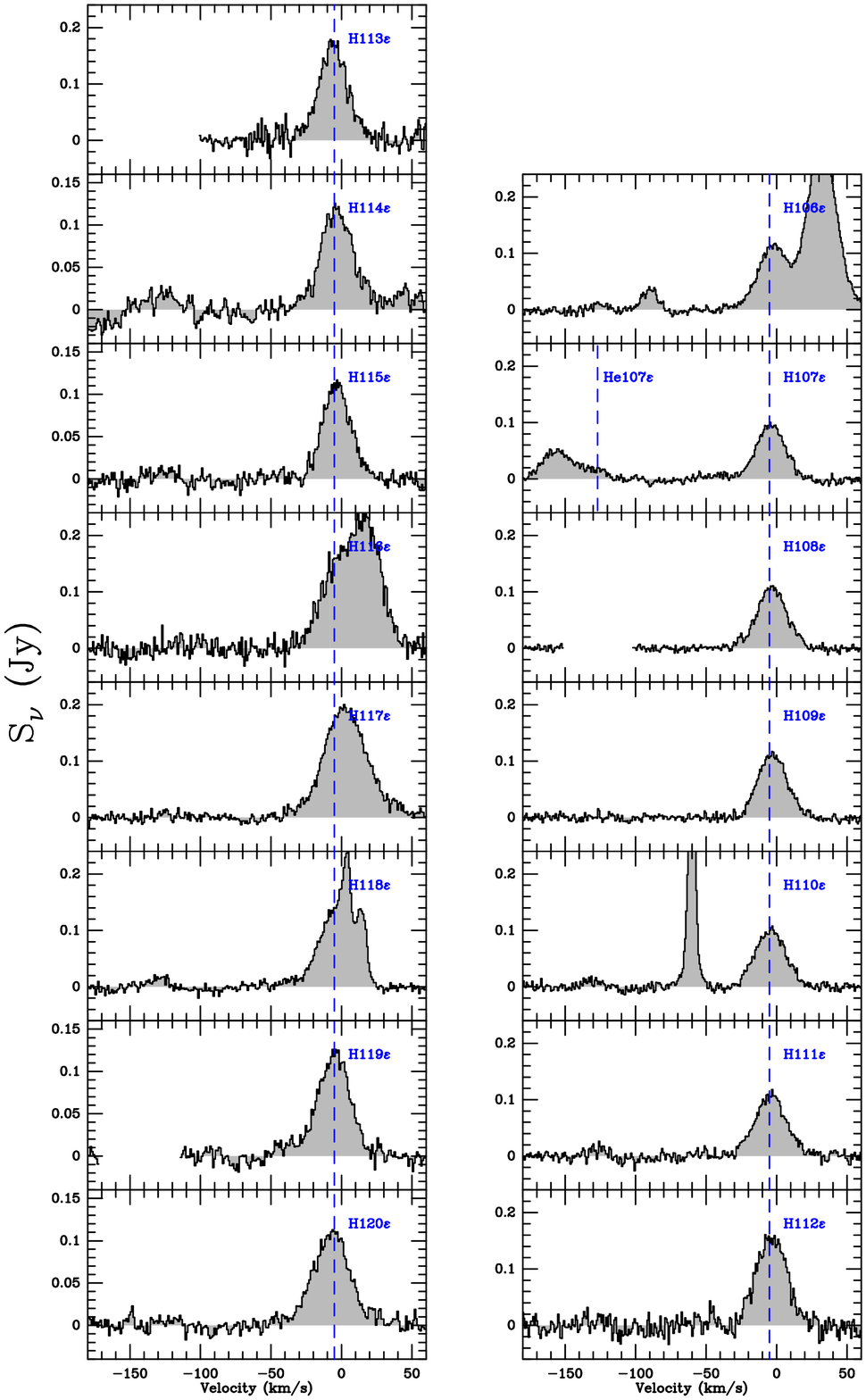}
\caption{{The observed H$\varepsilon$ and He$\varepsilon$ transitions indicated by dashed lines with $n$ and $\Delta n$ also given. The velocity scale refers to the respective H$\varepsilon$ line in each panel.} \label{Fig:epsilon}}
\end{figure*}

\begin{figure*}[!htbp]
\centering
\includegraphics[width = 0.8\textwidth]{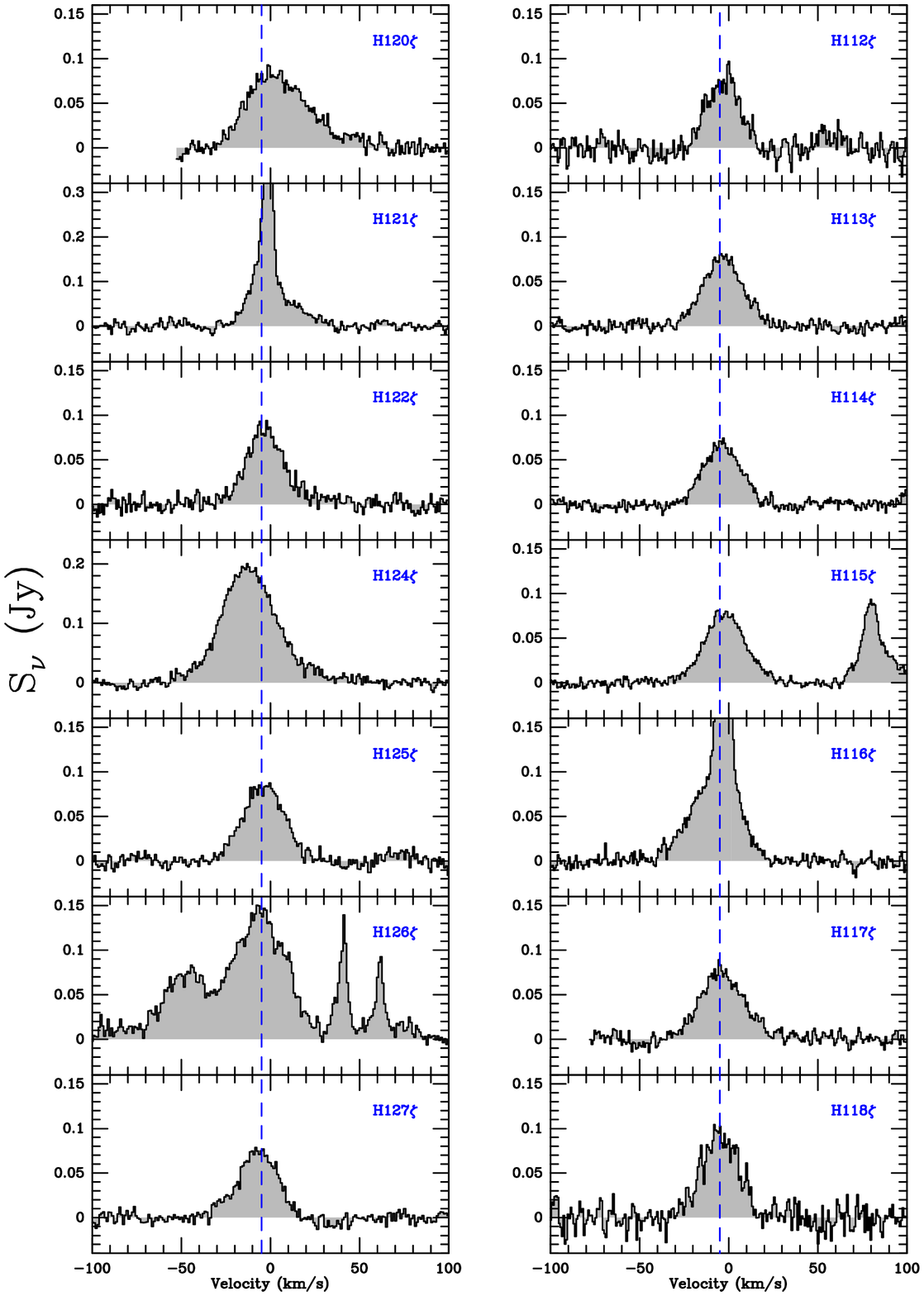}
\caption{{The observed H$\zeta$ transitions indicated by dashed lines with $n$ and $\Delta n$ also given. The velocity scale refers to the respective H$\zeta$ line in each panel.} \label{Fig:zeta}}
\end{figure*}

\begin{figure*}[!htbp]
\centering
\includegraphics[width = 0.8\textwidth]{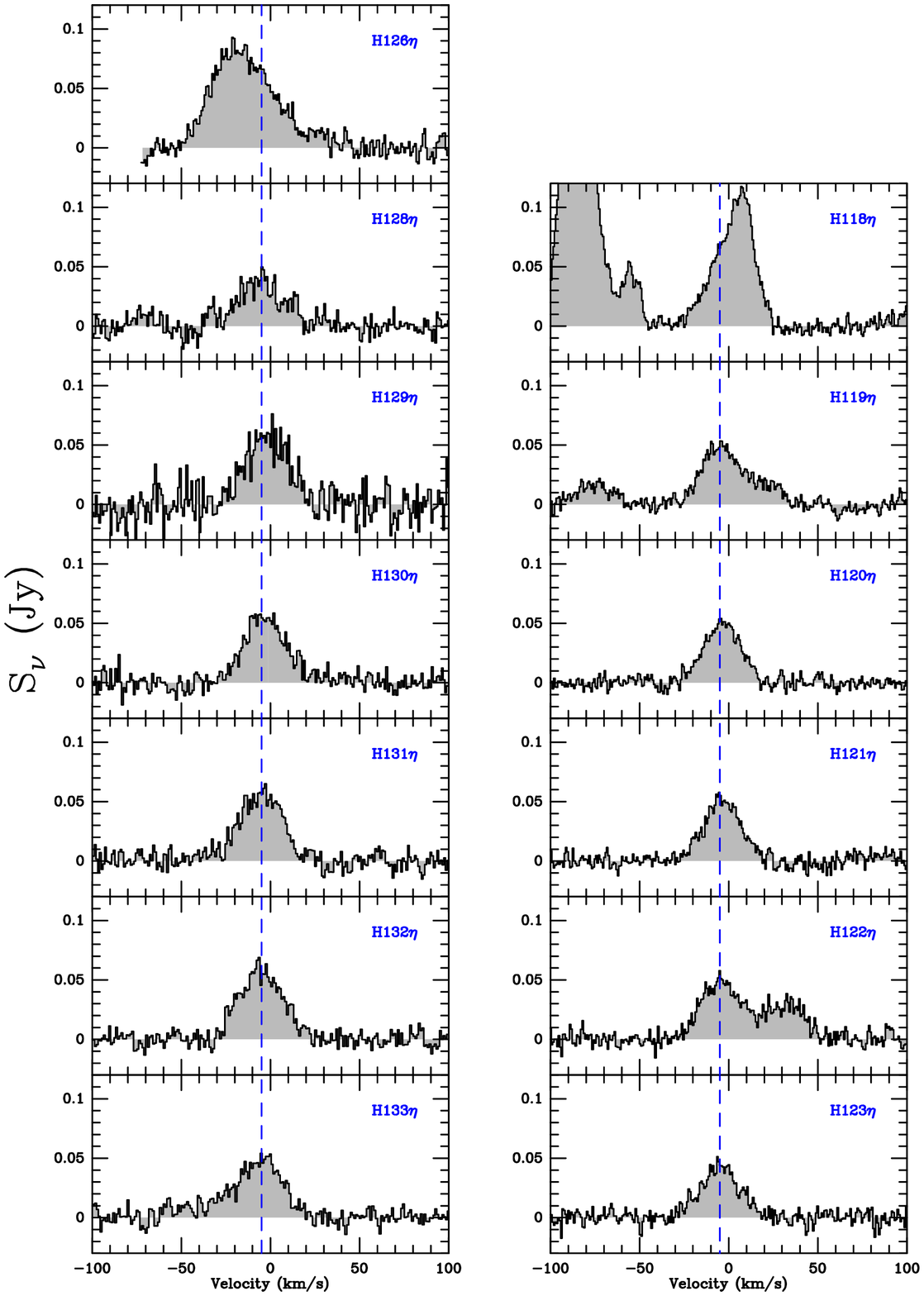}
\caption{{The observed H$\eta$ transitions indicated by dashed lines with $n$ and $\Delta n$ also given. The velocity scale refers to the respective H$\eta$ line in each panel.} \label{Fig:eta}}
\end{figure*}

\clearpage

\begin{figure*}[!htbp]
\centering
\includegraphics[width = 0.8\textwidth]{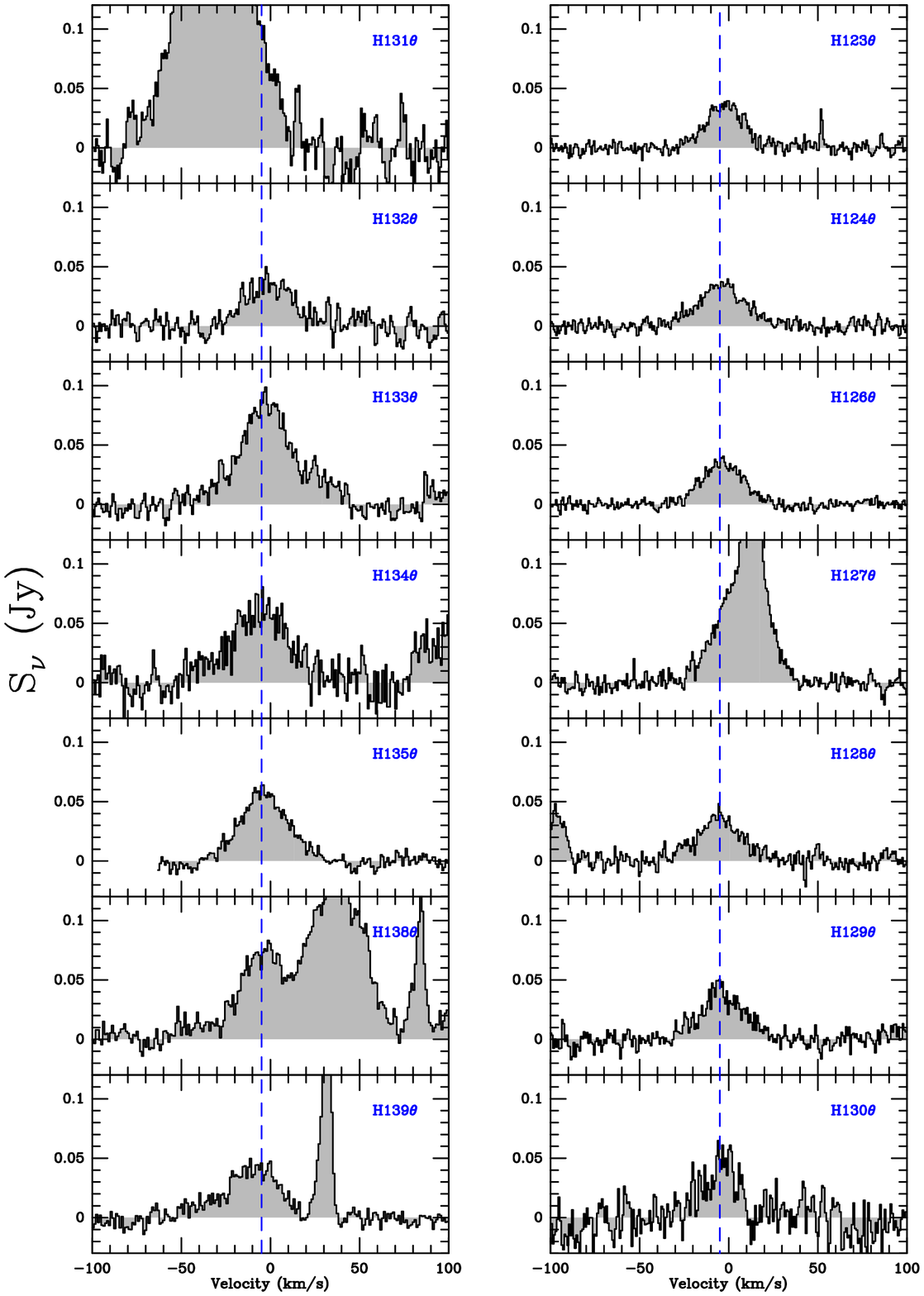}
\caption{{The observed H$\theta$ transitions indicated by dashed lines with $n$ and $\Delta n$ also given. The velocity scale refers to the respective H$\theta$ line in each panel.} \label{Fig:theta}}
\end{figure*}

\begin{figure*}[!htbp]
\centering
\includegraphics[width = 0.8\textwidth]{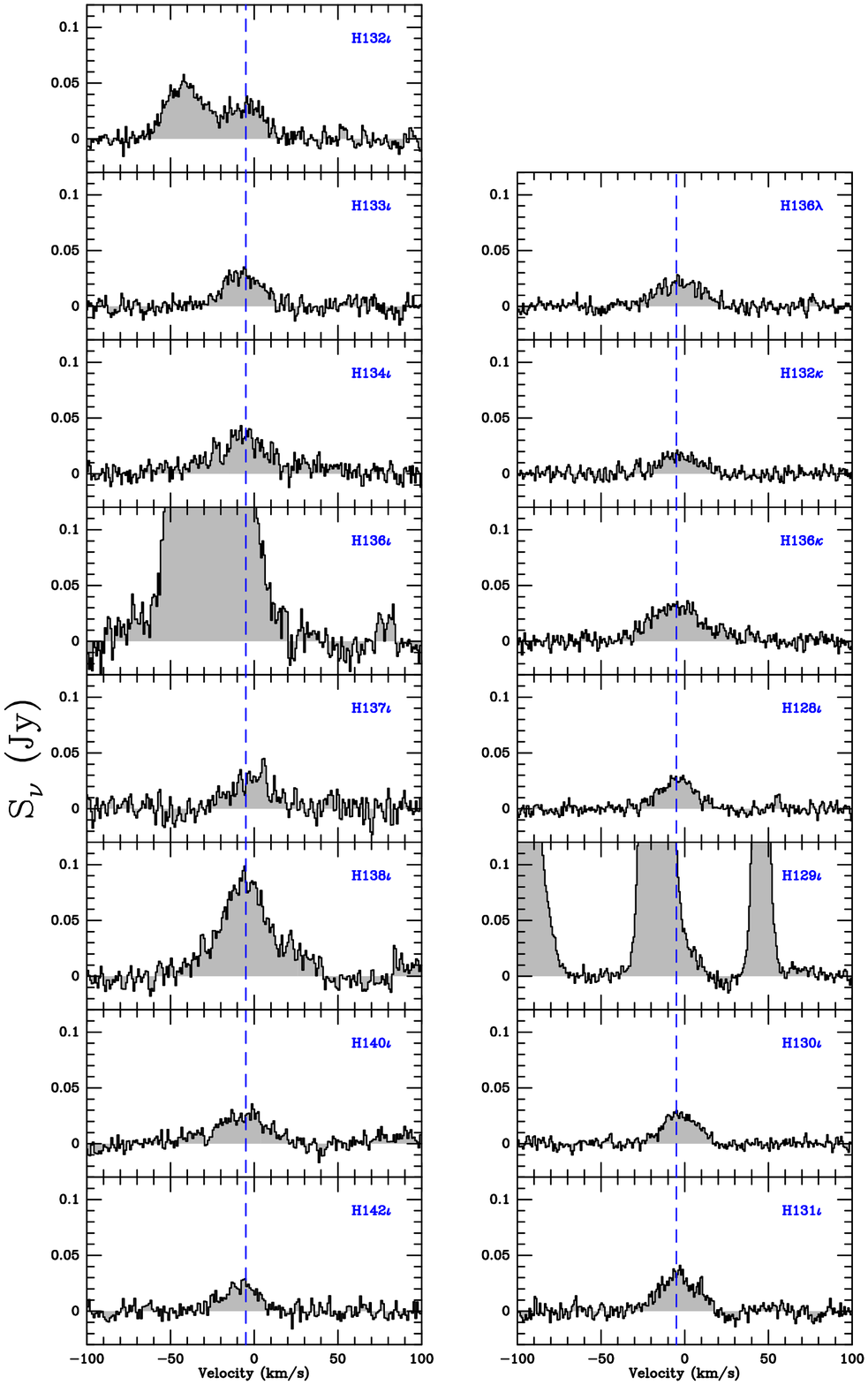}
\caption{{The observed H$\iota$, H$\kappa$ and H$\lambda$ transitions indicated by dashed lines with $n$ and $\Delta n$ also given. The velocity scale refers to the respective H$\iota$, H$\kappa$ and H$\lambda$ lines in each panel.} \label{Fig:iota}}
\end{figure*}

\begin{figure*}[!htbp]
\centering
\includegraphics[width = 0.35\textwidth]{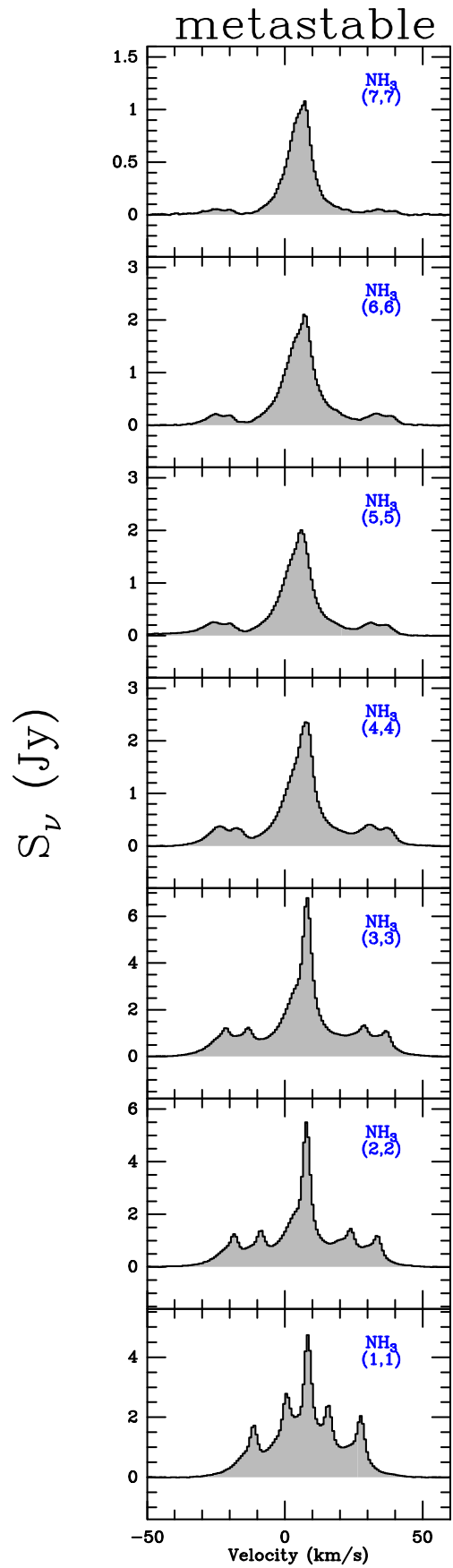}
\caption{{The observed NH$_{3}$ transitions with quantum numbers indicated in the upper right of each panel.} \label{Fig:nh3}}
\end{figure*}

\begin{figure*}[!htbp]
\centering
\includegraphics[width = 0.85 \textwidth]{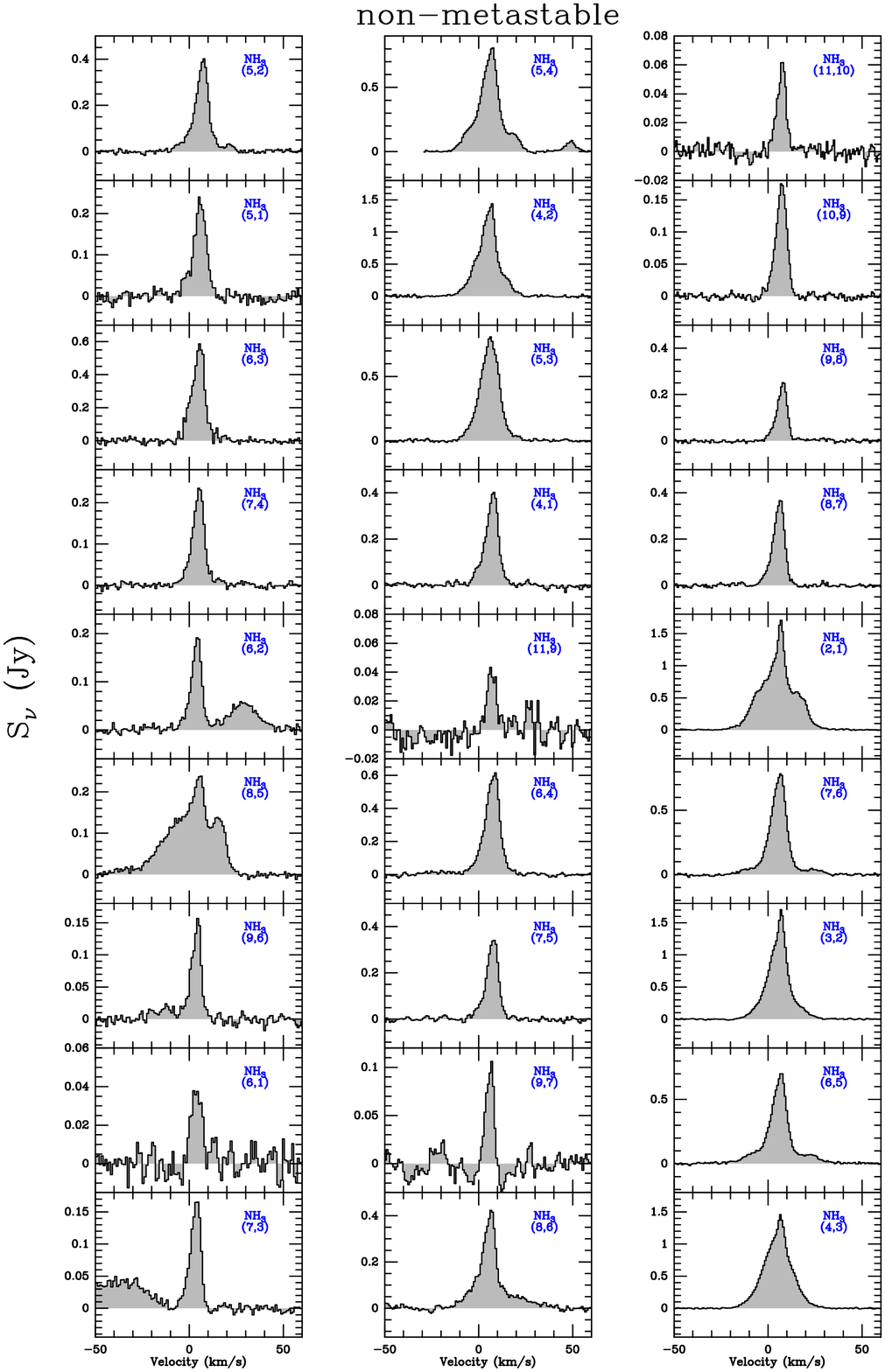}
\centerline{{Fig. \ref{Fig:nh3}. --- Continued.}}
\end{figure*}

\begin{figure*}[!htbp]
\centering
\includegraphics[width = 0.6 \textwidth]{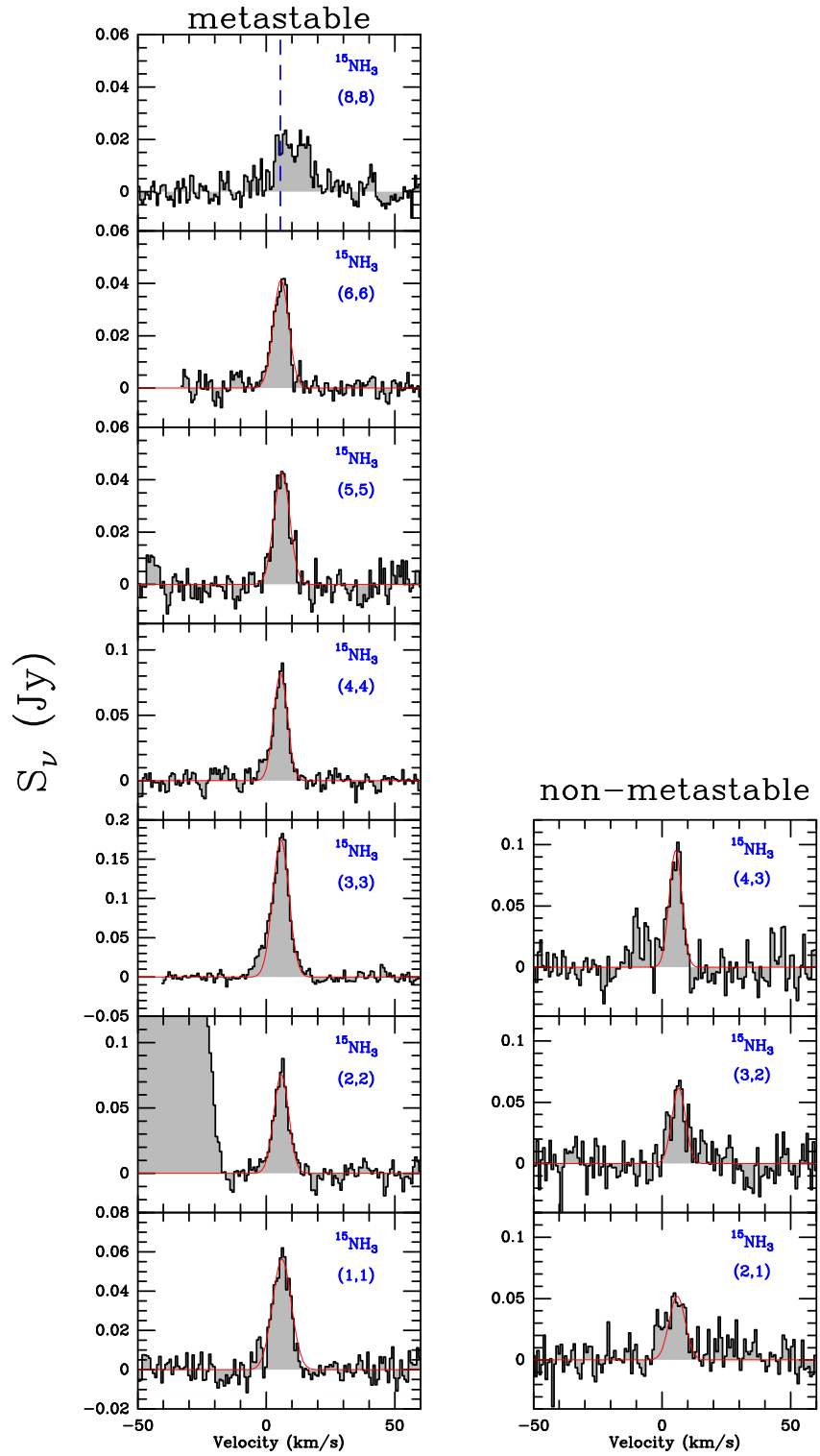}
\caption{{The observed $^{15}$NH$_{3}$ and NH$_{2}$D transitions with a one-component Gaussian fit shown (red lines). Quantum numbers are indicated in the upper right of each panel.} \label{Fig:nh32}}
\end{figure*}

\begin{figure*}[!htbp]
\centering
\includegraphics[width = 0.45 \textwidth]{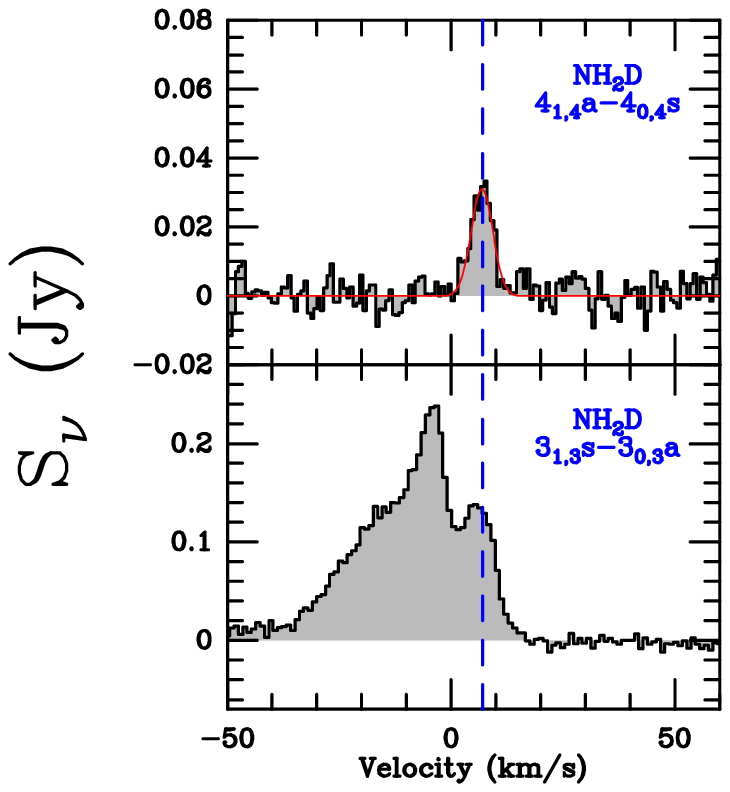}
\centerline{{Fig. \ref{Fig:nh32}. --- Continued.}}
\end{figure*}

\begin{figure*}[!htbp]
\centering
\includegraphics[width = 0.9 \textwidth]{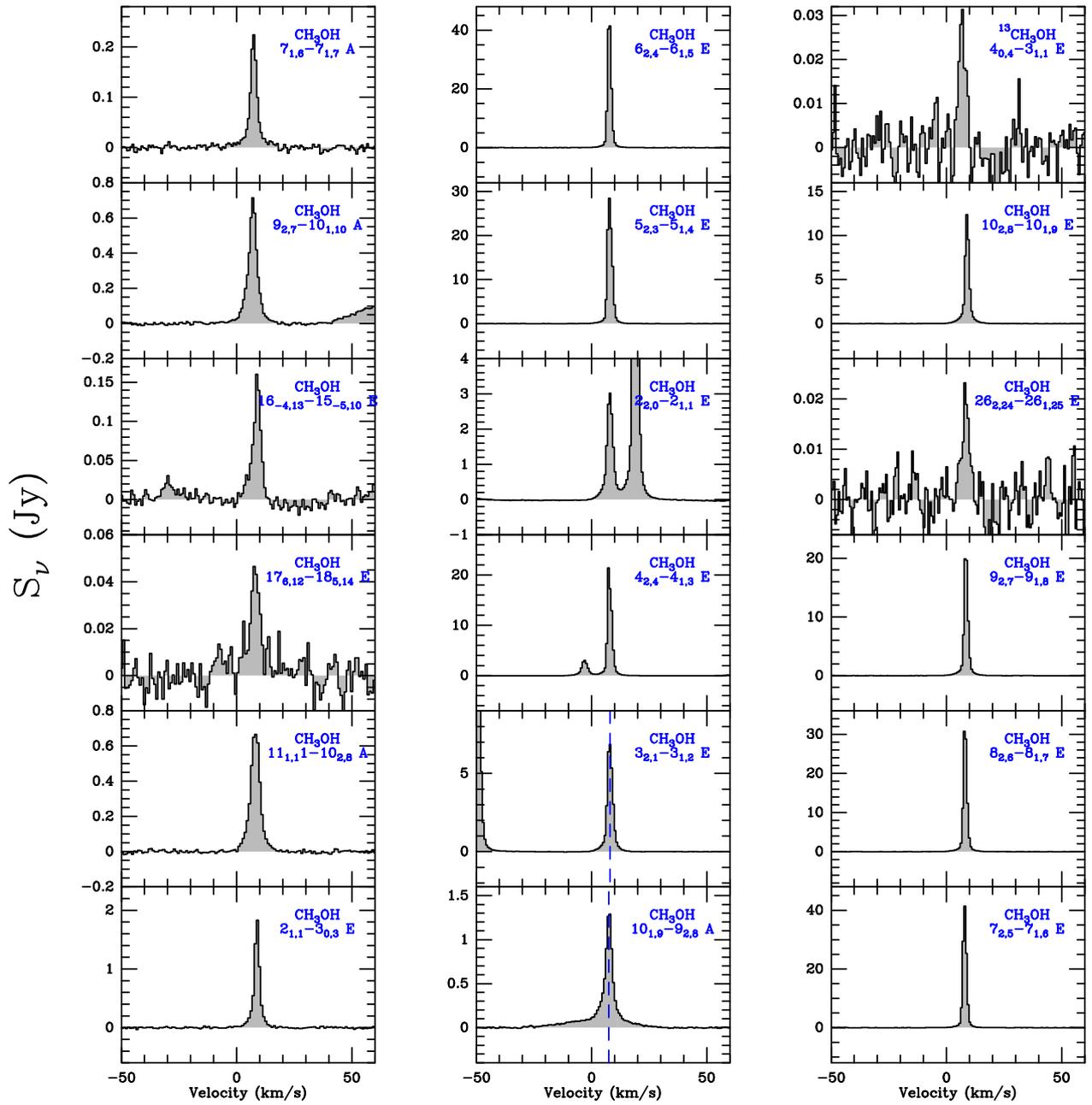}
\caption{{The observed CH$_{3}$OH and $^{13}$CH$_{3}$OH transitions with a one-component Gaussian fit shown (red lines). Species and quantum numbers are given in the upper right of each panel.} \label{Fig:ch3oh}}
\end{figure*}

\begin{figure*}[!htbp]
\centering
\includegraphics[width = 0.4 \textwidth]{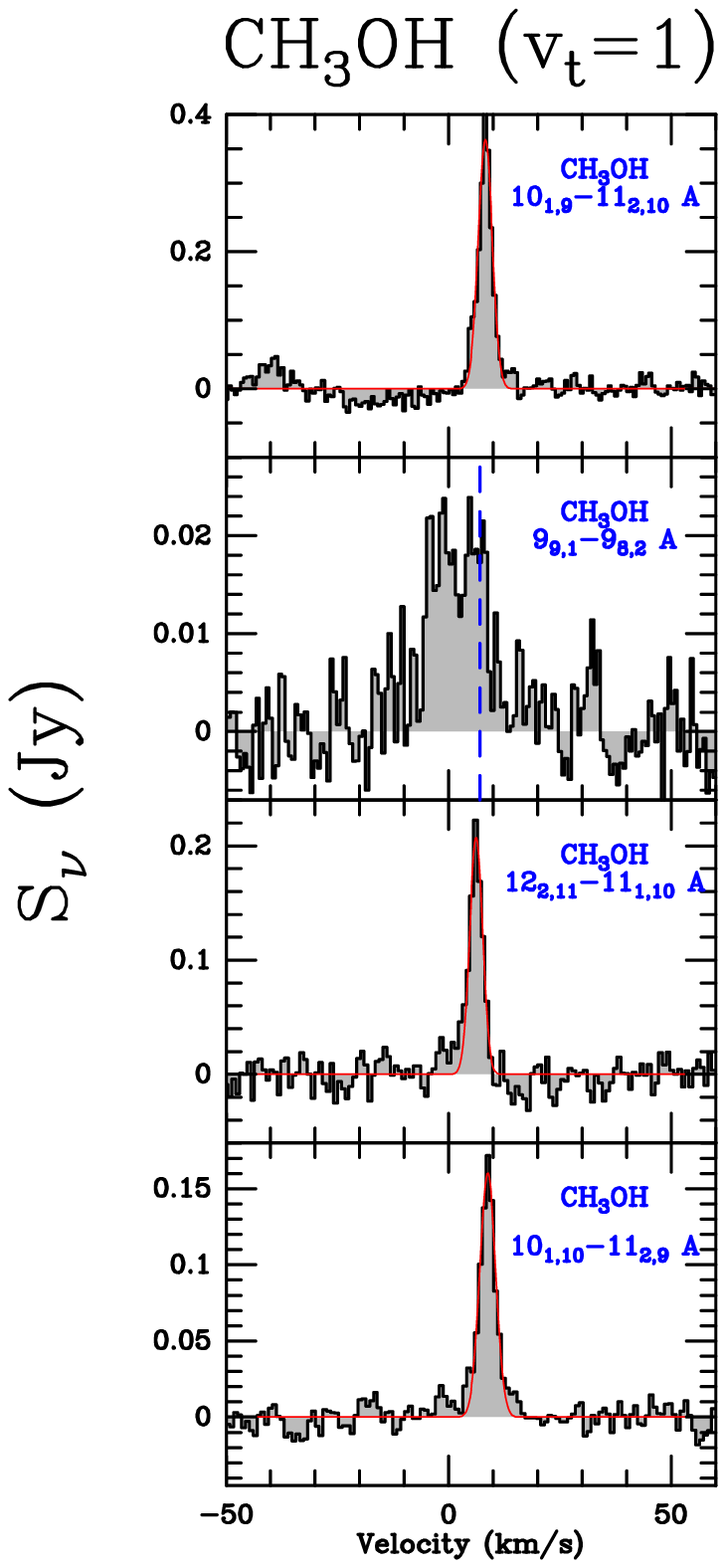}
\centerline{Fig. \ref{Fig:ch3oh}. --- Continued.}
\end{figure*}

\begin{figure*}[!htbp]
\centering
\includegraphics[width = 0.3 \textwidth]{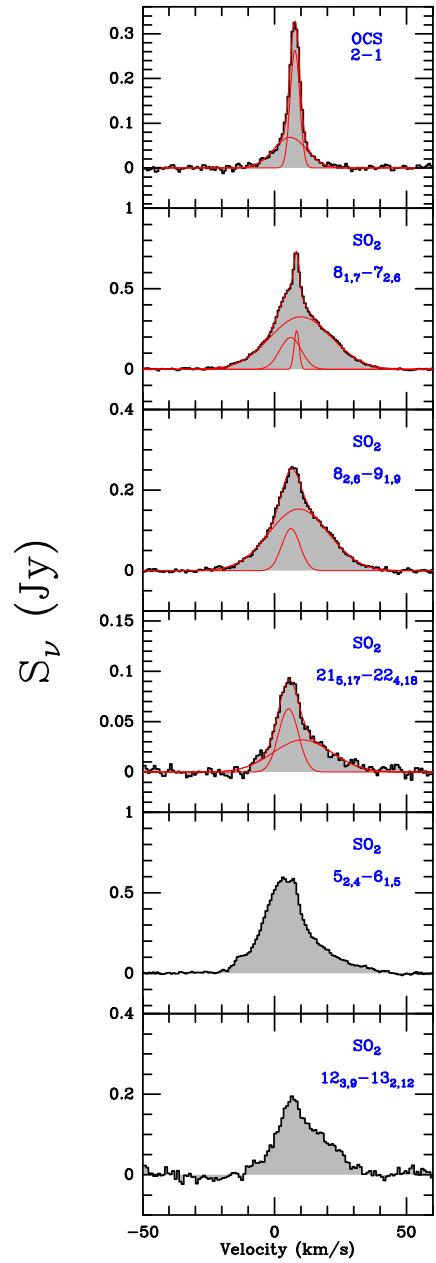}
\caption{{The observed SO$_{2}$ and OCS transitions (black lines) with a two- or three-component Gaussian fit shown together with the individual Gaussian components (red lines). Species and quantum numbers are given in the upper right of each panel. Note that SO$_{2}$ (5$_{2,4}$--6$_{1,5}$) is blended with He$65\alpha$ at 23413.8 MHz and SO$_{2}$ (12$_{3,9}$--13$_{2,12}$) is blended with He$107\delta$ at 20333.8 MHz.} \label{Fig:so2}}
\end{figure*}

\begin{figure*}[!htbp]
\centering
\includegraphics[width = 0.6 \textwidth]{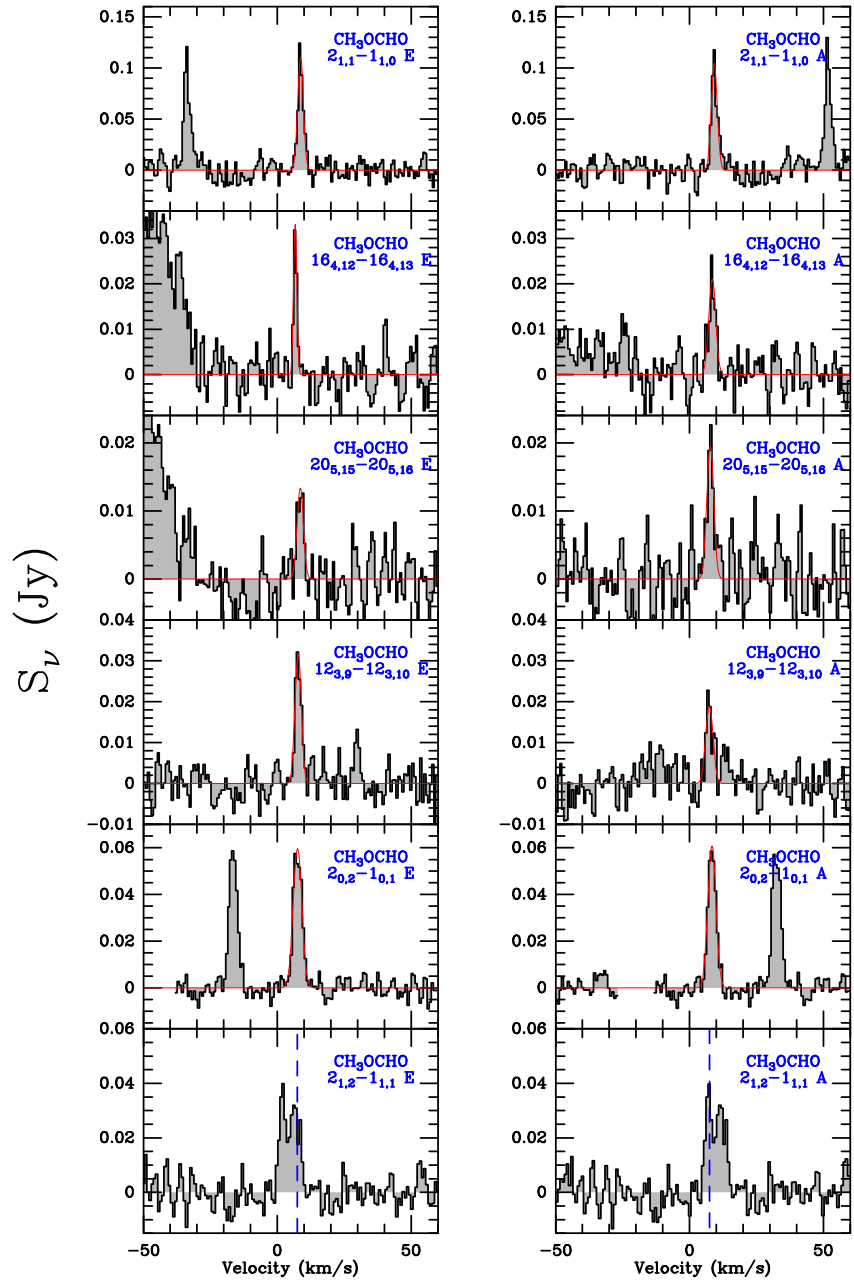}
\caption{{The observed CH$_{3}$OCHO transitions with one Gaussian fit shown (red lines). Species and quantum numbers are given in the upper right of each panel. In the CH$_{3}$OCHO (2$_{1,2}$--1$_{1,1}$ E) and CH$_{3}$OCHO (2$_{1,2}$--1$_{1,1}$ A) panels, the blue dashed lines represent the systemic velocities.} \label{Fig:ch3ocho}}
\end{figure*}

\begin{figure*}[!htbp]
\centering
\includegraphics[width = 0.8 \textwidth]{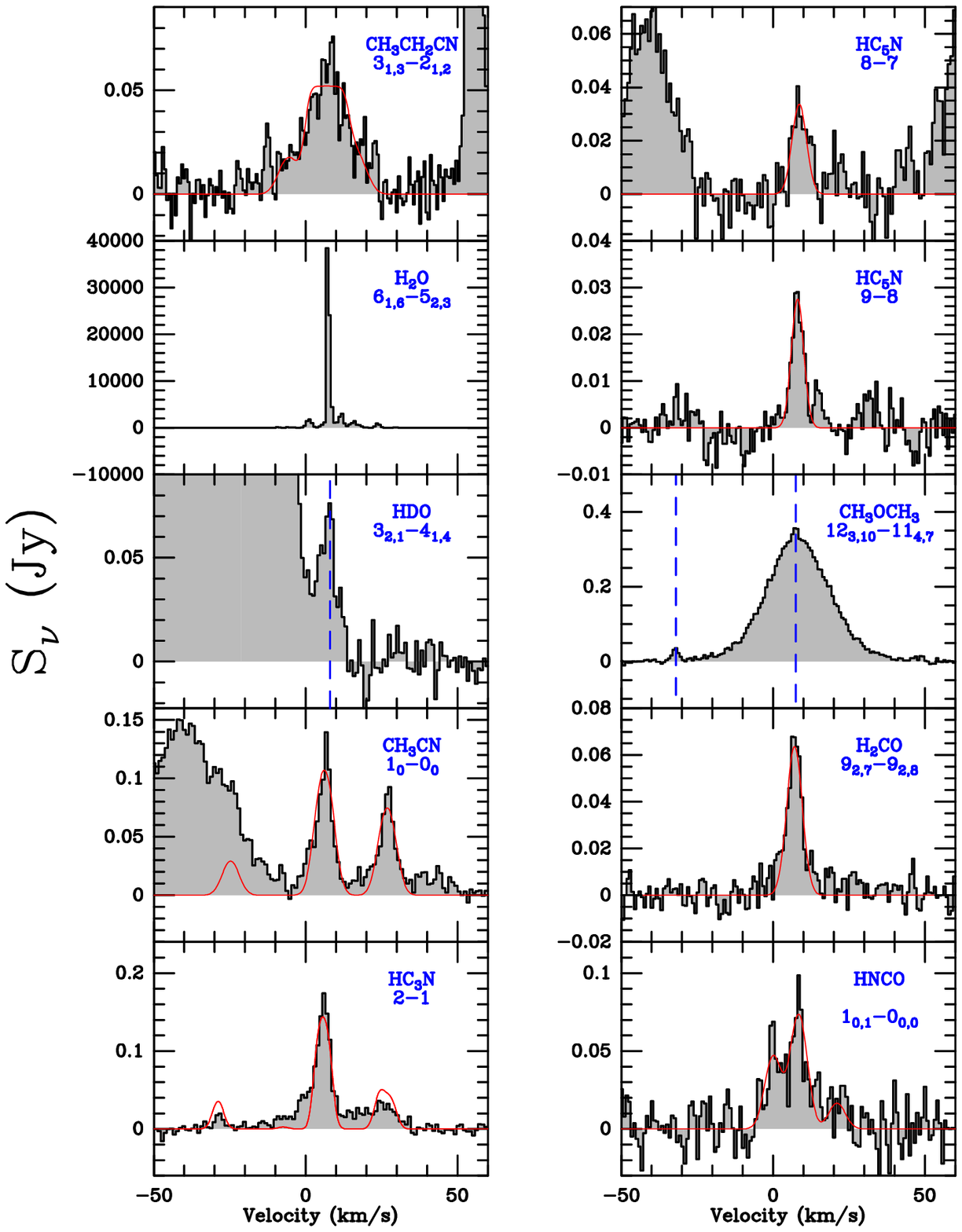}
\caption{{The observed HC$_{3}$N, CH$_{3}$CN, HDO, H$_{2}$O, HNCO, H$_{2}$CO, CH$_{3}$OCH$_{3}$, HC$_{5}$N, and CH$_{3}$CH$_{2}$CN transitions. Species and quantum numbers are indicated in the upper right of each panel. For HC$_{3}$N, CH$_{3}$CN, HNCO, and CH$_{3}$CH$_{2}$CN, the spectra are fitted with the HFS method indicated by red lines. For the HDO and CH$_{3}$OCH$_{3}$ transitions which are blended, their systemic velocity is indicated by blue dashed lines.} \label{Fig:one}}
\end{figure*}
\clearpage

\clearpage
\section{The ALMA maps of detected molecules and comparisons with other studies}\label{app.c}
\begin{figure*}[!htbp]
\centering
\includegraphics[width = 0.8 \textwidth]{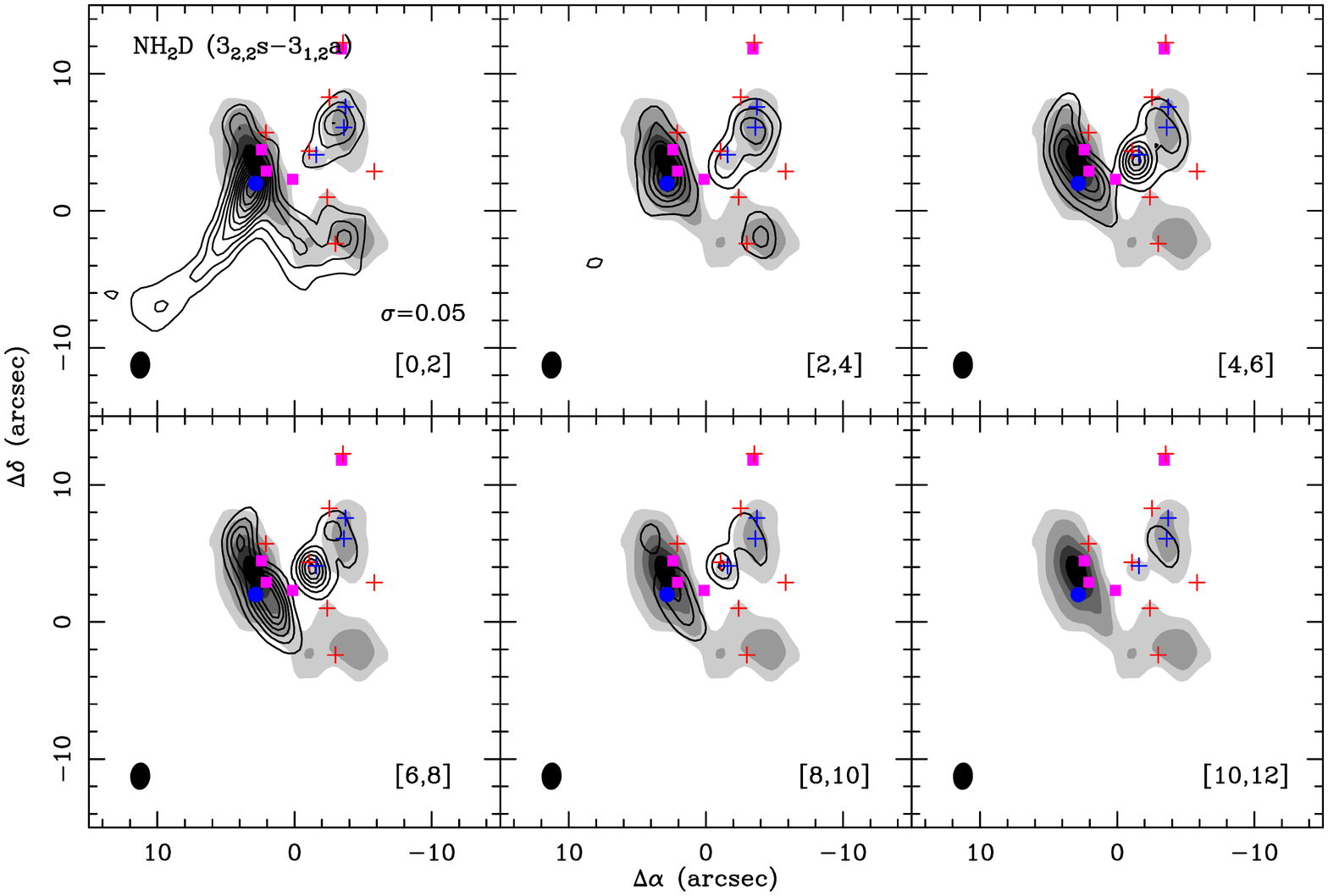}
\includegraphics[width = 0.8 \textwidth]{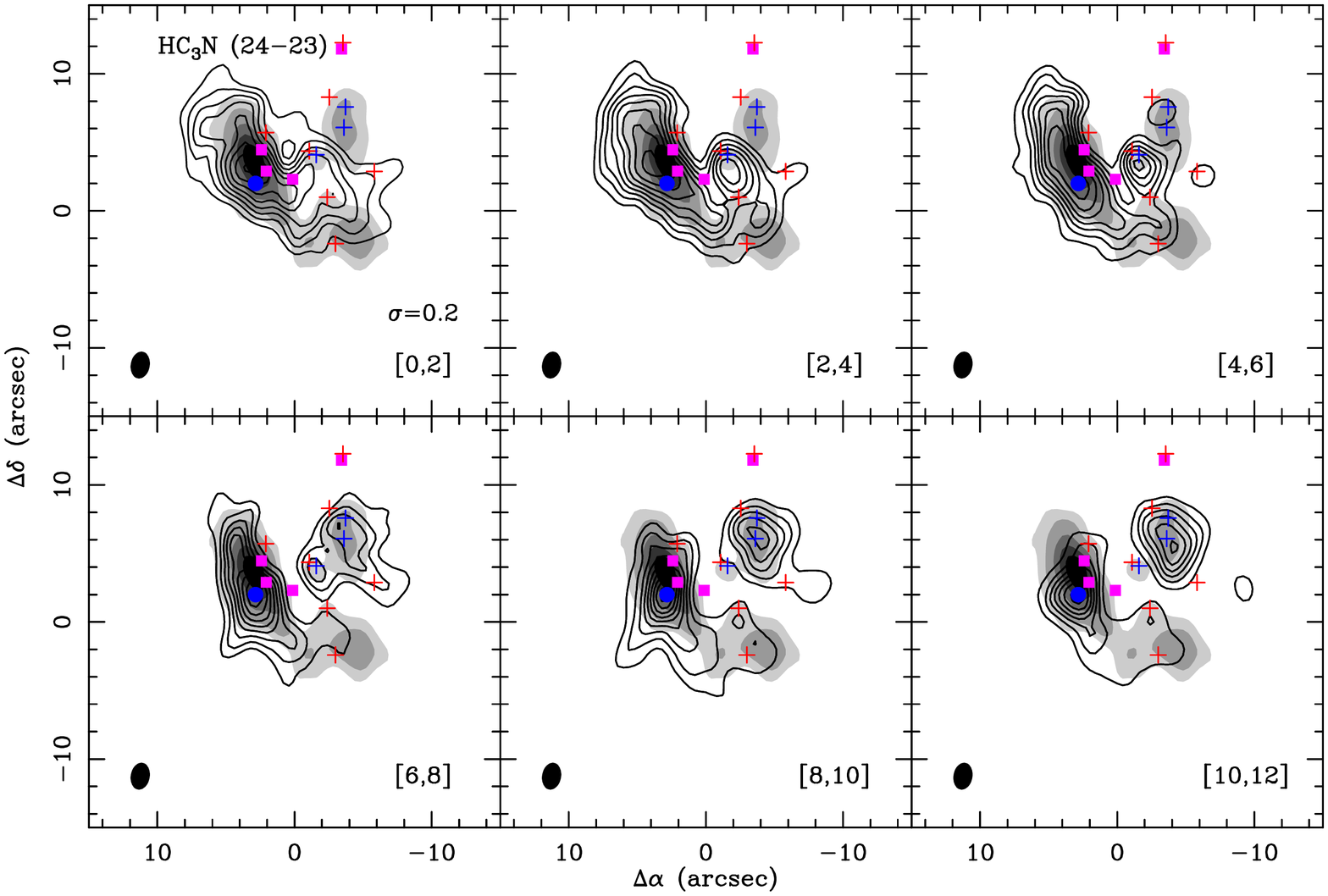}
\caption{{Molecular line channel maps (contours) overlaid on the 230 GHz continuum map (grey). Grey shadings of the continuum image are 10\%, 20\%, 40\%, 60\%, 80\% of the peak intensity of 1.406 Jy~beam$^{-1}$. The contour levels of the molecular line images start at 5~$\sigma$ and continue in steps of 5~$\sigma$, where the $\sigma$ value for each transition is shown in the first panel in units of Jy~beam$^{-1}$. The dotted contours are the negative features with the same contour absolute levels as the positive ones in each panel. The symbols are the same as in Fig.~\ref{Fig:alma-dust}. The corresponding molecular transitions are indicated in the upper left of the first panel. The velocity range is given in the lower right of each panel in \kms. The synthesized beams of the molecular line images are shown in the lower left of each panel. The (0, 0) position in each panel is ($\alpha_{\rm J2000}$, $\delta_{\rm J2000}$) = (05$^{\rm h}$35$^{\rm m}$14.350$^{\rm s}$, $-$05\degree22\arcmin35.00\arcsec).} \label{Fig:cmap}}
\end{figure*}

We make use of the ALMA line survey archival data to study the distribution of molecules detected by our 1.3 cm line survey. The transitions used are listed in Table~\ref{Tab:alma}. Fig.~\ref{Fig:cmap} shows the channel maps of these transitions. Note that the first [0,2]~\kms\, panel of NH$_{2}$D ($3_{2,2}s-3_{1,2}a$) is contaminated by H$_{2}$CO ($9_{1,8}-9_{1,9}$) at 216268.7 MHz.

Recently, \citet{2015arXiv150408012F} studied Orion KL using combined SMA and IRAM 30m data, which covers parts of the frequency range of the ALMA-SV data. Here, we make a brief comparison between these data. By comparing the dust continuum emission in both datasets, mm2 in \citet{2015arXiv150408012F} is resolved into MM4, MM5, and MM6 by ALMA. On the other hand, the dust continuum sources SR, NE, OF1N, and OF1S in \citet{2015arXiv150408012F} are not detected in the ALMA-SV data. A comparsion of OCS (18--17) demonstrates that \citet{2015arXiv150408012F} show more extended emission while the ALMA-SV data display more sub-structures. That is because the ALMA-SV data have a higher spatial resolution but lack short spacing and have a smaller primary beam.

We also estimate the flux loss by comparing the ALMA-SV data with those obtained with the IRAM-30m single dish telescope. We make use of HC$_{3}$N (24--23) from \citet{2013AA...559A..51E}, OCS (18--17) from \citet{2010AA...517A..96T}, and SO$_{2}$ ($11_{5,7}-12_{4,8}$) from \citet{2013AA...556A.143E}. Based on the telescope information\footnote[11]{http://www.iram.es/IRAMES/mainWiki/Iram30mEfficiencies}, we use a forward efficiency of 94\%, a main beam efficiency of 63\%, and a conversion factor from brightness temperature to flux of 7.5 Jy/K to derive the total flux observed by the IRAM-30m telescope. Meanwhile, the IRAM-30m telescope has a spatial resolution of $\sim$12\arcsec, covering most emitting regions in Orion KL, so the IRAM-30m data can be considered to represent the total flux densities of Orion KL. By integrating the whole regions of the ALMA-SV data, we find that the ALMA-SV data can account for 89\% of the HC$_{3}$N (24--23), 57\% of the OCS (18--17), and 44\% of the SO$_{2}$ ($11_{5,7}-12_{4,8}$) emission.

\begin{table*}[!hbt]
\caption{Transitions from the ALMA line survey..}\label{Tab:alma}             
\normalsize
\centering                                      
\begin{tabular}{ccc}          
\hline\hline                        
Frequencies                 & Transitions             & $E_{\rm u}/k$         \\
(MHz)                       &                         & (K)                 \\
\hline
\multicolumn{3}{l}{Orion KL}\\
\hline
216562.7                    & NH$_{2}$D (3$_{2,2}s-3_{1,2}a$)            &   120     \\
218324.7                    & HC$_{3}$N ($24-23$)                     &  131      \\
218222.2                    & H$_{2}$CO (3$_{0,3}-2_{0,2}$)             &  21        \\
218903.4                    & OCS ($18-17$)                          &  100       \\
218981.0                    & HNCO (10$_{1,10}-9_{1,9}$)                & 101        \\
223200.1                    & CH$_{3}$OCH$_{3}$ (8$_{2,7}-7_{1,6}$~$AE$) & 38         \\
229347.6                    & SO$_{2}$ (11$_{5,7}-12_{4,8}$)             & 122\\          
233827.5                    & CH$_{3}$CH$_{2}$CN ($8_{4,5}-7_{3,4}$)     &  33        \\ 
235190.6                    & CH$_{3}$OCHO (10$_{9,1}-10_{8,2}$~$A$)    & 86         \\
239064.3                    & CH$_{3}$CN (13$_{4}-12_{4}$)             & 195         \\
243915.8                    & CH$_{3}$OH (5$_{1,4}-4_{1,3}$~$A$)        &  50        \\
\hline
 \end{tabular}
 			        \normalsize
\end{table*}

\clearpage

\begin{figure*}[!htbp]
\centering
\includegraphics[width = 0.8 \textwidth]{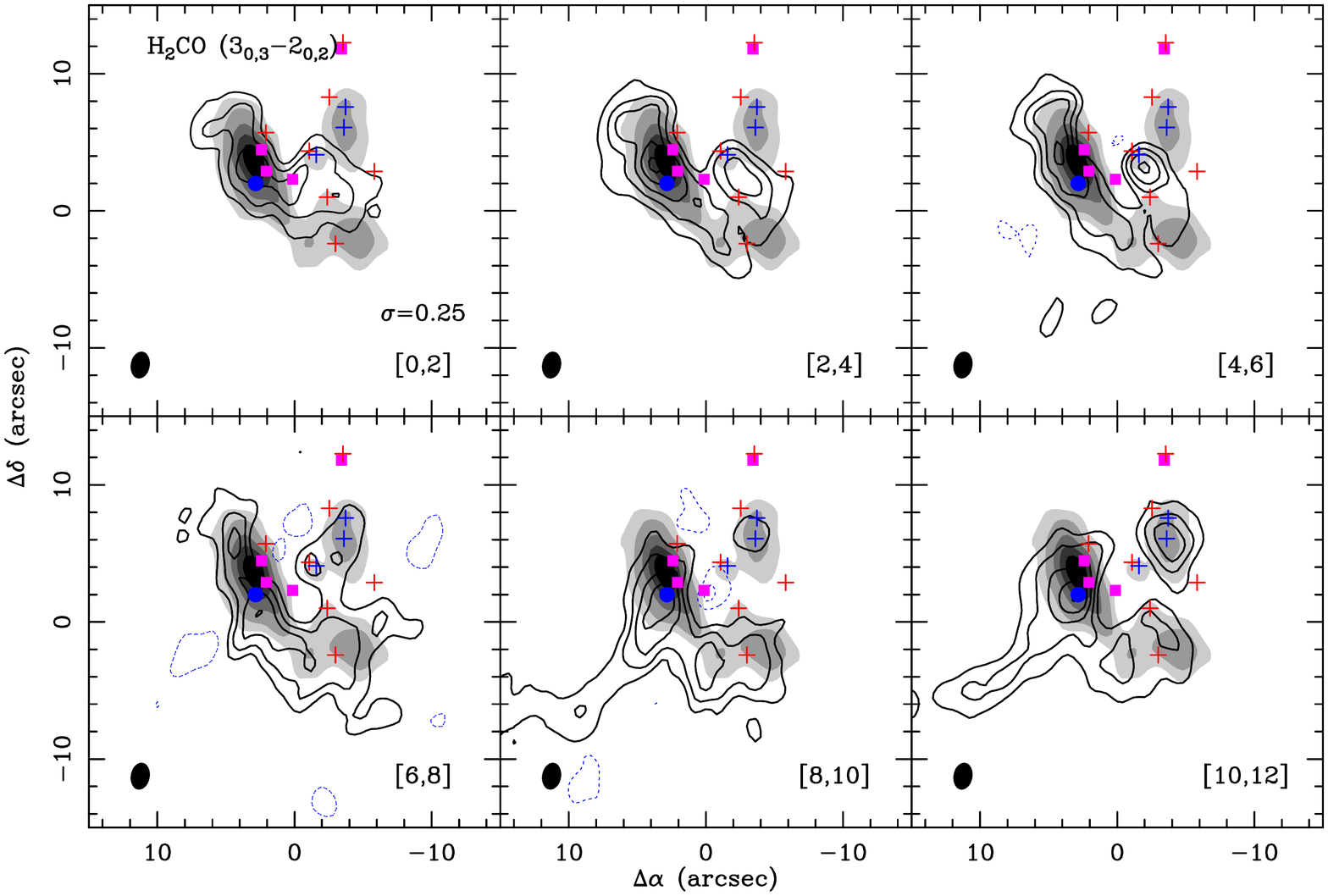}
\includegraphics[width = 0.8 \textwidth]{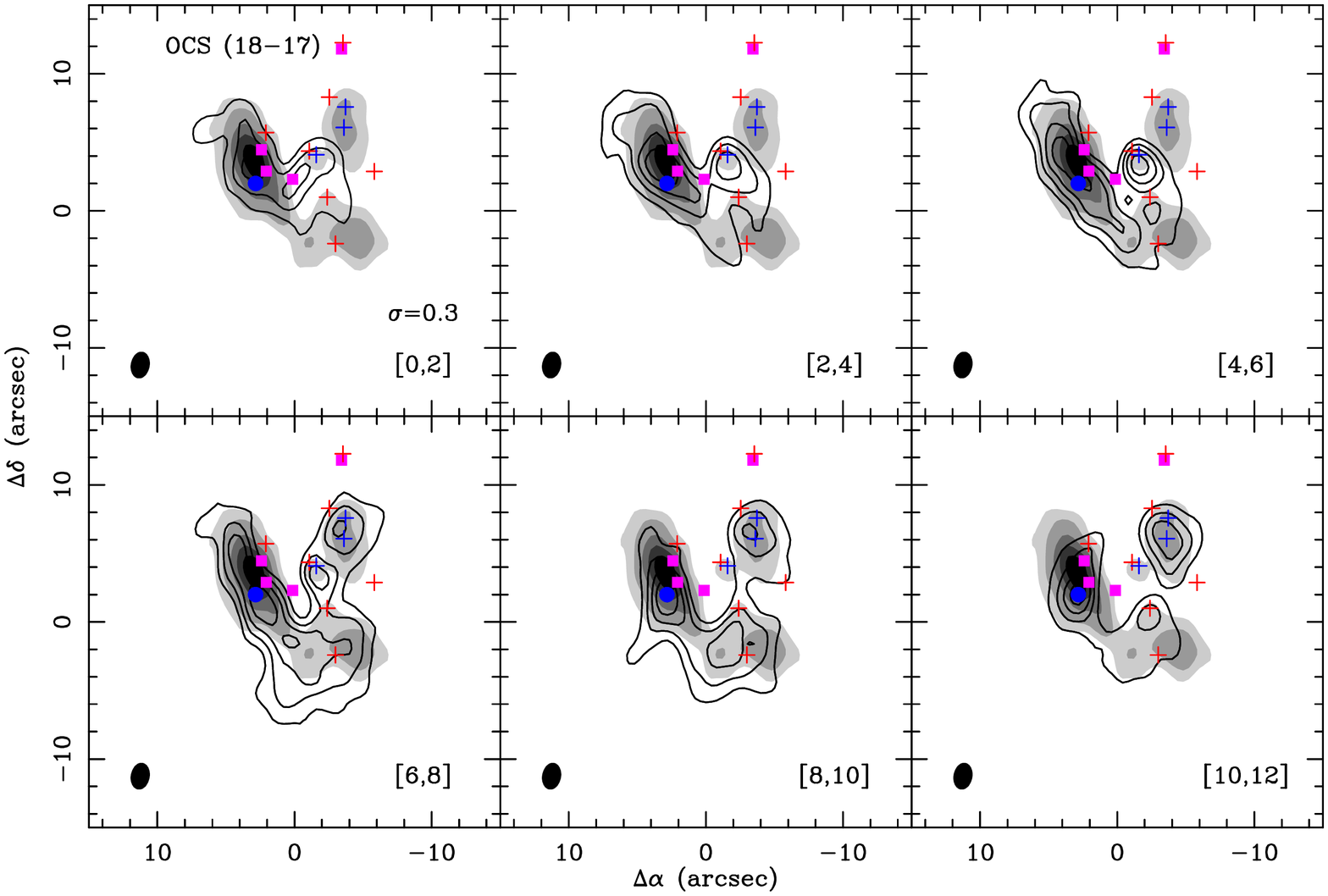}
\centerline{Fig. \ref{Fig:cmap}. --- Continued.}
\end{figure*}

\begin{figure*}[!htbp]
\centering
\includegraphics[width = 0.8 \textwidth]{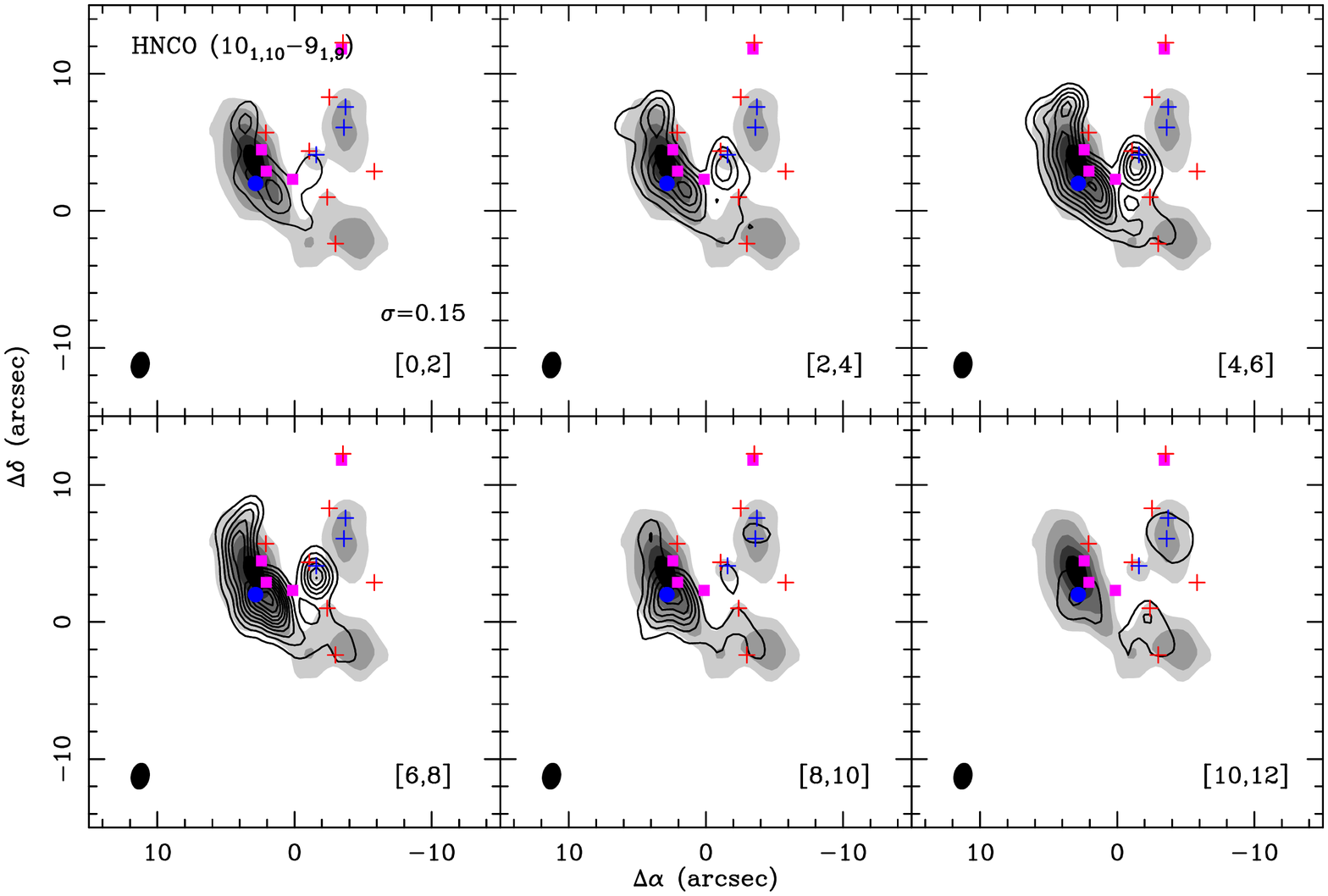}
\includegraphics[width = 0.8 \textwidth]{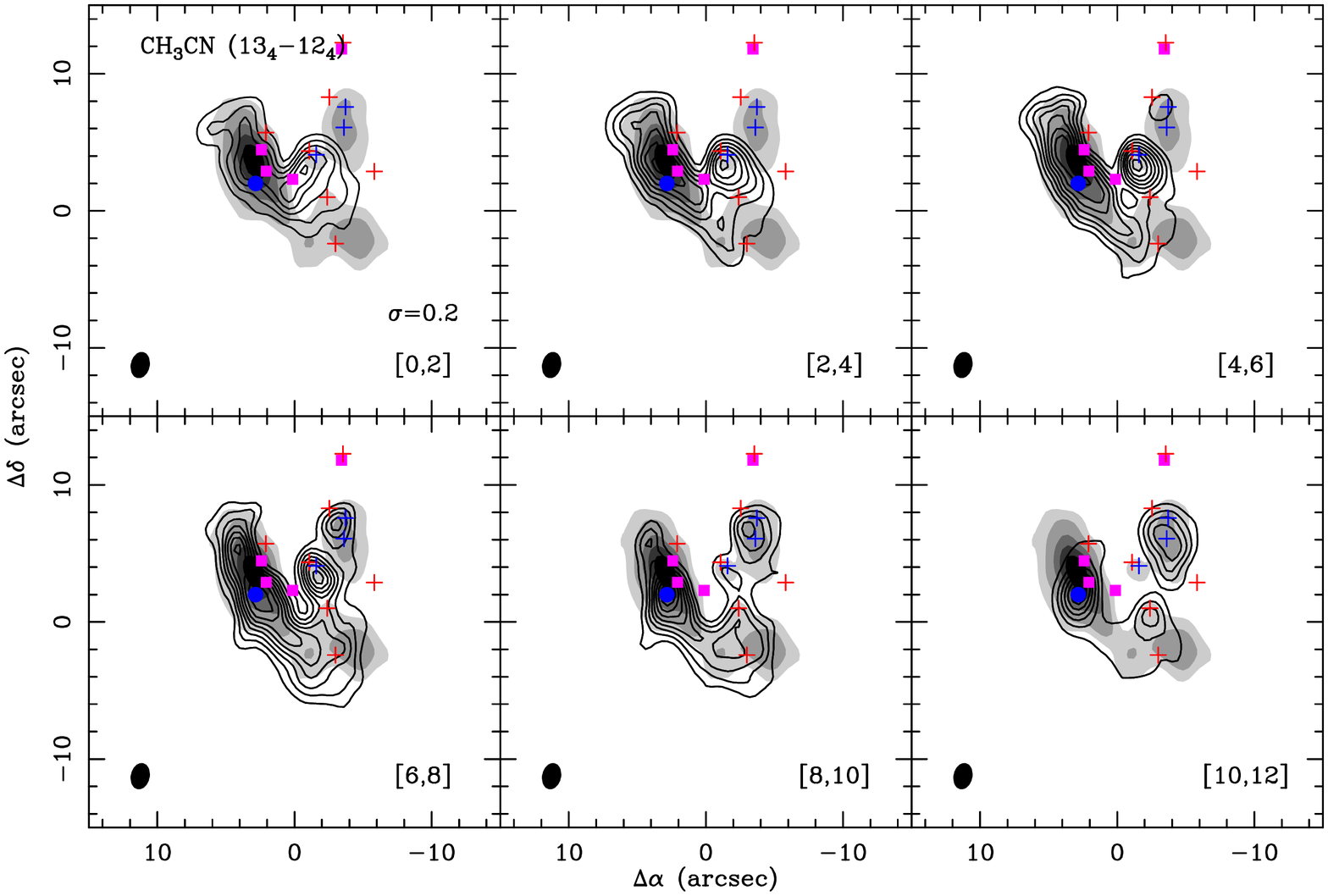}
\centerline{Fig. \ref{Fig:cmap}. --- Continued.}
\end{figure*}

\begin{figure*}[!htbp]
\centering
\includegraphics[width = 0.8 \textwidth]{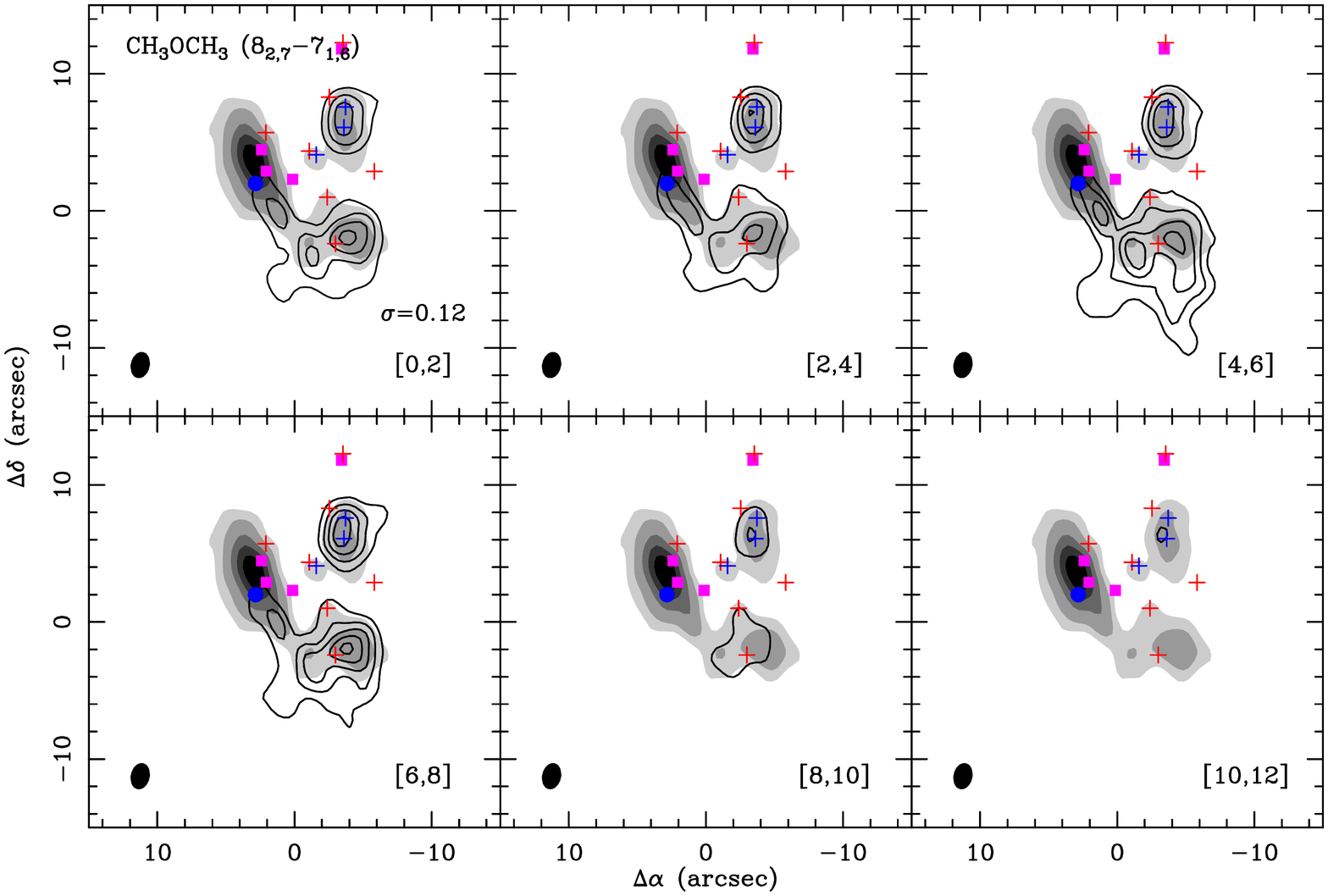}
\includegraphics[width = 0.8 \textwidth]{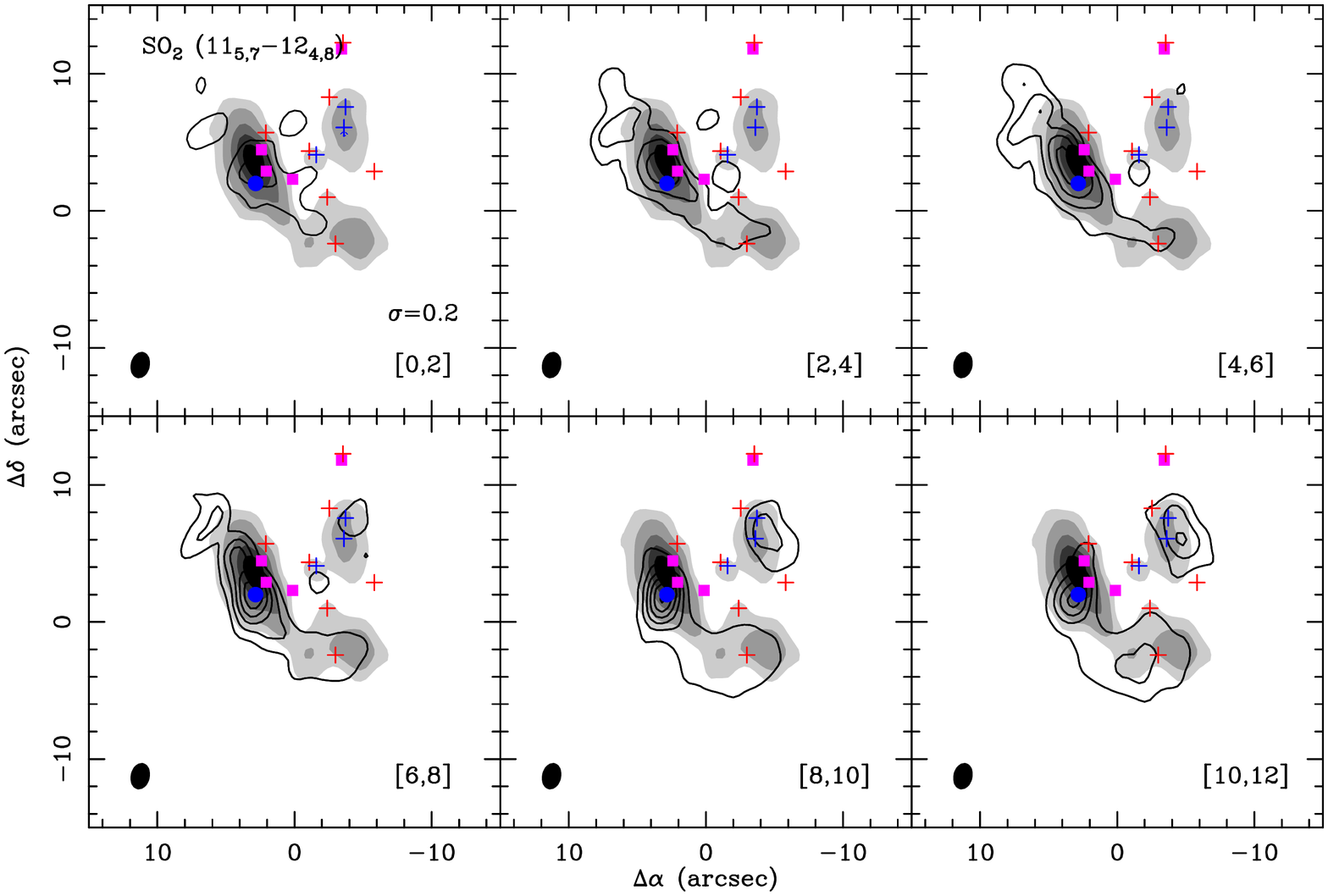}
\centerline{Fig. \ref{Fig:cmap}. --- Continued.}
\end{figure*}

\begin{figure*}[!htbp]
\centering
\includegraphics[width = 0.8 \textwidth]{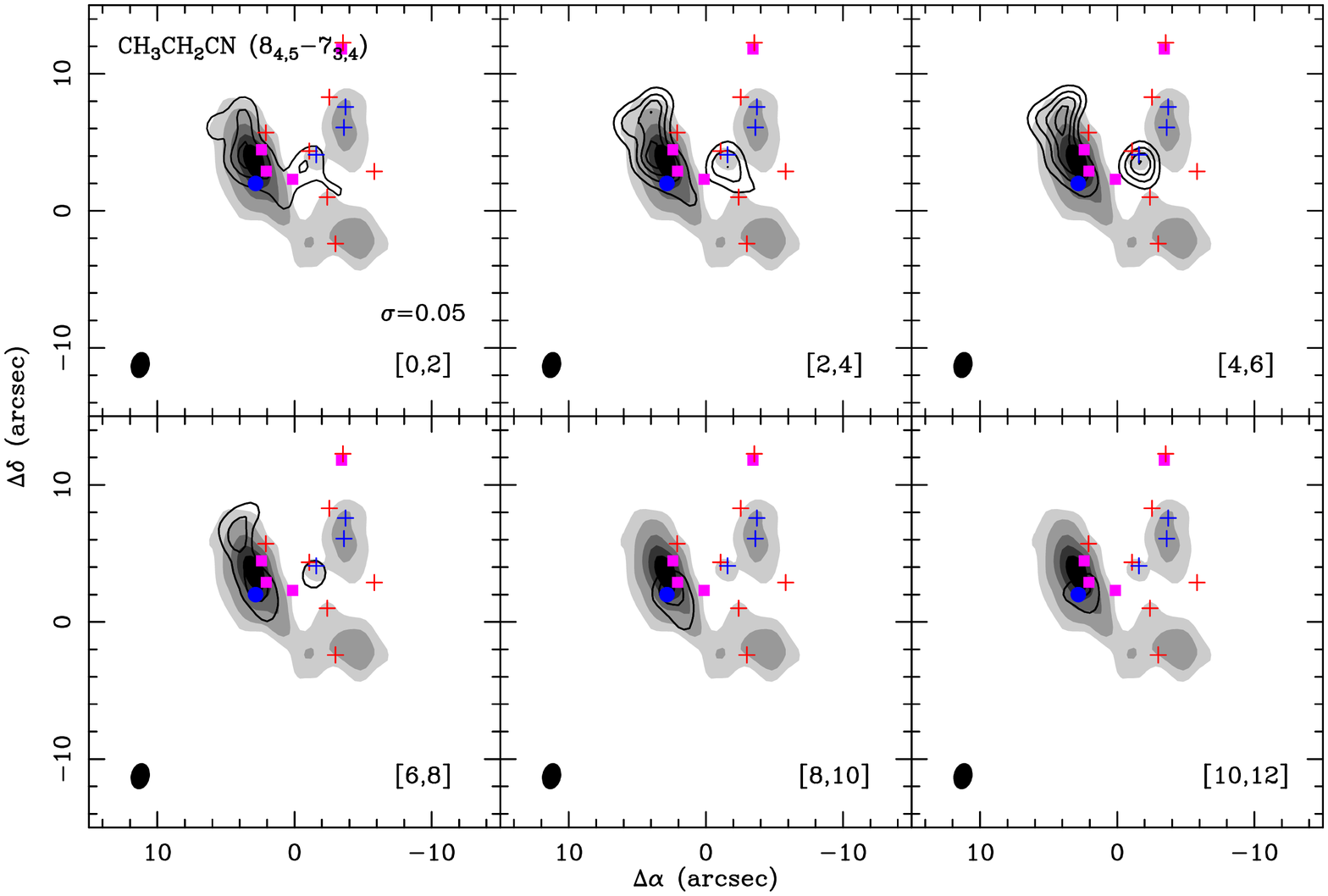}
\includegraphics[width = 0.8 \textwidth]{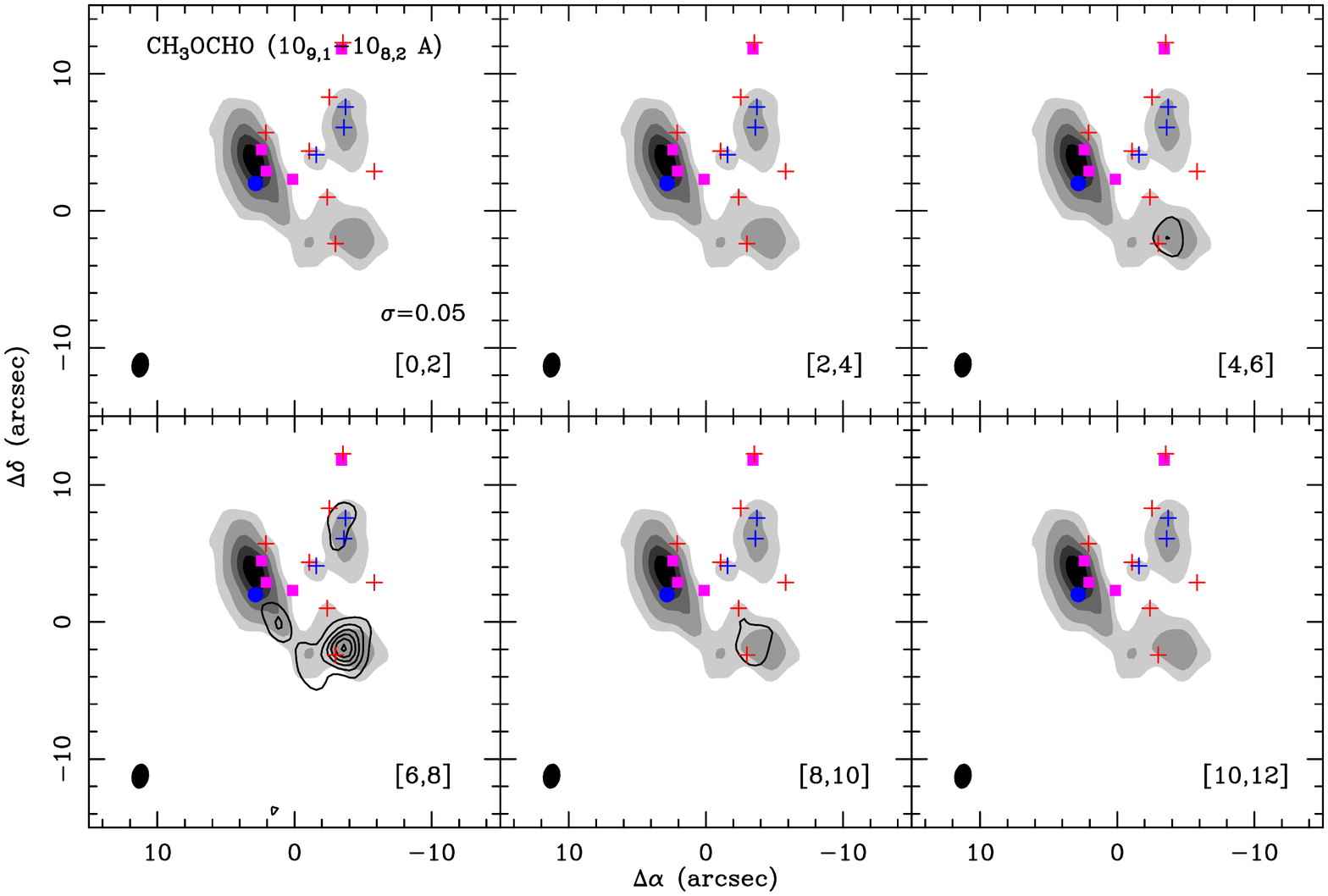}
\centerline{Fig. \ref{Fig:cmap}. --- Continued.}
\end{figure*}

\begin{figure*}[!htbp]
\centering
\includegraphics[width = 0.8 \textwidth]{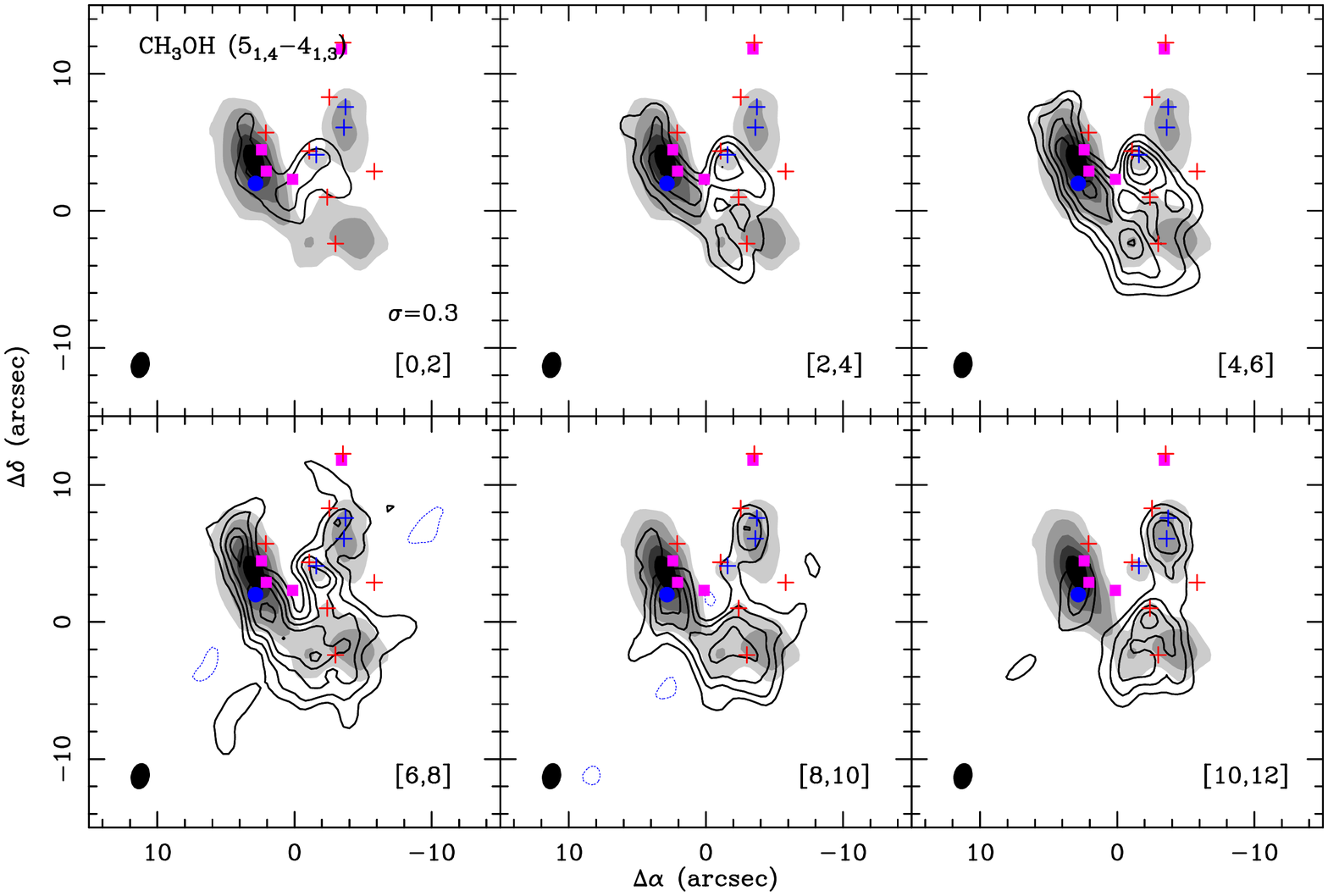}
\centerline{Fig. \ref{Fig:cmap}. --- Continued.}
\end{figure*}

\end{appendix}

\end{document}